\documentclass[fleqn,usenatbib,useAMS]{mnras}

\usepackage[T1]{fontenc}
\usepackage{ae,aecompl}
\usepackage{graphicx}	% Including figure files
\usepackage{amsmath}	% Advanced maths commands
\usepackage{amssymb}	% Extra maths symbols
\usepackage{amsfonts}
\usepackage[british]{babel}
\usepackage{multicol}        % Multi-column entries in tables
\usepackage{bm}		% Bold maths symbols, including upright Greek
\usepackage{pdflscape}	% Landscape pages
\usepackage{booktabs}

\title[Rotation curves with fuzzy and multistate SFDM]
{Rotation curves of high-resolution LSB and SPARC galaxies with fuzzy and
multistate (ultra-light boson) scalar field dark matter}

\author[T. Bernal et al.]
       {T. Bernal,$^{1}$\thanks{E-mail: ac13341@chapingo.mx}\thanks{Present address:
       Universidad Aut\'onoma Chapingo, km. 38.5 Carretera M\'exico-Texcoco, 56230, Texcoco,
       Estado de M\'exico, M\'exico}
       L.~M. Fern\'andez-Hern\'andez,$^{1}$\thanks{E-mail:
       lfernandez@fis.cinvestav.mx}
       T. Matos$^{2}$\thanks{E-mail: tmatos@fis.cinvestav.mx}\thanks{Part of the Instituto
       Avanzado de Cosmolog\'{\i}a (IAC) Collaboration}
       and  M.~A. Rodr\'{\i}guez-Meza$^{1}$\thanks{Email:
       marioalberto.rodriguez@inin.gob.mx}\thanks{Part of the Instituto
       Avanzado de Cosmolog\'{\i}a (IAC) Collaboration}\\
       $^{1}$Departamento de F\'{\i}sica, Instituto Nacional de
       Investigaciones Nucleares, AP 18-1027, Ciudad de M\'{e}xico
       11801, M\'{e}xico\\
	   $^{2}$Departamento de F\'{\i}sica, Centro de Investigaci\'on y de
       Estudios Avanzados del IPN, AP
       14-740, Ciudad de M\'exico 07000, M\'exico
       }

\date{\today}
\pubyear{2017}

\begin{document}
\label{firstpage}
\pagerange{\pageref{firstpage}--\pageref{lastpage}}
\maketitle

%%%%%%%%%%%%%%%% ABSTRACT %%%%%%%%%%%%%%%%%%
\begin{abstract}
  Cold dark matter (CDM) has shown to be an excellent candidate for the dark
matter (DM) of the Universe at large scales, however it presents some challenges
at the galactic level. The scalar field dark matter (SFDM), also called fuzzy,
wave, Bose-Einstein condensate or ultra-light axion DM, is identical to CDM at
cosmological scales but different at the galactic ones. SFDM forms core halos,
it has a natural cut-off in its matter power spectrum and it predicts
well-formed galaxies at high redshifts. In this work we reproduce the rotation
curves of high-resolution low surface brightness (LSB) and SPARC galaxies with
two SFDM profiles: (1)~The soliton+NFW profile in the fuzzy DM (FDM) model,
arising empirically from cosmological simulations of real, non-interacting
scalar field (SF) at zero temperature, and (2)~the multistate SFDM (mSFDM)
profile, an exact solution to the Einstein-Klein-Gordon equations for a real,
self-interacting SF, with finite temperature into the SF potential, introducing
several quantum states as a realistic model for a SFDM halo. From the fits with
the soliton+NFW profile, we obtained for the boson mass $0.212< m_\psi/(10^{-23}
\mathrm{eV}/c^2)<27.0$ and for the core radius $0.326< r_c/\mathrm{kpc}<8.96$.
From the combined analysis with the LSB galaxies, we obtained $m_\psi = 0.554
\times10^{-23}\mathrm{eV}$, a result in tension with the severe cosmological
constraints. Also, we show the analytical mSFDM model fits the observations as
well as or better than the empirical soliton+NFW profile, and it reproduces
naturally the wiggles present in some galaxies, being a theoretically motivated
framework additional or alternative to the FDM profile.
\end{abstract}


%%%%%%%%%%%%%%%%%% INTRO %%%%%%%%%%%%%%%%%%
\section{Introduction}
\label{sec:Intro}

  Dark matter (DM) was postulated first in order to explain the rotation curves of
disk galaxies and the observed velocity dispersions in galaxy clusters, as well
as the observational mass-to-light ratios in galaxies and clusters of galaxies
\citep{Zwicky:1933,Zwicky:1937,Smith:1936,Rubin:1980,Rubin:1983}. Later, its
necessity was evident to explain the gravitational lenses, the structure
formation in the early Universe, the acoustic baryonic oscillations, the
power spectrum of galaxies, among other astrophysical and cosmological
phenomena \citep[see e.g.][]{Bertone:2004,Bennett:2013}. Very recently, the space mission
\textit{Planck} has obtained the most precise map of the total matter-energy
content of the Universe, setting the contribution of dark matter to
$\sim 26\%$, meanwhile the baryonic matter is only $\sim 5\%$ and the
cosmological constant $\Lambda$ or dark energy is $\sim 69\%$
\citep{Planck:2016}.

  The most accepted DM model is the
cold dark matter (CDM), which is very successful at reproducing the
observations at cosmological scales; however, at the galactic level it faces
some problems \citep[see e.g.][]{Weinberg:2015}. One of the difficulties is the
so-called `cusp/core' problem, since from CDM $N$-body simulations the DM
halos are assembled with the `cuspy' Navarro-Frenk-White density profile
\citep{Navarro:1997}, which is proportional to $r^{-1}$ at small radii and to
$r^{-3}$ at large distances, meanwhile many observations suggest a constant
central density or `core' profile, e.g.
in rotation curves of galaxies \citep{Moore:1999,deBlok:2001,
McGaugh:2007,McGaugh:2016} and dwarf spheroidal (dSph) galaxies
\citep{Klypin:1999,Kroupa:2010,Boylan-Kolchin:2011,Walker:2012,Pawlowski:2014}.
Another issue is the CDM prediction of too many satellite halos around big
galaxies like the Milky Way \citep{Sawala:2015} which have not been observed.
Aside, it fails to reproduce the phase-space distribution of satellites around
the Milky Way and Andromeda galaxies \citep{Pawlowski:2012,Ibata:2013,
Ibata:2014} and the internal dynamics in tidal dwarf galaxies
\citep{Gentile:2007,Kroupa:2012}. Finally, another issue might lie in the early
formation of big galaxies, since the CDM model predicts big galaxies were
formed hierarchically from halos less massive than $10^{12} M_\odot$, as is
the typical case, but there are some recent observations of massive galaxies at
very high redshifts $z$ ($5 \leq z < 6$) \citep{Caputi:2015}.

  Such problems at the galactic scales might be solved by introducing the
baryonic physics to the simulations, which can be relevant at the centers of the
galaxies and galaxy clusters through the inclusion, for example, of star formation,
supernova explosions, stellar winds, active galactic nuclei, dynamical friction,
etc. \citep[see e.g.][]{El-Zant:2001,El-Zant:2004,Ma:2004,Nipoti:2004,Romano-Diaz:2008,
DelPopolo:2009,Governato:2010,DelPopolo:2012,DelPopolo:2012b,Teyssier:2013,
Pontzen:2012,Madau:2014,DiCintio:2014,Pontzen:2014,Nipoti:2015,DelPopolo:2017}.
From a phenomenological point of view, empirical CDM density profiles have
been proposed in order to explain the observations, e.g. the Burkert profile
\citep{Burkert:1995,Salucci:2000} and the generalized NFW profile
\citep{Zhao:1996}. Aside, there are some alternatives to CDM seeking to solve
the discrepancies at galactic scales without appealing to baryonic processes,
for instance, warm dark matter \citep{zavala:2009,navarro10,Lovell:2011},
self-interacting dark matter \citep{spergel:2000,yoshida:2000,dave:2001,Elbert:2015,
Robles:2017} and scalar field dark matter (SFDM) \citep[see][and references
therein]{Matos-Guzman:1998, Magana:2012,Suarez:2014,Matos-Robles:2016,
Marsh:2015b,Ostriker:2016}.

  We are interested in the SFDM model, studied within some special cases and
named differently depending on the authors. The motivation is the natural
solution emerging from this model to the CDM problems mentioned before. The
SFDM model assumes DM is a spin$-0$ scalar field (SF) $\psi$, with a
typical ultra-light mass $m_\psi$ $\sim$ $10^{-23}-10^{-21} \mathrm{eV}/c^2$, which might
include self-interactions. The
first time when this idea was mentioned was in \citet{Ruffini:1983}; since then
the idea was rediscovered several times using different names \citep[see e.g.][]
{Membrado:1989,Spergel:1989,Sin:1992,Ji:1994,Lee:1995,
Matos-Guzman:1998,Sahni:1999,Peebles:2000,Goodman:2000,Matos-Urena:2000,
Matos-Urena:2001,Hu:2000,Wetterich:2001,Arbey:2001a,Boehmer:2007,
Vazquez-Magana:2008,Woo:2009,Lundgren:2010,Bray:2010,Marsh-Ferreira:2010,Robles:2013,
Schive:2014dra} and more recently in \citet{Ostriker:2016}. However, the first
systematic study of this idea started in 1998 by \citet{Guzman:1999,
Matos-Guzman:1998}, showing that the observed rotation curves of disk galaxies
can be reproduced by the SFDM model, and the cosmology was studied for the first
time in \citet{Matos:1999,Matos-Urena:2000}. Other systematic studies of the
SFDM model were performed by \citet{Arbey:2001a,Arbey:2001b} and more recently
in \citet{Marsh-Ferreira:2010,Schive:2014dra,Marsh:2015b}.

  In \citet{Matos:1999,Matos-Urena:2000}, using the amount of
satellite galaxies observed in the vicinity of the Milky Way, it was found that
the mass of the SF should be $m_\psi$ $\sim$ $10^{-22} \mathrm{eV}/c^2$
\citep[see also][for the estimation of the SF mass from a Jeans' stability
analysis, $m_\psi$ $\sim$ $10^{-22} \mathrm{eV}/c^2$]{Hdez-Rguez:2011,Rguez-Meza:2017}.
With this ultra-light mass, \citet{Alcubierre:2001ea} found, through numerical
simulations, that the gravitational collapse of a SF configuration forms stable
objects with mass $M$ of the order of a galaxy halo, $M$ $\sim$ $10^{12} M_\odot$.
Also, with the characteristic ultra-light SF mass, the bosons condense very
early in the Universe at critical condensation temperatures of $\mathrm{TeV}$
\citep{Matos-Urena:2001}, making up Bose-Einstein condensates (BEC) interpreted as
the DM halos. One important property of the SFDM is that, at
cosmological scales, it behaves as dust and reproduces the same observations
as well as CDM: the cosmic microwave background (CMB) and the
mass power spectrum at large scales \citep{RodriguezMontoya:2010,Hlozek:2014,
Schive:2015}. Recently, \citet{Schive:2014dra,Schive:2014hza} have run high-resolution
cosmological simulations of SFDM and reproduced the same results.

  In a series of papers \citep{Matos:2007,Bernal:2008,Robles:2012uy} it was
shown that the SFDM forms core halos \citep[see also][]{Harko:2011xw}. Further
features of the SFDM have been analyzed, for example, the gravitational
lensing \citep{Nunez:2010,Robles-lensing:2013}, the $\psi^4$-SF
potential \citep[e.g.][]{Matos:2011,Matos:2016}, multistate SF solutions, i.e.
bosons in excited states outside the ground energy level \citep[see e.g.][]{BernalA:2010,Robles:2013,
Martinez-Medina:2015,Bernal:2016}. Numerical simulations of galaxy formation
were performed \citep{Martinez-Medina:2014hca,Martinez-Medina:2015jra},
where the characteristic spiral arms and bars of a disk galaxy were easily
generated. It was also shown that satellite galaxies are stable around SFDM
halos \citep{Robles:2015}. The hydrodynamical version of the SF equations of
motion were first shown in \citet{Hdez-Rguez:2011,Rguez-Meza:2017}, where it was
demostrated that vorticity is present in this `fluid'. Also a study of
high-resolution rotation curves of spiral galaxies in the BEC-DM and
flat-oscillaton models has been done \citep{Fdez-Hdez:2017}. And there are some
works investigating the possibility that galactic ultra-light boson halos may
host and even collapse into the supermassive black-holes observed in many
galaxies \citep{Chavanis:2016,Avilez:2017}.

  As mentioned before, the SFDM model has many variants, depending on the
specific characteristics studied by different authors: the SF might be real or
complex and possess a self-interaction, it might take into account also the
temperature of the SF, by introducing a suitable SF potential. The SFDM model
has been named as fuzzy DM (FDM) \citep{Hu:2000}, wave DM (WDM or $\psi$DM)
\citep{Bray:2012,Schive:2014dra}, ultra-light axion (ULA) DM \citep{Hlozek:2014},
Bose-Einstein condensate (BEC) DM \citep{Boehmer:2007,Fdez-Hdez:2017},
multistate SFDM (mSFDM) \citep{Robles:2013}, etc.

  In the BEC-DM model \citep{Boehmer:2007} it is assumed a dominant self-interaction
between the particles (Thomas-Fermi limit); in this approximation all the bosons lie in the
ground state at zero temperature; the mean boson mass from the
analytic density profile is $m_\psi$ $\sim$ $10^{-6} \mathrm{eV}/c^2$. This
model has been widely studied, showing up sharp
discrepancies in dwarf galaxies \citep{Diez-Tejedor:2014}, disk galaxies
\citep{Boehmer:2007,Robles:2012uy,Fdez-Hdez:2017} and galaxy clusters
\citep{Bernal:2016}, ruling out this approximation as a realistic SFDM model.

  The first model studied in the present work is the fuzzy, wave or ultra-light axion DM,
which share the same characteristics: a
quadratic non-thermal, non-interacting potential with ultra-light masses,
$m_\psi$ $\sim$ $10^{-23}-10^{-21} \mathrm{eV}/c^2$.
From the cosmological simulation by \citet{Schive:2014dra,Schive:2014hza}, an
empirical density profile was obtained for this model, composed by a
coupling of the asymptotic NFW decline with an inner soliton-like profile
from the SFDM (`soliton+NFW' profile) \citep{Schive:2014dra,Marsh-Pop:2015}.

  Secondly, we investigate the multistate SFDM model \citep{Robles:2013}, that considers
the SF is thermal at the very early Universe, interacting with the radiation
and the rest of matter. By introducing a self-interacting,
temperature-corrected SF potential, the authors obtained an exact analytic
solution for the SF density, including the ground and excited quantum states.
This model has been proved successful in fitting
the observations of dwarf spheroidal galaxies \citep{Martinez-Medina:2015},
the rotation curves of disk galaxies \citep{Robles:2013}, the strong
gravitational lensing \citep{Robles-lensing:2013} and the dynamical masses from X-ray
observations of clusters of galaxies \citep{Bernal:2016}. Observe that
both, the soliton+NFW and the mSFDM profiles are approaches of the same model
for galactic applications, the difference lies in that the first one comes from
numerical simulations, and the second one comes from approximations of the field
equations but with finte temperature in the SF potential.

  In this work, we fit the observed rotation curves of 18 high-resolution
low surface brightness (LSB) galaxies, where only the DM is taken into account,
4 representative SPARC (Spitzer Photometry \& Accurate Rotation Curves) galaxies
and 2 other NGC galaxies used in \citet{Robles:2013} (the last 6 galaxies with
high-resolution photometric information).
For the galaxies with baryonic data, \citet{McGaugh:2016,Lelli-McGaugh:2016b} analyzed 153
galaxies with the gas and stars information from the SPARC database \citep{Lelli:2016}, and found
an empirical radial acceleration relation between the observed acceleration
from the rotation curves of the galaxies and the acceleration from the baryonic
component, showing a deviation at the value $g^\dag=1.2 \times 10^{-10}
\mathrm{m/s}^2$. This result could suggest that baryons are the source of the
gravitational potential, at least at small radii. As \citet{McGaugh:2016}
pointed out, such relation can be explained as the end product of galaxy
formation processes (including the baryonic matter), new DM physics or the result of a modified
gravity law. In the CDM paradigm, \citet{Ludlow:2016} introduced different stellar and AGN 
feedback processes to explain the observed acceleration relation in well-resolved
galaxies from the EAGLE simulation, showing that the
empirical relation can be `accommodated' within the model.
However, instead to look for complicated and diverse baryonic processes, in
this work we test the two SFDM models with the observed rotation
curves of some representative SPARC galaxies, showing that the model is consistent
with the data. The results given here will give us a clue for the 
next generation of numerical simulations. The soliton+NFW profile does not contain
finite temperature while the mSFDM profile is not a numerical result.
The best SFDM profile will be the one that takes these two ingredients into account.

  The idea is to analyze statistically both approximations: the study can give
us a clue for whether the scalar field is thermal or not, or if the
self-interaction between SF particles can be important, assuming the scalar
field is the dark matter of the galaxies. In order to do so, this work is
organized as follows: In Section~\ref{sec:SFDM} we briefly explain the SFDM model
and the two different approaches to be compared; in Section~\ref{section:Samples}
we describe the galaxies' samples used to fit the SFDM models; in
Section~\ref{sec:Results} we present the results, and in Section~\ref{sec:Conclusions}
we discuss the results and present our conclusions.

%%%%%%%%%%%% SCALAR FIELD DARK MATTER %%%%%%%%%%%%%
\section{Scalar Field Dark Matter}
\label{sec:SFDM}

  From particle physics motivation, spin-0 scalar fields are the simplest
bosonic particles, described by the Klein-Gordon (KG) equation:
\begin{equation}
	\Box \psi - \frac{\mathrm{d}V(\psi)}{\mathrm{d}\psi} = 0 ,
\label{eq:KG}
\end{equation}
for the real or complex scalar field $\psi$ and the SF potential $V(\psi)$;
from hereafter we use the notation $\hbar=1=c$, for $c$ the speed of light and
$\hbar$ the reduced Planck's constant. The scalar fields
arise in cosmology to explain the inflationary epoch at the very early Universe and
as alternatives to explain the accelerated expansion. They can
describe also compact objects, like boson stars and supermassive black holes. As
alternative to the CDM paradigm, it is proposed that DM halos are
constituted by ultra-light scalar fields, for which the complete
Einstein-Klein-Gordon equations reduce to the Schr\"odinger-Poisson system in
the Newtonian limit.

  Until now, there is not an agreement on the correct form of the potential
$V(\psi)$ for DM applications and some of them have been proposed in order to apply
the SFDM theory to different astronomical and cosmological situations
\citep[see e.g.][]{Suarez:2014}. A general potential is given by
\begin{equation}
	V(\psi) = \frac{1}{2} m_\psi^2 \psi^2 + \frac{1}{4} \lambda \psi^4 \, ,
\label{eq:double-pot}
\end{equation}
the so-called `double-well' potential, which includes the mass $m_\psi$ of the SF and a
self-interaction $\lambda$. For positive $m_\psi^2$ the potential has a single
minimum at $\psi=0$ and the potential is $Z_2$ invariant; for $-m_\psi^2$ the
potential has two minima and the $Z_2$ symmetry appears spontaneously
broken, being of great interest for physical situations (see
Subsection~\ref{section:FT-SFDM}). A recent study of large scale structure formation
with the last potential has been developed in \citet{Suarez:2016}. For this
potential, in the hydrodynamic approach, it is possible to write an equation of
state of the SF, resulting in a polytrope of index $n=1$.

  In the case $\lambda=0$, the potential~\eqref{eq:double-pot} reduces to
\begin{equation}
	V(\psi)= \frac{1}{2} m_\psi^2 \psi^2 ,
\label{eq:psi2}
\end{equation}
which has been widely used to model the SFDM in the Universe, as discussed in
the next Subsection.

%%%%%%%%%%%%%%%%%%%% SCHIVE + NFW %%%%%%%%%%%%%%%%%%%%%%%%%%%%%%%%%%%
\subsection{Fuzzy (wave) dark matter}
\label{section:Schive+NFW}

  The fuzzy DM (FDM) \citep{Hu:2000}, wave DM (WDM or $\psi$DM) \citep{Bray:2012,
Schive:2014dra} or ultra-light axion (ULA) DM \citep{Hlozek:2014} (here we name
it as \citet{Hu:2000}, FDM), considers the SFDM
is described by the potential~\eqref{eq:psi2}, for ultra-light masses $m_\psi$
$\sim$ $10^{-23}-10^{-21} \mathrm{eV}$ and null self-interaction $\lambda=0$.
As showed in \citet{Matos-Urena:2001} \citep[see also][]{Schive:2014dra,Urena-Gonzalez:2016},
this approximation presents a cut-off in the power spectrum which suppresses the
small scale structure formation below the de Broglie wavelength $\lambda_\mathrm{deB}$ 
(corresponding to halo masses $M$$<$$10^8 M_\odot$
for $m_\psi$ $\sim$ $10^{-22} \mathrm{eV}$), as a result of the quantum
properties of the model (Heisenberg uncertainty principle), solving in this way
the small scale structure overproduction in CDM. This approach assumes the
SFDM is at zero temperature, implying all the bosons are in the ground state,
i.e. the lowest energy level with no nodes. Also, \citet{Suarez:2011} showed
that, within this approach, the evolution of
perturbations of the SFDM model is identical to $\Lambda$CDM and
\citet{RodriguezMontoya:2010} showed the SFDM model is consistent with
the acoustic peaks of the CMB for a boson mass $m_\psi$
$\sim$ $10^{-22} \mathrm{eV}$.

  Nevertheless, \citet{RindlerDaller:2013} found that this model is not
consistent with the big bang nucleosynthesis (BBN) constraints for the boson
mass $m_\psi$ for any model with $\lambda=0$, at 1$\sigma$-confidence
level (CL), making necessary to introduce a self-interaction into the
SF potential. Although their results allow to alleviate such restrictions at
2$\sigma$-CL, their work is a motivation to study a SF potential including a
non-null self-interaction (see Subsection~\ref{section:FT-SFDM}).

  \citet{Schive:2014dra,Schive:2014hza} run a high-resolution cosmological simulation
based on the dynamics of FDM in the Newtonian limit, governed by the
Schr\"odinger-Poisson system \citep[see also][]{Schwabe:2016,Mocz-Robles:2017}.
From the simulation, \citet{Schive:2014dra} derived an empirical density profile
for the DM halos, consisting of a soliton-like core in every system, prominent
before a transition radius, embedded in a NFW density halo dominant at large
radii. The motivation of this proposal is the expected loss of phase
coherence of the FDM waves at large distances from the center of
distribution, expecting a transition from the soliton to the NFW profile
\citep{Marsh-Pop:2015}. Such profile can be approximated by
\begin{equation}
	\rho_{\mathrm{FDM}} (r) = \Theta \left( r_\epsilon - r \right)
	\rho_\mathrm{sol}(r) + \Theta \left( r - r_\epsilon \right)
	\rho_\mathrm{NFW}(r) \, ,
\label{eq:rho-schive}
\end{equation}
where $\Theta$ is a step function, $r_\epsilon$ is the transition radius
where the density changes from the soliton profile \citep{Schive:2014dra}
\begin{equation}
   \rho_\mathrm{sol} (r) = \frac{\rho_c}
   {[1 + 0.091 (r/r_c)^2]^8} \, ,
\label{eq:rho-core}
\end{equation}
to the NFW profile \citep{Navarro:1997}
\begin{equation}
	\rho_\mathrm{NFW}(r) = \frac{\rho_s}{(r/r_s)(1+r/r_s)^2} .
\label{nfw-rho}
\end{equation}
In equation~\eqref{eq:rho-core}, $\rho_c:=1.9 (m_\psi/10^{-23} \mathrm{eV})^{-2}
(r_c/\mathrm{kpc})^{-4} M_\odot \mathrm{pc}^{-3}$ is the central
soliton density, $m_\psi$ the boson mass and $r_c$ the half-light radius of
the soliton-like region. In equation~\eqref{nfw-rho}, $\rho_s$ is related to
the density of the Universe at the moment the halo collapsed and $r_s$ is a
scale radius. The total density profile~\eqref{eq:rho-schive} has five free
parameters ($m_\psi$, $r_c$, $r_\epsilon$, $\rho_s$, $r_s$), which are
reduced to four asking for continuity of the function at the transition radius
$r_\epsilon$ (see Appendix~\ref{appendix}). From the cosmological simulations, 
\citet{Schive:2014dra} found that the transition radius $r_\epsilon$ is, in
general, $r_\epsilon>3r_c$.

  There are several constraints to the boson mass $m_\psi$ in the FDM model: In
the cosmological context, from CMB and galaxy clustering data,\citet{Hlozek:2014}
constrained the mass $m_\psi > 10^{-24} \mathrm{eV}$.  From the galaxy UV-luminosity
function and reionization constraints, \citet{Bozek:2014} obtained $m_\psi >
10^{-23} \mathrm{eV}$, and from the high-redshift luminosity
function of galaxies, \citet{Schive:2015} get $m_\psi > 1.2 \times 10^{-22}
\mathrm{eV}$. From Lyman-$\alpha$ observations, \citet{Sarkar:2016} obtained
$m_\psi > 10^{-23} \mathrm{eV}$, and very recently, \citet{Irsic:2017,Armengaud:2017}
derived $m_\psi \gtrsim 2\times10^{-21} \mathrm{eV}$, providing a severe constraint
to smaller masses.

  From astrophysical observations, the study of the effect of tidal forces
on the cold clumps in Ursa Minor, \citet{Lora:2012} found $m_\psi \sim (0.3-1)
\times 10^{-22} \mathrm{eV}$, and in Sextants, \citet{Lora:2014} found $m_\psi
\sim (0.12-8) \times 10^{-22} \mathrm{eV}$.  From ultra-faint dwarf spheroidal (dSph) galaxies,
\citet{Calabrese:2016} obtained $m_\psi \sim (3.7-5.6) \times 10^{-22} \mathrm{eV}$.
For the classical dSph galaxies in the Milky Way,
\citet{Marsh-Pop:2015} constrained $m_\psi$ using the soliton+NFW density
profile~\eqref{eq:rho-schive}, through stellar populations in Fornax
and Sculptor and found $m_\psi < 1.1 \times 10^{-22} \mathrm{eV}$.
\citet{Chen-Schive:2016} applied the Jeans analysis to the kinematic data of
the eight dSphs to constrain $m_\psi$, where, for all the dSphs except for Fornax, as
the transition radii are outer the half-light radii $r_c$, they took into
account the soliton core profile~\eqref{eq:rho-core} only, deriving a mass
$m_\psi=1.18 \times 10^{-22} \mathrm{eV}$; for Fornax, they applied the
complete density profile~\eqref{eq:rho-schive} and found a larger
$m_\psi=1.79 \times 10^{-22} \mathrm{eV}$ and smaller $r_c$, with respect to
the soliton-only model results. \citet{Gonzalez-Marsh:2016}
constrained $m_\psi$ with the complete soliton+NFW
profile~\eqref{eq:rho-schive}, using kinematic mock data of
Fornax and Sculptor, that include the stellar components of the galaxies,
finding core radii $r_c>1.5\ \mathrm{kpc}$ and $r_c>1.2\ \mathrm{kpc}$,
respectively, and $m_\psi< 4 \times 10^{-23} \mathrm{eV}$;
the authors propose that such small boson mass might be the
result of baryonic feedback processes present in the dSph galaxies.
Studying a set of LSB rotation curves, \citet{Garcia-Aspeitia:2015}
showed that FDM gives values of $M\sim 10^7 M_\odot$ within $300 \ \mathrm{pc}$,
which is consistent with results found in \citet{Strigari:2008} for satellite
galaxies in the Milky Way.

And very recently, \citet{Urena-Robles:2017} proposed a way to determine the
boson mass using the results from the mass discrepancy-acceleration relation
presented in \citet{McGaugh:2016,Lelli:2016} from the observations in the
SPARC catalog, that implies a universal maximum acceleration for all the
galaxy halos, and found a relation between the soliton core radius $r_c$, the
boson mass $m_\psi$ and the central surface density of the galaxies. With
their analysis with the soliton profile, they found that the dwarf galaxies in
the Milky-Way and Andromeda 
are consistent with a boson mass $m_\psi \approx 1.2 \times 10^{-21} \mathrm{eV}$,
in agreement with the cosmological constraint $m_\psi > 10^{-23} \mathrm{eV}$.

Therefore, we have the lower limit $m_\psi=10^{-23} \mathrm{eV}$
\citep{Bozek:2014,Sarkar:2016}, and from the latest works with Lyman-$\alpha$
data the limit $m_\psi \approx 2 \times 10^{-21} \mathrm{eV}$
\citep{Irsic:2017,Armengaud:2017}. Taking into account the last result, there
would be a tension with many of the constraints mentioned before, coming from
diverse observations. At this point, there is not an agreement on the correct
mass of the ultra-light boson, and in this article, we explore the viability
of the FDM model in galaxies of many sizes to report the boson mass needed to
reproduce the observed rotation curves.

  For the LSB and SPARC galaxies analyzed in the present work, we assume the
soliton+NFW profile~\eqref{eq:rho-schive} and expect the transition radius
$r_\epsilon$ is well inside the radius of the last observed point (spanning a
range $R_\mathrm{max}\sim1-30 \ \mathrm{kpc}$, for the maximum radius of luminous
matter), thus we apply the complete density profile. A continuity condition
between the soliton and NFW profiles at the transition radius
$r_\epsilon$ is imposed, assuming an abrupt transition between the soliton and
the external NFW halo, as found in the cosmological simulations
\citep{Schive:2014dra,Schwabe:2016,Mocz-Robles:2017}. With this condition, it
is possible to write the NFW parameter $\rho_s$ in function of the other four
free parameters, thus we fit only $m_\psi$, $r_c$, $r_\epsilon$ and $r_s$. The
complete expressions are written in Appendix~\ref{appendix}.

%%%%%%%%%%%%%%%%%%%%% FINITE TEMPERATURE SFDM %%%%%%%%%%%%%%%%%%%%%%%%%%%%%
\subsection{Multistate scalar field dark matter}
\label{section:FT-SFDM}

  As a step forward on the study of the SFDM model, \citet{Robles:2013}
considered the general double-well potential~\eqref{eq:double-pot}, with a non-null
self-interaction term, plus one-loop finite-temperature $T$ corrections
\citep{Kolb:1994}:
\begin{equation}
	V(\psi) = - \frac{1}{2} m_\psi^2 \psi^2 + \frac{1}{4} \lambda \psi^4 +
	\frac{1}{8} \lambda \psi^2 T^2 - \frac{\pi^2}{90} T^4 \, ,
\label{eq:ft-potential}
\end{equation}
in units $k_\mathrm{B}=1$ for the Boltzmann constant and where the $Z_2$
symmetry appears spontaneously broken (with the term $-m_\psi^2$). Such
self-interacting, finite-temperature corrected potential is motivated by the
theoretical expectation of modern particle physics that at high temperatures
symmetries spontaneously broken today were restored. Symmetry breaking studies
are of great interest for diverse physical situations, assuming the Universe
has underwent phase transitions during its evolution \citep[see e.g.][and
references therein]{Kolb:1994}, as for example, the inflationary era. With this
motivation in mind, \citet{Robles:2013} considered that, at the very early
Universe, the initial SF fluctuations from inflation interact with
the rest of matter and radiation at very high temperatures $T \gg T_c$, for
$T_c$ the critical temperature where $\psi$$=$0 is a minimum of the potential and
the symmetry is restored. In this case, the SF is
embedded in a thermal bath at finite temperature $T$ at very early
times\footnote{In equation~\eqref{eq:ft-potential}, the term $\propto \psi^2
T^2$ appears due to the interaction of the SF with the
thermal bath and the term $\propto T^4$ is due to the thermal bath only.}. As
the Universe expands, the temperature decreases and eventually the SF decouples
from the rest of matter, evolving independently. As the
potential~\eqref{eq:ft-potential} depends on the 4th-power with respect to
$T$, as the temperature continues decreasing, the SF goes through a $Z_2$
spontaneous symmetry breaking (SSB), which turns a minimum of the potential to
a maximum, that increases the amplitude of the initial SF perturbations
forming the initial galaxy halos \citep[see also][for the study of the SSB of
a charged complex SF]{Matos-Rguez:2014}.

  For cosmological applications, \citet{Robles:2013} assumed the SSB at
$T$$=$$T_c$ takes place in the radiation dominated era. They found
that at that moment, the SF fluctuations can start growing in the linear regime
for $T$$<$$T_c$, until they reach a new stable minimum. Then obtained the perturbed
Einstein-Klein-Gordon evolution equations for the SF perturbation, $\delta\psi$,
in a Friedmann-Lema\^{i}tre-Robertson-Walker background space-time. Under
these assumptions, the galactic halos could have been formed almost at the same time
of the SSB and with similar masses, $M\sim 10^{12} M_\odot$ (for
$m_\psi \sim 10^{-22} \mathrm{eV}$), which is the typical mass of a galaxy like
ours. Later on, such halos can enter in the non-linear regime, merging and
constituting larger structures, hierarchically, just like in the CDM model.

  Under the linear approximation to describe the evolution of a galaxy halo, in
the Newtonian regime, the exact solution to the perturbation
equations for the temperature-corrected potential~\eqref{eq:ft-potential}
is found as \citep{Robles:2013}
\begin{equation}
	\rho^j_\mathrm{mSFDM}(r) = \rho^j_0 \left[ \frac{\sin(k_j r)}{(k_j r)}
	\right]^2 ,
\label{T-density}
\end{equation}
whose mass distribution is given by
\begin{equation}
   M^j_\mathrm{mSFDM}(r) = \frac{4 \pi \rho_0}{k_j^2} \frac{r}{2} \left[ 1-
	\frac{\sin(2 k_j r)}{(2 k_j r)} \right] .
\label{T-mass}
\end{equation}
This solution is naturally core. Thus, the finite-temperature corrected
potential~\eqref{eq:ft-potential} implies
the existence of different excitation states $j$ as solution to the SF
perturbation equation, i.e. the bosons are thermally distributed in the ground
and excited states at higher energy levels, and the general solution allows a
configuration out of the fully condensate system. In the last equation,
$j=1,2,3,...$ is the number of the excited state required to fit the mass
distribution, $\rho^j_0 = \rho^j_\mathrm{mSFDM}(0)$ is the central density and
the radius $R$ of the SF configuration is fixed through the condition
$\rho^j_\mathrm{mSFDM}(R)=0$, i.e. $k_j R = j \pi$.

  Notice that when using the analytic profile~\eqref{T-density}
there is a degeneracy in the function, since if we keep the ratio $j/R$ constant, the form of the density profile does not change, only the state $j$ and radius $R$ become bigger. The radius $R$ is chosen as the minimum radius required to fit the last observational
point and it is defined as the radius of the halo, even though the realistic
SFDM halo goes to infinity. In this approximation, to fit diverse observations,
the oscillations for $r>R$ are not taken into account given that they decay very
quickly with $r$.

  Now, as the solution to the SF perturbation equation is linear, it is
possible to have combinations of excited states; this means that the total
density $\rho_\mathrm{mSFDM}$ can be written as the sum of the densities in the
different excited states \citep{Robles:2013}:
\begin{equation}
 \rho_\mathrm{mSFDM}(r) = \sum_j \rho^j_0 \left[ \frac{\sin(j \pi r / R)}
	{(j \pi r / R)} \right]^2 .
\label{densitytotal}
\end{equation}
The last equation corresponds to the multistate SFDM (mSFDM) model. It is worth
noting that the radius $R$ in the last equation is the same for all the excited
states present in the configuration and, as explained before, it is defined as the radius
of the mSFDM halo. Such profile has at least three free parameters for one state
and every additional excited state adds two parameters more, $j$ and $\rho_0^j$.

This theoretically motivated SFDM profile has a big disadvantage: the physical
parameters of the SF, like the boson mass $m_\psi$ and self-interaction parameter
$\lambda$, are degenerated with other quantities, like the critical temperature
$T_c$ at the SSB, the temperature of the halo at the formation time, etc., thus
it is not possible to constrain the properties of the boson particle with only
one independent observation. There are some works trying to constrain the
self-interaction for a $\psi^4$-potential \citep[see e.g.][]{RindlerDaller:2013,
Li:2016,Suarez:2016}. However, the mSFDM model has been proved successful in
fitting the observations of dwarf spheroidal galaxies \citep{Martinez-Medina:2015},
the rotation curves of galaxies \citep{Robles:2013}, the dynamical masses from
X-ray observations of clusters of galaxies \citep{Bernal:2016}, in predicting the
size of the Einstein radius of lensed galaxies by strong gravitational lensing
\citep{Robles-lensing:2013} and the survivability of satellite SFDM halos orbiting
around a Milky Way-like galaxy \citep{Robles:2015}. At this point, the mSFDM profile
can be used as a useful fitting function and along with or as an analytic
alternative to the empirical soliton+NFW density profile, as disscused below.

%%%%%%%%%%%%%%%% OBSERVATIONS %%%%%%%%%%%%%%
\section{Galaxies' samples}
\label{section:Samples}

\subsection{High-resolution LSB galaxies}
\label{hr-lsb}

  We are using two sets of galaxy rotation curves according to the availability of
photometric data. For the low surface brightness (LSB) galaxies, in which the dark
matter is the dominant component, we are using the observed high-resolution
rotation curves of 18 LSB galaxies reported in \citet{deBlok:2001}. In that work,
the authors considered the visible data contribution, classifying the galaxies
according with the availability of photometric data and using three models: the
minimum disk model, where the dark matter is the principal component in the halos
making zero all the visible components for the galaxies with photometry; a
constant $M/L_*$ model, and the maximum disk model, that considers the galaxies
with photometric data only.

  For the first part of the work, we used the set of the LSB galaxies 
without photometric data, where the DM is the dominant component. This set is
described in Table~\ref{tab:gal-info}, where we present some important
characteristics of each galaxy. 

\begin{table*}
\caption{High-resolution LSB galaxies. In this Table we show the characteristics
of the 18 high-resolution LSB galaxies \citep{deBlok:2001} used in the present work.
The columns read: (1)~The name of the galaxy, (2)~Morphology, (3)~Absolute magnitude,
(4)~Maximum observational radius, (5)~Maximum velocity of the rotation curve and
(6)~Distance to the galaxy.}
\label{tab:gal-info}
\begin{tabular}{lccccc}
\hline
Galaxy	&	Morphology	&	$M_\mathrm{abs}$	&	$R_\mathrm{max}$
&	$V_\mathrm{max}$	& $D$ \\
Name	&				&	$(\mathrm{mag})$	&	$(\mathrm{kpc})$	&	$(\mathrm{km/s})$   & $(\mathrm{Mpc})$    \\
(1)		&	(2)			&	(3)			&	(4)			&	(5)	&	       (6)  	\\
\hline
ESO-LV 014-0040  & Spiral				& -21.6      & 29.2     & 263	& 212	\\
ESO-LV 084-0411  & Edge-on				& -18.1       & 8.9     & 61 	&  80   \\
ESO-LV 120-0211  & Fuzzy Magellanic bar	& -15.6       & 3.5     & 25    & 15	\\
ESO-LV 187-0510  & Irregular spiral, flocculent & -16.5& 3.0	& 40    & 18  \\
ESO-LV 206-0140  & Spiral				& -19.2      & 11.6     & 118   & 60   \\
ESO-LV 302-0120  & Spiral, hint of bar?	& -19.1      & 11.0     & 86    & 69  \\
ESO-LV 305-0090  & Barred spiral		& -17.3      & 4.8      & 54    &  11  \\
ESO-LV 425-0180  & Spiral				& -20.5      & 14.4     & 145   & 86   \\
ESO-LV 488-0490  & Inclined Magellanic bar& -16.8    & 6.0      & 97    & 22   \\
F730-V1			 & Spiral				& ...        & 11.9     & 145   & 144 \\
UGC 4115         & Im 				& -12.4      & 1.0      & 40    & 3.2  \\
UGC 11454        & Fuzzy spiral, small core &$-18.6^a$&11.9     & 152 & 91  \\
UGC 11557        &  	SBdms & -20.0  & 6.2		& 95    & 22  \\
UGC 11583        & dI & $-14.0^a$   & 1.5     & 36    & 5  \\
UGC 11616        & Sc     &$-20.3^a$ & 9.6        & 143   & 73 \\
UGC 11648        & I            & $-21.0^a$  & 12.7     & 145   & 48 \\
UGC 11748        & Sbc &$-22.9^a$& 21.0 & 242   & 73 \\
UGC 11819        & dG            & $-20.3^a$  & 11.7     & 153   & 60 \\
\hline
\end{tabular}
%\newline
\flushleft
$^a$Zwicky magnitude 17.
\end{table*}

\subsection{NGC galaxies with photometric data}
\label{sparc}

  In the second place we selected, as representative systems of the set of
  galaxies with photometric data, the three sample
galaxies reported in \citet{McGaugh:2016} (NGC 7814, 6503, 3741) from the SPARC (Spitzer
Photometry \& Accurate Rotation Curves) database~\citep{Lelli:2016}, and the
three sample galaxies used in \citet{Robles:2013} (NGC 1003, data from SPARC; NGC
1560, data from \citet{deBlok:2001}; NGC 6946, data from \citet{McGaugh:2005}).
The characteristics of these galaxies are shown in Table~\ref{tab:gal-photo}.

\begin{table}
\caption{Galaxies with photometric information. The same as Table~\ref{tab:gal-info}
for the galaxies with photometric data.}
\label{tab:gal-photo}
\begin{tabular}{lccccc}
\hline
Galaxy	&	Morphology	&	$M_\mathrm{abs}$	&	$R_\mathrm{max}$
&	$V_\mathrm{max}$	& $D$ \\
Name	&				&	$(\mathrm{mag})$	&	$(\mathrm{kpc})$	&
$(\mathrm{km/s})$   & $(\mathrm{Mpc})$     \\
(1)		&	(2)	&	(3)	&	(4)	&	(5)	&	(6) \\
\hline
NGC 7814$^a$	& Sabl & -20.15$^d$ 	& 5.03	& 218.9	& 14.40 \\
NGC 6503$^a$	& Scd & -17.7$^e$	& 3.04	& 116.3	& 6.26  \\
NGC 3741$^a$	& Im	& -13.13$^f$	& 5.14	& 50.1	& 3.21	\\
NGC 1003$^a$	& SA(s)cd	& -19.2	& 31.3 	& 115	& 11.8 \\
NGC 1560$^b$	& Sd	& -15.9	& 8.3	& 78	& 3.0 \\
NGC 6946$^c$	& SABcd	&	---	&	---	& 224.3	& 10.1\\
\hline
\end{tabular}
\newline
$^a$Data from the SPARC database \citep{Lelli:2016}.\newline
$^b$Data from \citet{deBlok:2001}.\newline
$^c$Data from \citet{McGaugh:2005}.\newline
$^d$\citet{Monachesi:2016-NGC7814}.\newline
$^e$\citet{Koda:2015-NGC6503}.\newline
$^f$\citet{Dutta:2009-NGC3741}.
\end{table}

  The SPARC database contains the near-infrared photometry (tracing the stellar
mass distribution) and high-resolution HI/H$\alpha$ rotation curves (tracing the gravitational
potential out to large radii) of 175 disk galaxies, deriving with high
resolution the stellar and gas components of the galaxies \citep{Lelli:2016}.
The sample represents a wide range in morphological types, stellar masses,
surface brightnesses and gas fractions, and the mass models are reconstructed
from the observed distributions of stars and gas for different characteristic
radii and values of the stellar mass-to-light ratio.
The database includes
low surface brightness (LSB) galaxies, in major part gas-dominated dwarf
galaxies; high surface brightness (HSB) galaxies, in major part
bulge-dominated spiral galaxies; and intermediate surface brightness (ISB) galaxies,
which are disk-dominated systems.

  Usually, in LSB galaxies the DM component is dominant even at
small radii, whilst in HSB galaxies the contribution of the stellar and gas
components to the total mass is important at the central regions.
However, \citet{McGaugh:2016} found that, even when the dark matter is dominant
at the inner regions for many galaxies, the observed acceleration in 153
galaxies of all types (LSB, ISB and HSB galaxies) from the SPARC database, strongly
correlates with the acceleration from the baryonic matter, showing a mass
discrepancy at the value $g^\dag = 1.2 \times 10^{-10}\ \mathrm{m/s}^2$.
In this work, we investigate the consistency of the two SFDM models
with the observational rotation curves of four representative SPARC galaxies, including the 
baryonic information.

\subsection{Statistical calibration method}
\label{mcmc-method}

  The galaxy rotation curve is given by
\begin{equation}
	V(r) = \sqrt{\frac{G M_T(r)}{r}} ,
\end{equation}
where $M_T = M_\mathrm{bulge} + M_\mathrm{disk} + M_\mathrm{gas} +
M_\mathrm{DM}$, for $M_\mathrm{DM}$ the dark matter contribution depending on the
density profile used (soliton+NFW or mSFDM). In the case of the LSB
galaxies (Subsection~\ref{hr-lsb}), $M_T = M_\mathrm{DM}$, and for the galaxies with
photometric information (Subsection~\ref{sparc}) we
considered the different baryonic components (bulge, disk-stars and gas), if given.

  In order to constrain the free parameters of every SFDM model, we used the
Markov Chain Monte Carlo (MCMC) method \citep{Gamerman:1997}, through a
maximization of the likelihood function $\mathcal{L}(\mathbf{p})$ given by
\begin{equation}
	\mathcal{L}({\bf p}) = \frac{1}{(2 \pi)^{N/2}
	|{\bf C}|^{1/2}} \exp{\left ( - \frac{{\bf \Delta}^{T}
	{\bf C}^{-1} {\bf \Delta}}{2} \right )} ,
\label{eq:likelihood}
\end{equation}
where $\mathbf{p}$ is the vector of parameters, $N$ the number of
observational points for each galaxy, ${\bf \Delta} = V_\mathrm{obs}(r_i) -
V_\mathrm{model}(r_i,\mathbf{p})$, for $V_\mathrm{obs}$ the observational
circular velocity at the radius $r_i$ and $V_\mathrm{model}$ the derived total velocity for a
given SFDM model computed in the same position where $V_\mathrm{obs}$ was
measured, and $\mathbf{C}$ a diagonal matrix.

  We sample the parameter space from uniform prior ranges with two Markov
chains and tested the convergence of the fit with the Gelman-Rubin convergence
criterion ($\mathcal{R}-1 < 0.1$) \citep{Gelman-Rubin:1992}. The fitting
parameters and $1 \sigma$ and $2 \sigma$ confidence levels (CL) are computed from the
Markov chains with 30\% as burn-in.

  Simultaneously, we used the data analysis software \textit{ROOT}
\citep{ROOT:1997}, to minimize the $\chi^2$ errors to obtain the best fit from the observations:
\begin{equation}
	\chi_\mathrm{red}^2 = \frac{1}{N-N_p} \sum_{i=1}^N \left( \frac{V_\mathrm{obs}(r_i) -
	V_\mathrm{model}(r_i,\mathbf{p})}{\sigma_i} \right)^2 \, ,
\end{equation}
where $N$ is the number of data points, $N_p$ the number of parameters and $\sigma_i $
the error in the measurement of $V_\mathrm{obs}(r_i)$.

%%%%%%%%%%%%%%%% RESULTS %%%%%%%%%%%%%%%%
\section{Results}
\label{sec:Results}

  In this Section we show the results of the fits with the two SFDM models. In
Subsection~\ref{subsec:Schive+NFW} we present the results for the FDM
model with the soliton+NFW density profile~\eqref{eq:rho-schive}, separately
for the high-resolution LSB galaxies (Subsection~\ref{subsubsec:lsb-1})
and for the galaxies with photometric information
(Subsection~\ref{subsubsec:sparc-1}). In~\ref{subsubsec:comb-lsb} we show the
results of the combined analysis of the LSB galaxies to determine the single
boson mass needed to fit all the rotation curves together. In
Subsection~\ref{subsec:Multistates} we show the results for the mSFDM model with
the density profile~\eqref{densitytotal}, also separating the fits with the
high-resolution LSB galaxies (Subsection~\ref{subsubsec:lsb-2}) and the
galaxies with photometric data (Subsection~\ref{subsubsec:sparc-2}).

%%%%%%%%%%%%%%%% SCHIVE+NFW %%%%%%%%%%%%%%%%
\subsection{Soliton+NFW fits}
\label{subsec:Schive+NFW}

\subsubsection{High-resolution LSB galaxies}
\label{subsubsec:lsb-1}

  Table~\ref{tab:results2} shows the results for the 18 high-resolution
LSB galaxies reported in \citet{deBlok:2001}, for the FDM model with the soliton+NFW density
profile~\eqref{eq:rho-schive}. As explained in Appendix~\ref{appendix}, we
restricted the density profile asking for continuity of the
function~\eqref{eq:rho-schive} at the transition radius $r_\epsilon$, as an abrupt
transition between the soliton and NFW regions, as showed in the cosmological
simulations~\citep{Schive:2014dra,Schive:2014hza,Schwabe:2016,Mocz-Robles:2017}. With this
restriction, we have four free parameters: the central density $\rho_c$, the
soliton core radius $r_c$, the transition radius $r_\epsilon$, and the NFW
characteristic radius $r_s$. For these quantities we report the $1\sigma$
errors from the MCMC method used, also the $\chi_\mathrm{red}^2$
errors for each galaxy. From $\rho_c$ and $r_c$ we obtained the boson mass
$m_\psi$, and with the four free parameters the
NFW density parameter $\rho_s$ can be computed. Additionally,
we derive the virial radius $r_{200}$ and concentration parameter
$c:=r_{200}/r_s,$\footnote{$r_{200}$ is the radius where the
density is 200 times the critical density of the Universe, and defines
the halo radius. In the CDM framework,
concentrations are strongly correlated with the halo formation epoch,
and must be in agreement with the $N$-body simulations.} shown in the same Table.

\begin{table*}
\caption{Soliton+NFW density profile in high-resolution LSB galaxies. In this
Table we show the resulting fitting parameters $\rho_c$, $r_c$, $r_\epsilon$ and
$r_s$ for the FDM model with the soliton+NFW density
profile~\eqref{eq:rho-schive}, and the resulting boson mass $m_\psi$, all the
quantities $\pm 1\sigma$ errors from the MCMC method used. We also show the
resulting ratio $r_\epsilon/r_c$, the $r_{200}$ radius from the NFW halo,
concentration parameter $c$ and $\chi^2_\mathrm{red}$ errors for the 18
high-resolution LSB galaxies in \citet{deBlok:2001}.}
\label{tab:results2}
\begin{tabular}{lccccccccc}
    \hline
    Galaxy	&	$\rho_c$	&	$r_c$	&	$r_\epsilon$	&	$r_\epsilon/r_c$	&	$m_\psi$	&	$r_s$	&	$r_{200}$	&	$c$	&	$\chi^2_\mathrm{red}$ \\
    	&	$(10^{-2} M_\odot/\mathrm{pc}^3)$	&	$(\mathrm{kpc})$	&	$(\mathrm{kpc})$	&	&	$(10^{-23}\mathrm{eV})$	&	$(\mathrm{kpc})$	&		$(\mathrm{kpc})$	&	&	\\
\hline
ESO 014-0040 & $36.8\pm 8.9$ 			& $2.38^{+0.42}_{-0.94}$ 	& $1.59^{+0.21}_{-0.68}$	& 0.67	& $0.50^{+0.13}_{-0.32}$	& $15.3^{+2.6}_{-3.8}$     			  	& 287 	& 18.8  		 	& 6.22  \\
ESO 084-0411 & $0.483^{+0.068}_{-0.11}$ 	& $8.96^{+0.94}_{-3.9}$ 	& $7.7^{+1.3}_{-2.7}$ 		& 0.86	& $0.33^{+0.12}_{-0.25}$	& $1.56^{+0.29}_{-1.5}\times 10^{-5}$ 	& 123 	& $7.90\times 10^6$	& 0.878	\\
ESO 120-0211 & $2.70^{+0.60}_{-1.2}$ 	& $1.59^{+0.18}_{-0.83}$	& $1.22^{+0.15}_{-0.47}$	& 0.77	& $5.2^{+1.9}_{-4.5}$ 		& $1.47^{+0.19}_{-1.5}\times 10^{-3}$ 	& 30.3	& $2.06\times 10^4$	& 1.03  \\
ESO 187-0510 & $3.95^{+0.57}_{-1.0}$ 	& $1.59^{+0.21}_{-0.42}$	& $2.17^{+0.29}_{-0.99}$	& 1.36	& $3.06\pm 0.97$ 			& $1.67^{+0.44}_{-1.7}\times 10^{-2}$ 	& 42.6	& $2.55\times 10^3$ & 0.44  \\
ESO 206-0140 & $20.4^{+4.5}_{-5.7}$ 		& $1.98^{+0.31}_{-0.93}$	& $1.10^{+0.21}_{-0.46}$	& 0.56	& $1.09^{+0.31}_{-0.86}$	& $6.0^{+1.3}_{-1.8}$  				  	& 123 	& 20.4 				& 3.13  \\
ESO 302-0120 & $3.55^{+0.57}_{-0.82}$ 	& $3.98^{+0.39}_{-1.3}$ 	& $4.25^{+0.46}_{-1.5}$ 	& 1.07	& $0.54^{+0.23}_{-0.32}$	& $1.74^{+0.48}_{-1.7}\times 10^{-3}$ 	& 106 	& $6.09\times 10^4$	& 0.58  \\
ESO 305-0090 & $2.38^{+0.35}_{-0.62}$ 	& $2.81^{+0.42}_{-0.68}$	& $4.34^{+0.62}_{-2.3}$ 	& 1.54	& $1.24^{+0.27}_{-0.49}$	& $7.6^{+1.8}_{-7.5}\times 10^{-3}$   	& 68.1	& $8.96\times 10^3$ & 0.17  \\
ESO 425-0180 & $106^{+51.0}_{-27.0}$ 	& $0.326^{+0.057}_{-0.22}$ 	& $0.151^{+0.021}_{-0.089}$ & 0.46	& $27.0^{+6.0}_{-30.0}$ 			& $21^{+7}_{-10}$					  	& 205 	& 9.78 				& 5.25  \\
ESO 488-0490 & $7.15^{+0.96}_{-1.4}$ 	& $2.66^{+0.22}_{-0.45}$  	& $4.13^{+0.55}_{-1.8}$ 	& 1.55	& $0.77^{+0.17}_{-0.12}$	& $1.08^{+0.37}_{-1.1}\times 10^{-4}$ 	& 108 	& 1.00$\times 10^6$	& 0.83  \\
F730-V1     & $23.6^{+4.6}_{-6.7}$ 		& $1.75^{+0.31}_{-0.67}$	& $1.16^{+0.19}_{-0.55}$	& 0.66	& $1.16^{+0.27}_{-0.77}$	& $9.3^{+2.1}_{-3.0}$  				  	& 160 	& 17.2 				& 5.72  \\
UGC 4115    & $15.5^{+1.8}_{-3.3}$ 		& $0.96^{+0.18}_{-0.34}$	& $0.94^{+0.16}_{-0.34}$	& 0.98	& $4.62^{+0.93}_{-2.8}$ 	& $1.90^{+0.48}_{-0.91}\times 10^{-6}$	& 46.7	& 2.46$\times 10^7$ & 0.084 \\
UGC 11454   & $17.3^{+3.0}_{-5.1}$ 		& $2.56^{+0.47}_{-1.2}$		& $1.29^{+0.25}_{-0.51}$	& 0.50	& $0.72^{+0.21}_{-0.55}$	& $12.6^{+2.0}_{-2.8}$      		  	& 185 	& 14.6 				& 9.09  \\
UGC 11557   & $1.48^{+0.17}_{-0.33}$ 	& $6.7^{+1.0}_{-2.9}$ 		& $6.7^{+1.1}_{-2.9}$ 		& 1.00	& $0.335^{+0.089}_{-0.25}$	& $1.88^{+0.10}_{-0.89}\times 10^{-3}$	& 129 	& $6.88\times 10^4$	& 0.389 \\
UGC 11583   & $10.2^{+1.6}_{-3.2}$ 		& $0.95^{+0.15}_{-0.32}$	& $1.04^{+0.14}_{-0.44}$ 	& 1.09	& $5.6^{+1.5}_{-3.1}$ 		& $2.27^{+0.17}_{-1.3}\times 10^{-6}$ 	& 42.3	& $1.86\times 10^7$ & 0.617 \\
UGC 11616   & $16.6^{+2.3}_{-3.8}$ 		& $2.51^{+0.40}_{-1.0}$ 	& $1.48^{+0.28}_{-0.50}$	& 0.59	& $0.69^{+0.20}_{-0.49}$	& $7.2^{+1.7}_{-2.5}$   			  	& 142 	& 19.8 				& 3.49  \\
UGC 11648   & $29.5^{+3.6}_{-10.0}$ 	& $1.01\pm 0.19$			& $1.72\pm 0.31$			& 1.70	& $2.73^{+0.37}_{-0.88}$	& $54^{+10}_{-20}$        			  	& 285 	& 5.28 				& 1.08  \\
UGC 11748   & $86.0^{+12.0}_{-25.0}$ 	& $2.63^{+0.52}_{-1.1}$		& $1.58^{+0.27}_{-0.38}$ 	& 0.60	& $0.283^{+0.085}_{-0.20}$	& $2.24\pm 0.32$   					  	& 170 	& 76.0  			& 11.6  \\
UGC 11819   & $6.35^{+0.40}_{-0.44}$ 	& $4.46^{+0.19}_{-0.28}$	& $6.97^{+0.84}_{-1.8}$ 	& 1.56	& $0.278^{+0.025}_{-0.019}$	& $1.72^{+0.25}_{-0.71}\times 10^{-4}$	& 175 	& $1.01\times 10^6$ & 0.408 \\
\hline
\end{tabular}
\end{table*}

  From the best fit for each galaxy reported in Table~\ref{tab:results2}, we notice
that the ratio between the transition and core radii is in the interval
$0.46 \leq r_\epsilon/r_c \leq 1.70$. This means that, in many cases, the soliton
contribution is overlapped with the NFW halo in order to fit the observations. In
all cases, the values of $r_\epsilon$ do not correspond to $r_\epsilon>3r_c$, as
found in the cosmological simulations \citep{Schive:2014dra}. Thus, the soliton
contribution is not as prominent as in the cosmological simulations, according to
the best fits in the LSB galaxies.

For the 18 galaxies, the values for the boson mass are in the range
$0.278 \leq m_\psi/10^{-23} \mathrm{eV} \leq 27.0$, and core radius within $0.326
\leq r_c / \mathrm{kpc} \leq 8.96$. The wide range in core radii is consistent
with the variety of galaxy sizes. It most be noticed that the half of the
galaxies in the sample have resulting boson masses $m_\psi < 10^{-23} \mathrm{eV}$,
in disagreement with the strongest cosmological constraints
\citep{Bozek:2014,Sarkar:2016}. For those rotation curves, we tried to restrict
the boson mass to be at least $m_\psi=10^{-23} \mathrm{eV}$, but it was not
possible to fit the data. In these cases, the FDM model is in tension with the
LSB observations. And as we show in Subsection~\ref{subsubsec:sparc-1}, the low
masses $m_\psi < 10^{-23} \mathrm{eV}$ persist in the analysis of the NGC
galaxies with photometric data.

\begin{figure*}
 \begin{center}
 $\begin{array}{cccc}
\includegraphics[width=0.255\textwidth]{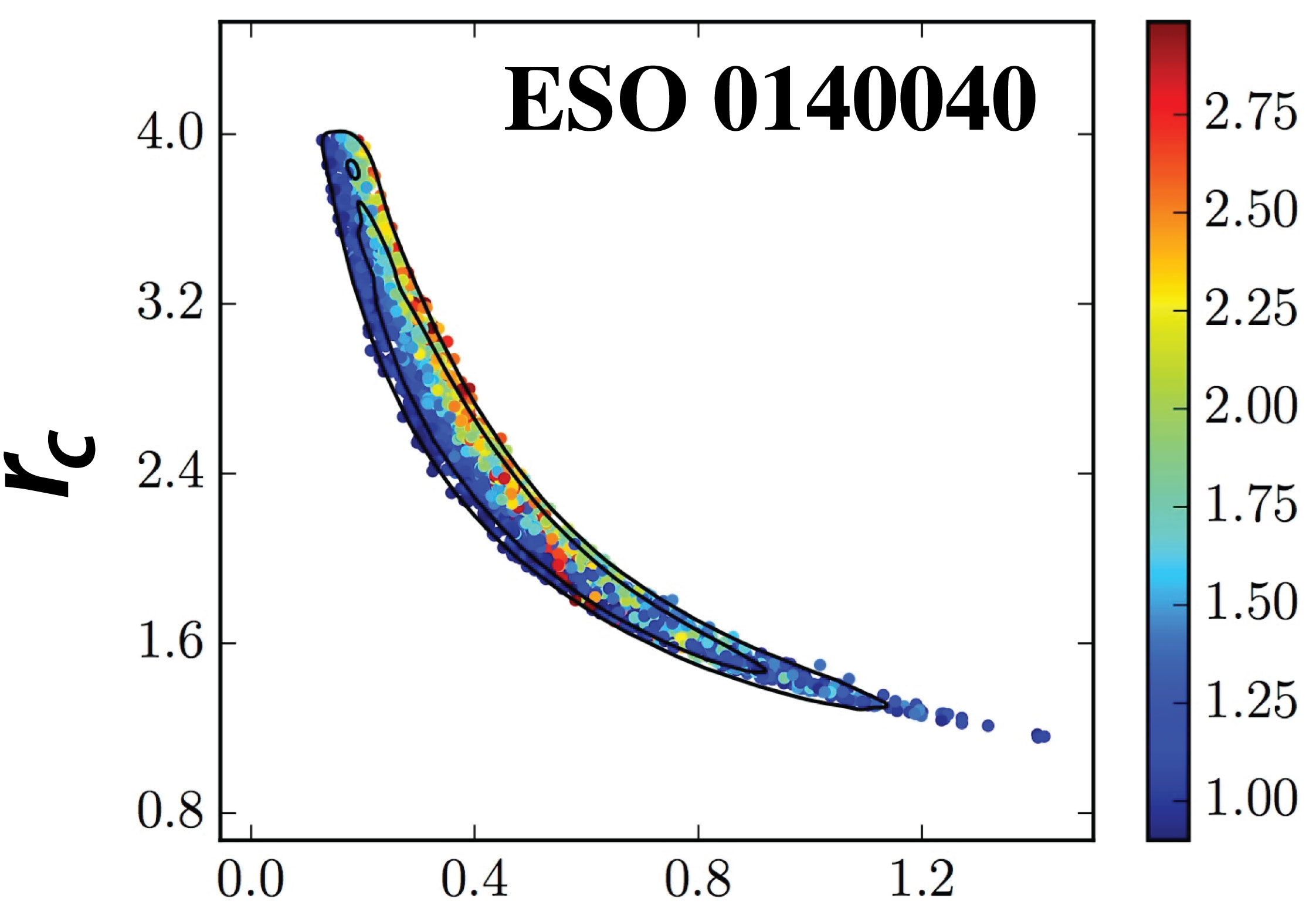} &
\includegraphics[width=0.225\textwidth]{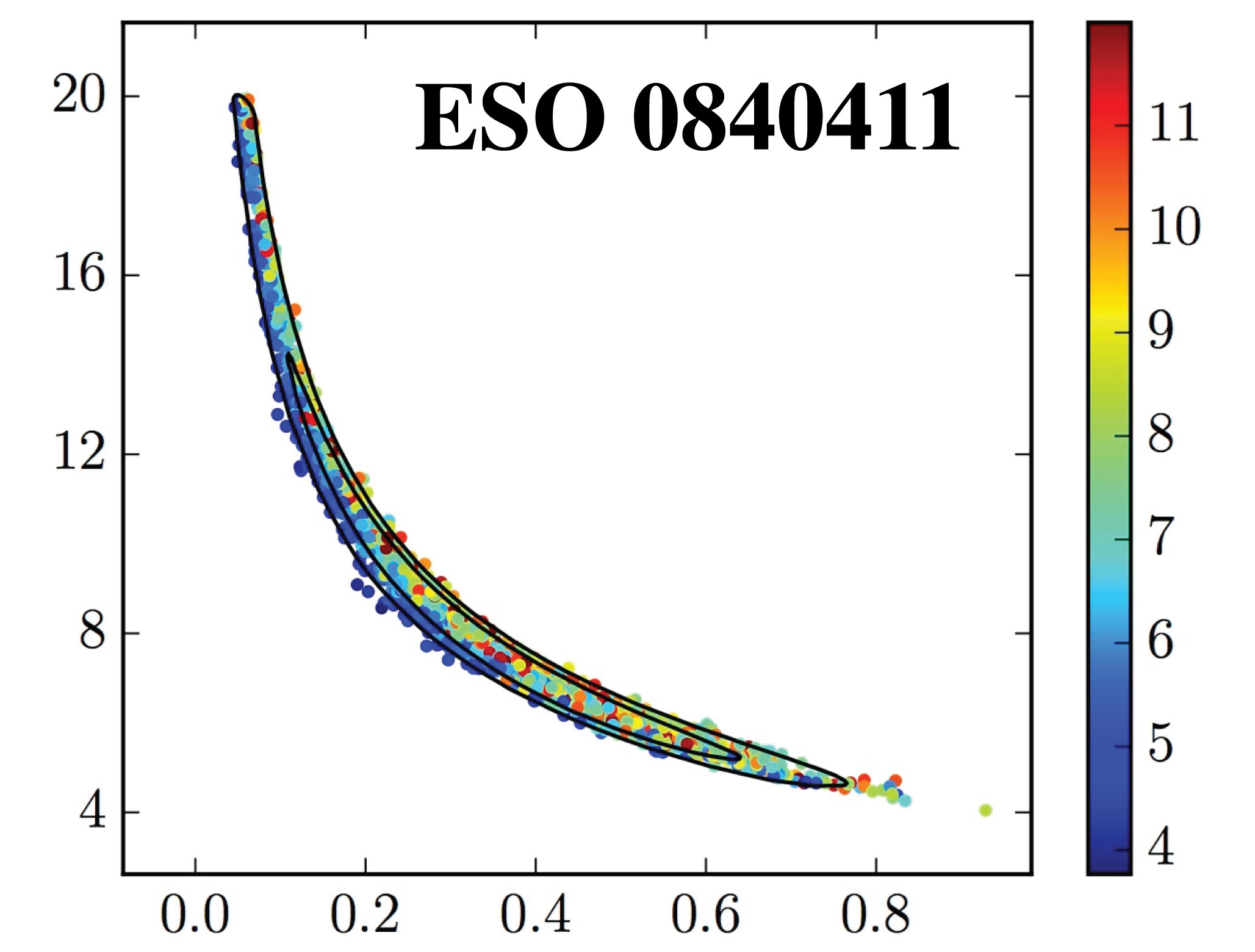} &
\includegraphics[width=0.235\textwidth]{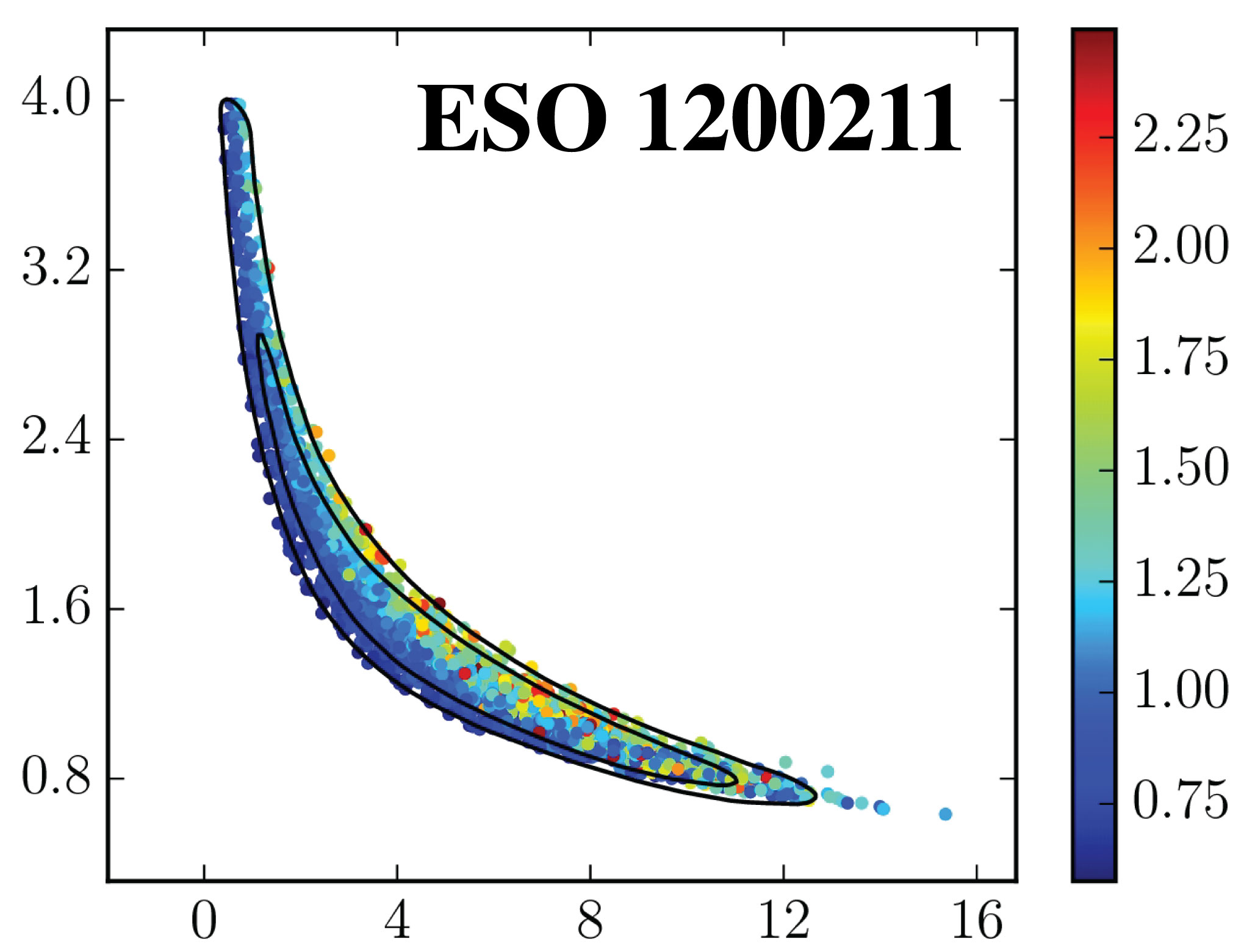} &
\includegraphics[width=0.245\textwidth]{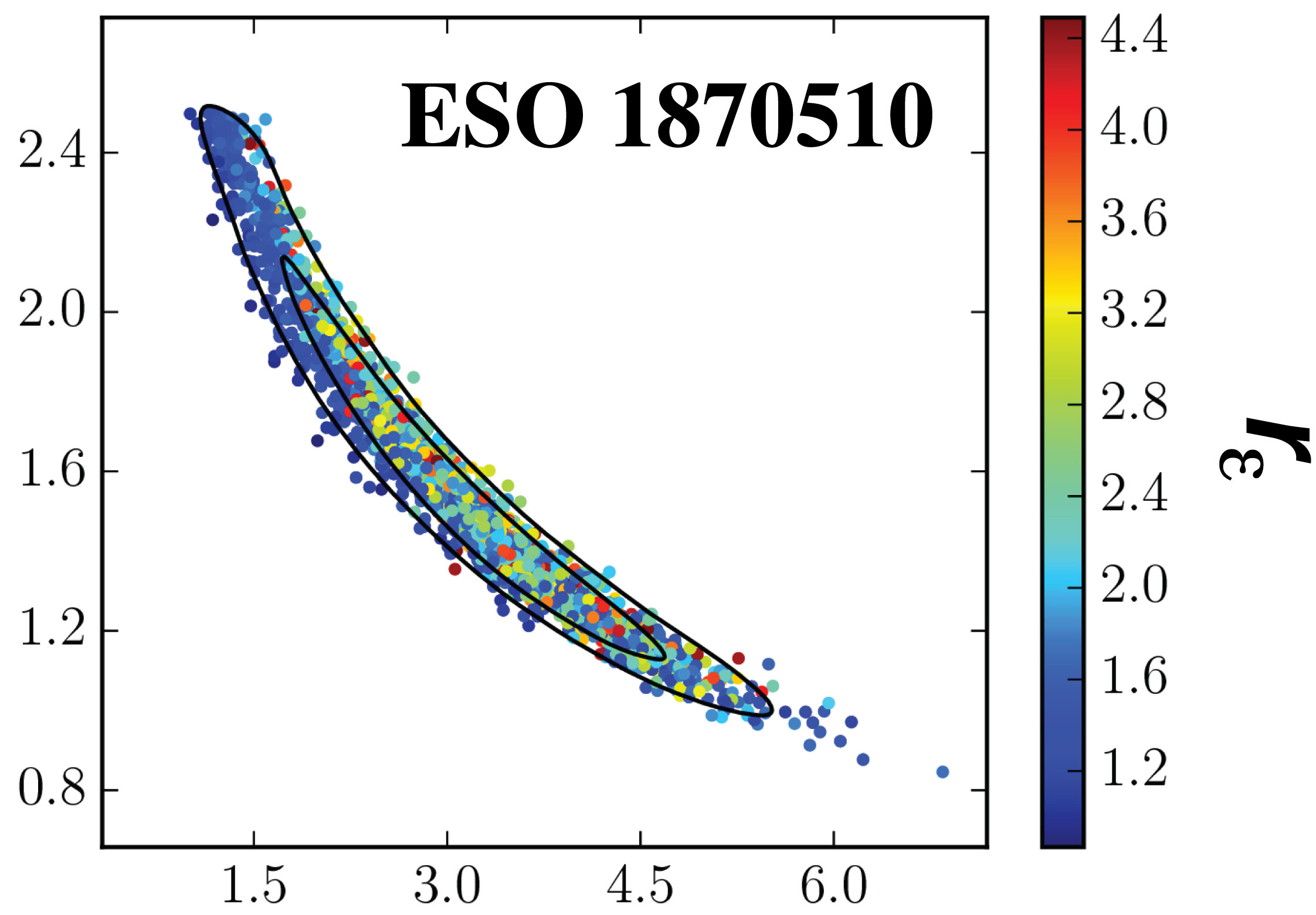} \\
\includegraphics[width=0.25\textwidth]{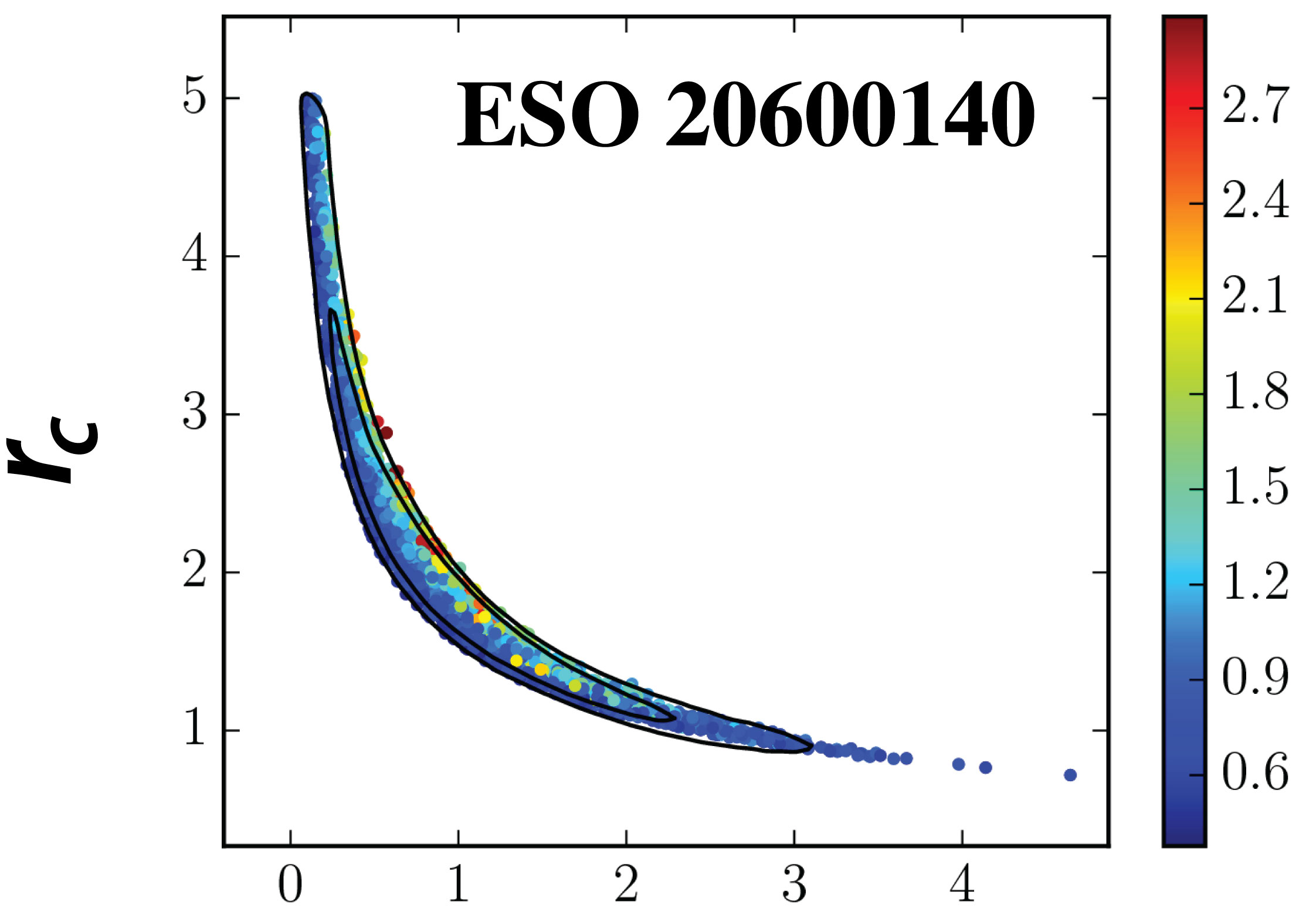} &
\includegraphics[width=0.23\textwidth]{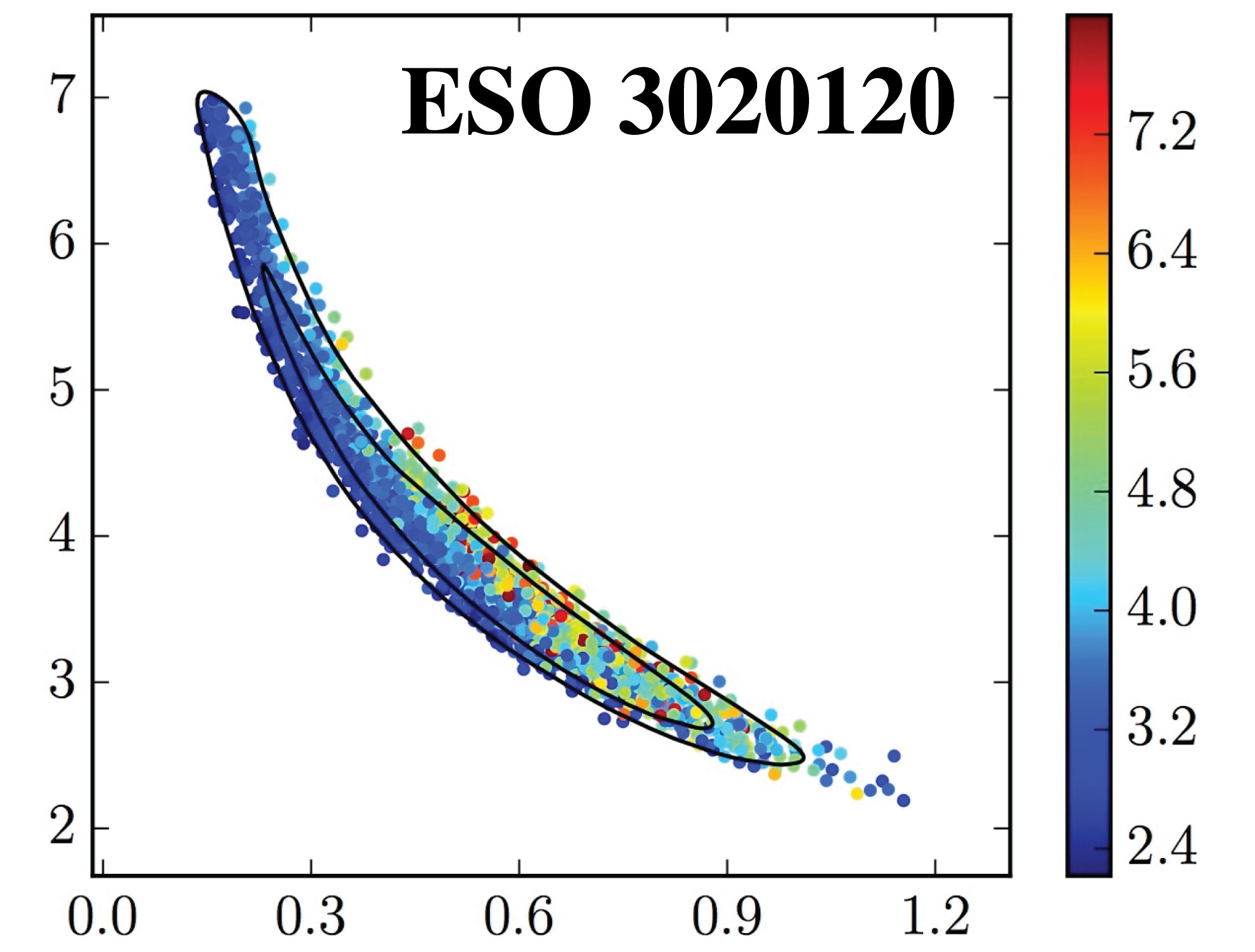} &
\includegraphics[width=0.23\textwidth]{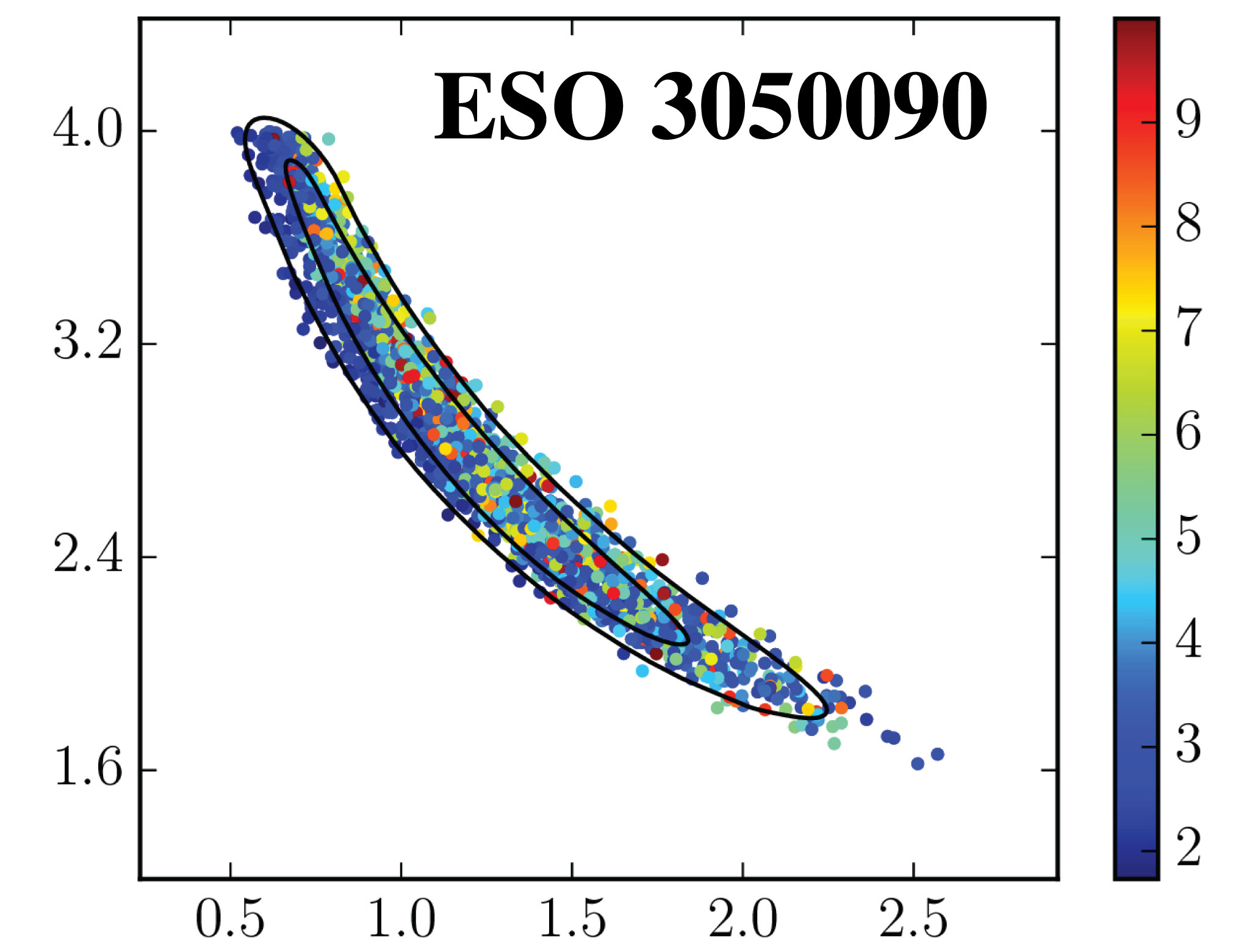} &
\includegraphics[width=0.25\textwidth]{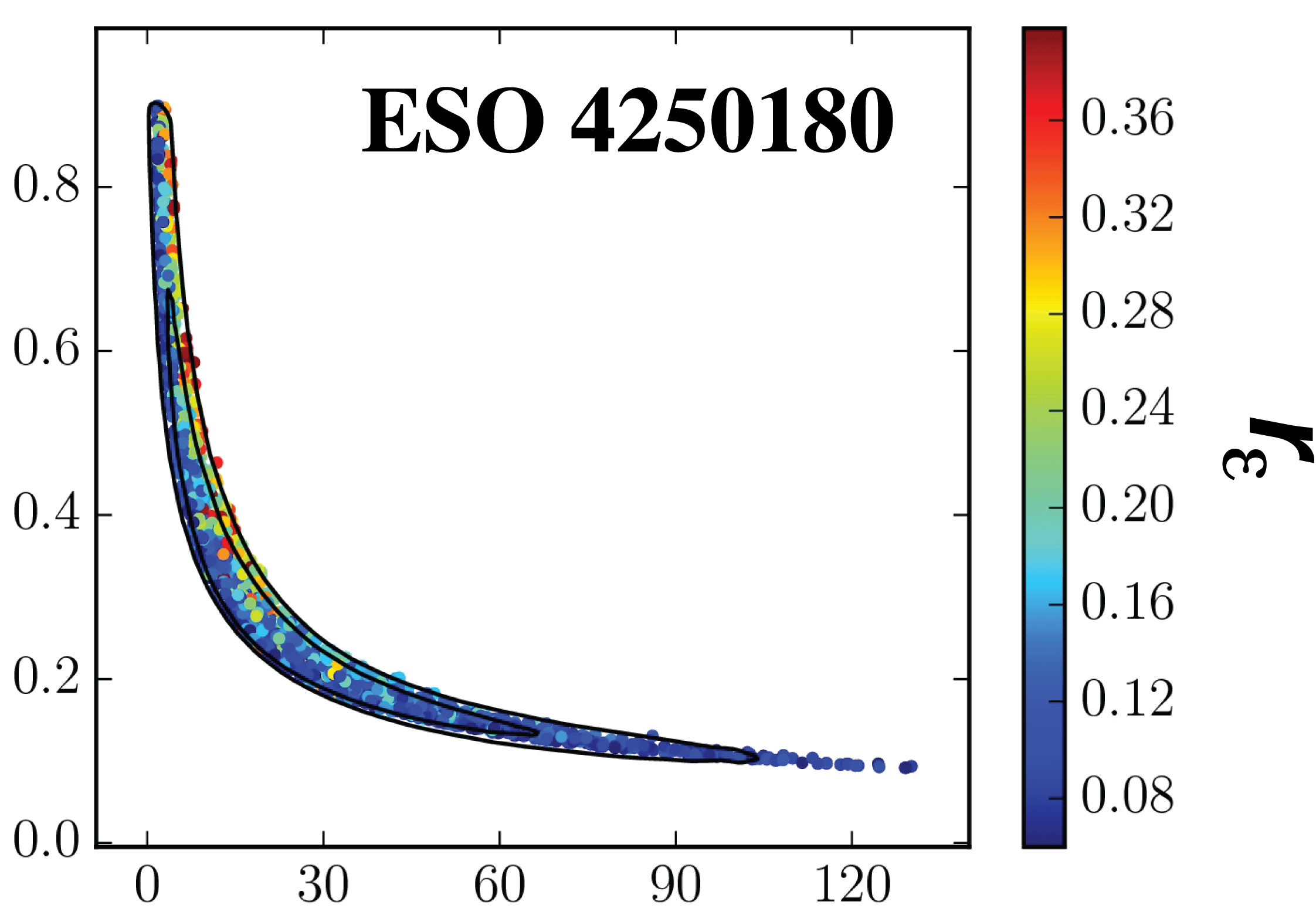} \\
\includegraphics[width=0.25\textwidth]{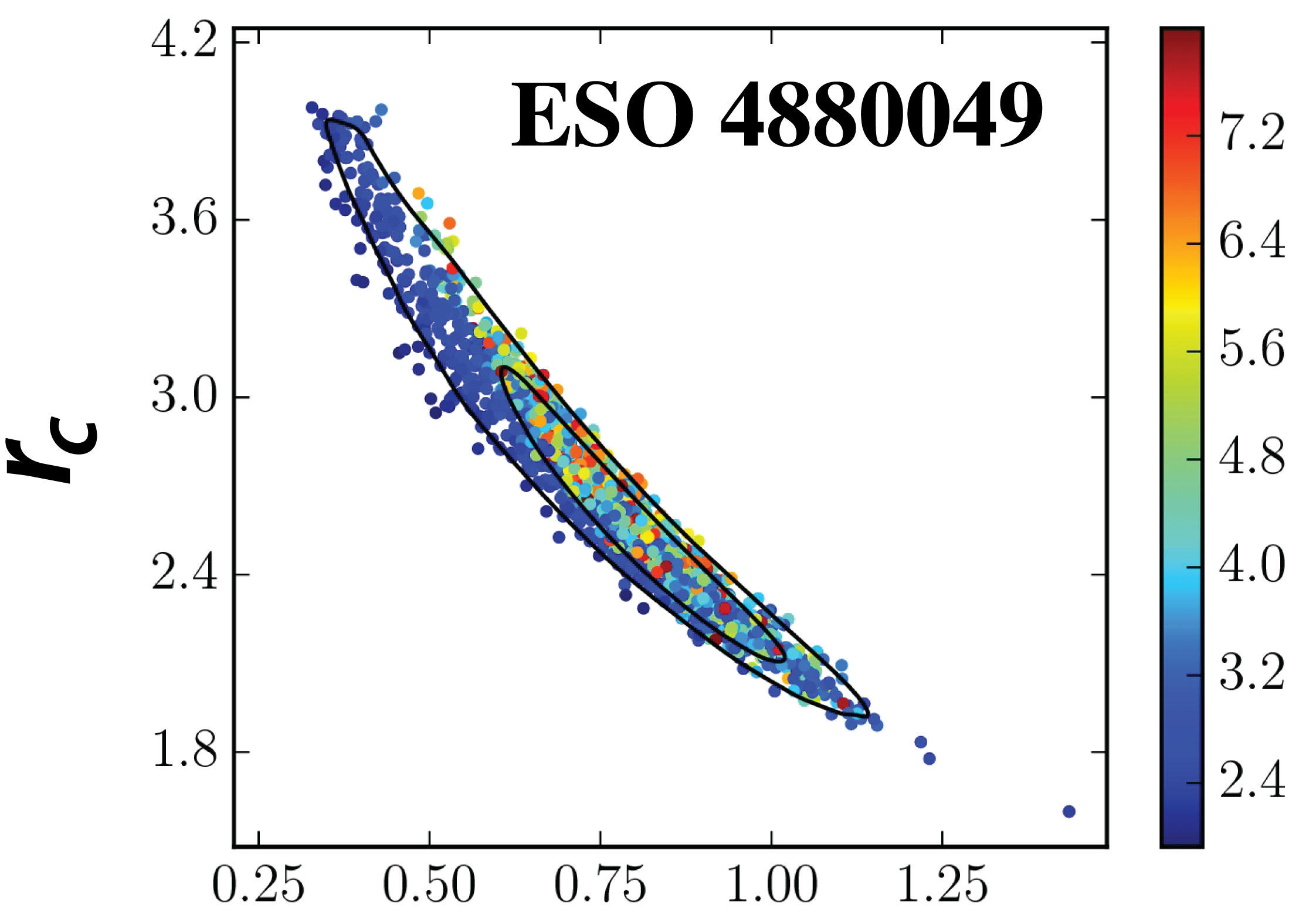} &
\includegraphics[width=0.23\textwidth]{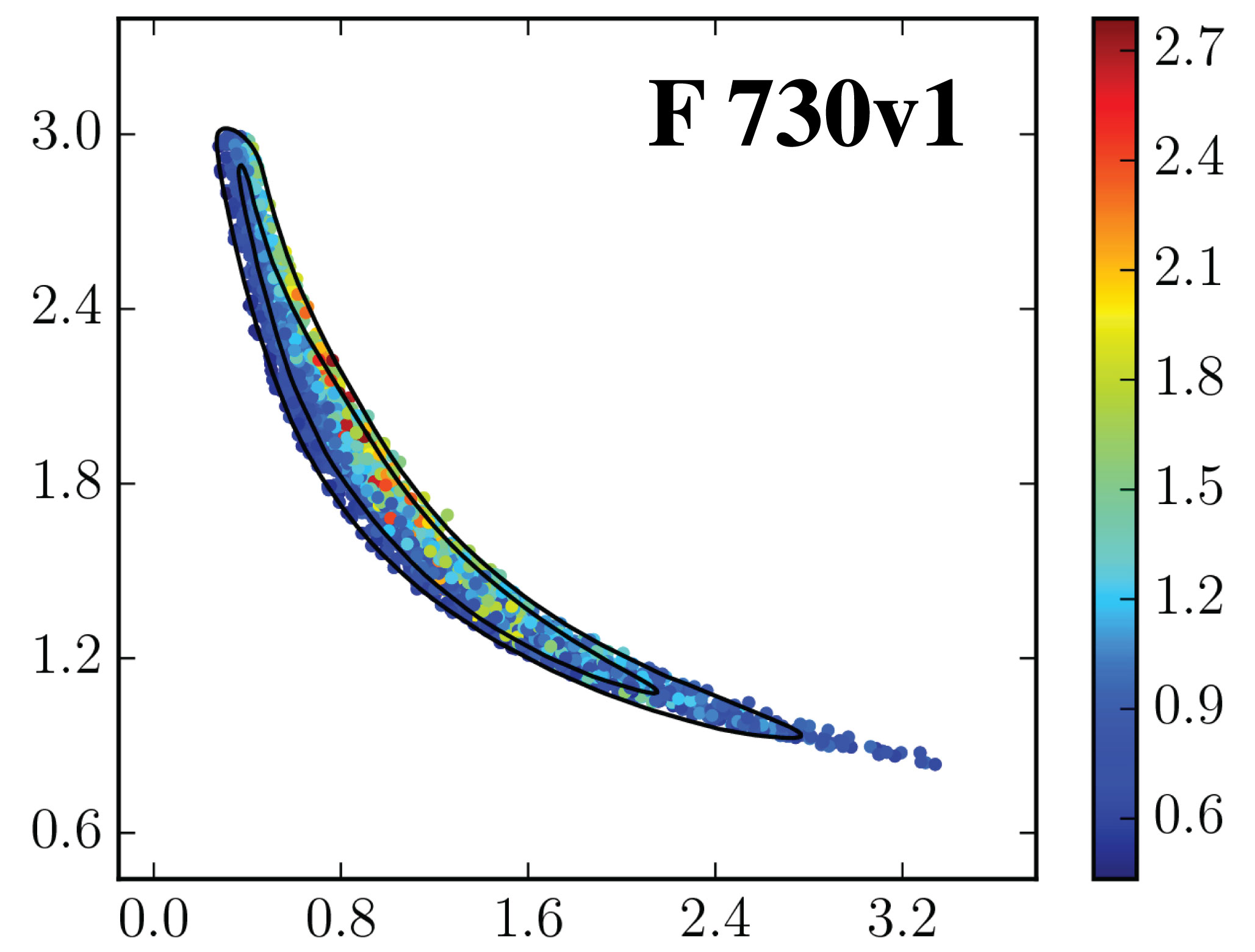} &
\includegraphics[width=0.24\textwidth]{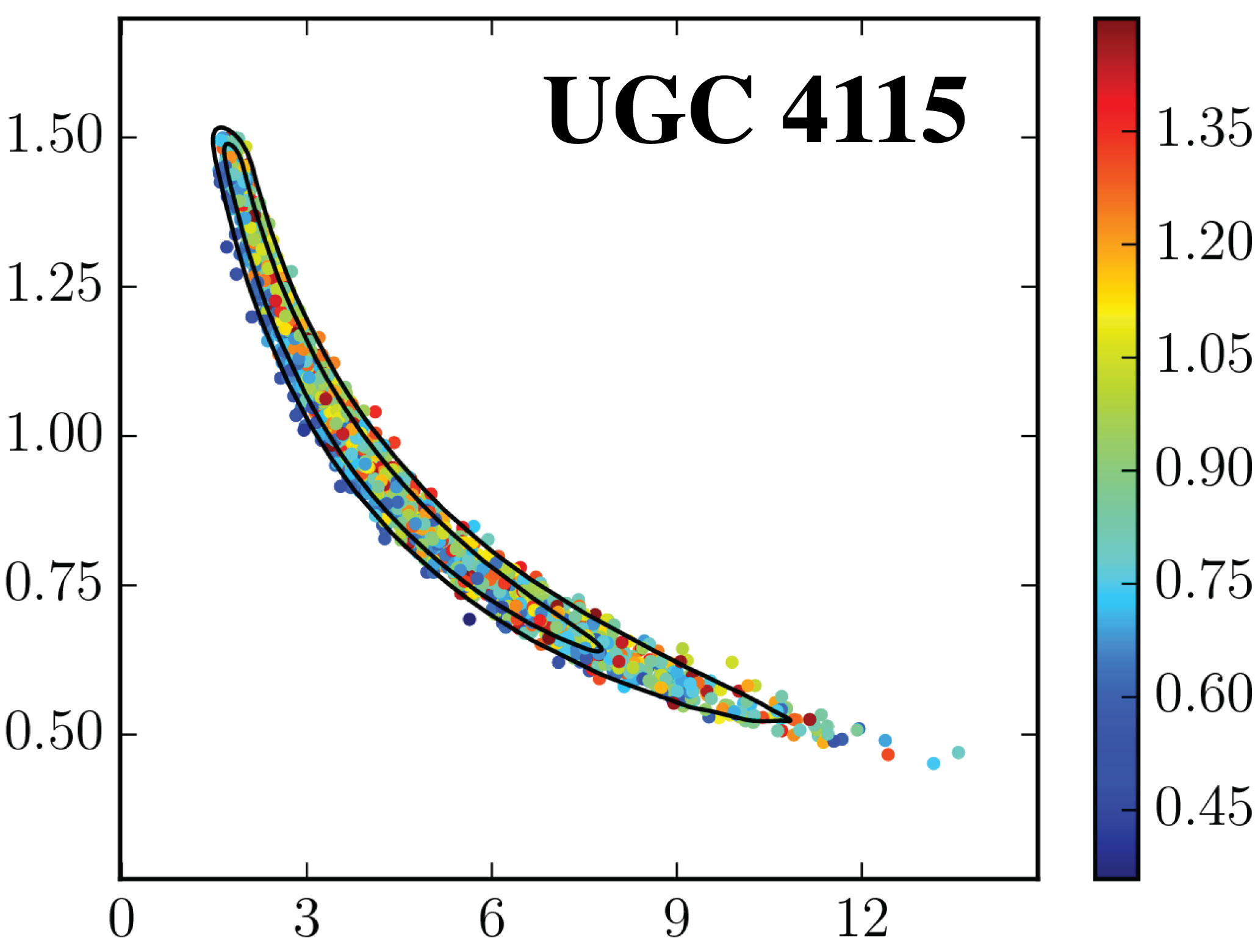}  &
\includegraphics[width=0.25\textwidth]{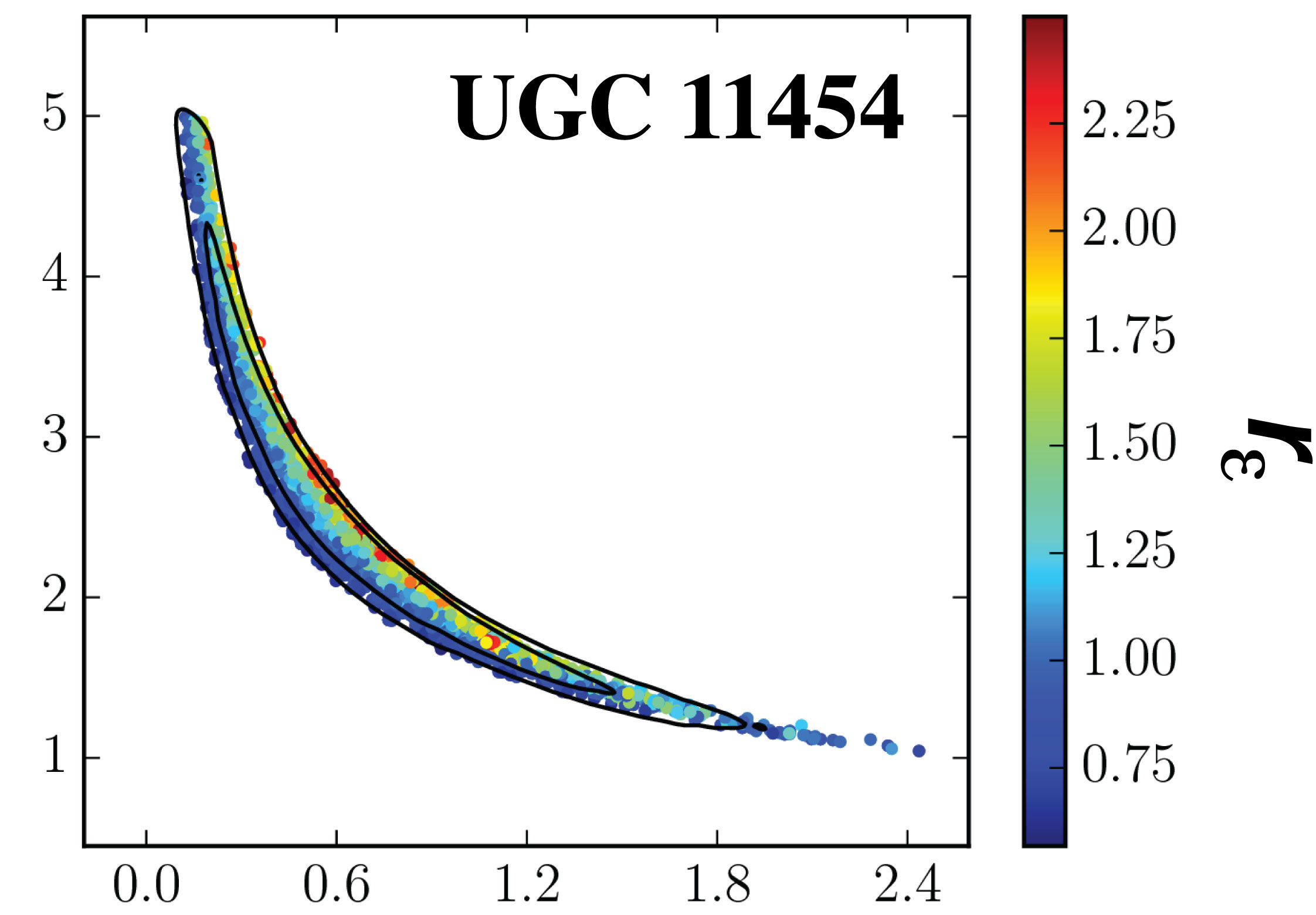} \\
\includegraphics[width=0.245\textwidth]{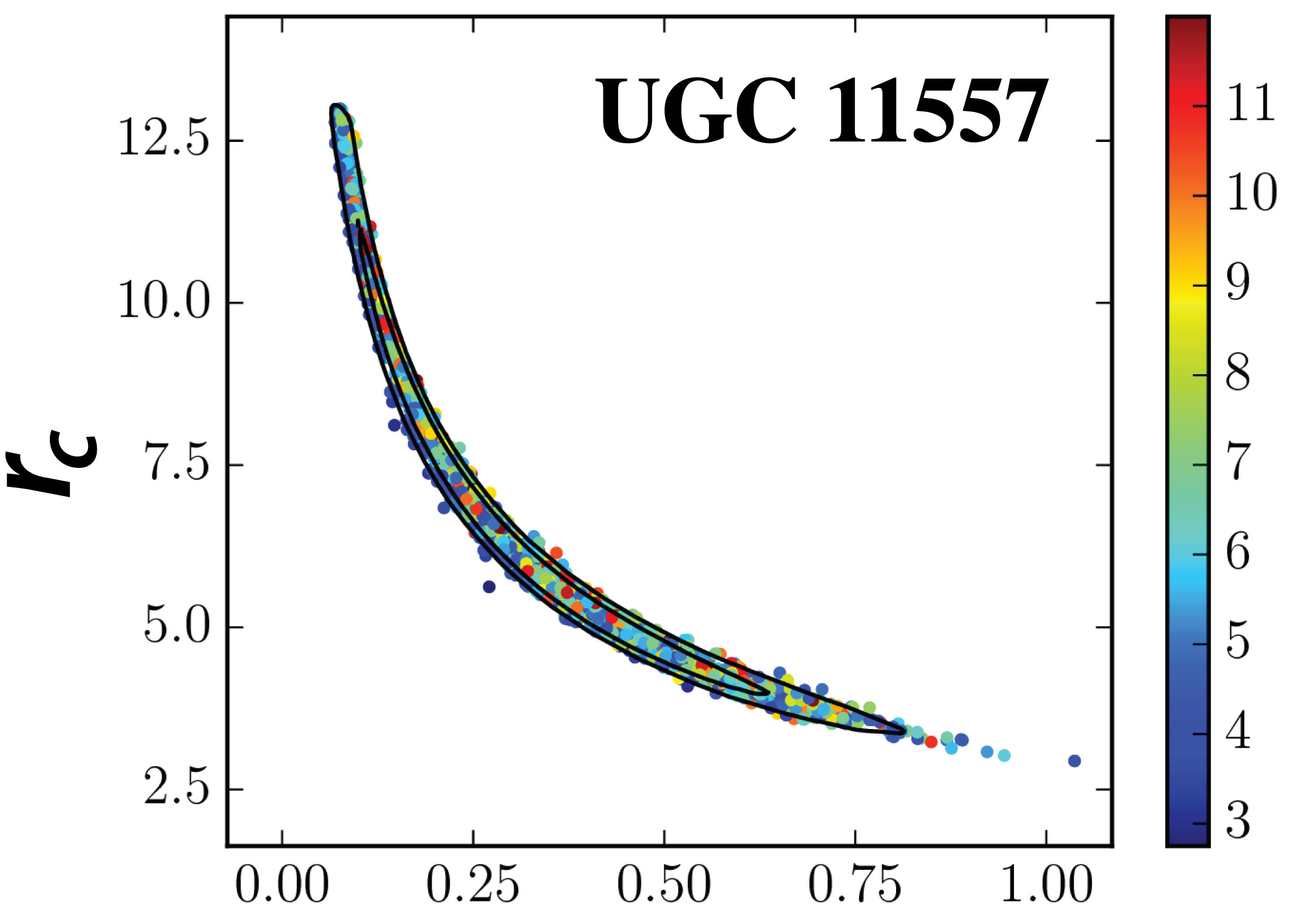} &
\includegraphics[width=0.235\textwidth]{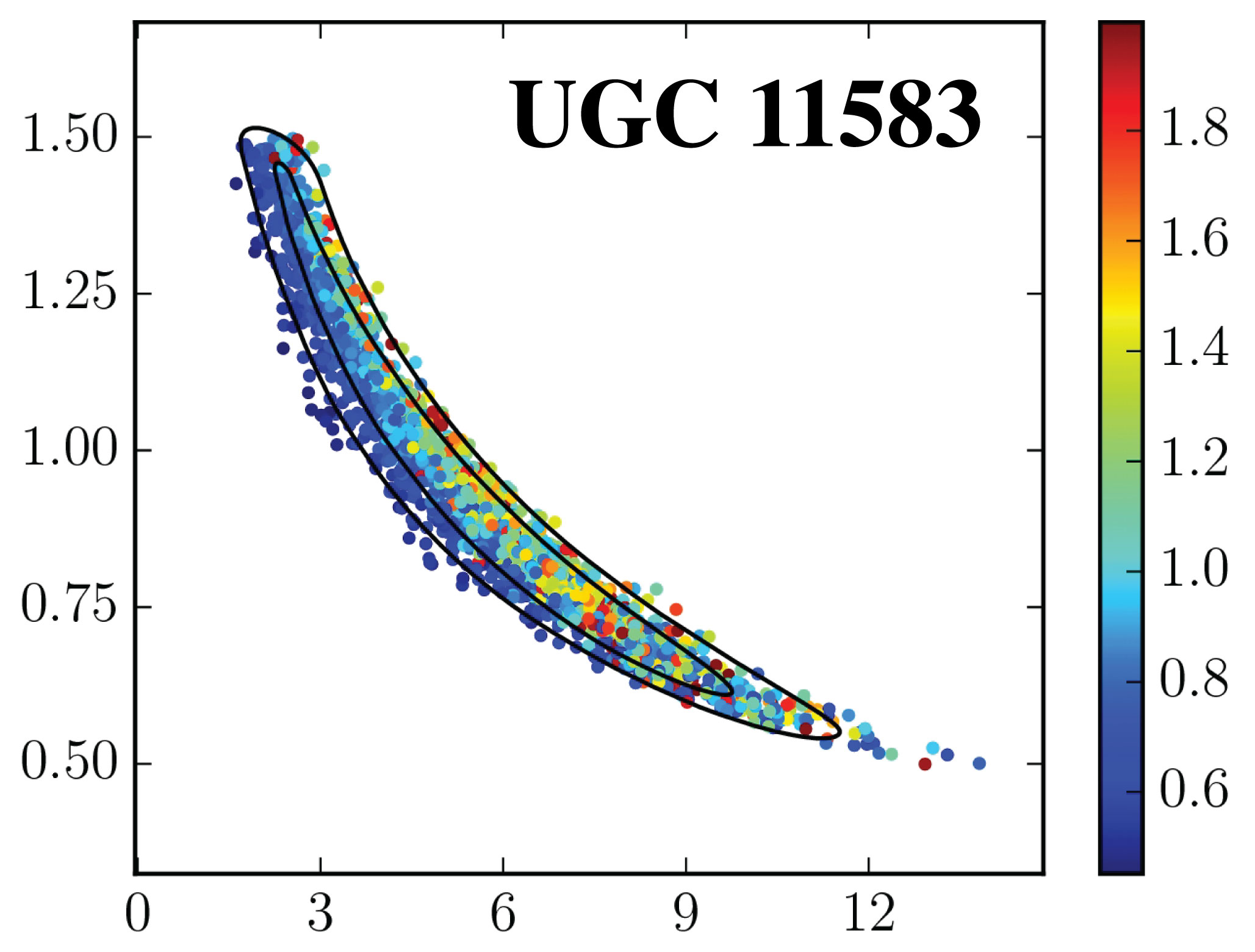} &
\includegraphics[width=0.23\textwidth]{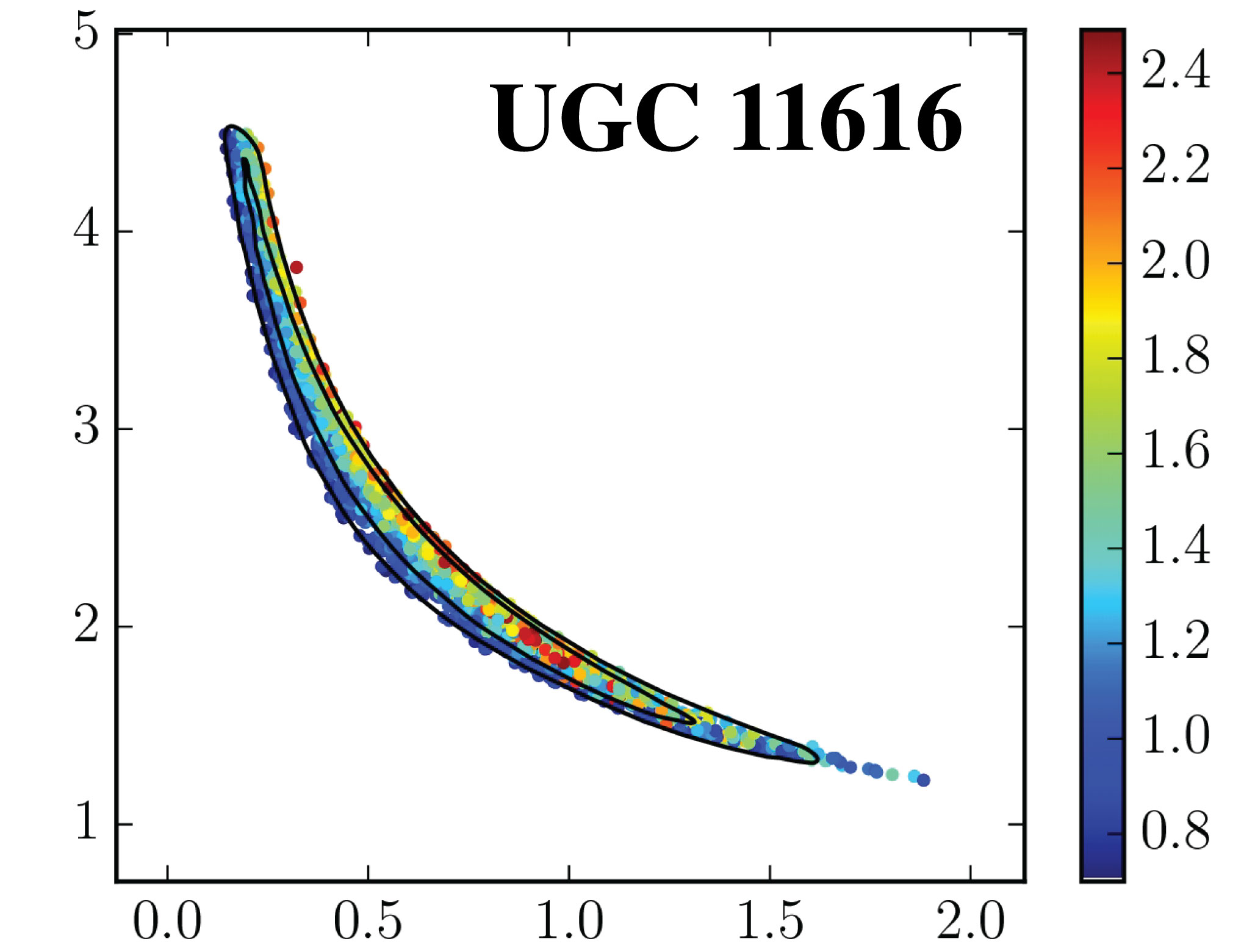} &
\includegraphics[width=0.25\textwidth]{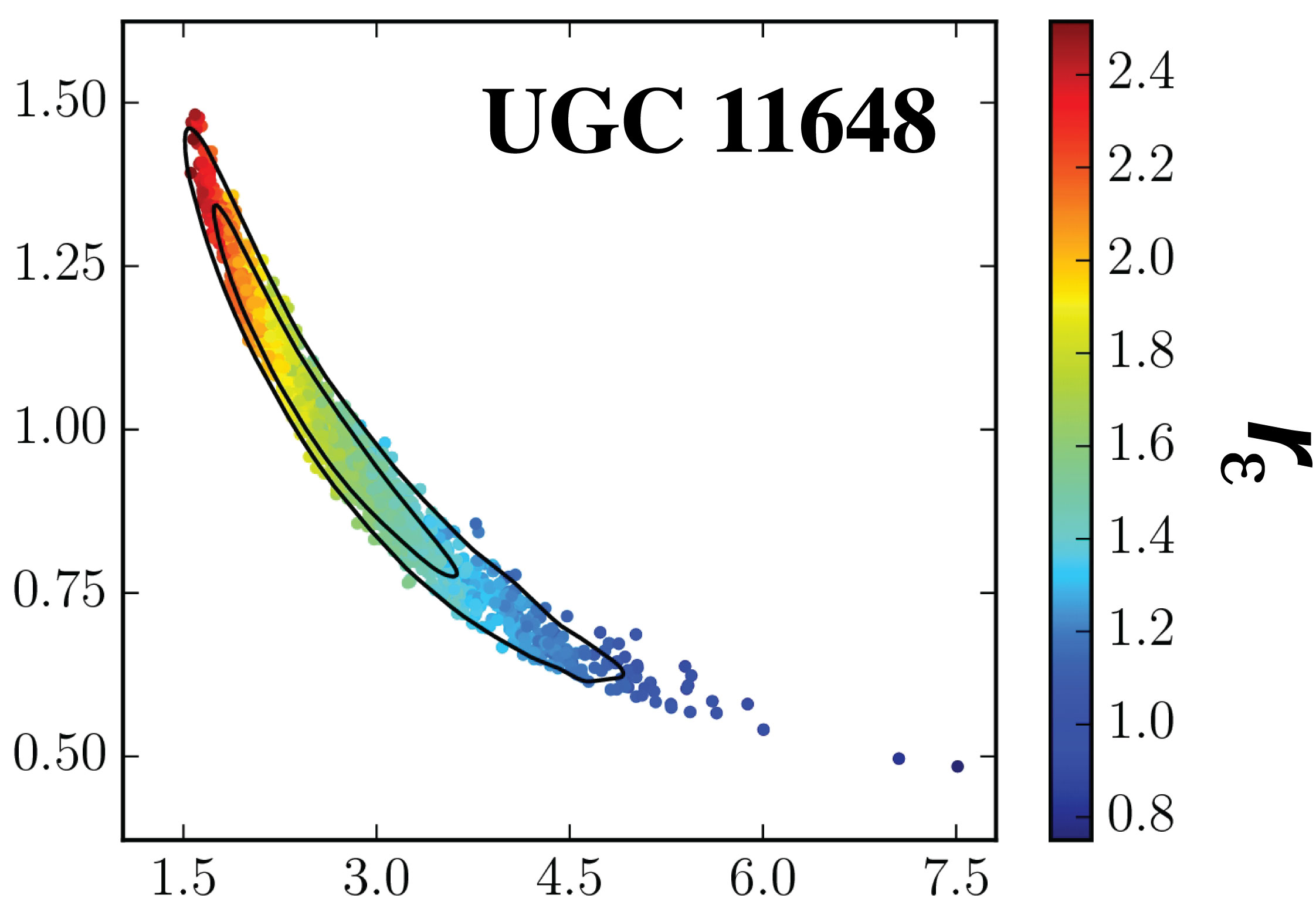} \\
\includegraphics[width=0.24\textwidth]{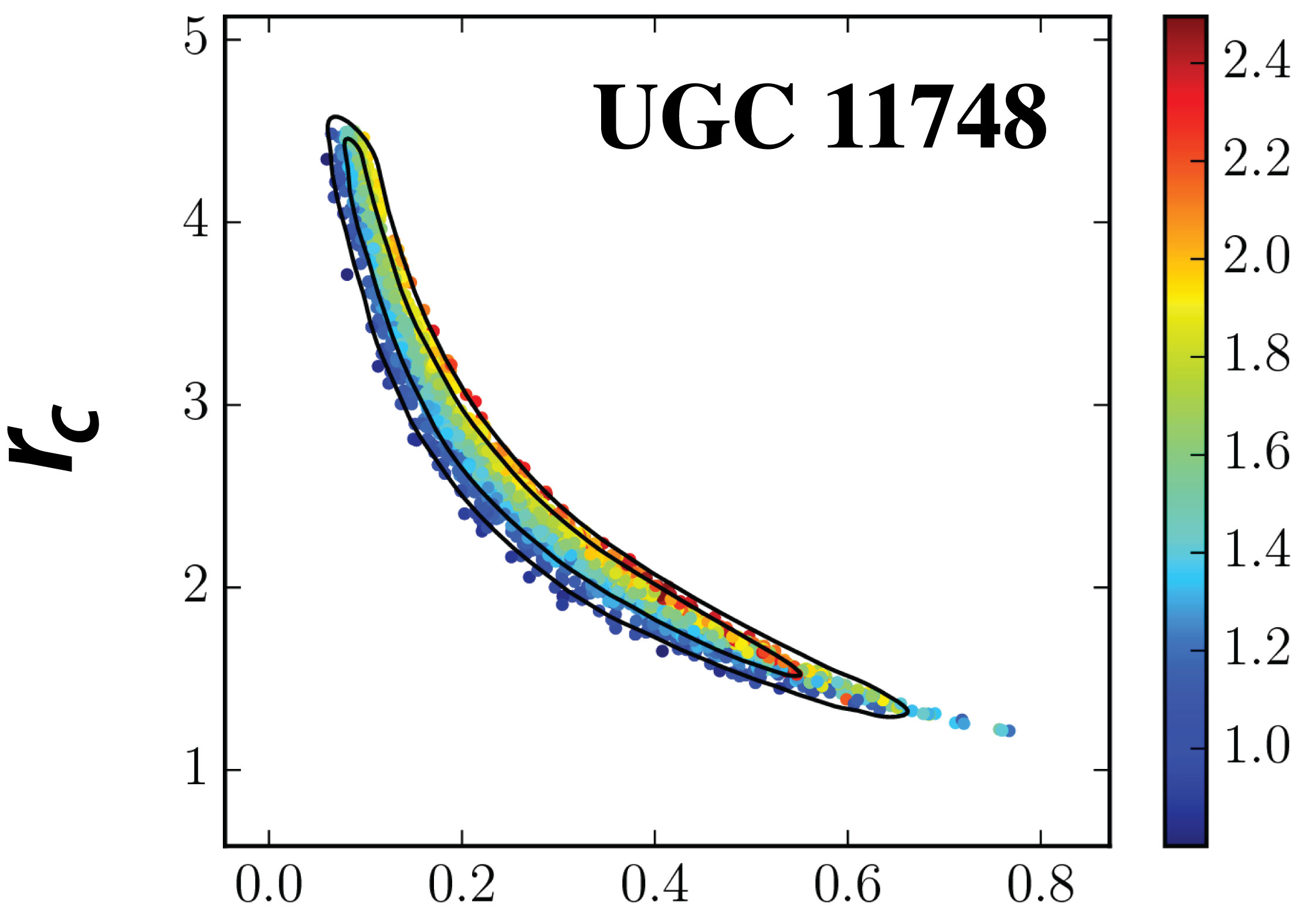} &
\includegraphics[width=0.235\textwidth]{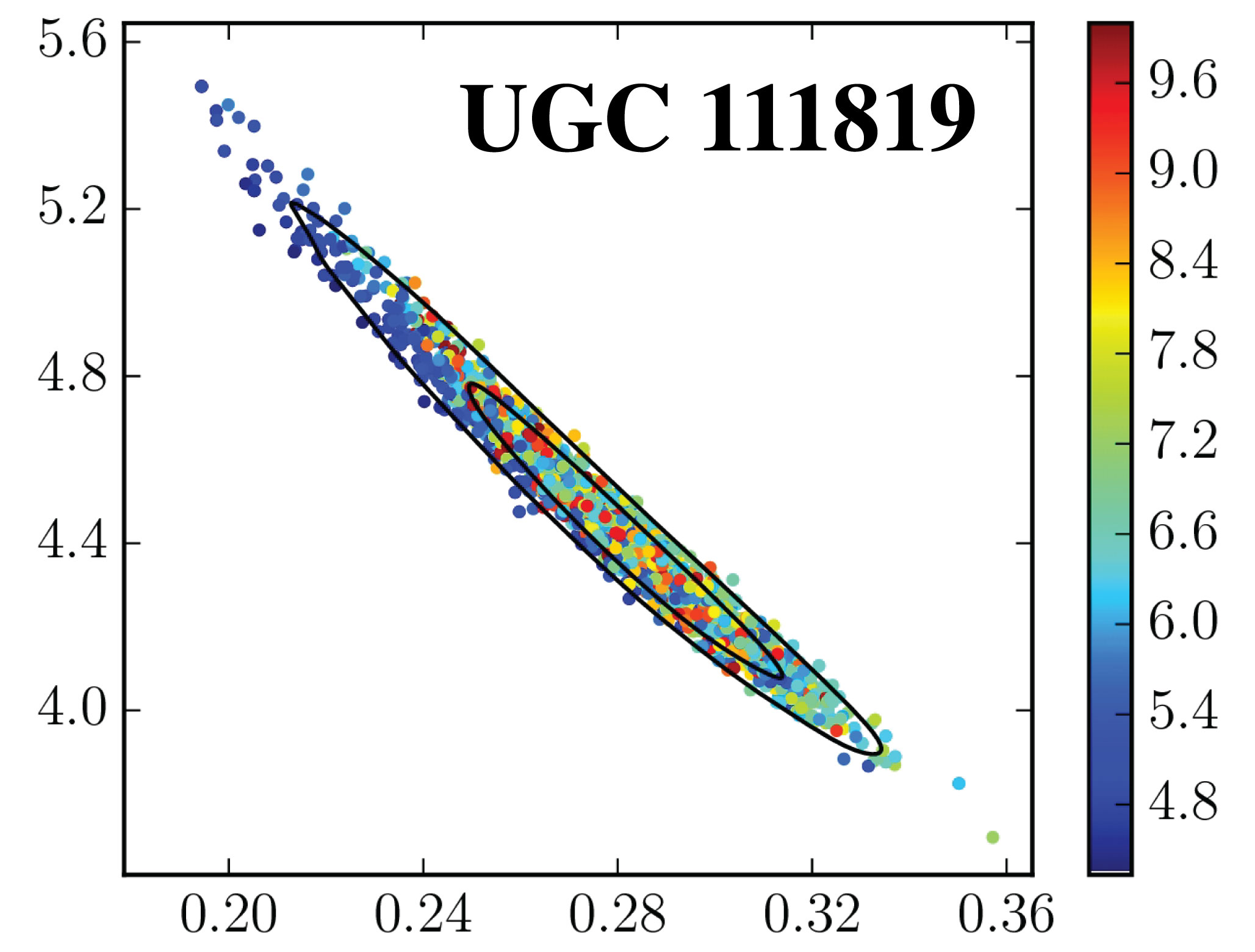} &
\includegraphics[width=0.24\textwidth]{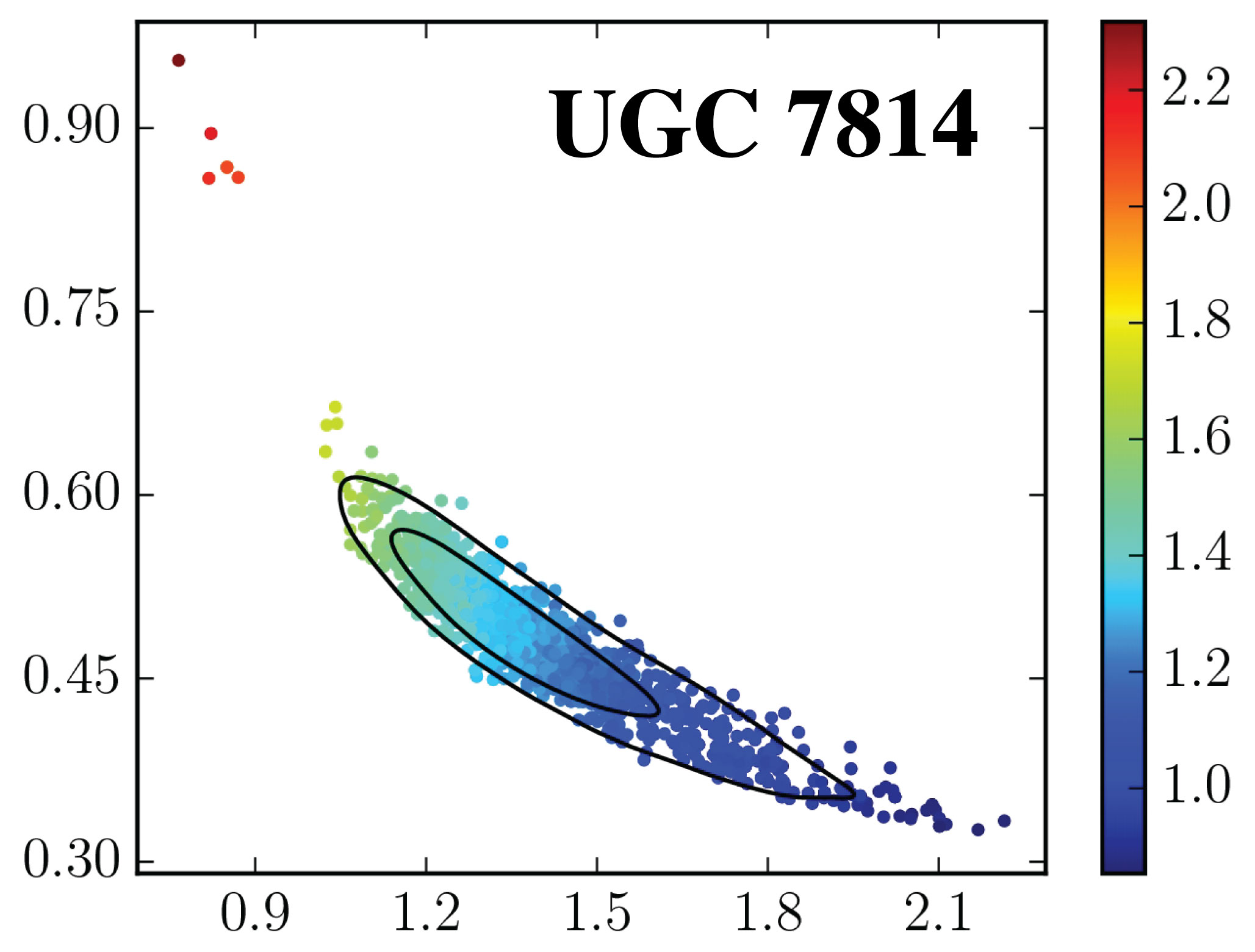} &
\includegraphics[width=0.245\textwidth]{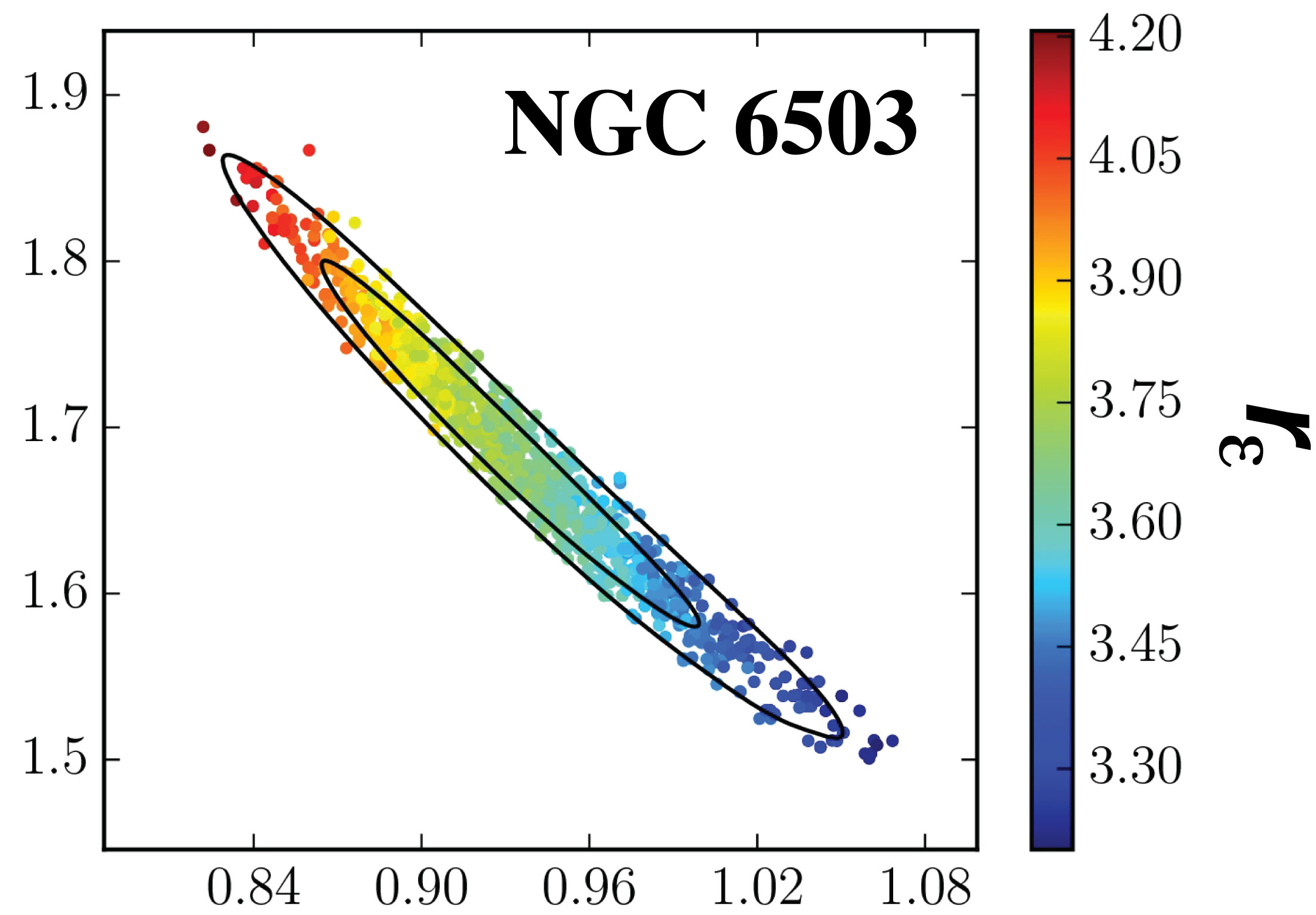} \\
\includegraphics[width=0.255\textwidth]{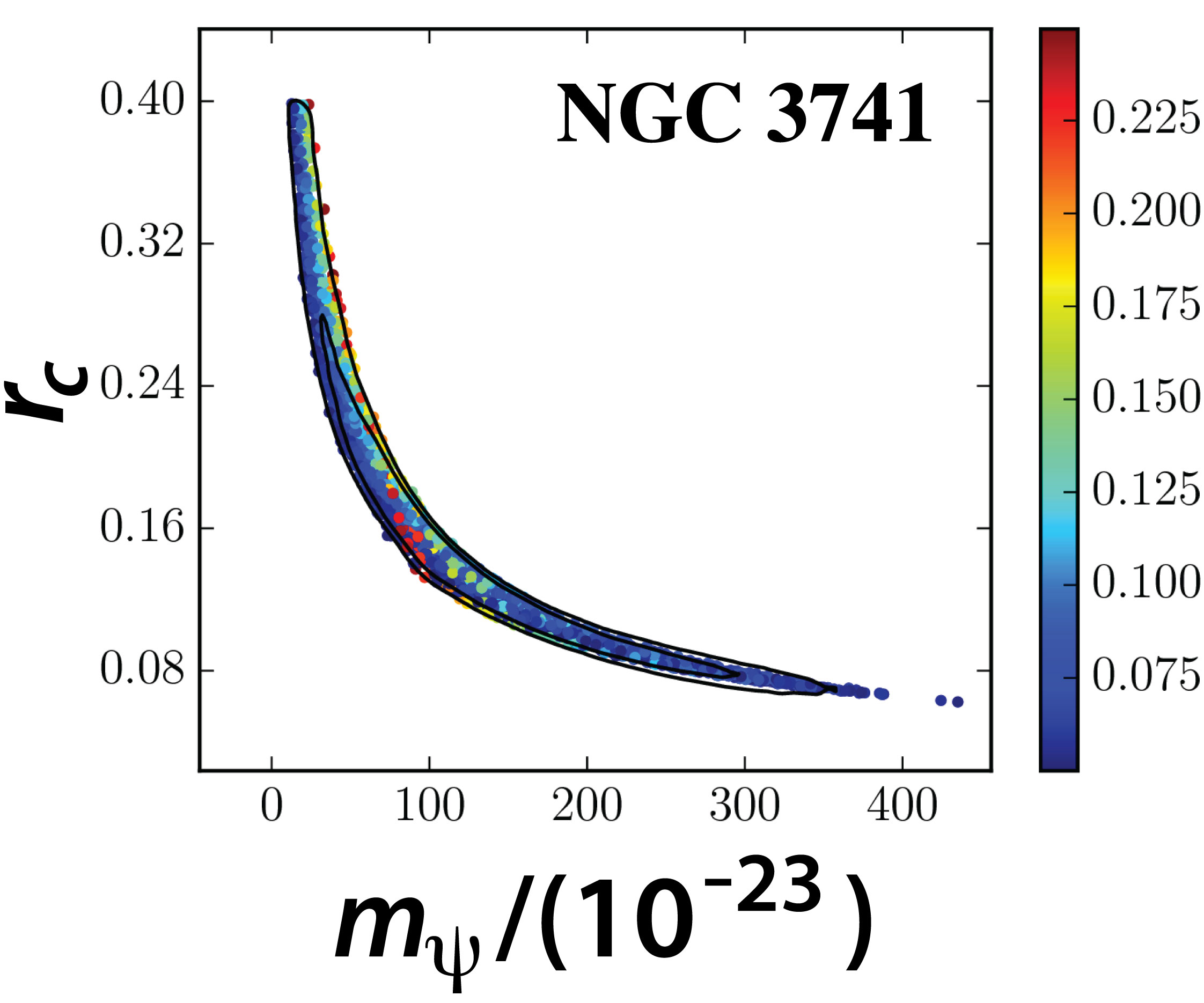} &
\includegraphics[width=0.235\textwidth]{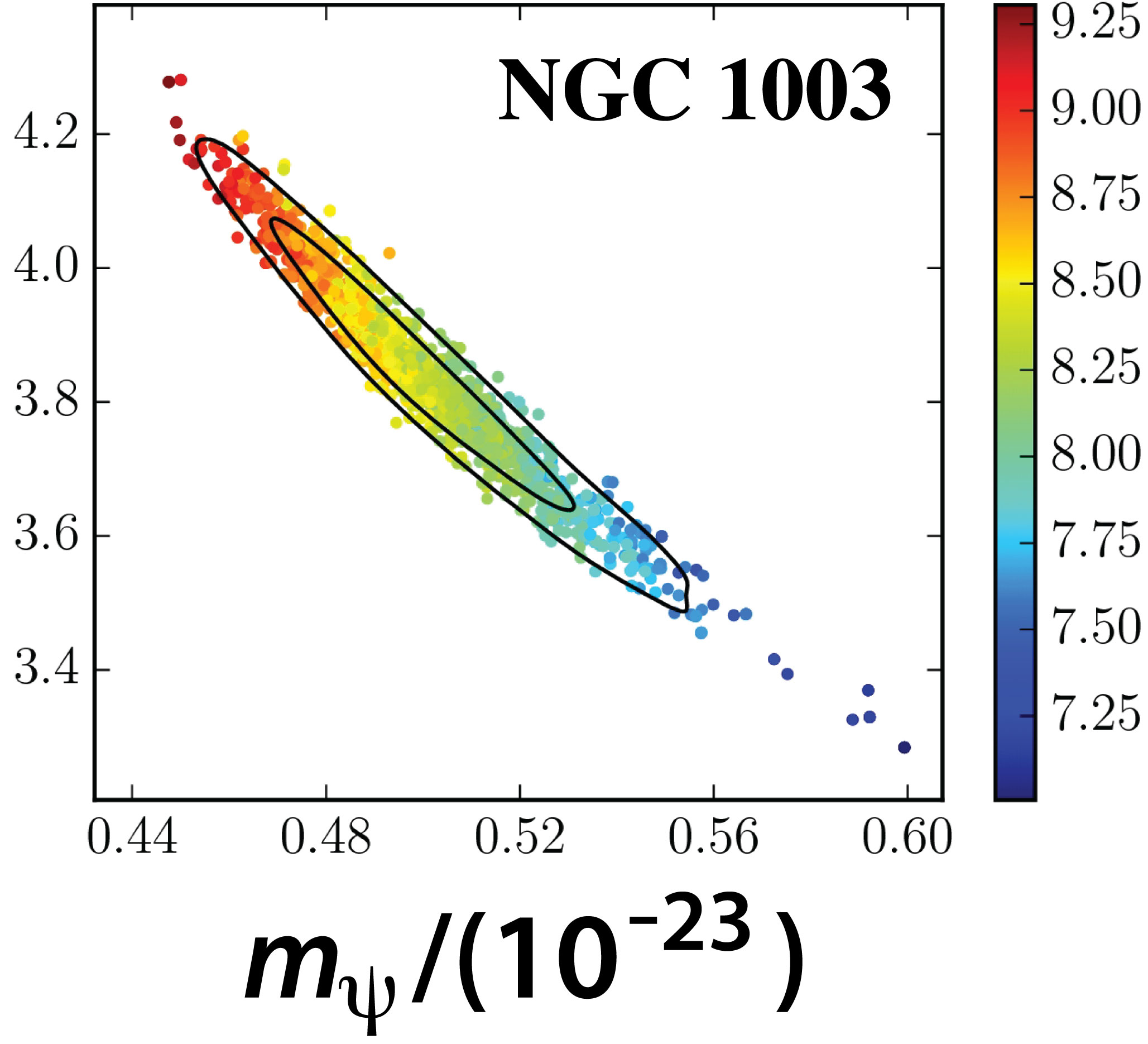} &
\includegraphics[width=0.225\textwidth]{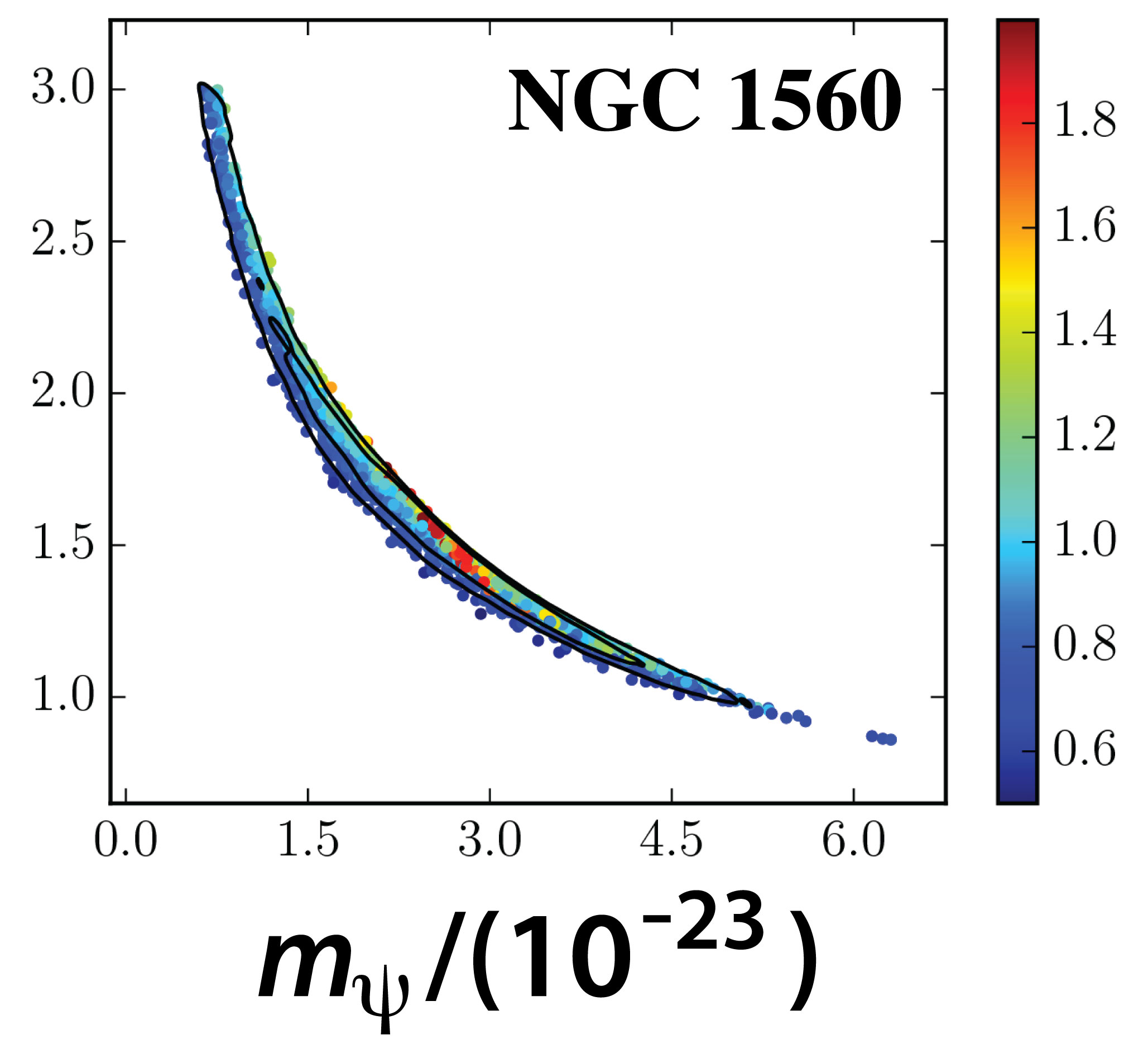} &
\includegraphics[width=0.245\textwidth]{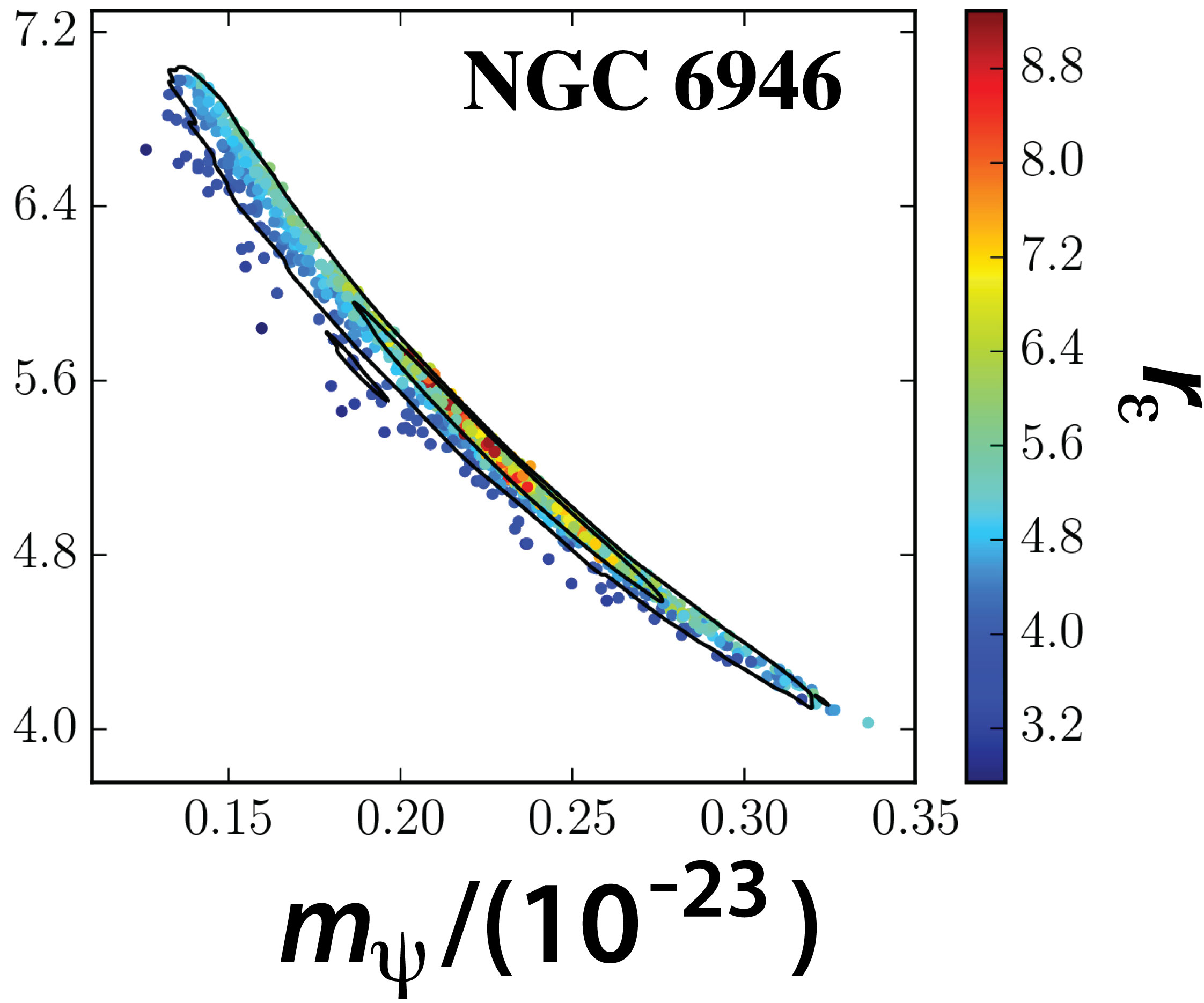}
 \end{array}$
 \caption{Posterior distributions for the parameters $m_\psi(10^{-23}
\mathrm{eV})$ and $r_c(\mathrm{kpc})$, with the transition radius $r_\epsilon
(\mathrm{kpc})$ as the scatter plot, for the high-resolution LSB and NGC
galaxies in the FDM model. The contours are the $1\sigma$ and
$2\sigma$ confidence regions. The plots were obtained with the GetDist 0.2.6
Package.}
 \label{fig:ellipses-Schive}
 \end{center}
\end{figure*}

	Fig.~\ref{fig:ellipses-Schive} shows the $1\sigma$ and $2\sigma$
contours for the parameters $m_\psi$ and $r_c$ from the MCMC analysis, and the
transition radius $r_\epsilon$ is the scatter plot. From this figure, it can be
noticed the degeneracy in the boson mass $m_\psi$ between the galaxies,
appearing to be inconsistent with each other. However, the boson mass must be a
constant of the model, changing only its core radius to fit the observations. In
the next Subsection we report the results of the combined analysis for the 18
LSB galaxies, keeping the boson mass as constant.

\subsubsection{Combined analysis for the LSB galaxies}
\label{subsubsec:comb-lsb}

Taking into account that the mass of the ultra-light SF particle in the SFDM
model must be a constant, in this Subsection we report the results of a combined
analysis fitting all the LSB galaxies with a single mass $m_\psi$, leaving the
other parameters free for each galaxy. This analysis was performed minimizing
the $\chi^2$ errors for the whole data of the 18 LSB galaxies to obtain the
best-fit mass from the observations. Notice that we only include the LSB
galaxies, as a sample spanning a wide range in sizes and morphological types, to
lead with DM-only fits and not with the baryons. The results are shown in
Table~\ref{tab:LSB-comb}, where we report the free parameters $\rho_c$, $r_c$,
$r_\epsilon$ and $r_s$, the resulting constant boson mass $m_\psi$ from the
analysis, the ratio $r_\epsilon/r_c$, and the NFW density $\rho_s$, virial
radius $r_{200}$ and concentration parameter $c$.

\begin{table*}
\caption{Soliton+NFW density profile -- Combined analysis in high-resolution LSB
galaxies. In this Table we show the resulting fitting parameters $\rho_c$, $r_c$,
$r_\epsilon$ and $r_s$ for the FDM model with the soliton+NFW density
profile~\eqref{eq:rho-schive}. The resulting boson mass is $m_\psi=0.554\times
10^{-23}\mathrm{eV}$. The combined analysis was performed minimizing the $\chi^2$
errors for the whole sample, and we obtained $\chi_\mathrm{red}^2=1.208$. We
also show the resulting NFW density parameter $\rho_s$, the ratio $r_\epsilon/r_c$,
the $r_{200}$ radius from the NFW halo, concentration parameter $c$ and
total DM mass $M_{200}=M_\mathrm{FDM}(r_{200})$, for the 18 high-resolution LSB
galaxies in \citet{deBlok:2001}.}
\label{tab:LSB-comb}
\begin{tabular}{lcccccccccc}
\hline
    Galaxy	&	$\rho_c$	&	$r_c$	&	$r_\epsilon$	&	$r_\epsilon/r_c$	&	$m_\psi$	&	$r_s$	&	$\rho_s$	&	$c$	&	$r_{200}$	&	$M_{200}$ \\
    	&	$(10^{-2} M_\odot/\mathrm{pc}^3)$	&	$(\mathrm{kpc})$	&	$(\mathrm{kpc})$	&	&	$(10^{-23}\mathrm{eV})$	&	$(\mathrm{kpc})$	&($10^{-2} M_\odot/\mathrm{pc}^3$)	&	&	$(\mathrm{kpc})$	&	($10^{11} M_\odot$)	\\
\hline
ESO 014-0040 & 50.2 	& 1.87 & 2.09 & 1.1  & 0.554 & 7.87 	& 9.07  & 29.3 & 231	& 13.5	\\
ESO 084-0411 & 0.542 	& 5.82 & 5.79 & 1.0  & 0.554 & 4.42 	& 1.90  & 16.0 & 70.6 	& 0.36	\\
ESO 120-0211 & 3.09 	& 3.76 & 0.54 & 0.15 & 0.554 & 1.18 	& 3.02  & 19.2 & 22.5 	& 0.010	\\
ESO 187-0510 & 4.04 	& 3.52 & 1.06 & 0.30 & 0.554 & 0.73 	& 33.0 	& 47.8 & 34.7 	& 0.040	\\
ESO 206-0140 & 7.78 	& 2.99 & 3.58 & 1.2  & 0.554 & 3.28 	& 13.9 	& 34.5 & 113	& 1.55	\\
ESO 302-0120 & 3.67 	& 3.60 & 3.70 & 1.0  & 0.554 & 1.20 	& 90.8 	& 69.6 & 83.5 	& 0.57	\\
ESO 305-0090 & 1.82 	& 4.30 & 2.55 & 0.59 & 0.554 & 1.70 	& 13.4 	& 34.0 & 57.6 	& 0.20	\\
ESO 425-0180 & 10.9 	& 2.74 & 6.21 & 2.3  & 0.554 & 2.13 	& 22.9 	& 41.6 & 88.7 	& 0.82	\\
ESO 488-0490 & 5.52 	& 3.26 & 3.06 & 0.94 & 0.554 & 1.13 	& 109	& 74.6 & 84.6 	& 0.60	\\
F730-V1      & 12.08 	& 2.68 & 2.64 & 0.99 & 0.554 & 4.62 	& 8.63  & 28.8 & 133	& 2.57	\\
UGC 4115     & 12.0 	& 2.68 & 1.26 & 0.47 & 0.554 & 1.53 	& 28.0 	& 44.9 & 68.8 	& 0.35	\\
UGC 11454    & 10.7 	& 2.76 & 2.36 & 0.86 & 0.554 & 9.31 	& 2.54  & 17.9 & 167	& 5.12	\\
UGC 11557    & 1.70 	& 4.37 & 4.55 & 1.0  & 0.554 & 2.26 	& 14.6 	& 35.1 & 79.4 	& 0.50	\\
UGC 11583    & 4.65 	& 3.40 & 1.77 & 0.52 & 0.554 & 0.86 	& 73.2 	& 64.3 & 55.3 	& 0.17	\\
UGC 11616    & 11.8 	& 2.69 & 4.59 & 1.7  & 0.554 & 4.21 	& 8.60  & 28.7 & 121	& 1.95	\\
UGC 11648    & 9.86 	& 2.82 & 1.34 & 0.48 & 0.554 & 11.5   	& 1.21  & 13.4 & 154	& 4.02	\\
UGC 11748    & 102 		& 1.57 & 2.13 & 1.4  & 0.554 & 2.60 	& 80.3 	& 66.5 & 173	& 5.65	\\
UGC 11819    & 10.3 	& 2.78 & 2.98 & 1.1  & 0.554 & 8.98		& 2.75  & 18.5 & 166	& 5.04	\\
\hline
\end{tabular}
\end{table*}

The best-fit mass obtained in the combined analysis is $m_\psi = 0.554 \times
10^{-23} \mathrm{eV}$, with a total reduced error $\chi_\mathrm{red}^2=1.208$.
Unfortunately, the computation time of the combined analysis was too
expensive and we were not able to obtain the confidence level for the FDM boson
mass. However, we believe that it is not too large given the errors obtained in
the individual analysis of these galaxies (see Table~\ref{tab:results2}) and the
value of the combined $\chi^2_\mathrm{red}$ we obtained here. Thus, it is
possible to fit all the galaxies together but with a very small mass that does
not meet the cosmological constraint imposed to the FDM model
\citep{Bozek:2014,Sarkar:2016}. If we impose the condition $m_\psi>10^{-23}
\mathrm{eV}$ into the fitting method, it is not possible to fit all the rotation
curves in a combined analysis. These results set a tension between the FDM model
and the observations.
  
In this case, the soliton core radii are in the interval $1.57 \leq r_c/
\mathrm{kpc} \leq 5.82$ and ratio between the transition and core radii within
$0.30 \leq r_\epsilon/r_c \leq 2.3$. Again, the best fits do not correspond to
$r_\epsilon>3r_c$, as expected in the cosmological simulations \citep{Schive:2014dra}.
In both cases, the individual and combined analysis, the soliton contribution
required to fit the observations is not as prominent as in the simulations.

\begin{figure*}
\begin{center}$
\begin{array}{ccc}
\includegraphics[width=0.3\textwidth]{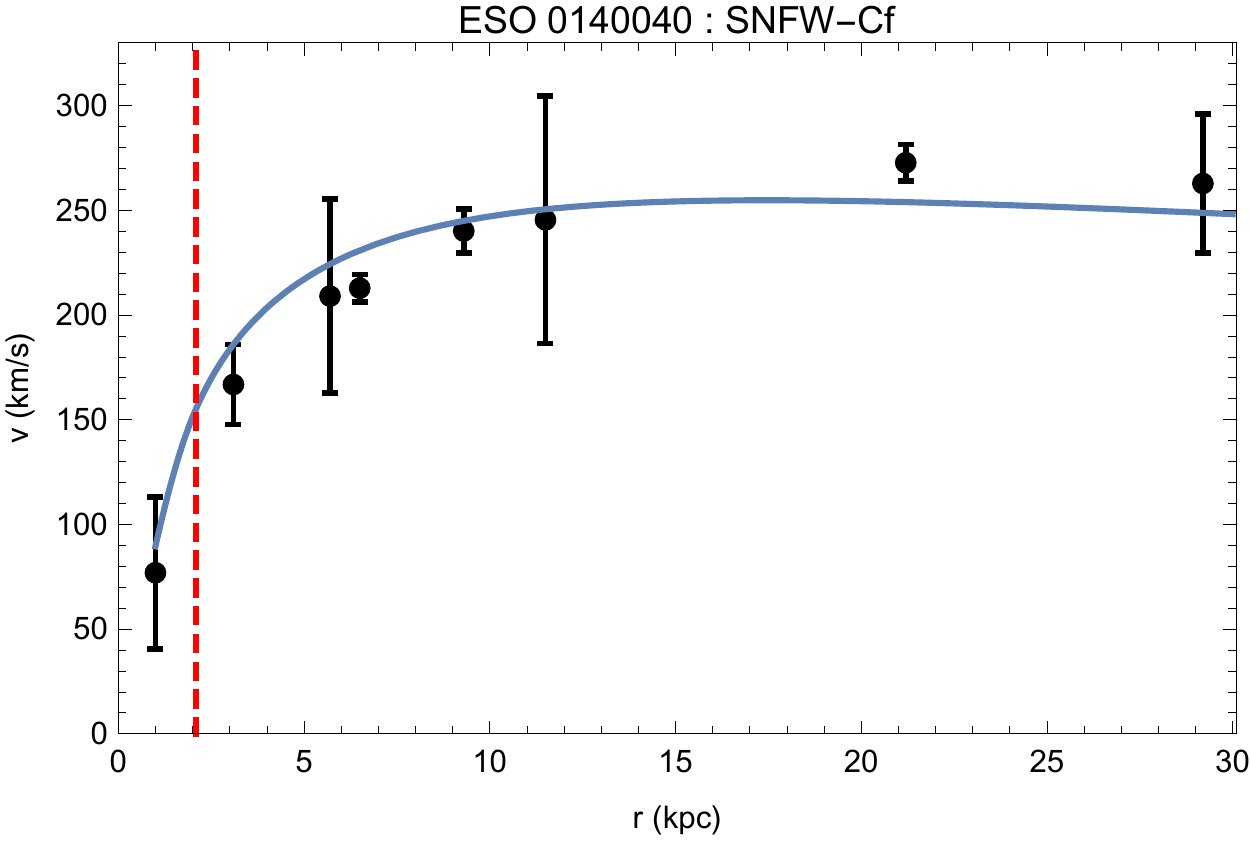} &
\includegraphics[width=0.3\textwidth]{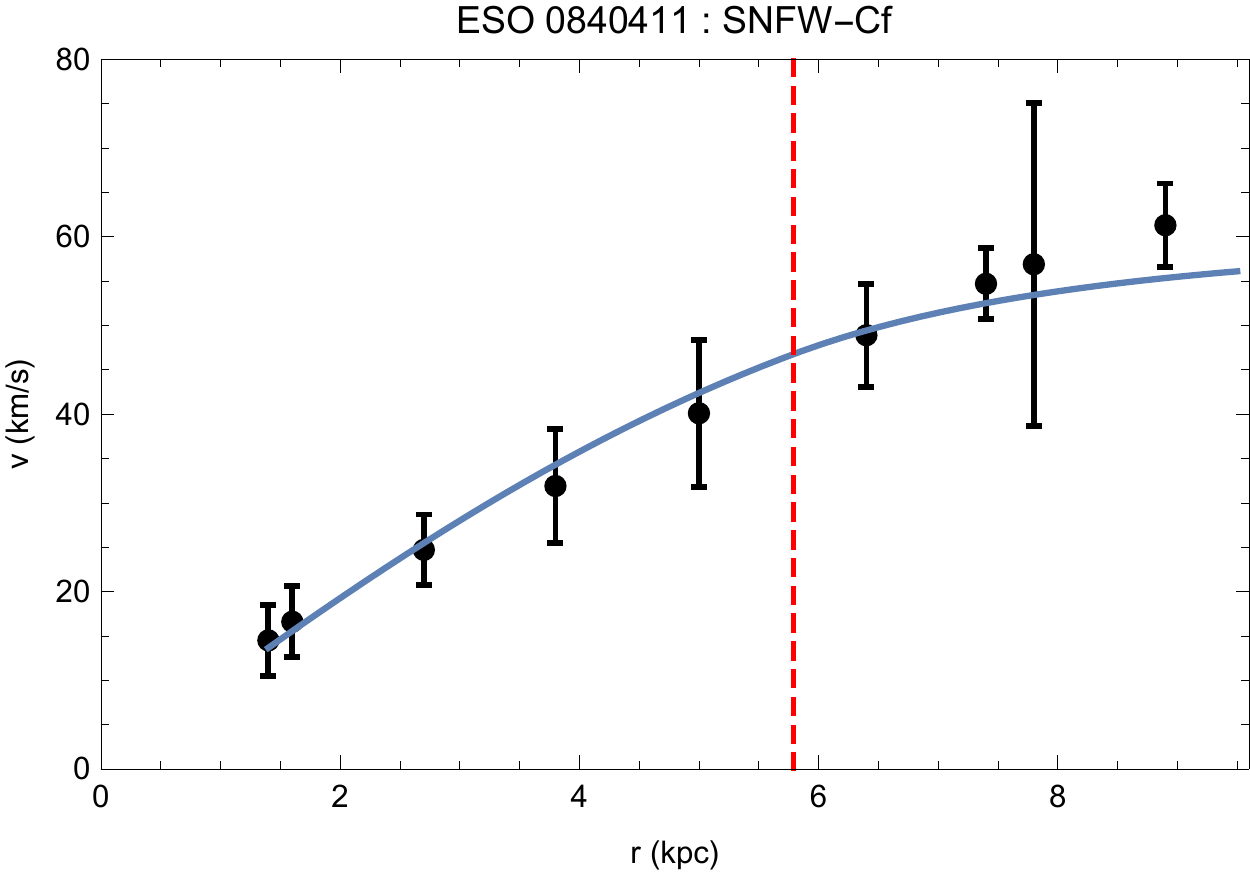}&
\includegraphics[width=0.3\textwidth]{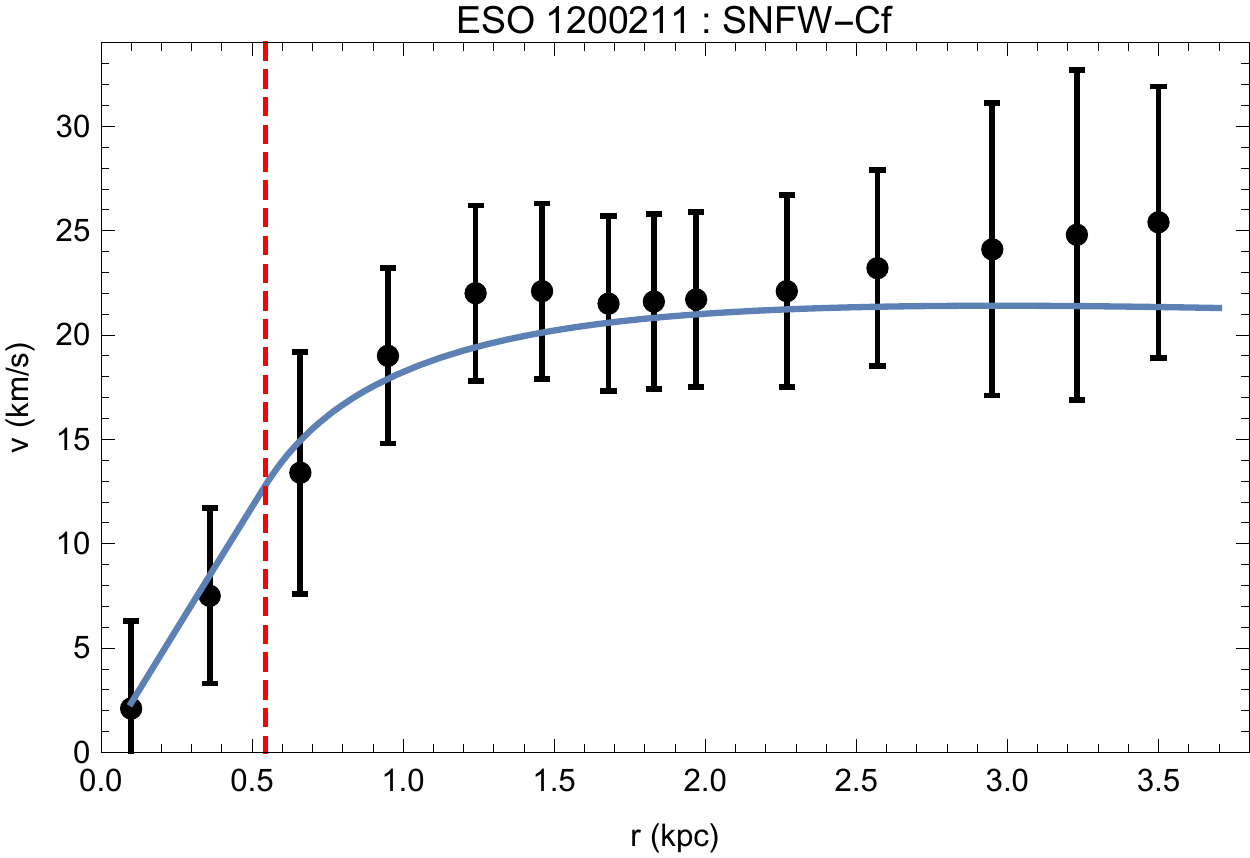} \\
\includegraphics[width=0.3\textwidth]{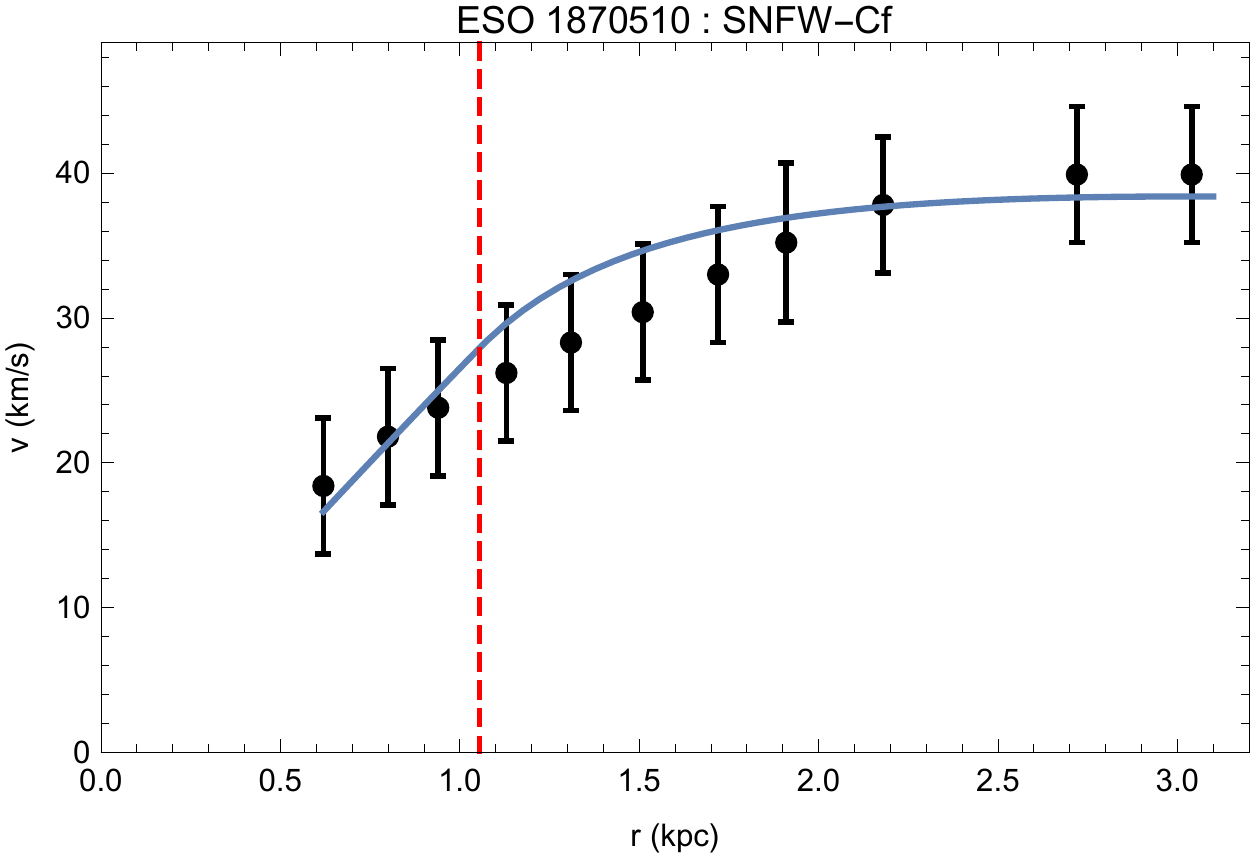}&
\includegraphics[width=0.3\textwidth]{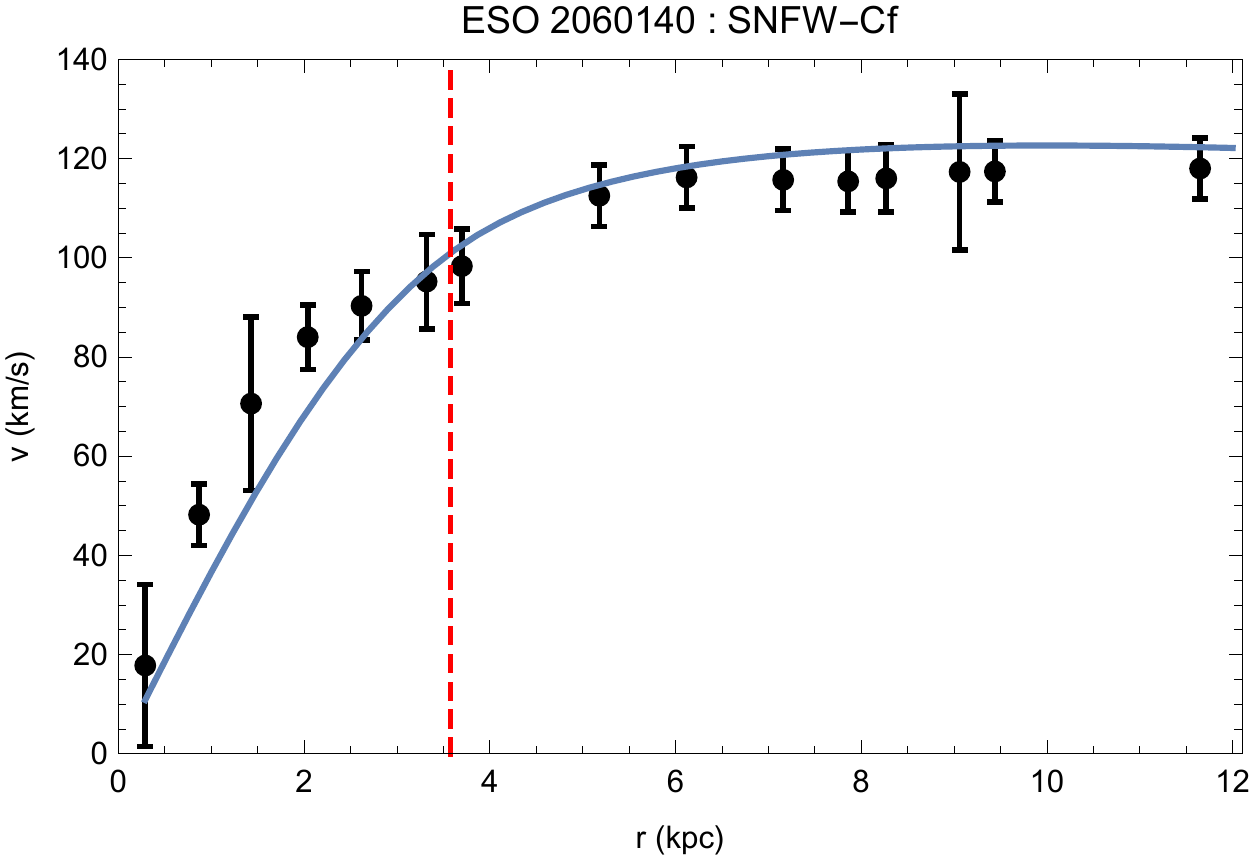} &
\includegraphics[width=0.3\textwidth]{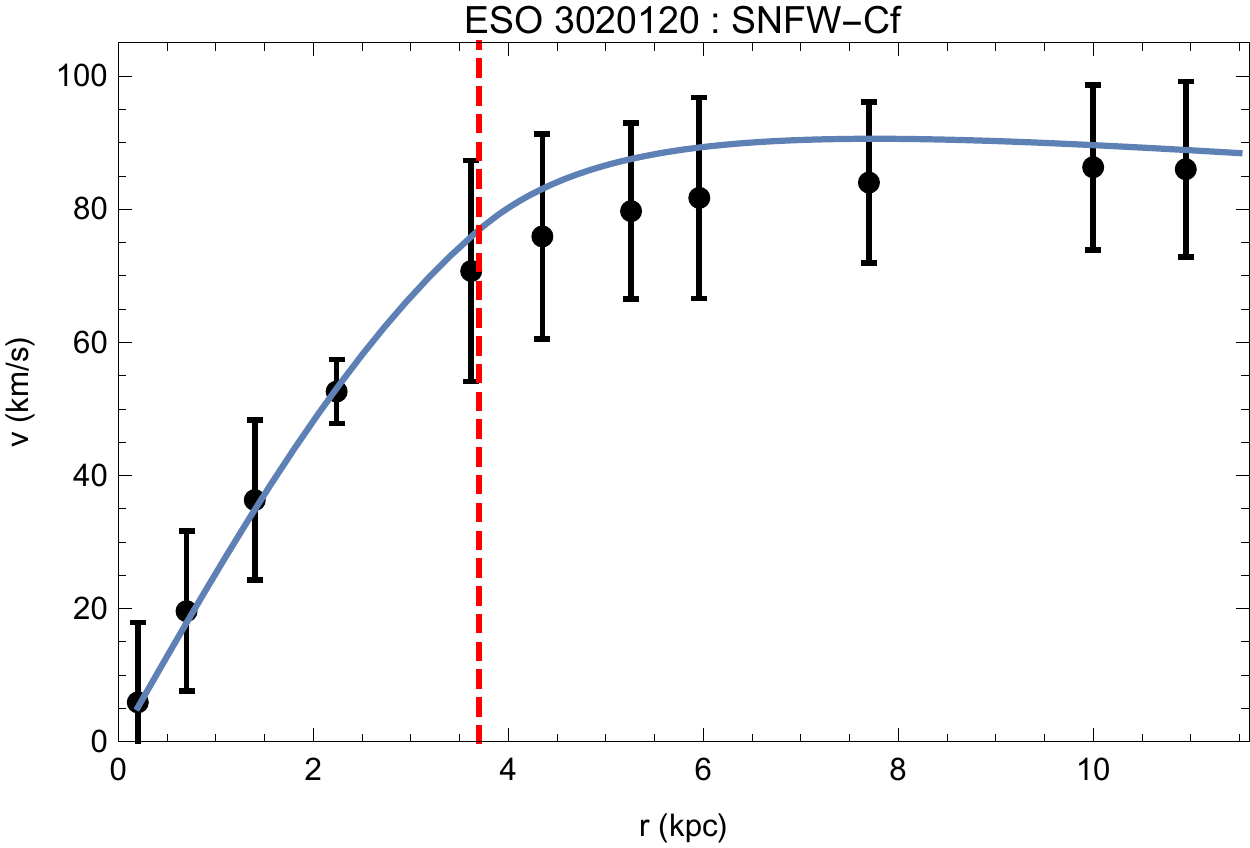} \\
\includegraphics[width=0.3\textwidth]{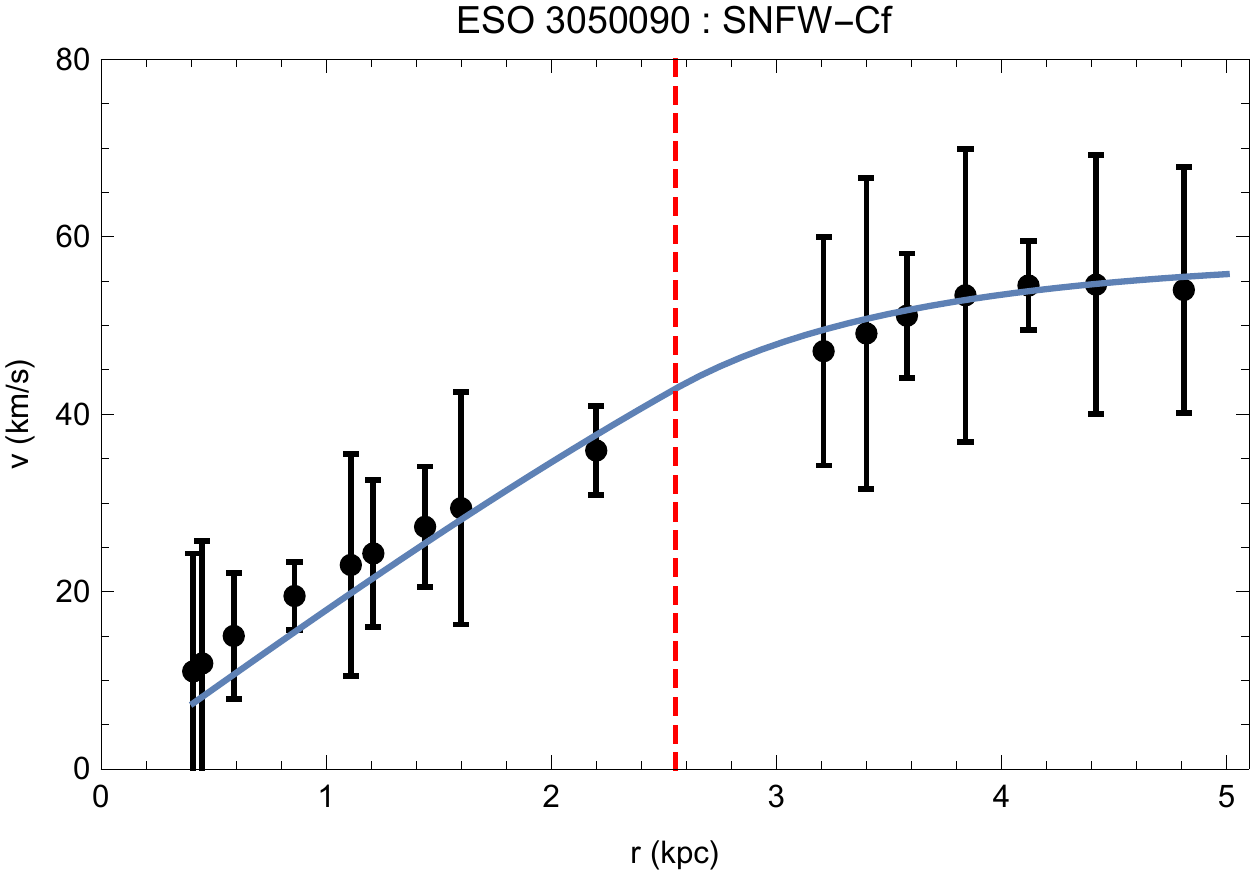} &
\includegraphics[width=0.3\textwidth]{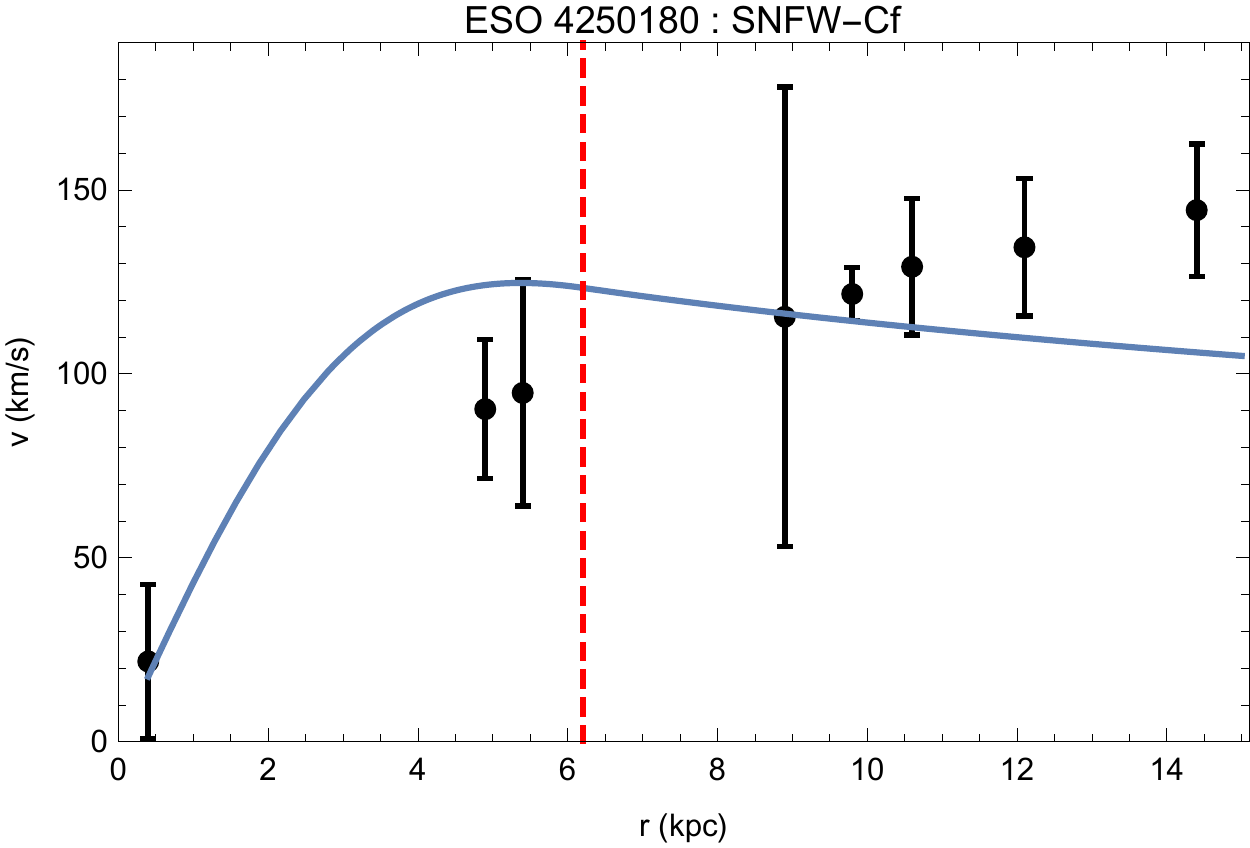}&
\includegraphics[width=0.3\textwidth]{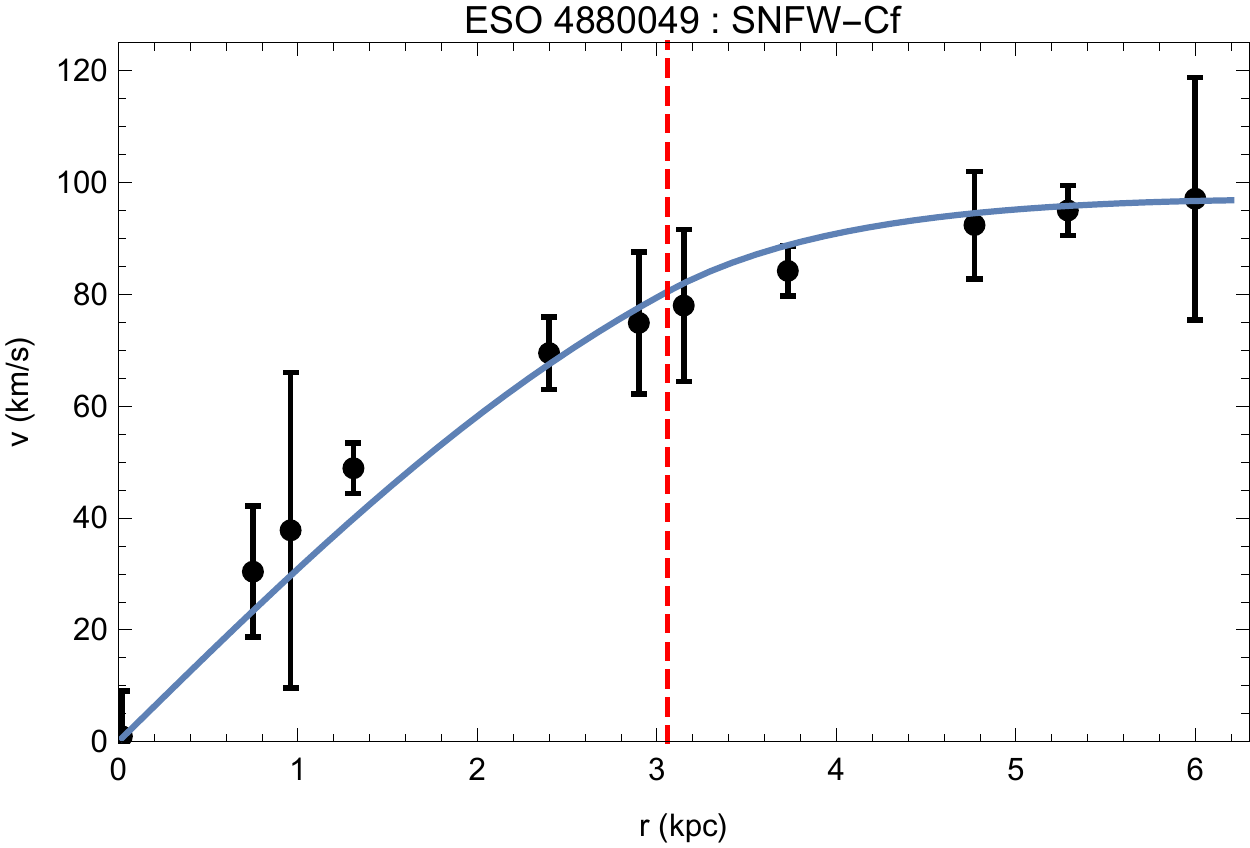} \\
\includegraphics[width=0.3\textwidth]{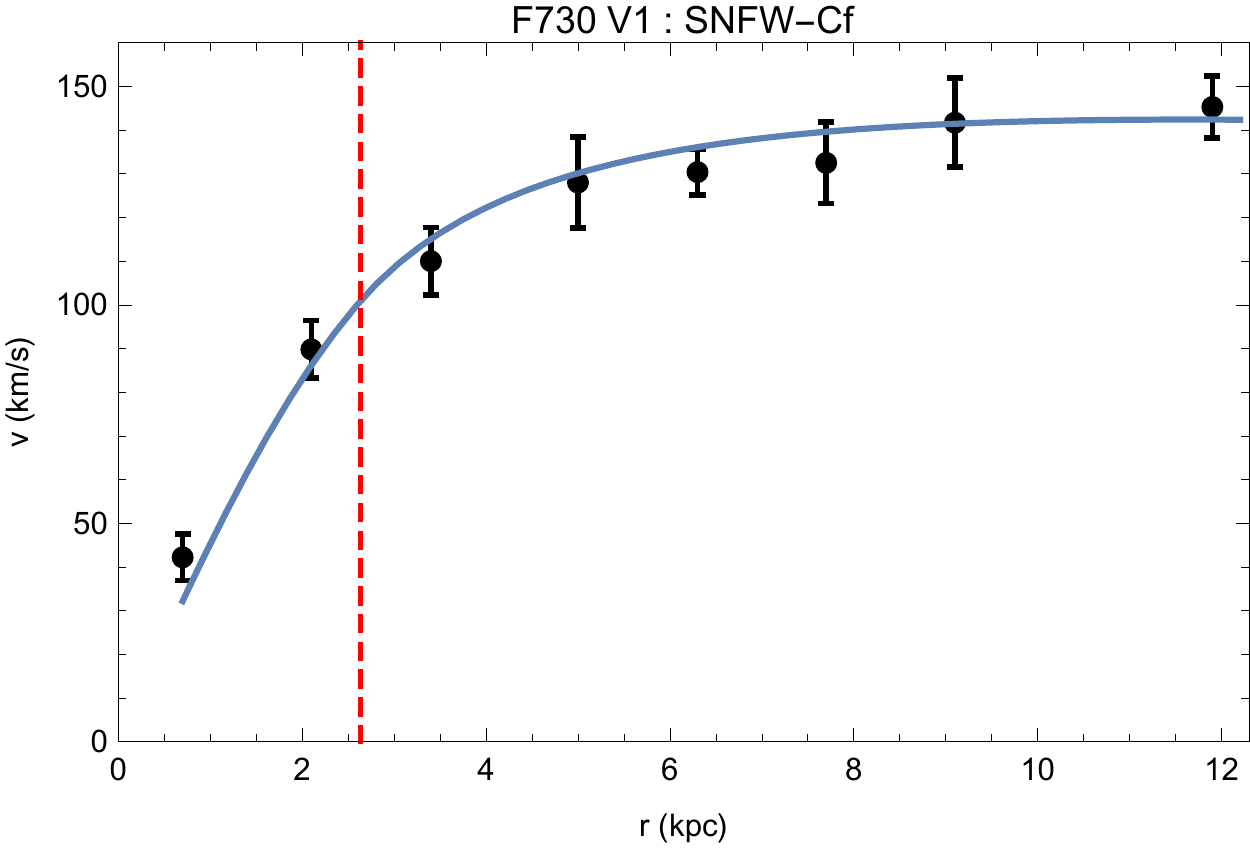}&
\includegraphics[width=0.3\textwidth]{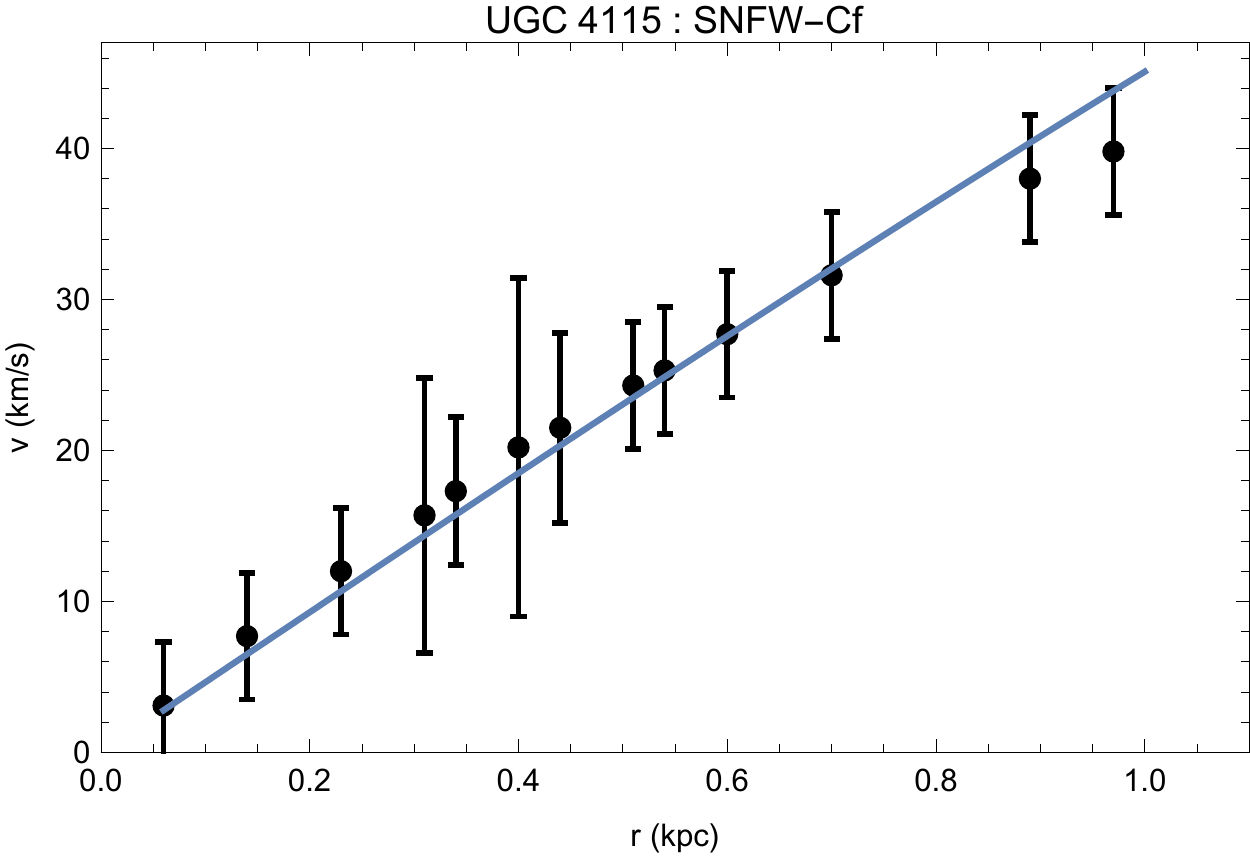} &
\includegraphics[width=0.3\textwidth]{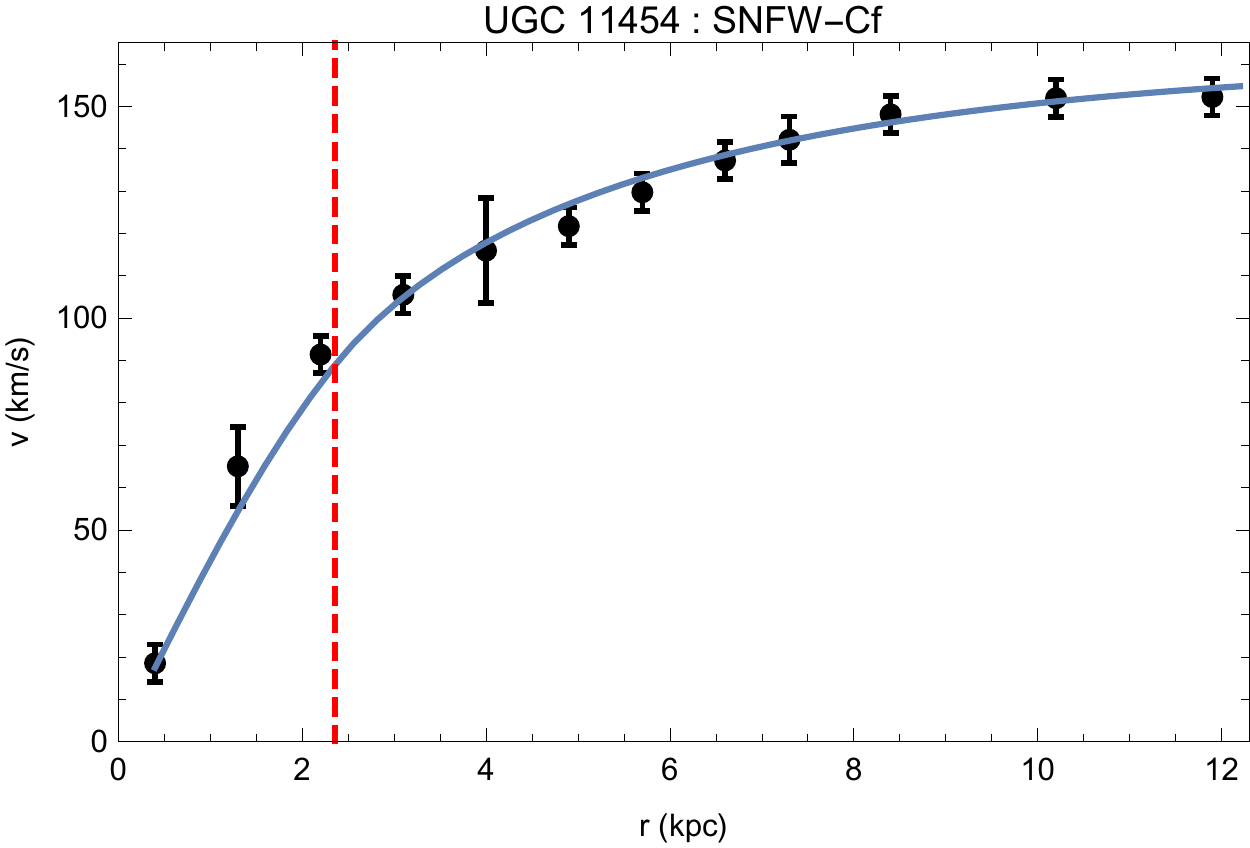} \\
\includegraphics[width=0.3\textwidth]{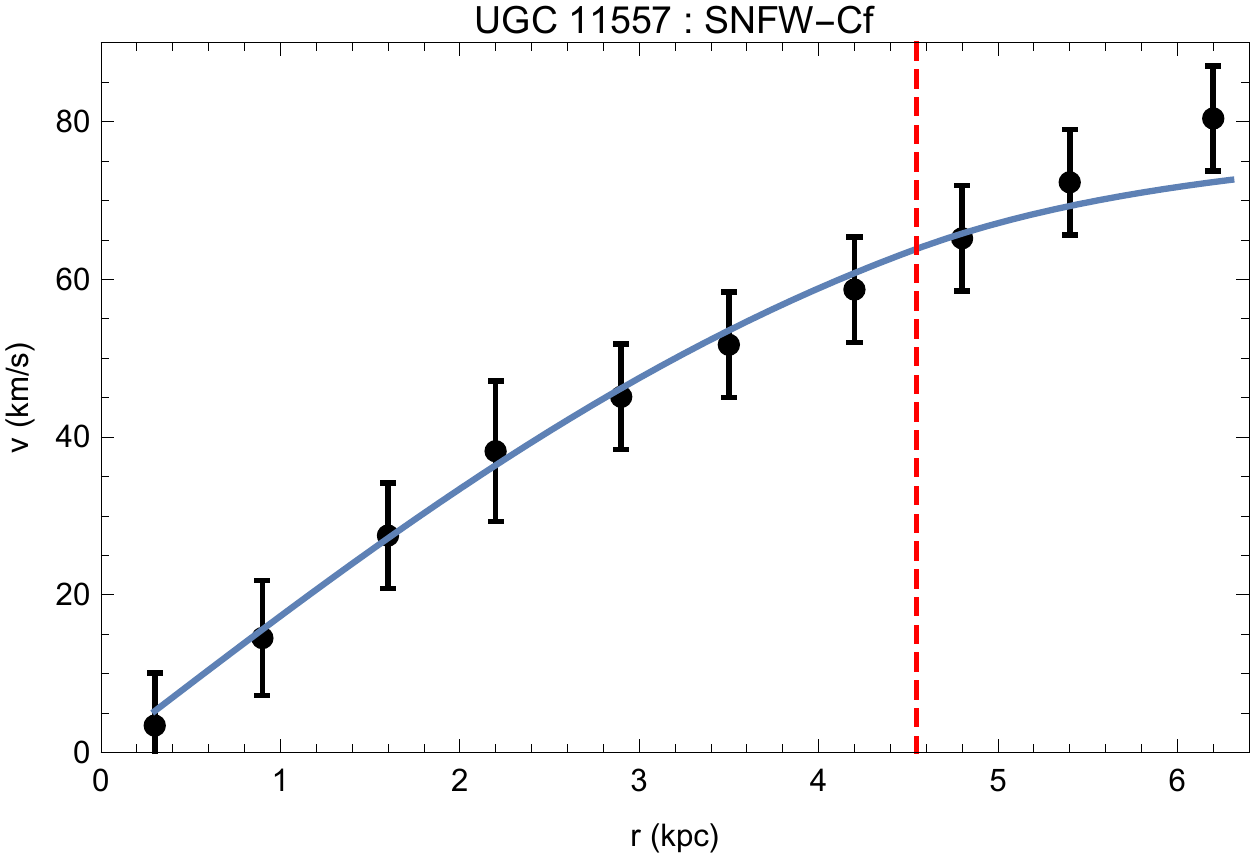} &
\includegraphics[width=0.3\textwidth]{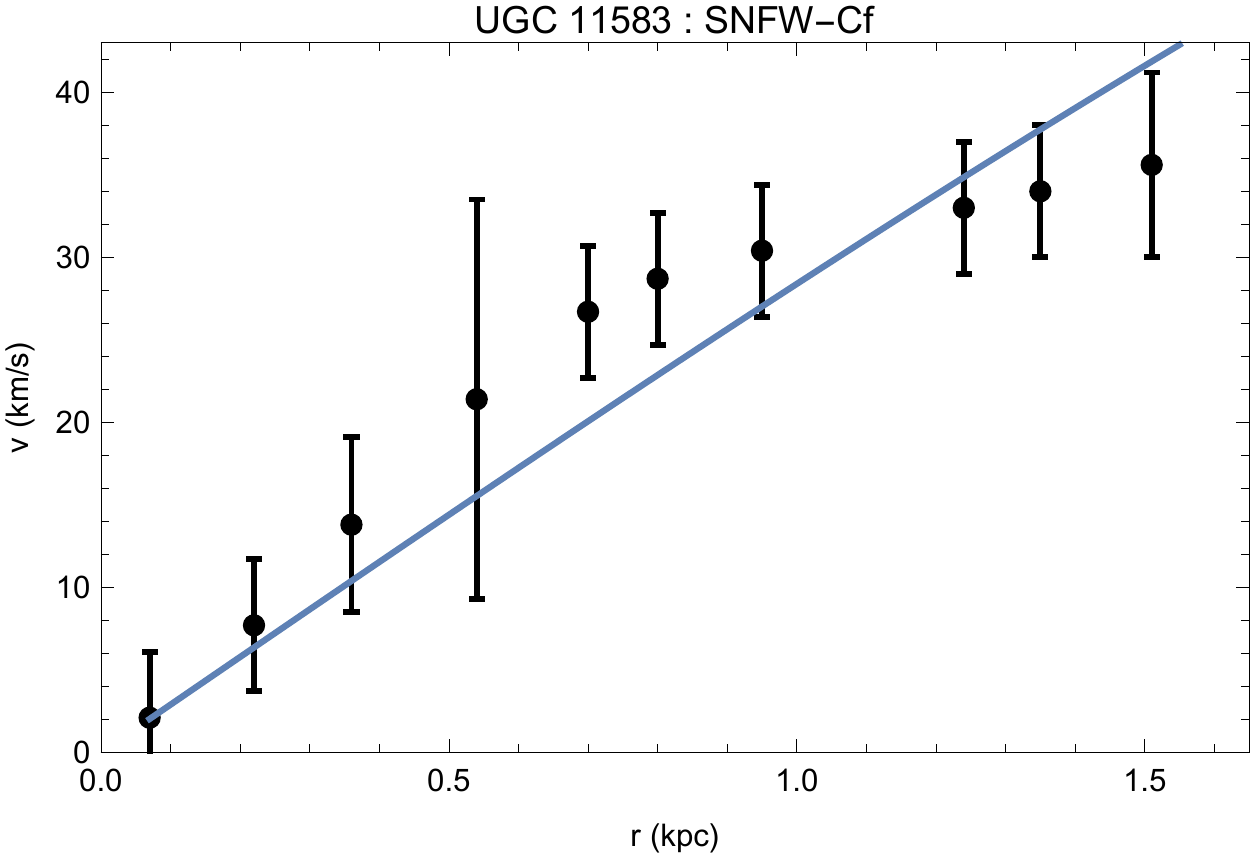} &
\includegraphics[width=0.3\textwidth]{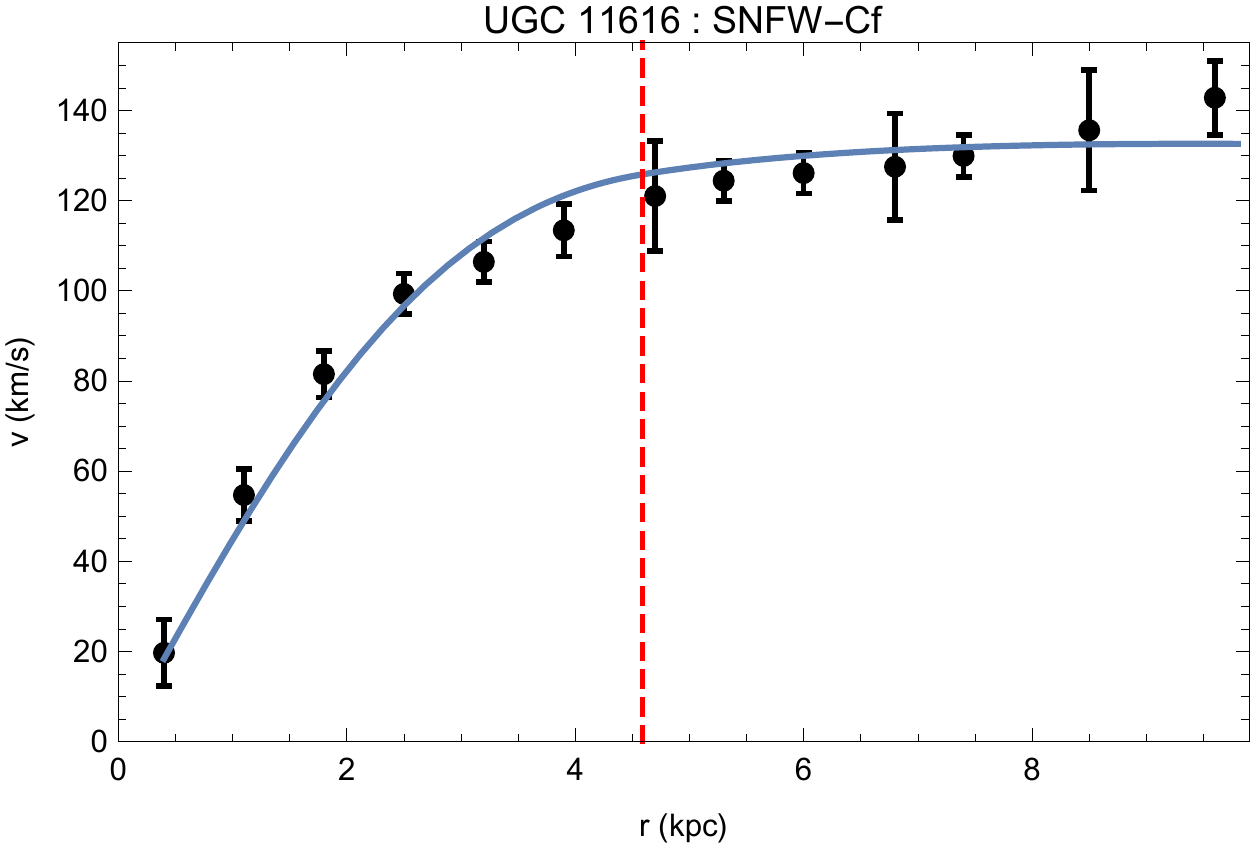}\\
\includegraphics[width=0.3\textwidth]{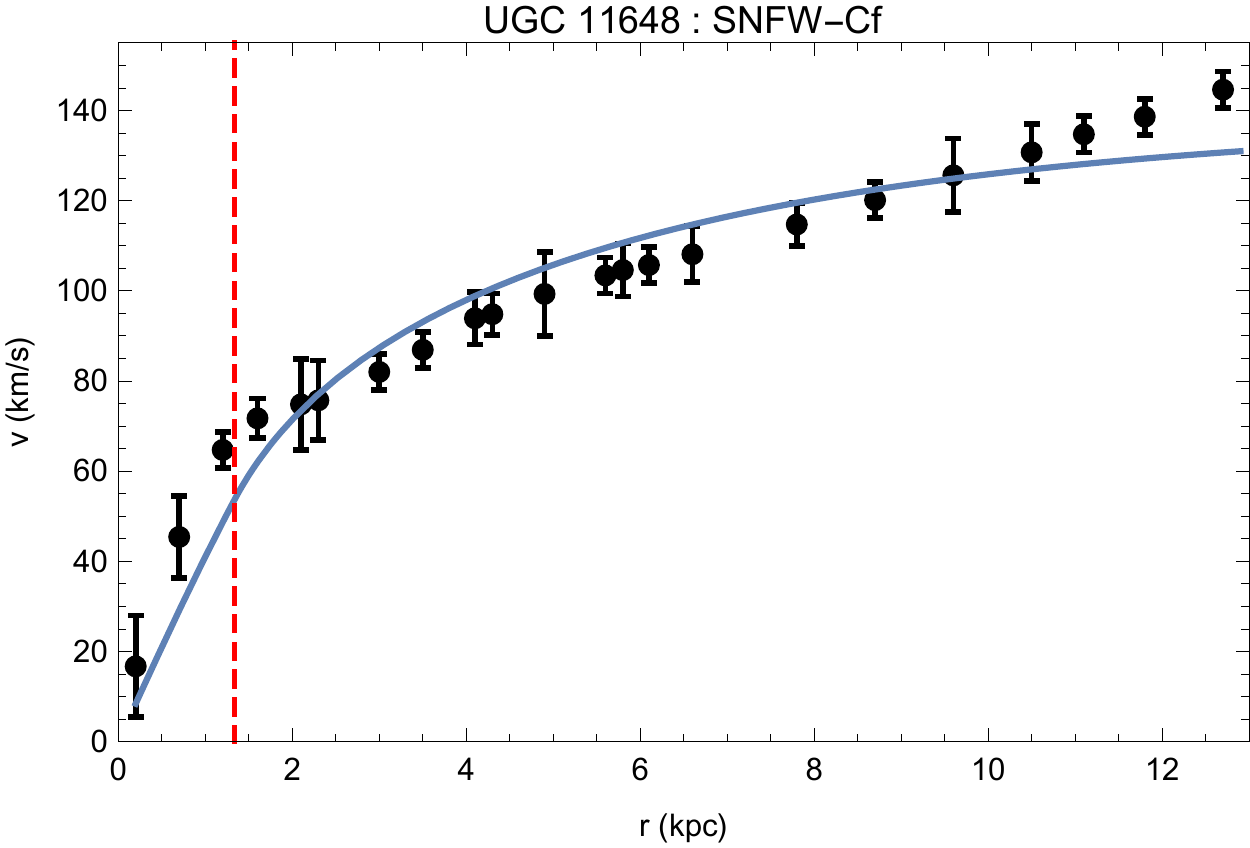} &
\includegraphics[width=0.3\textwidth]{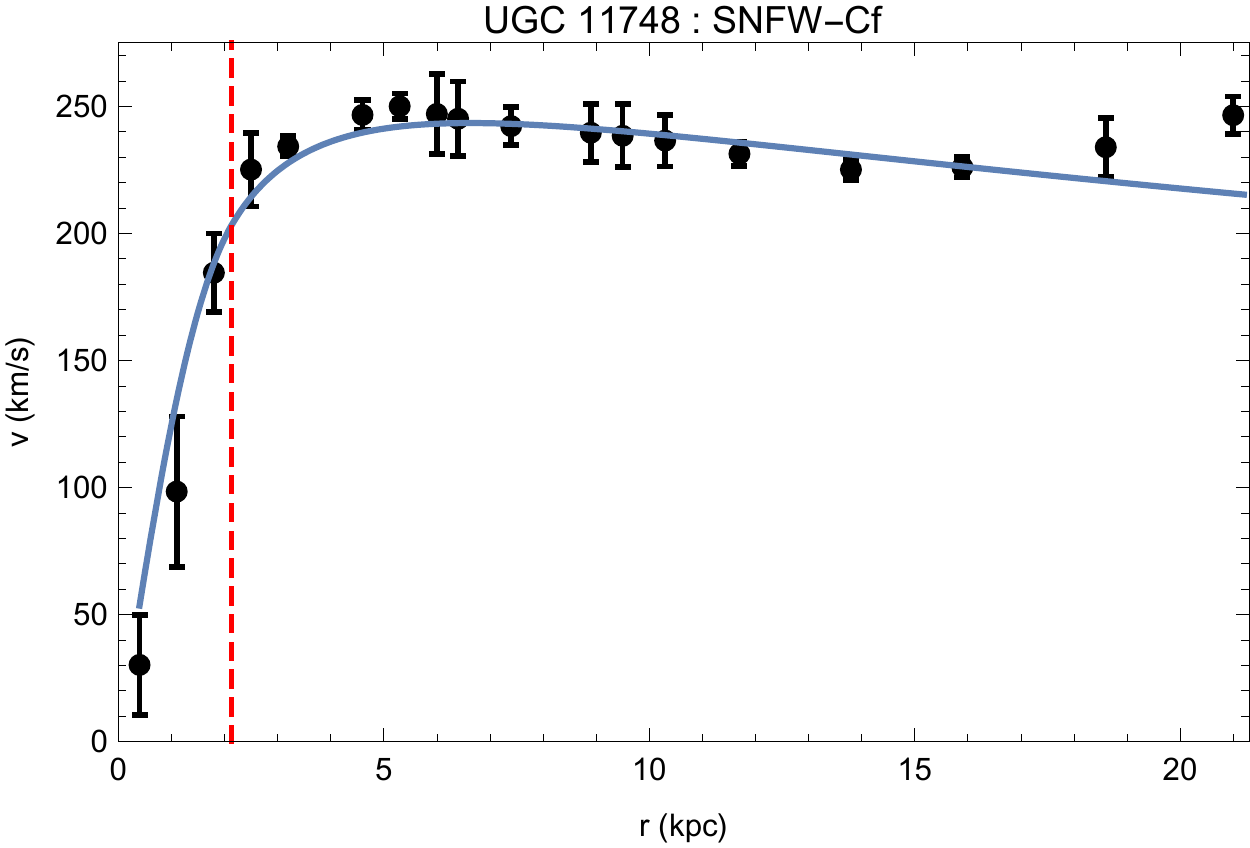}&
\includegraphics[width=0.3\textwidth]{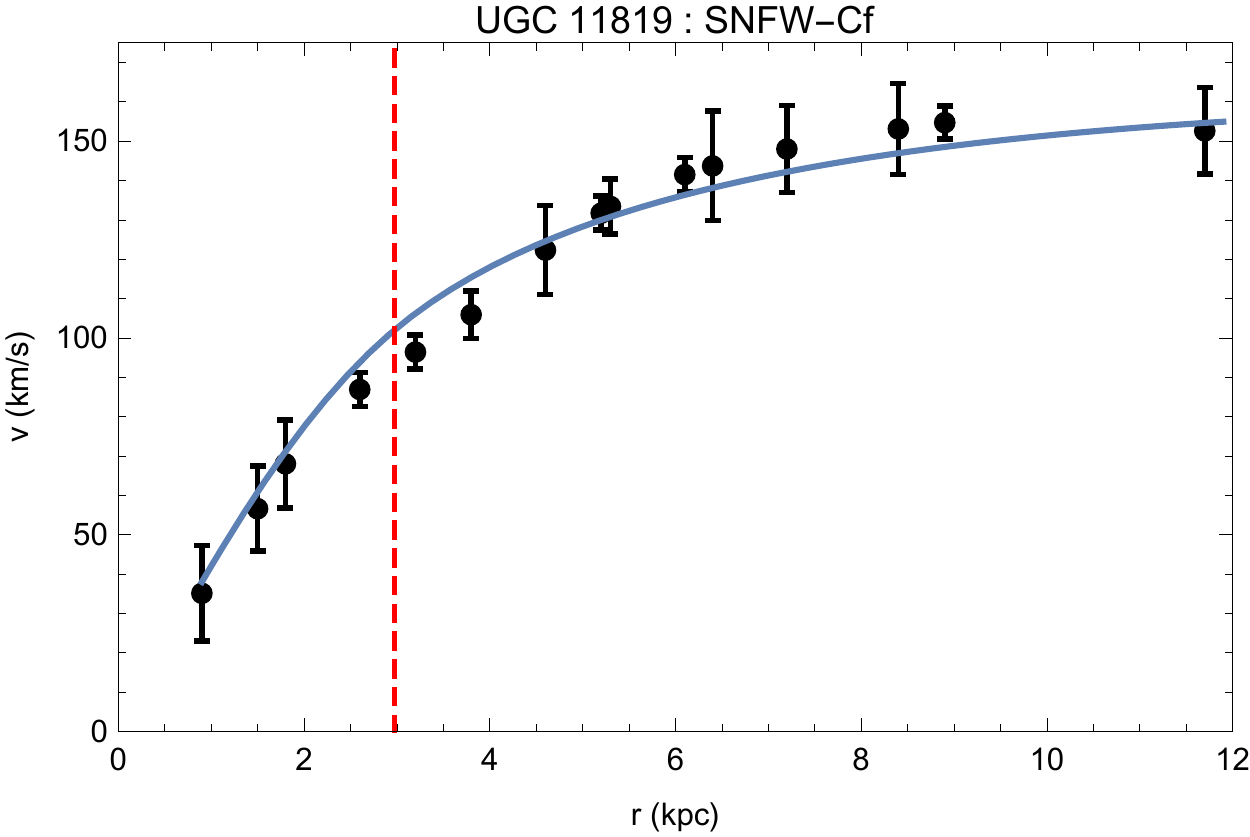}
\end{array}$
\end{center}
\caption{Best fits from the combined analysis for the 18 high-resolution LSB
galaxies, in the FDM model using the soliton+NFW profile. The corresponding fitting
parameters are shown in Table~\ref{tab:LSB-comb}, for a constant boson mass
$m_\psi = 0.554\times 10^{-23}\mathrm{eV}$. The vertical lines correspond to the
transition radii $r_\epsilon$. In all cases, the soliton contribution is
important in order to fit the whole rotation curve.}
\label{fig:Schive18-comb}
\end{figure*}

Fig.~\ref{fig:Schive18-comb} shows the resulting rotation curves for each galaxy
with the fitting parameters shown in Table~\ref{tab:LSB-comb}. The transition
radius $r_\epsilon$ is the vertical dashed line shown for all the galaxies.
Remarkably, with a constant boson mass the fits are good for almost all the
galaxies, given the large observational errors for some galaxies in the sample,
except for ESO 425-0180, UGC 11583, 11648 and 11748, and some regions in ESO
206-0140.

\subsubsection{NGC galaxies with photometric data}
\label{subsubsec:sparc-1}

  In Table~\ref{tab:results1} we show the fitting parameters for the three
representative galaxies in \citet{McGaugh:2016} (NGC 7814, 6503, 3741) from
the SPARC database \citep{Lelli:2016} and the three sample galaxies analyzed in
\citet{Robles:2013} (NGC 1003, data from SPARC; NGC 1560, data from
\citet{deBlok:2001}; NGC 6946, data from \citet{McGaugh:2005}), for the
FDM model with the soliton+NFW density profile. In the
top panel we show the results for the galaxies without the photometric
information, and on the bottom panel the results including
the disk, gas and/or bulge contribution, to notice the effect of the baryons on the
rotation curves. In each case, we report the fitting parameters $\rho_c$, $r_c$,
$r_\epsilon$ and $r_s$, with the resulting boson mass $m_\psi$, the ratio
$r_\epsilon/r_c$, the virial radius $r_{200}$, concentration parameter $c$,
the $\chi^2_\mathrm{red}$ error and $\pm 1\sigma$ uncertainties from the MCMC
method.

\begin{table*}
\caption{Soliton+NFW density profile in NGC galaxies with photometric data. The
top panel shows the DM-only fits and the bottom panel the DM+baryons analysis,
both for the same galaxies with high-resolution photometric information. Both
panels show the fitting parameters $\rho_c$, $r_c$, $r_\epsilon$ and $r_s$, for
the FDM model with the soliton+NFW density profile~\eqref{eq:rho-schive}, and the
resulting boson mass $m_\psi$; all the quantities $\pm 1\sigma$ errors from the
MCMC method used. We also show the resulting ratio $r_\epsilon/r_c$, the $r_{200}$
radius from the NFW halo, concentration parameter $c$ and $\chi^2_\mathrm{red}$
errors.}
\label{tab:results1}
\begin{tabular}{lccccccccc}
	\hline
    \multicolumn{10}{c}{DM-only fits} \\
    \hline
    Galaxy	&	$\rho_c$	&	$r_c$	&	$r_\epsilon$	&	$r_\epsilon/r_c$	&	$m_\psi$	&	$r_s$	&	$r_{200}$	&	$c$	&	$\chi^2_\mathrm{red}$ \\
    	&	$(10^{-2} M_\odot/\mathrm{pc}^3)$	&	$(\mathrm{kpc})$	&	$(\mathrm{kpc})$	&	&	$(10^{-23}\mathrm{eV})$	&	$(\mathrm{kpc})$	&	$(\mathrm{kpc})$	&	&	\\
\hline
NGC 7814	&	$1910^{+300}_{-510}$	& $0.485\pm 0.057$			&   $1.29^{+0.16}_{-0.14}$		&	2.7 	& $1.39^{+0.11}_{-0.20}$	&   $5.17^{+0.67}_{-0.93}$	&  168	&  32.4 &  2.48	\\
NGC 6503	&	$27.3^{+2.0}_{-2.4}$	& $1.686\pm 0.072$			&   $3.70 \pm 0.18$				&	2.2 	& $0.933^{+0.040}_{-0.049}$	&   $9.73^{+0.93}_{-1.0}$	&  116	&  11.9 &  1.12	\\
NGC 3741	&	$42.0\pm 12.0$			& $0.163^{+0.023}_{-0.091}$	&   $0.0962^{+0.0092}_{-0.047}$	&	0.59	& $129^{+40}_{-100}$		&   $15.7^{+2.5}_{-4.4}$	&  93.5	&  5.96 &  8.4	\\
NGC 1003	&	$3.51^{+0.23}_{-0.27}$	& $3.84\pm 0.15$			&   $8.36^{+0.36}_{-0.30}$		&	2.2 	& $0.501^{+0.018}_{-0.023}$	&   $128^{+30}_{-50}	$	&  223	&  1.74 &  1.29	\\
NGC 1560	&	$5.67^{+0.54}_{-0.90}$	& $1.64^{+0.16}_{-0.52}$	&   $1.07^{+0.13}_{-0.33}$		&	0.65	& $2.5\pm 1.0$				&   $12.7^{+1.3}_{-1.7}$	&  8.06	&  102	&  2.98	\\
NGC 6946	&	$4.87^{+0.31}_{-0.46}$	& $5.34^{+0.31}_{-0.57}$	&   $6.1^{+1.0}_{-1.6}$			&	1.1 	& $0.225^{+0.036}_{-0.024}$	&   $5.7^{+1.2}_{-1.5}$		&  157	&  27.5 &  15.4	\\
\hline
    \multicolumn{10}{c}{DM+baryons fits} \\
\hline
NGC 7814	&	$4.85^{+0.46}_{-0.68}$	&	$5.47\pm 0.40$			&   $9.0 \pm 1.0$			&	1.65	&	$0.212^{+0.016}_{-0.023}$	&	$21^{+6}_{-10}$			&  214	&  10.2 &  0.668\\
NGC 6503	&	$4.07^{+0.32}_{-0.40}$	&	$3.31\pm 0.20$			&   $5.59\pm 0.51$			&	1.69	&	$0.628^{+0.040}_{-0.056}$	&	$11.5^{+1.5}_{-2.0}$	&  114	&  9.95 &  1.18\\
NGC 3741	&	$5.7^{+1.1}_{-1.6}$		&	$0.759^{+0.098}_{-0.28}$&   $0.615^{+0.069}_{-0.22}$&	0.81	&	$12.1^{+3.9}_{-6.7}$  		&	$23^{+4}_{-10}$			&  101	&  4.39 &  2.9	\\
NGC 1003	&	$1.80^{+0.14}_{-0.19}$	&	$4.52\pm 0.24$			&   $8.76^{+0.50}_{-0.43}$	&	1.94	&	$0.507^{+0.027}_{-0.035}$ 	&	$159^{+40}_{-80}$		&  232	&  1.46 &  1.33	\\
NGC 1560	&	$4.22^{+0.46}_{-0.79}$	&	$1.74^{+0.22}_{-0.59}$	&   $1.10^{+0.14}_{-0.38}$	&	0.63	&	$2.65^{+0.85}_{-1.6}$	 	&	$13.7^{+1.8}_{-2.4}$	&  96.0	&  7.01 &  2.42	\\
NGC 6946	&	$2.99^{+0.21}_{-0.55}$	&	$4.56^{+0.59}_{-0.43}$	&   $8.0\pm 1.0$			&	1.75	&	$0.395^{+0.039}_{-0.077}$ 	&	$71^{+20}_{-30}$		&  242	&  3.40	&  2.03	\\
\hline
\end{tabular}
\end{table*}

\begin{figure*}
    \begin{center}
\includegraphics[width=0.5\textwidth]{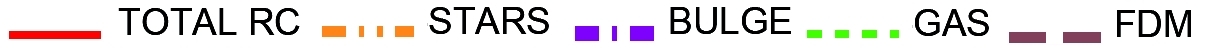}
    $\begin{array}{ccc}
\includegraphics[width=2.3in]{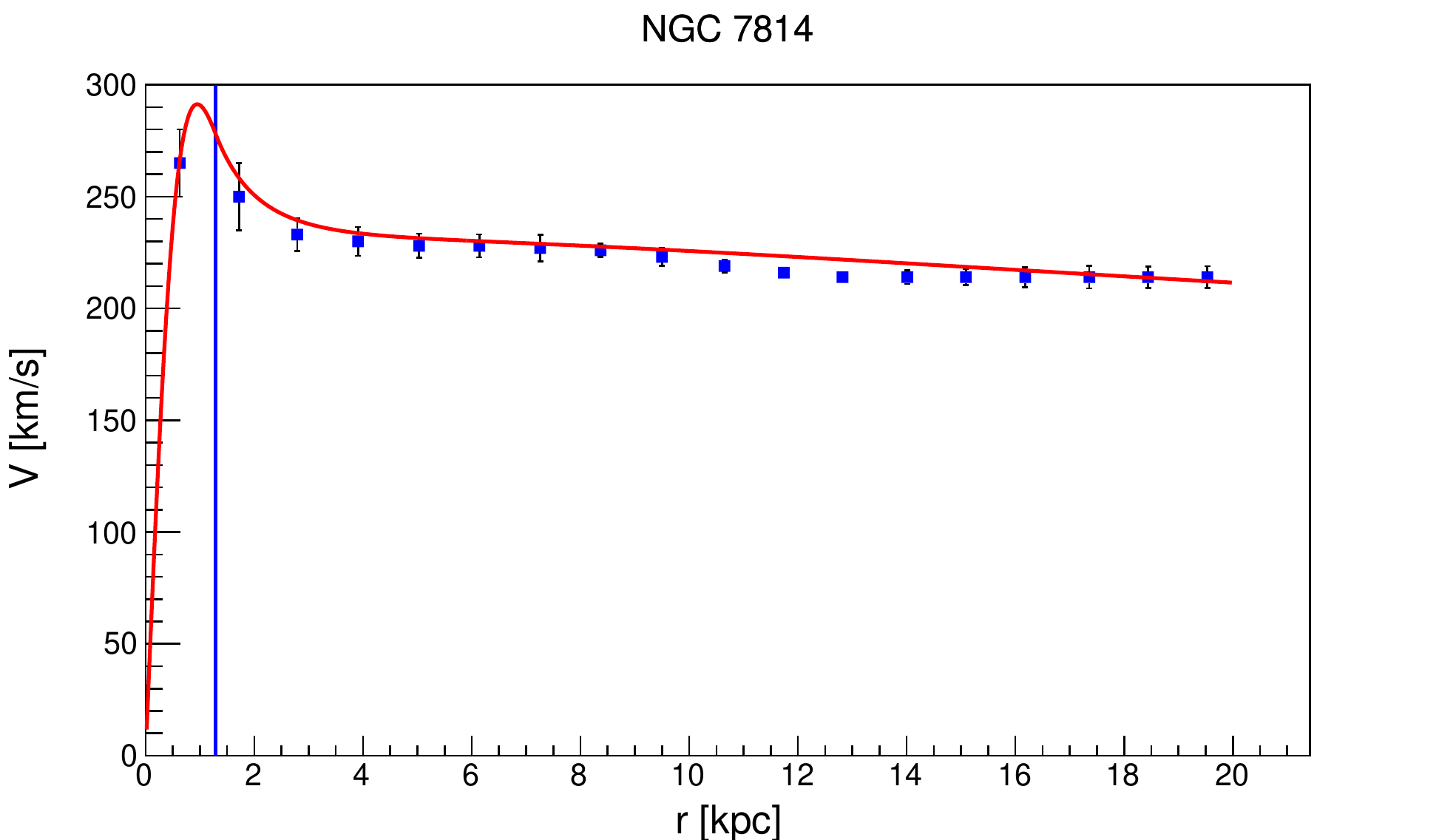} &
\includegraphics[width=2.3in]{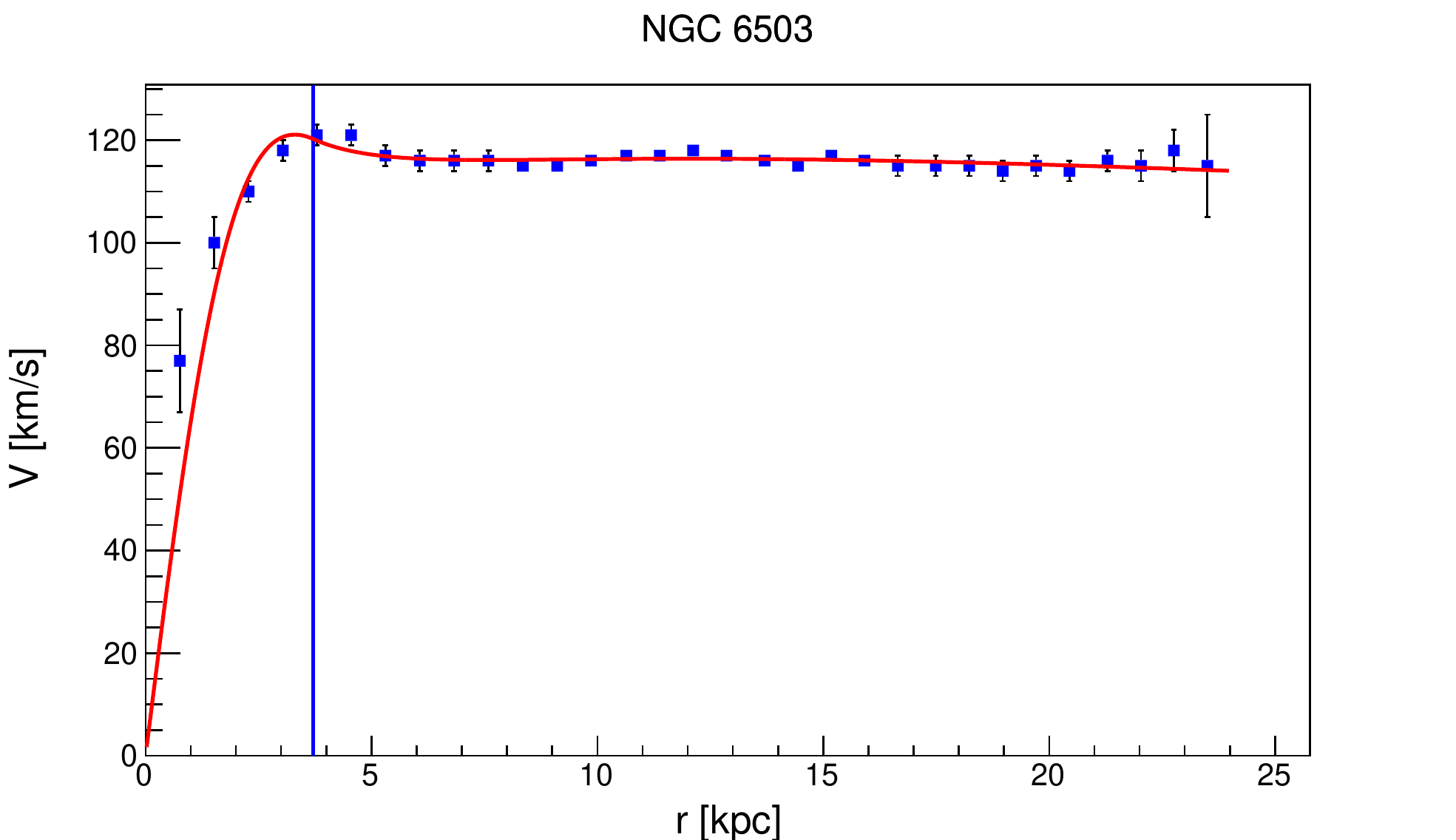} &
\includegraphics[width=2.3in]{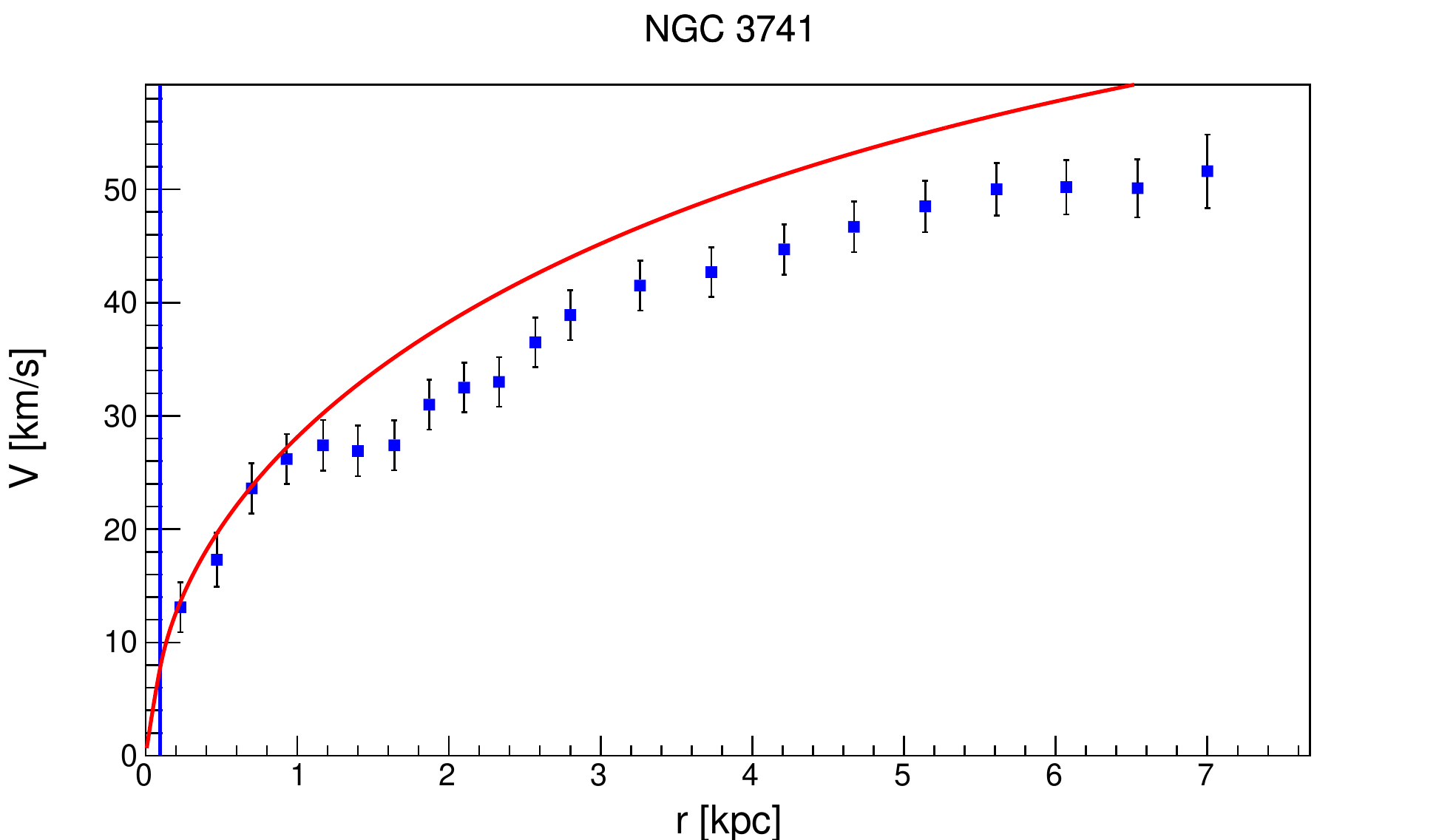} \\
\includegraphics[width=2.3in]{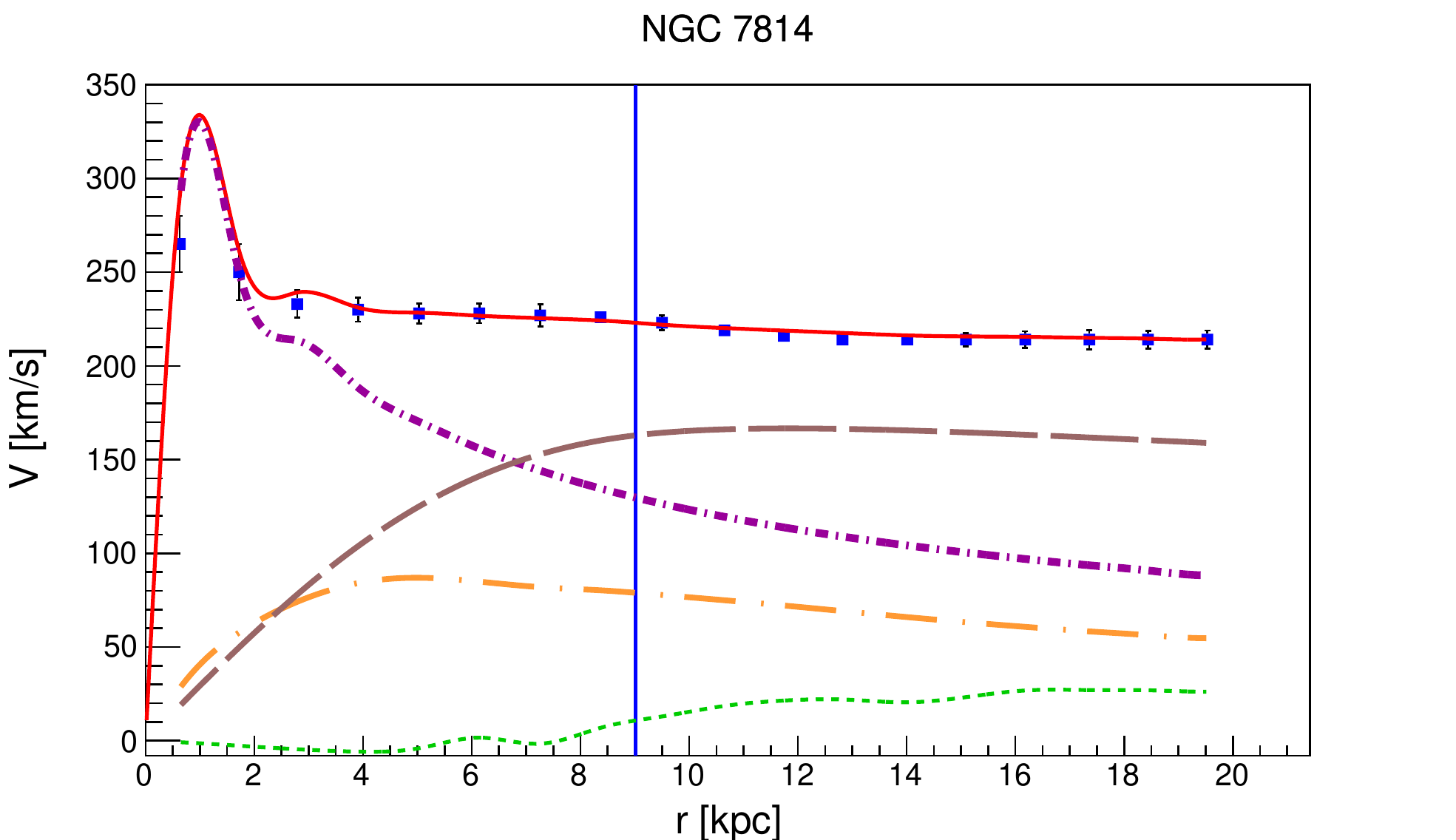} &
\includegraphics[width=2.3in]{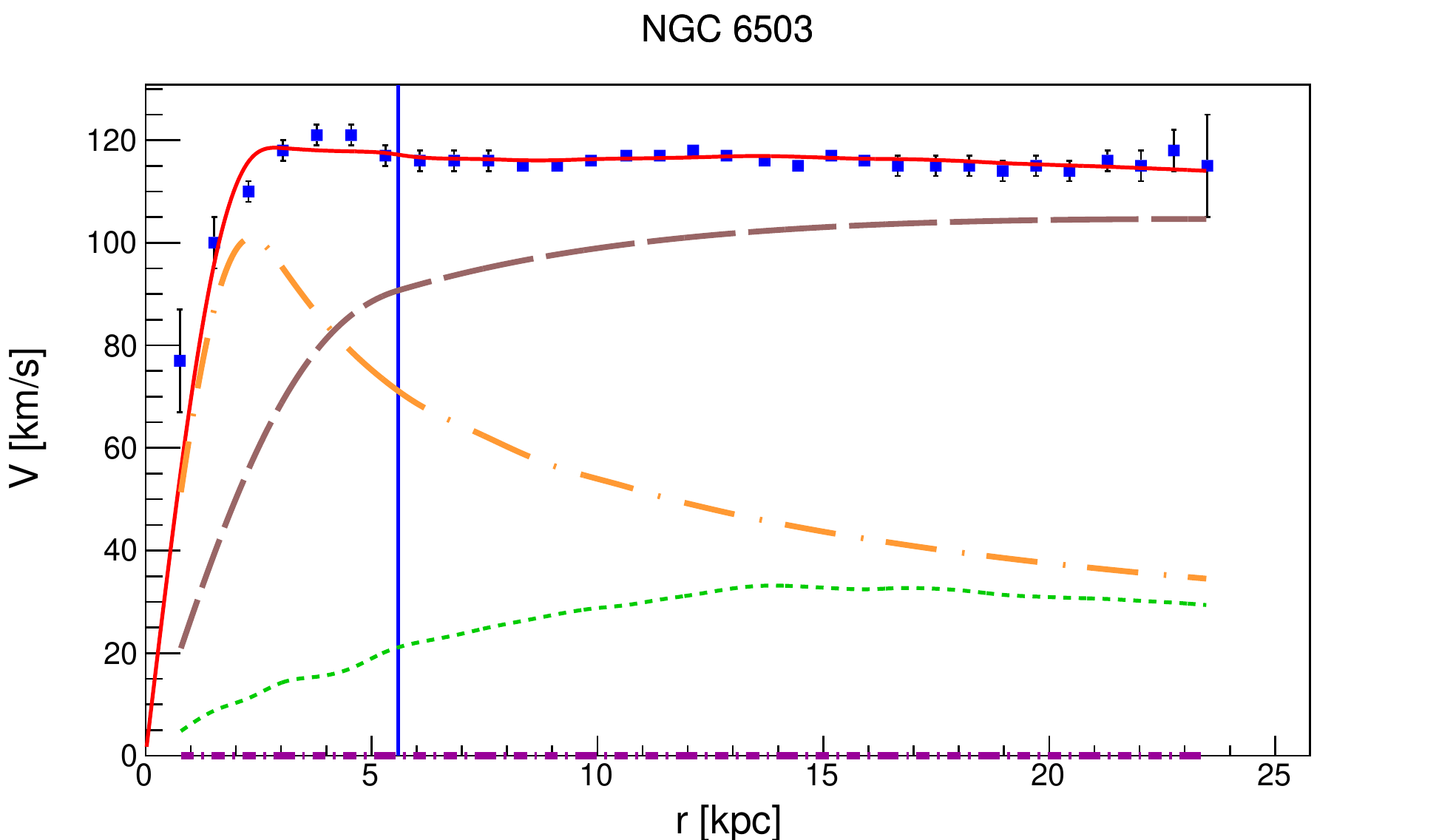} &
\includegraphics[width=2.3in]{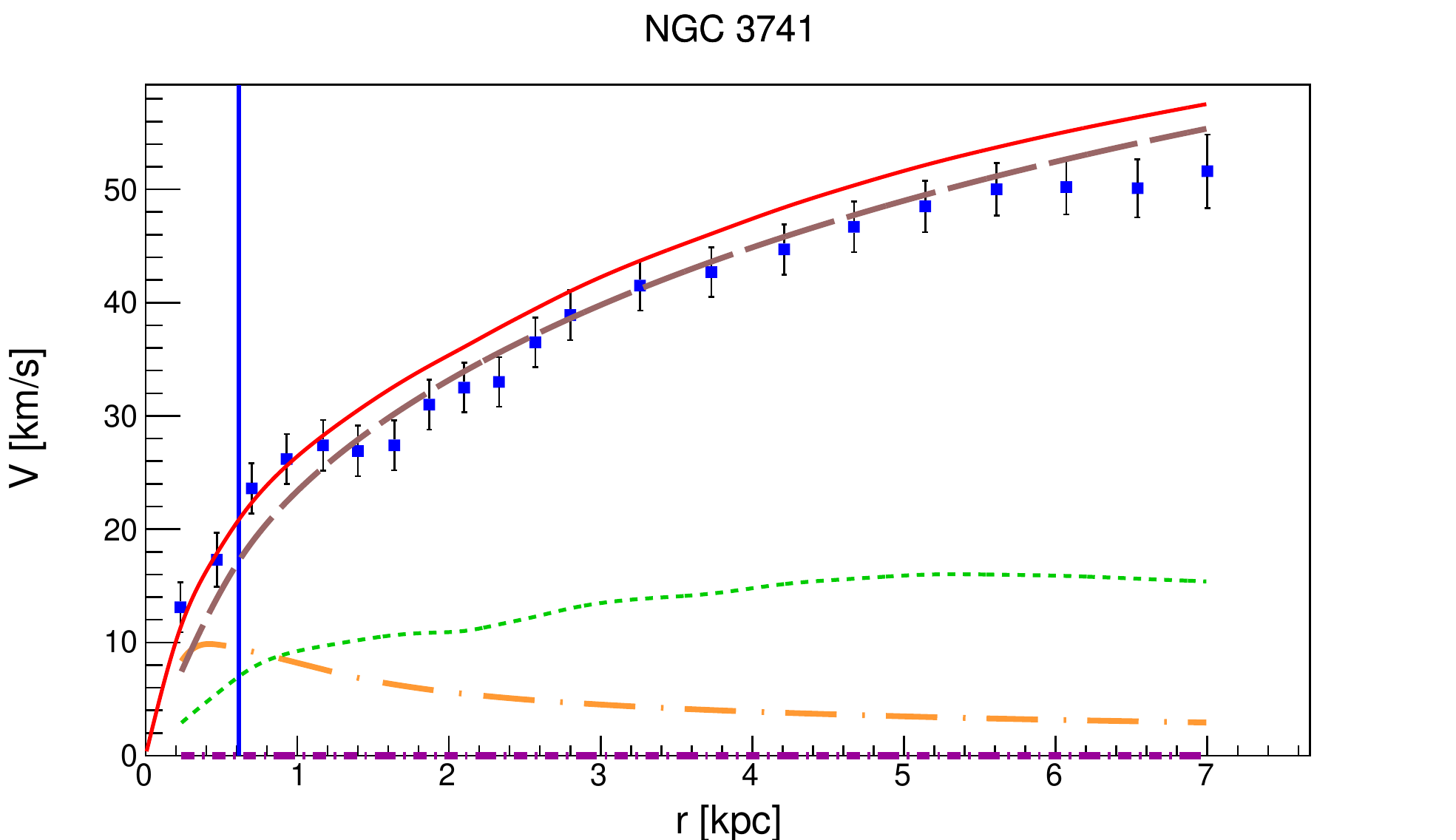} \\
\\
\\
\includegraphics[width=2.3in]{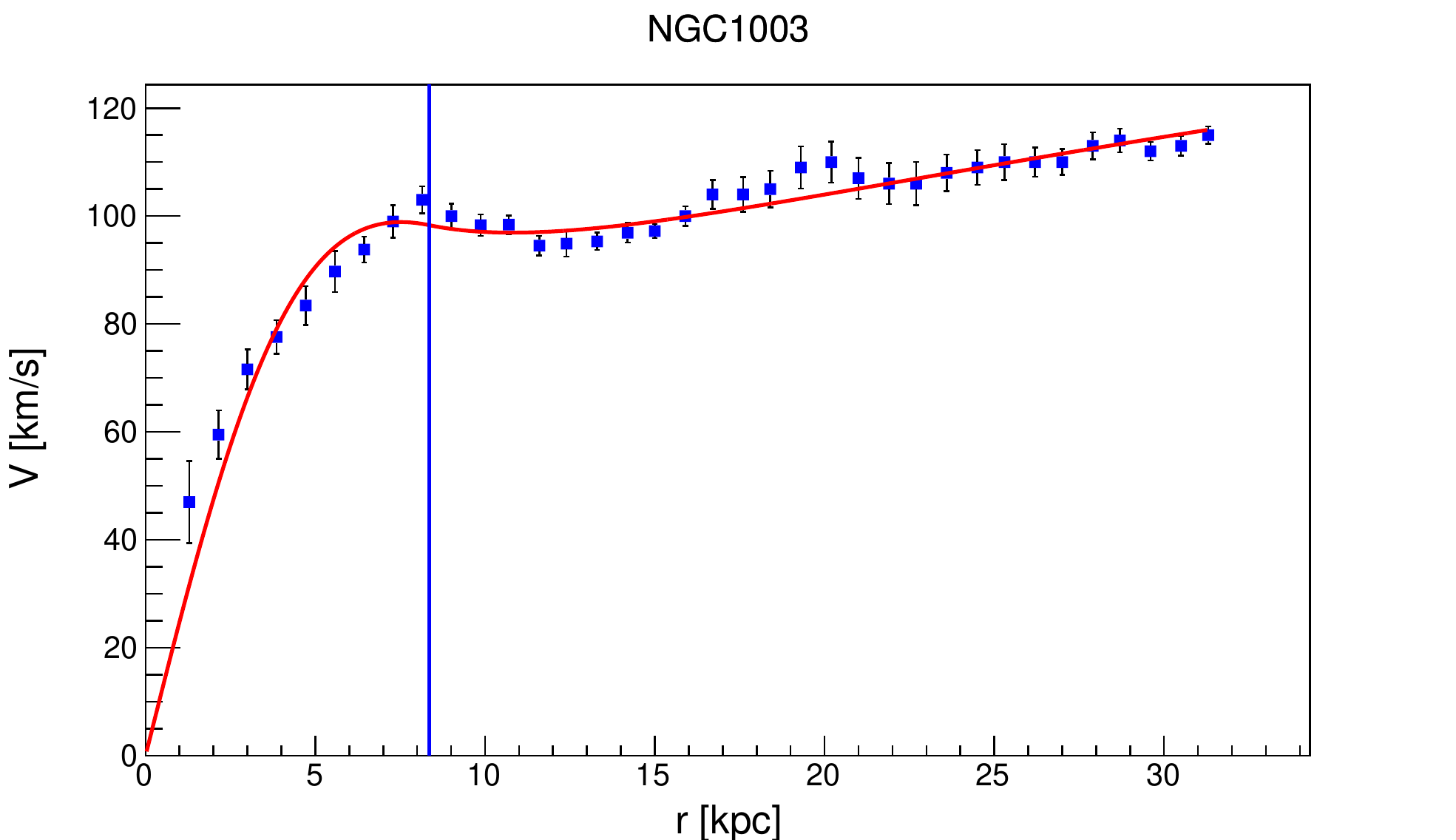} &
\includegraphics[width=2.3in]{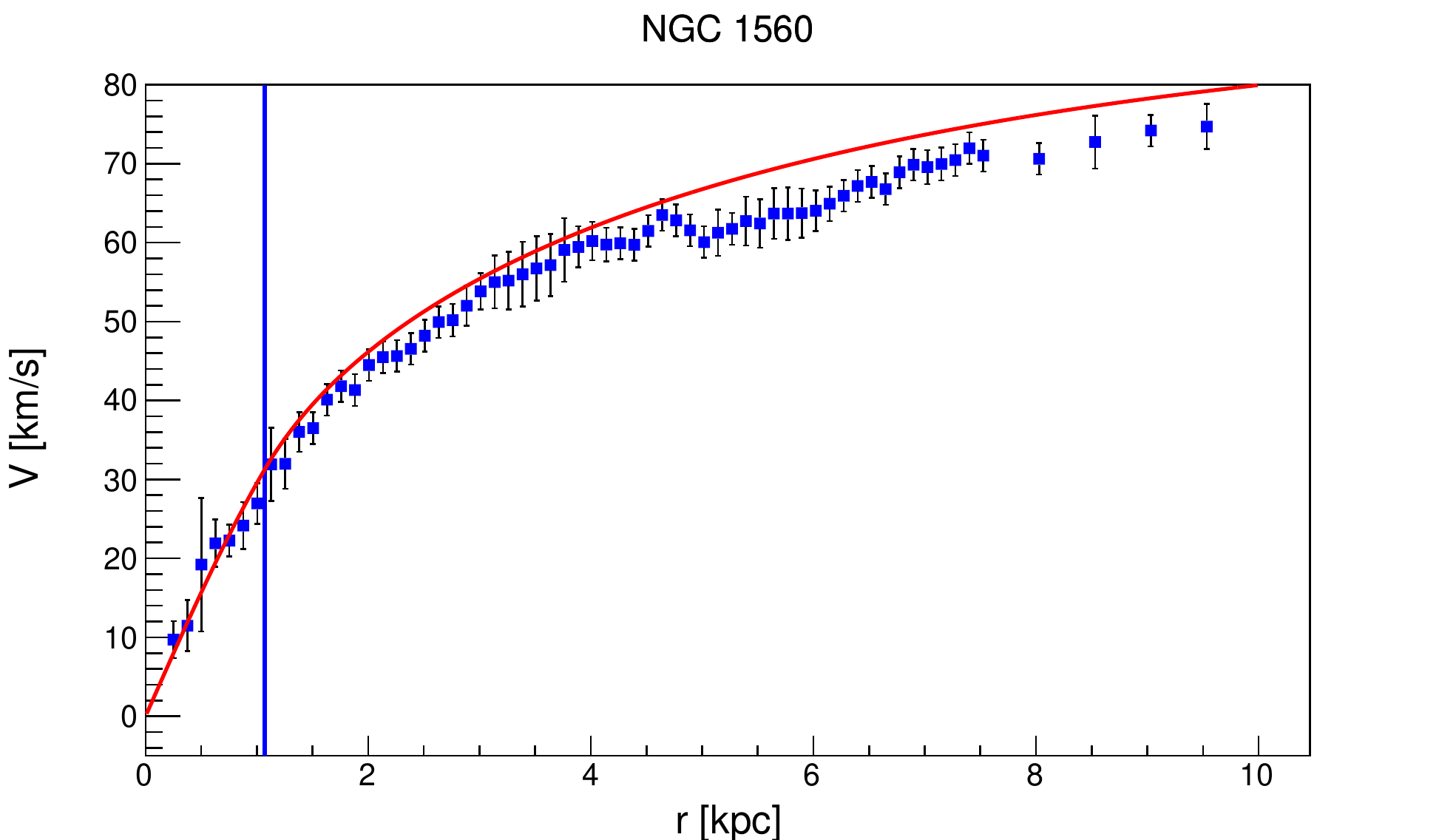} &
\includegraphics[width=2.3in]{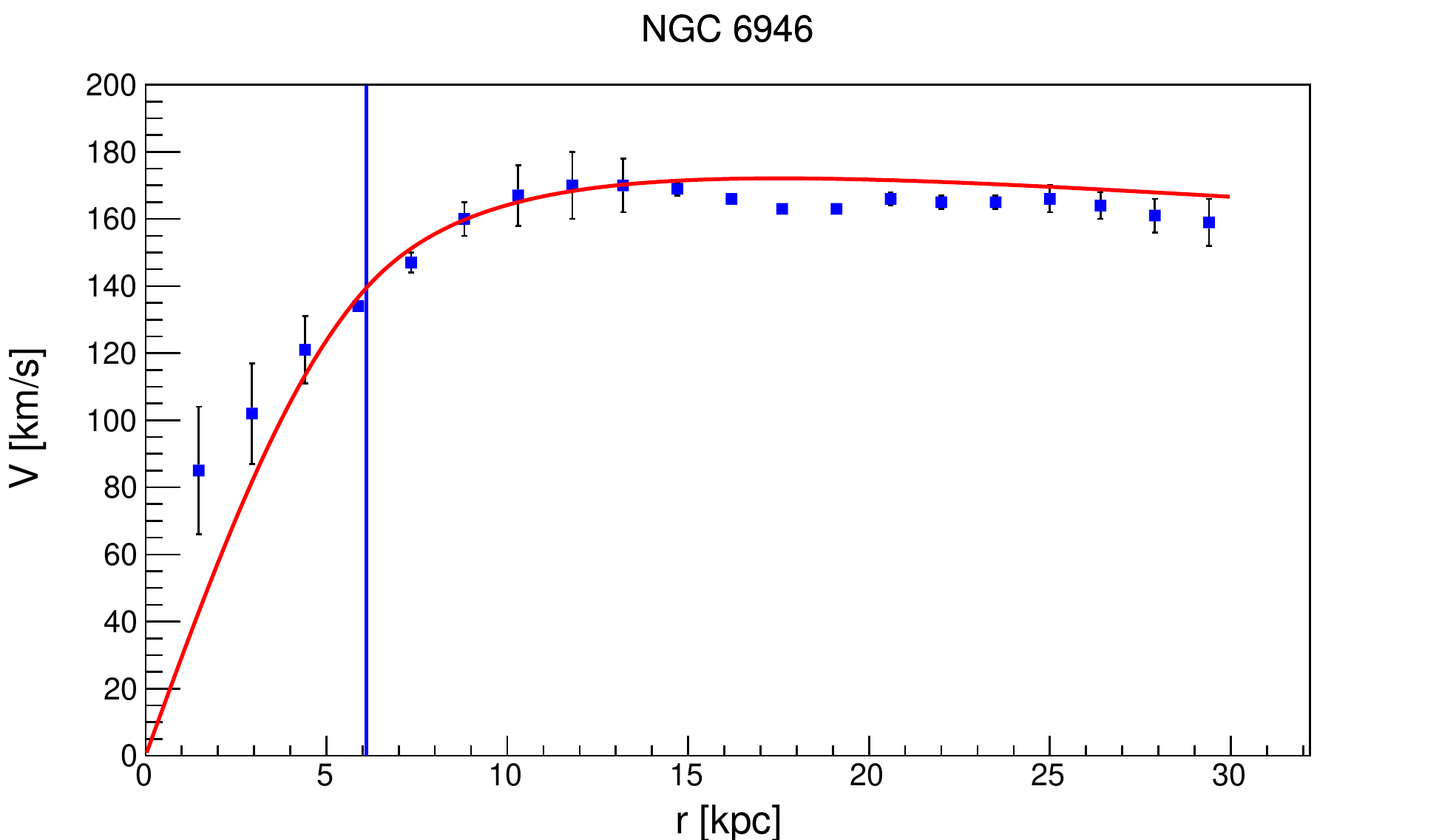} \\
\includegraphics[width=2.3in]{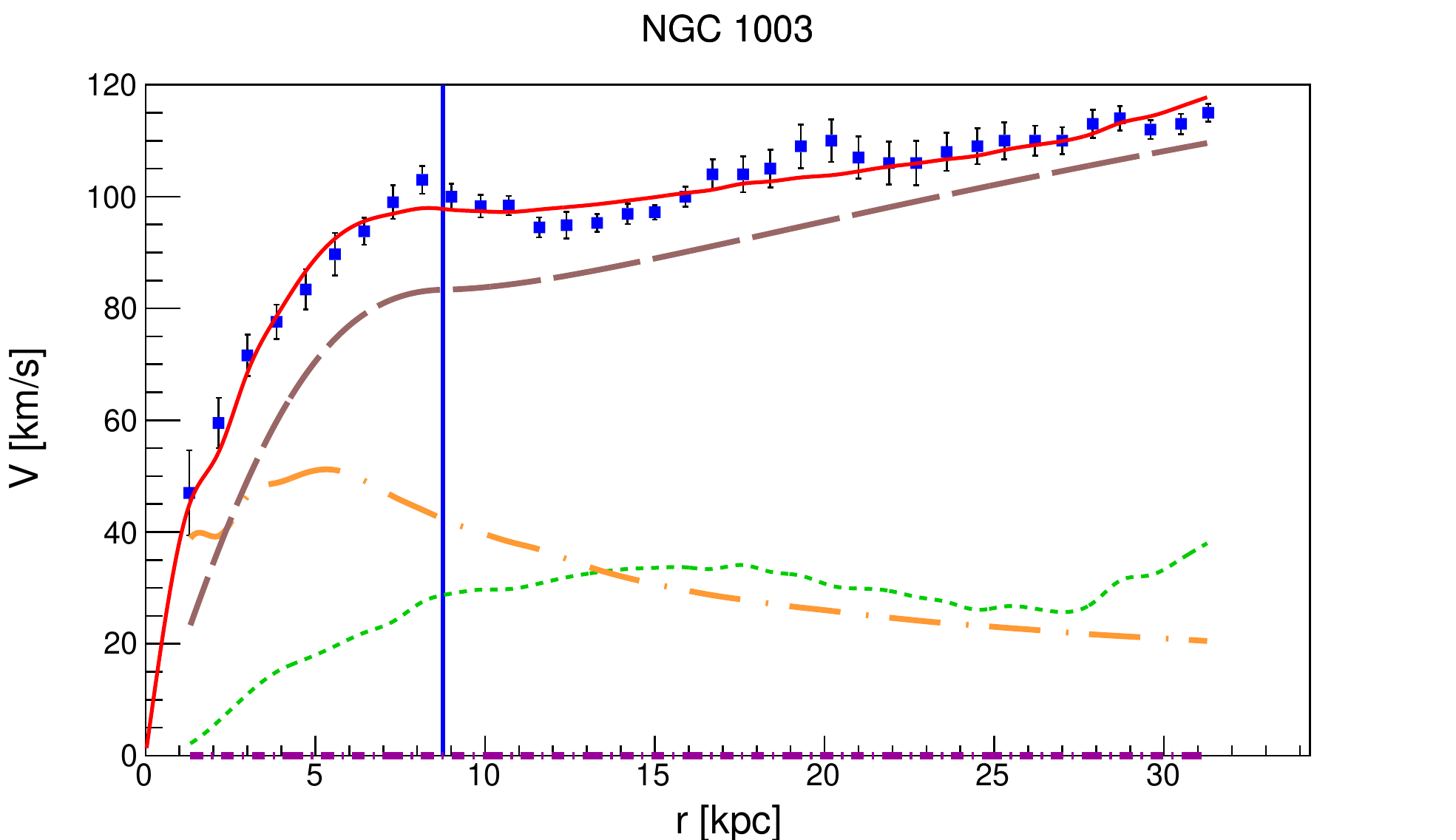} &
\includegraphics[width=2.3in]{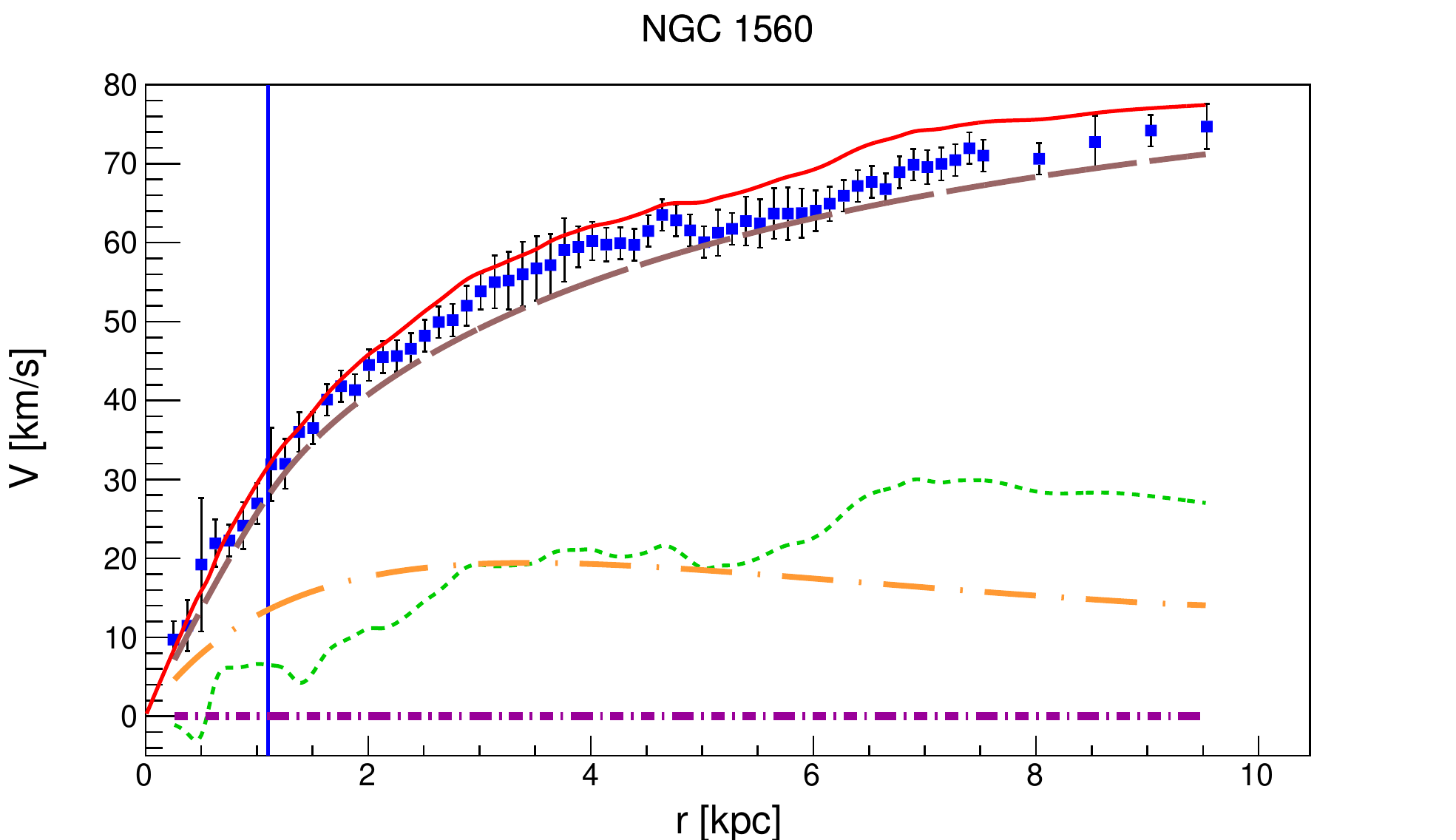} &
\includegraphics[width=2.3in]{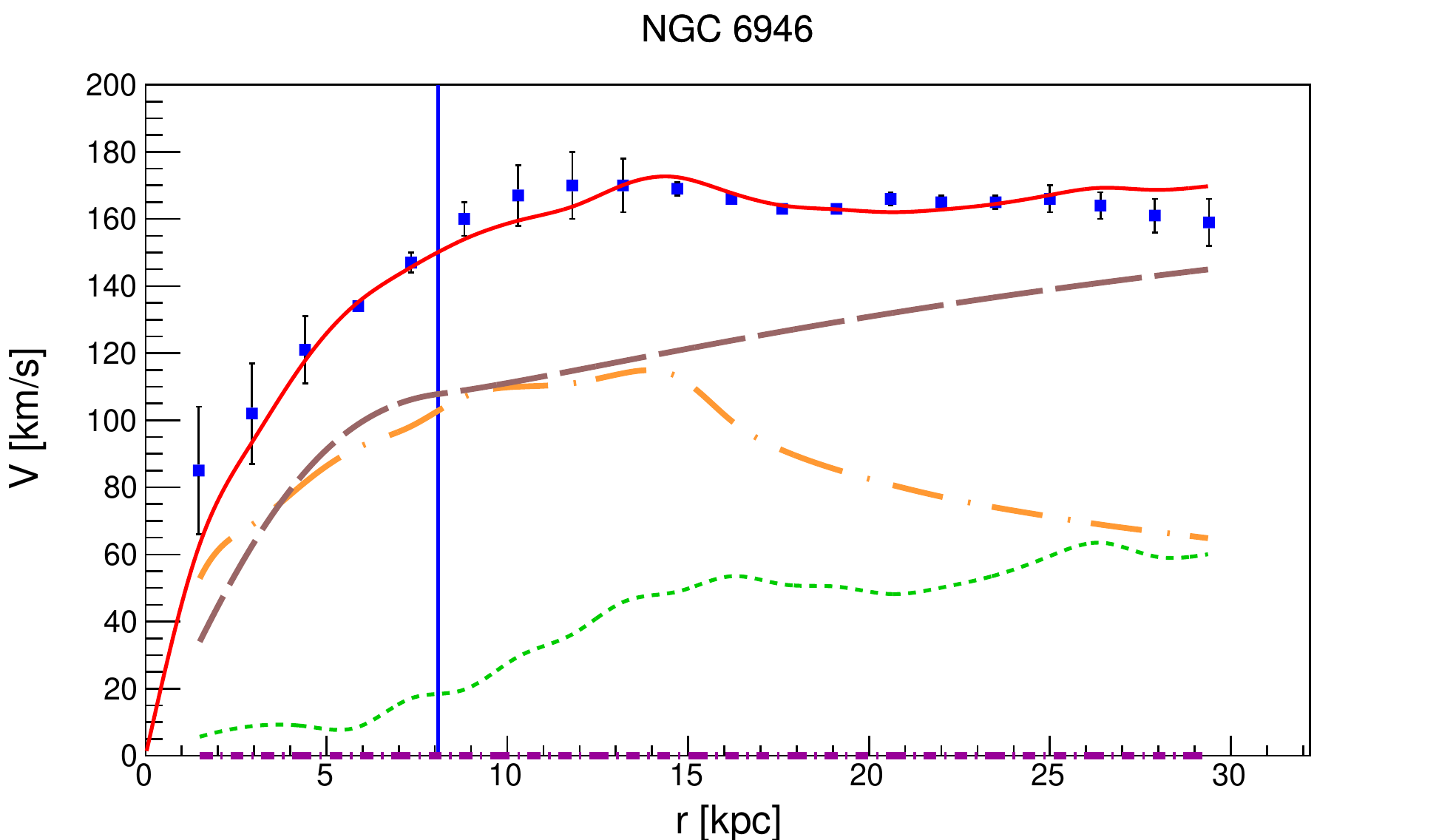}
    \end{array}$
    \end{center}
\caption{Best fits in the FDM model for the soliton+NFW profile for the NGC
galaxies with photometric data. The
fitting parameters are shown in Table~\ref{tab:results1}. We compare the
results without and taking into account the baryonic information. The
vertical lines are the corresponding transition radii $r_\epsilon$
from the MCMC method. For all the galaxies, the soliton contribution is less
concentrated at the inner radii once the baryonic data is taken into account.}
\label{fig1:Schive+NFW}
\end{figure*}

 In Fig.~\ref{fig1:Schive+NFW} we show the plots for the corresponding fitting
parameters presented in Table~\ref{tab:results1}, for the galaxies without and
with photometric data; the vertical lines show the corresponding transition
radius $r_\epsilon$ for each galaxy. Fig.~\ref{fig:ellipses-Schive} shows the $1
\sigma$ and $2\sigma$ contours for $m_\psi$ and $r_c$ for the NGC galaxies with
photometric data, where $r_\epsilon$ is the scatter plot.

We notice that, in general, the fits are improved with the baryons inclusion,
especially at the innermost regions. For NGC 7814, the bulge-dominated galaxy,
both fits (with and without photometric data) are very good; in this case the
soliton is capable to reproduce the bulge behavior. For NGC 6503 and NGC 6946,
both disk-dominated galaxies, the innermost regions are reproduced once the
baryons are included. We have the same for NGC 1003, but
as can be noticed, even when the fit is better, the soliton+NFW profile cannot
explain the oscillations from the high-resolution observations; in this case the
mSFDM profile is relevant \citep[see][and Subsection~\ref{subsec:Multistates}]
{Robles:2013}. Finally, for NGC 3741 and 1560, the total velocities overestimate
the observational points for the outer radii, even once the baryons have been
included.

The central soliton densities are larger in the DM-only analysis since the
soliton must fit the rotation curves at the innermost radii where the baryons
might contribute more, especially in NGC 7814. As can be easily seen from
Fig.~\ref{fig1:Schive+NFW}, we notice that in all cases the transition radii
increase once the baryons have been included, consistent with those densities.
The same occurs for the soliton core radii (except for NGC 6946). Therefore, the
soliton is less concentrated than in the DM-only study.

Finally, the boson masses for the DM+baryons analysis are in the range $0.212 \leq
m_\psi / 10^{-23}\mathrm{eV} \leq 12.1$, and core radii within $0.759 \leq r_c/
\mathrm{kpc} \leq 5.47$. We notice again that 4 of 6 galaxies have corresponding
boson masses $m_\psi<10^{-23}\mathrm{eV}$, a result that does not meet the
cosmological constraints \citep{Bozek:2014,Sarkar:2016}, setting a possible
tension between the FDM model and the rotation curves of these galaxies.

%%%%%%%%%%%%%%%% MULTISTATE SFDM %%%%%%%%%%%%%%%%
\subsection{Multistate SFDM fits}
\label{subsec:Multistates}

Here we present the results for the multistate SFDM model, with the density
profile~\eqref{densitytotal}. The reported combinations of excited states are the
ones that best reproduce the data.

The preferred state we are looking for is the ground state ($j=1$), and
combinations including it, since we expect the galaxy halos in the SFDM scenario
were formed from a BEC with this initial condition. As the halos evolve, a
realistic approximation describing the bosons in such configurations is to expect
that those particles reach higher energy levels due to galaxy and galaxy clusters
formation processes and their interaction with the baryonic matter. Within this
hypothesis, it is reasonable to expect lower excitation states, since they are
more easily attainable than the higher levels, and for the rotation
curves analyzed here, radii corresponding to lower excitation states are enough
to fit the observations. Therefore we preferred lower states to restrict the
freedom in $j$. Further numerical simulations are
needed to investigate the distribution of excitation states in diverse
environments and under different initial conditions, and their comparison with
observations.

Now, it is a well-known result that it is not possible to reproduce the whole
rotation curves of large galaxies with the ground state only, since the
corresponding velocity profile decreases very quickly after its maximum value
\citep{Boehmer:2007,Robles:2012uy,Guzman:2015jba}. As shown below, it was
possible to fit only some LSB galaxies with the single ground state $j=1$ and
none of the NGC galaxies. Furthermore, it is important to note that
in some cases, the resulting configurations do not display the ground state as a
dominant component. This might be the result of different formation histories
for each galaxy, including interactions with the baryonic matter resulting in
diverse multistate configurations.

\subsubsection{High-resolution LSB galaxies}
\label{subsubsec:lsb-2}

\begin{figure*}
    \begin{center}$
    \begin{array}{ccc}
\includegraphics[width=2.3in]{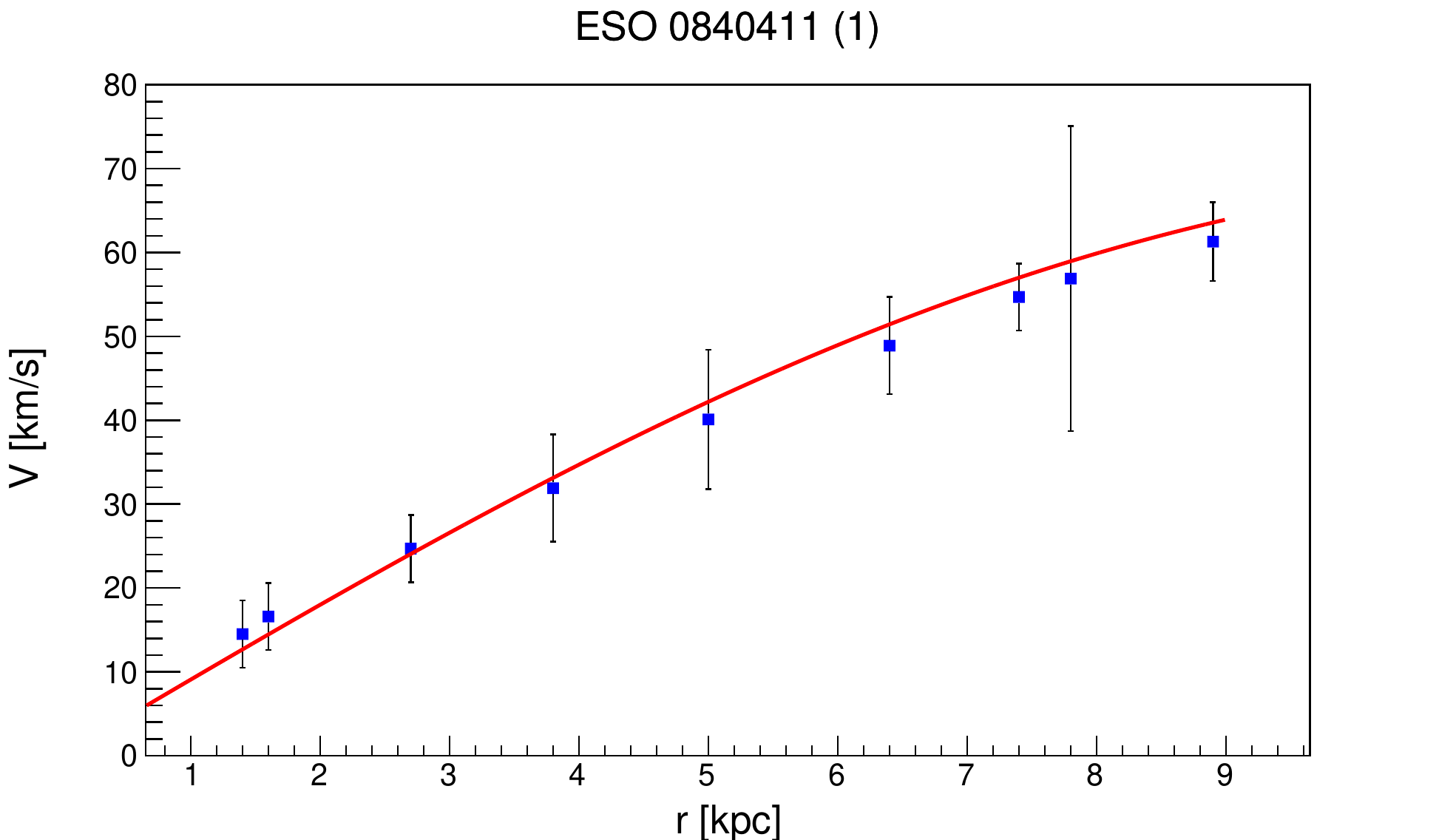}&
\includegraphics[width=2.3in]{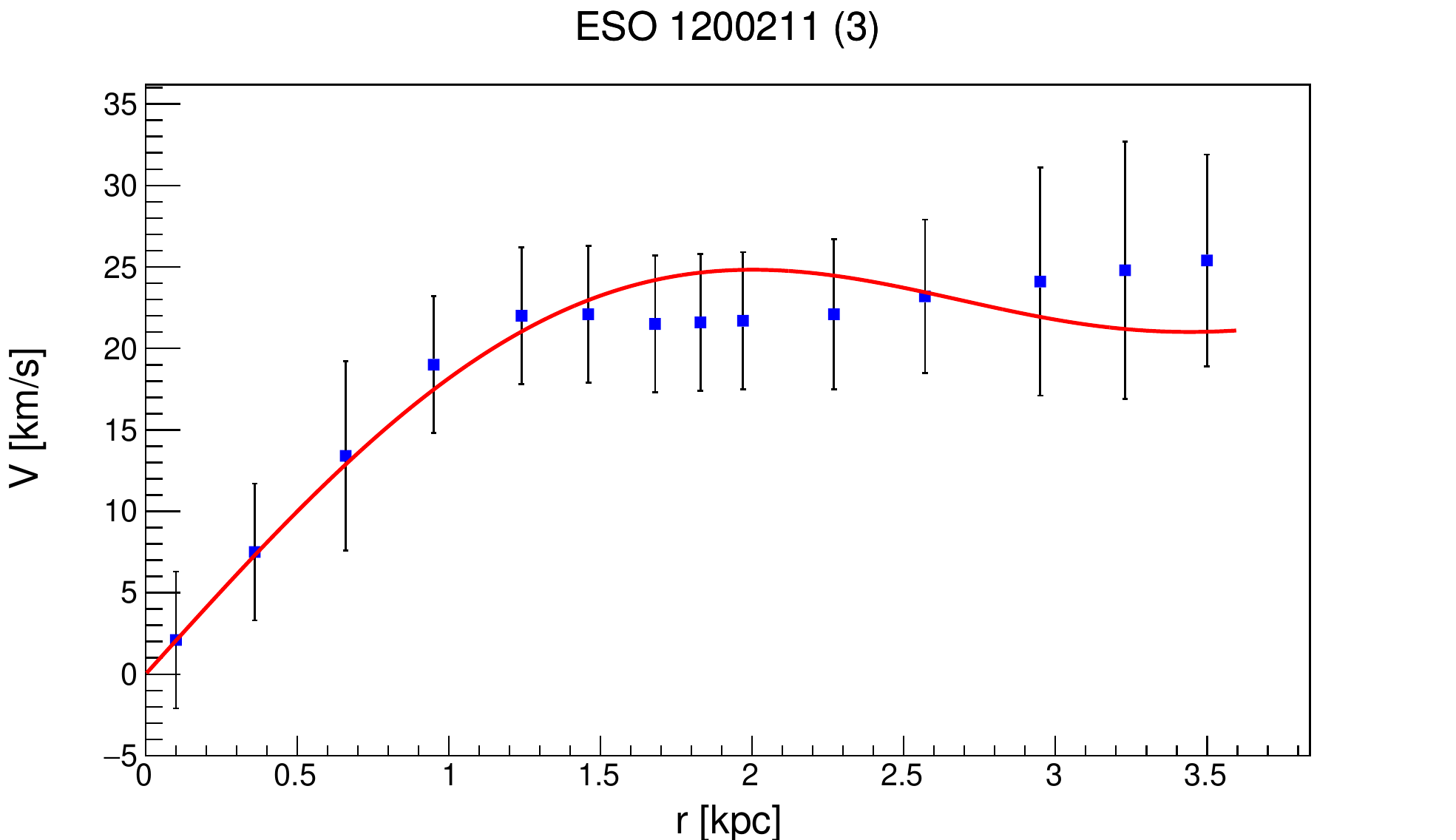}&
\includegraphics[width=2.3in]{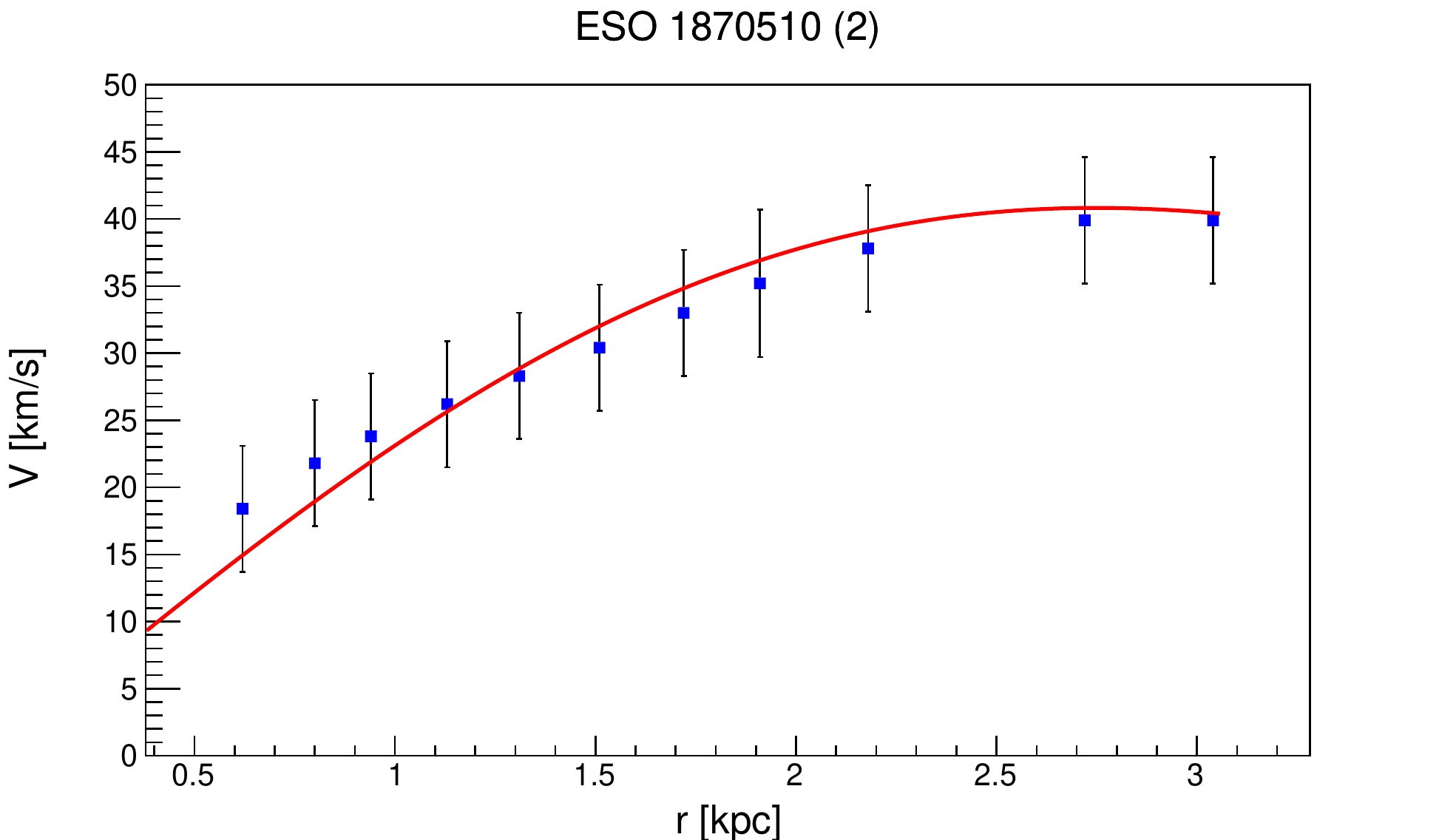}\\
\includegraphics[width=2.3in]{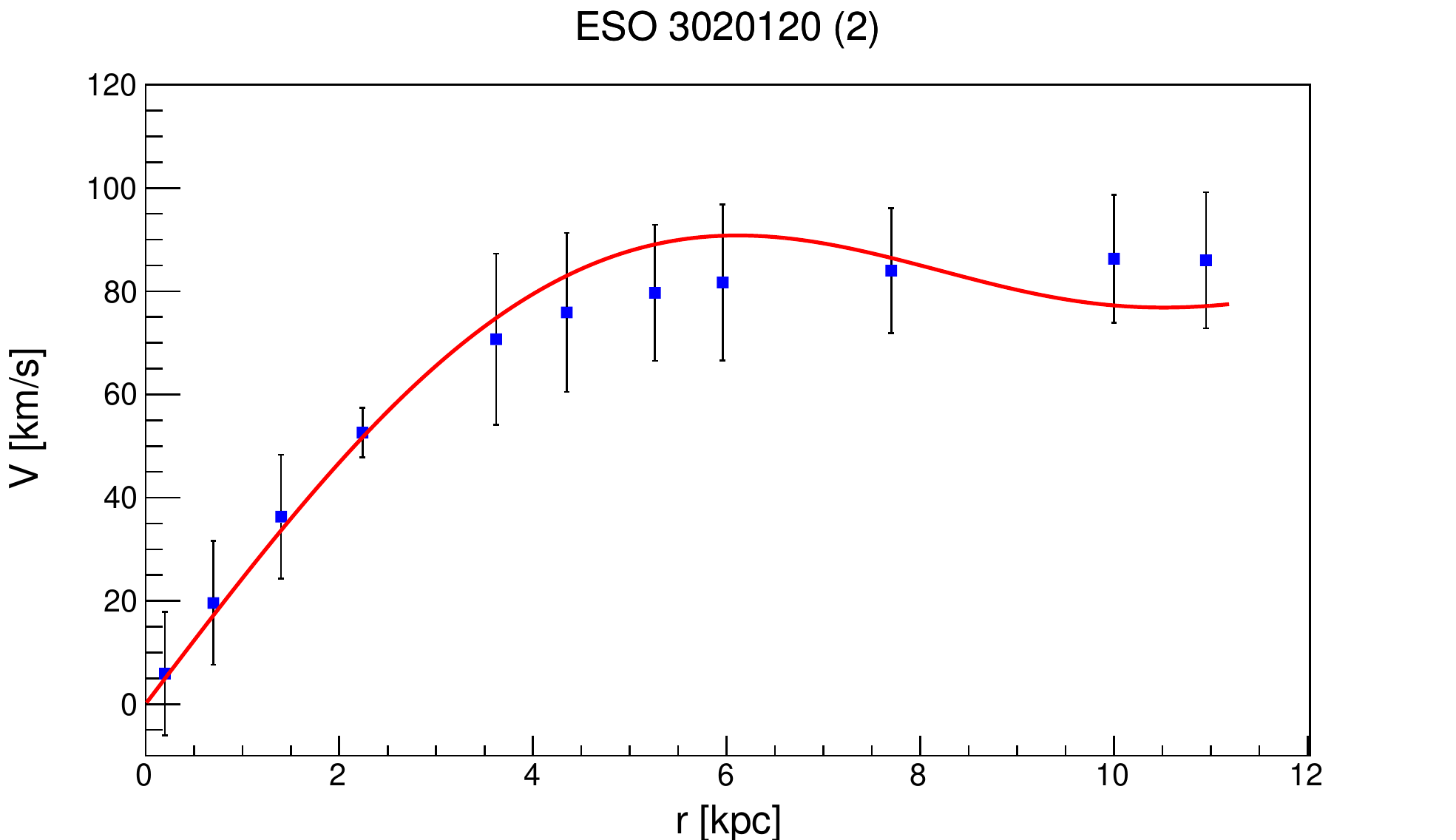}&
\includegraphics[width=2.3in]{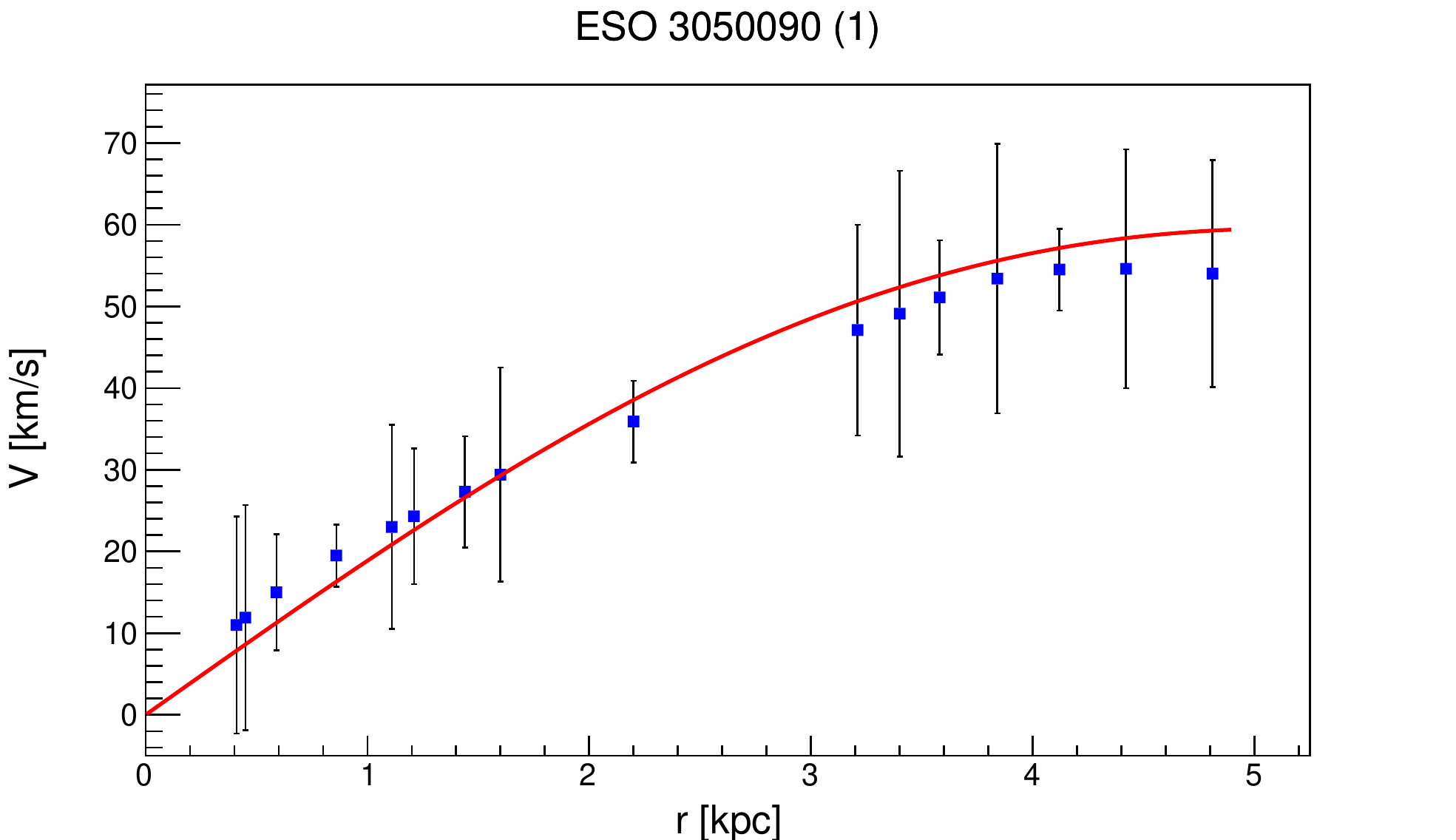}&
\includegraphics[width=2.3in]{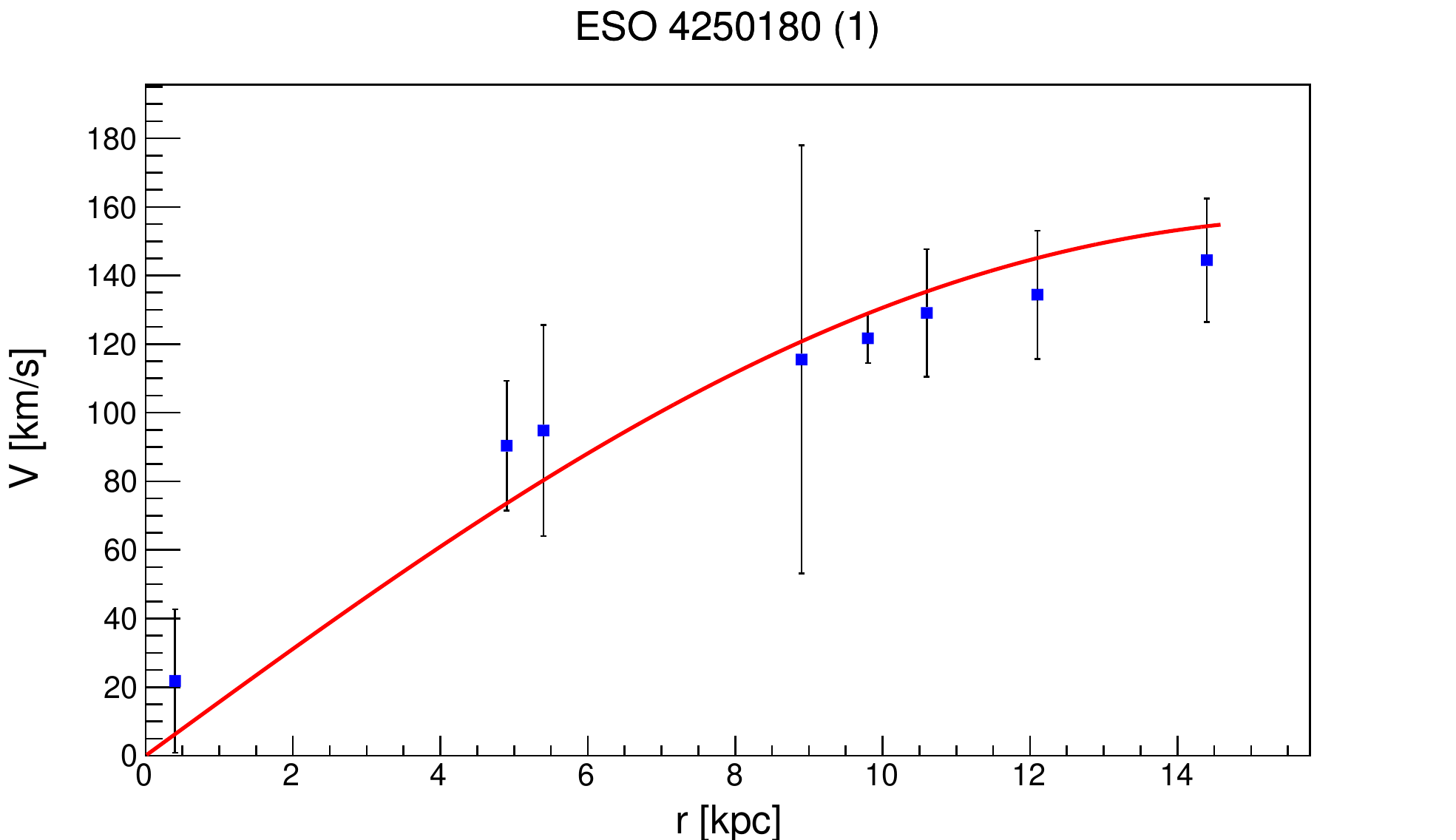}\\
\includegraphics[width=2.3in]{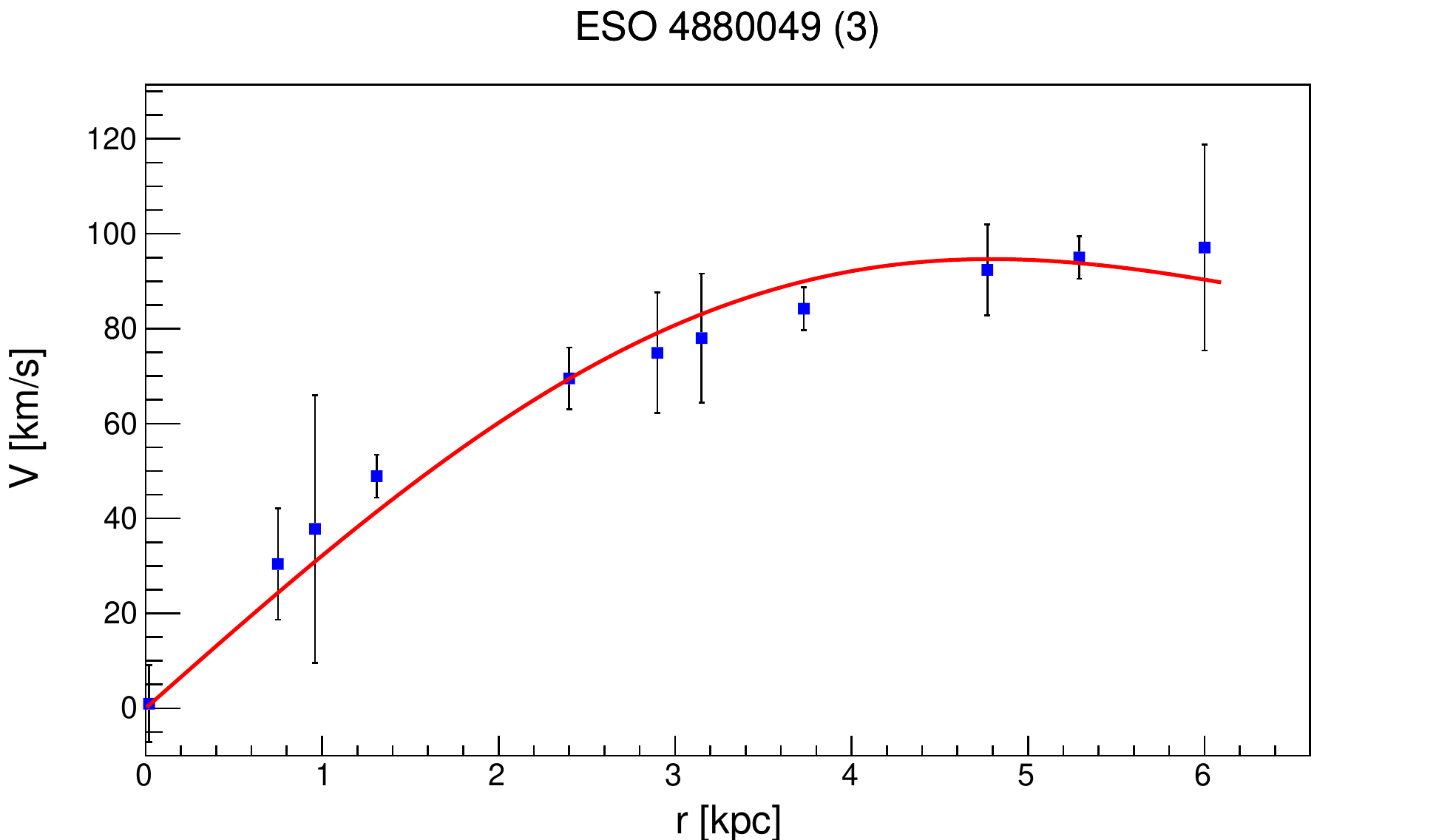}&
\includegraphics[width=2.3in]{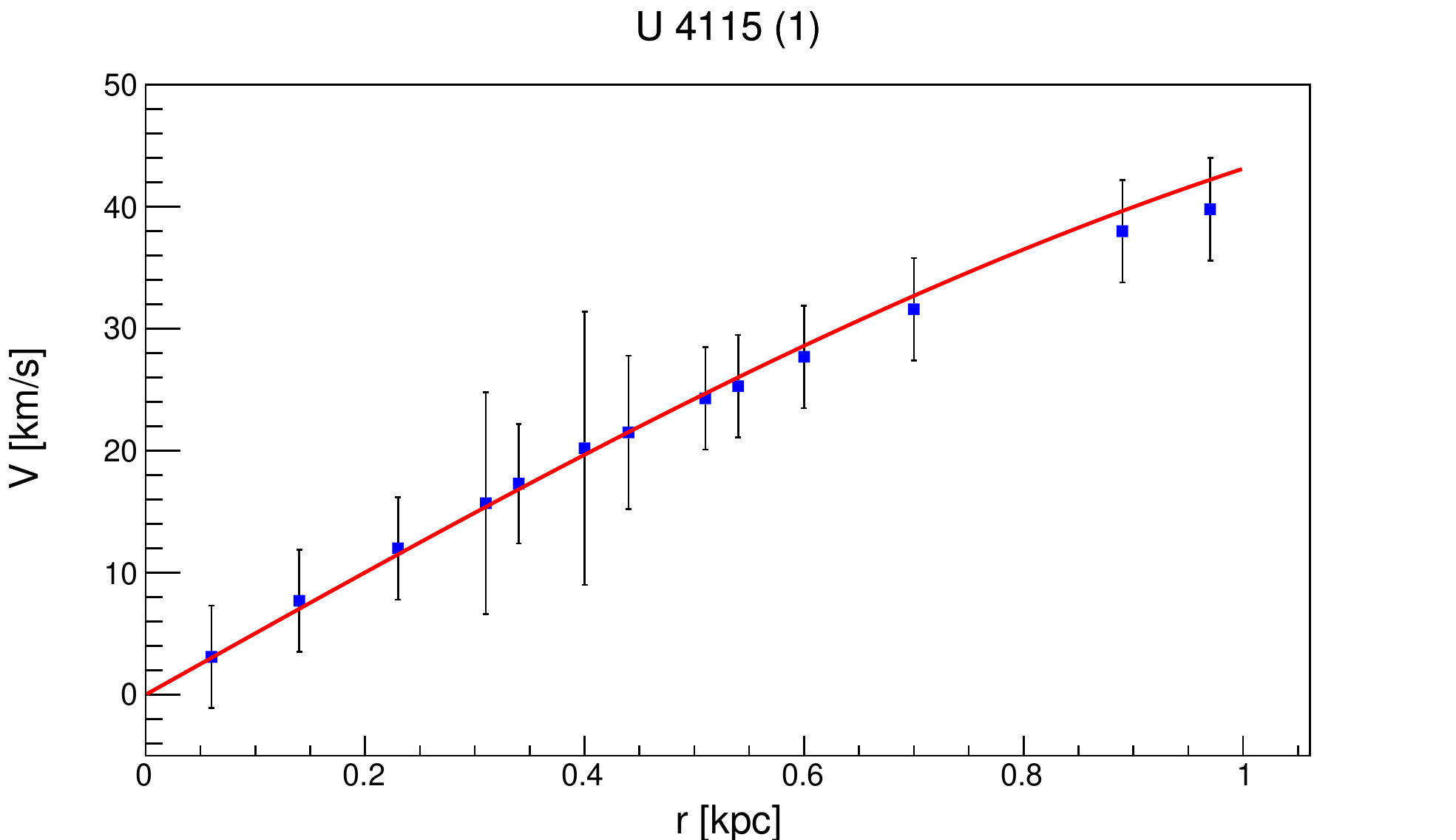} &
\includegraphics[width=2.3in]{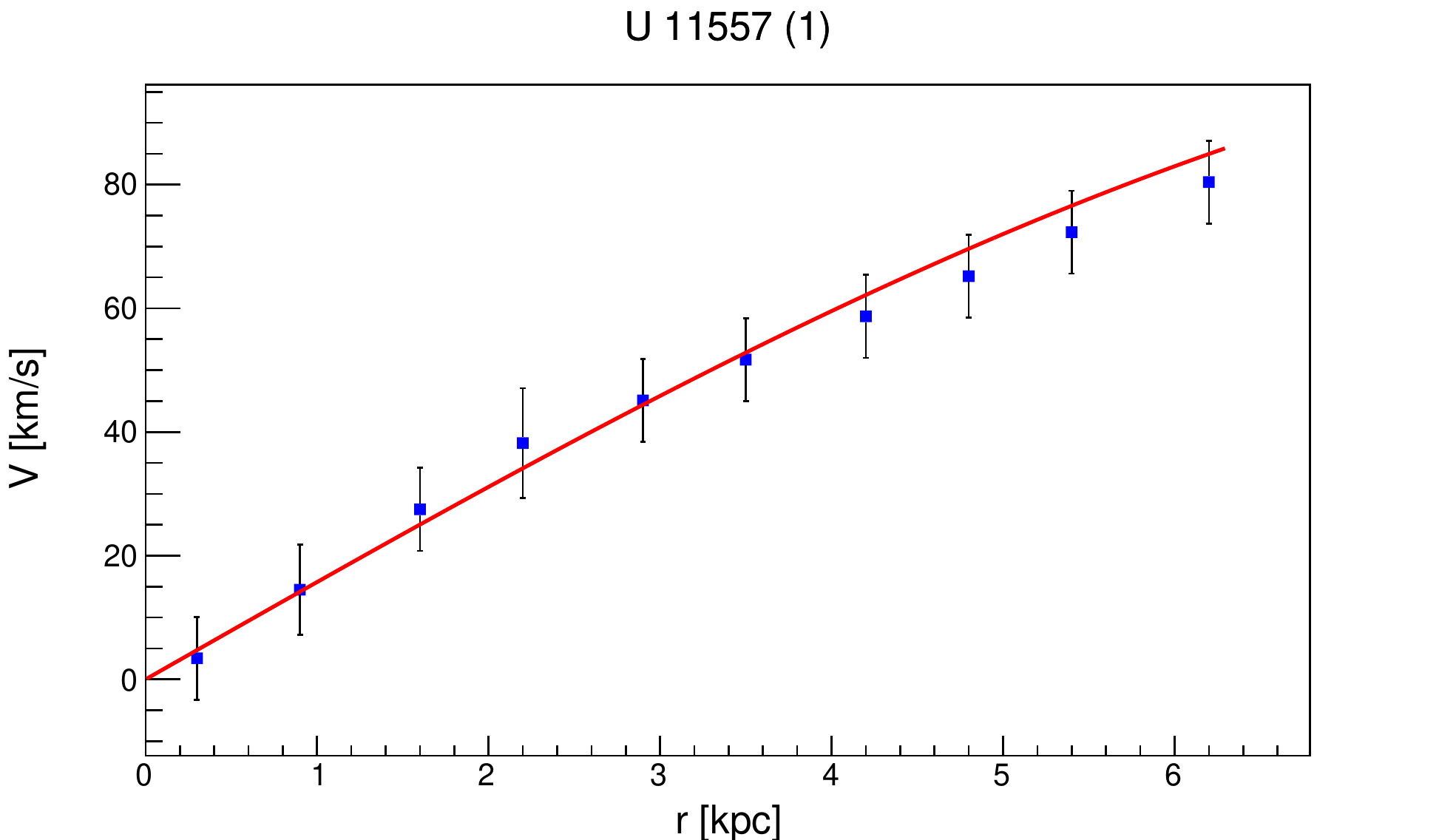}\\
\includegraphics[width=2.3in]{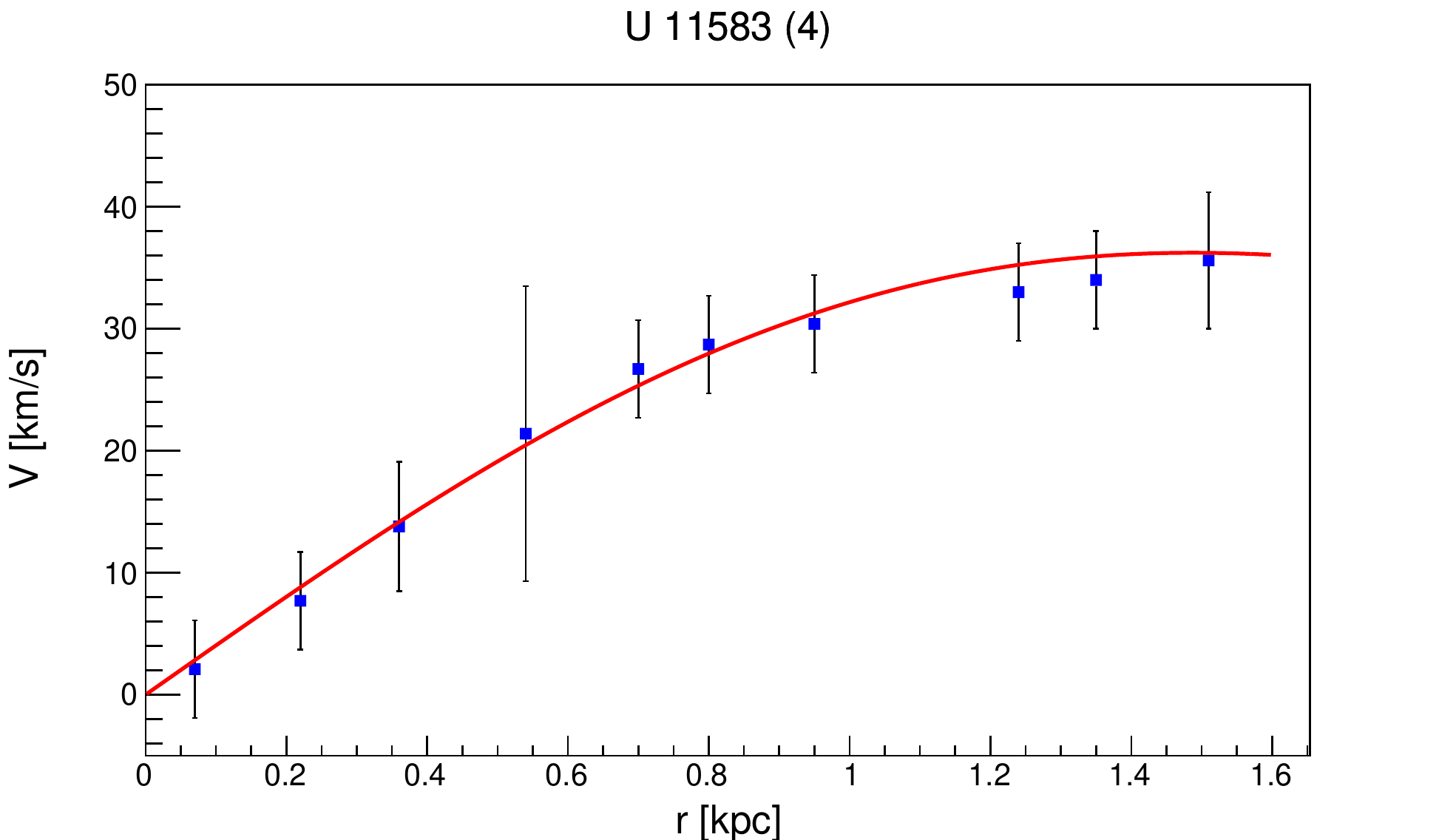}&
\includegraphics[width=2.3in]{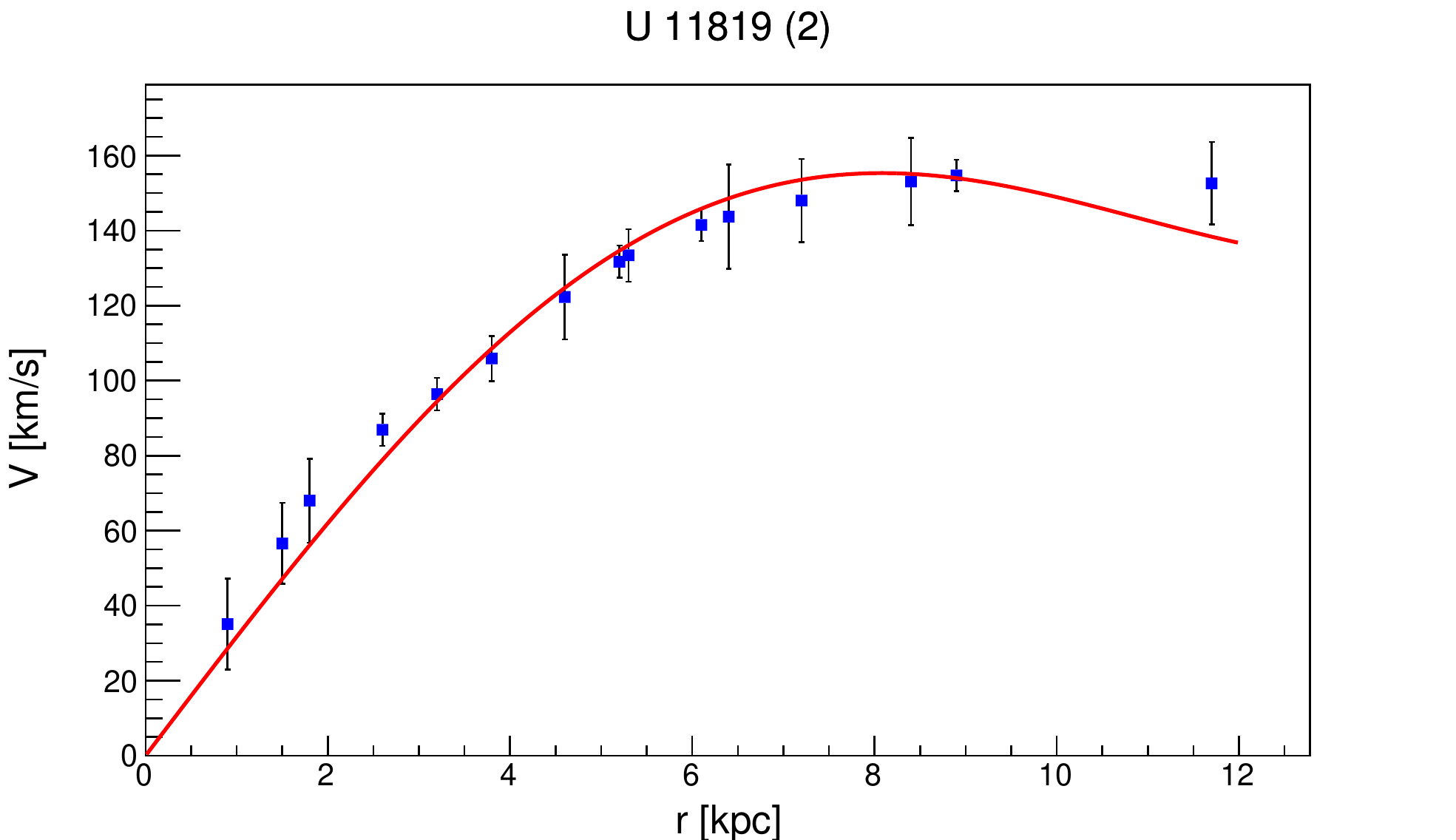}
 \end{array}$
\end{center}
\caption{Best-fit profiles for the high-resolution LSB galaxies reported in
Table~\ref{tab:mSFDM-1s}, in the mSFDM model with one excited state $j$.}
\label{fig:mSFDM-1s}
\end{figure*}

  In Table~\ref{tab:mSFDM-1s} we show the results for 11 of the 18
high-resolution LSB galaxies from \citet{deBlok:2001}, whose rotation curves can
be reproduced with one state only. From them, only 5 can be fitted with the
ground state $j=1$, and the other ones with an excited state $j>1$. We report the
fitting parameters $j$, $R$ and the central density for the $j$-th state,
$\rho_0^j$, all the quantities $\pm 1\sigma$ errors from the MCMC method used. We
report also the corresponding mSFDM mass $M_j = M_\mathrm{mSFDM}(R)$ and
$\chi^2_\mathrm{red}$ errors for each galaxy. In Fig.~\ref{fig:mSFDM-1s} we show
the corresponding rotation curves for these galaxies, showing that the fits are
well inside the observational errors.

\begin{table*}
\caption{Multistate SFDM with one excited state in high-resolution LSB galaxies.
In this Table we show the fitting parameters $R$ and $\rho_0^j$ for one excited
state $j$, $\pm1\sigma$ errors from the MCMC method used, in the multistate SFDM
model~\eqref{densitytotal} for 11 high-resolution LSB galaxies in \citet{deBlok:2001}.
We also report the resulting DM mass $M_j$ and $\chi^2_\mathrm{red}$ errors from
the fitting method.}
\label{tab:mSFDM-1s}
\begin{tabular}{lccccc}
    \hline
    Galaxy & $j$ & $R$ & $\rho_0^j$ & $M_j$ & $\chi^2_\mathrm{red}$ \\
     & & $(\mathrm{kpc})$ & $(10^{-2}M_\odot/\mathrm{pc}^3)$ & $(10^{10}M_\odot)$ & \\
\hline
ESO 084-0411   & 1 & $18.0^{+2.1}_{-5.4}$  & $0.461^{+0.058}_{-0.10}$ & 0.836  & 0.347 \\
ESO 120-0211   & 3 & $8.4^{+1.2}_{-2.1}$   & $2.36^{+0.47}_{-1.0}$    & 0.0359	& 0.3062 \\
ESO 187-0510   & 2 & $7.68^{+0.62}_{-1.2}$ &	$3.39^{+0.52}_{-0.64}$   & 0.116	& 0.266 \\
ESO 302-0120   & 2 & $17.1^{+2.2}_{-2.8}$  &	$3.38^{+0.60}_{-0.83}$   & 1.51  & 0.383 \\
ESO 305-0090   & 1 & $7.20^{+0.53}_{-1.9}$ &	$2.05^{+0.37}_{-0.52}$   & 0.393	& 0.202 \\
ESO 425-0180   & 1 & $23.3^{+2.2}_{-7.5}$  &	$1.37^{+0.21}_{-0.41}$   & 7.97  & 1.11 \\
ESO 488-0490   & 3 & $20.1^{+1.2}_{-1.9}$  &	$5.99^{+0.71}_{-0.88}$   & 1.14	& 0.866 \\
UGC 4115          & 1 & $2.52^{+0.35}_{-1.1}$ & $14.1^{+1.6}_{-3.0}$      & 0.0402  & 0.086 \\
UGC 11557         & 1 & $16.4^{+1.5}_{-7.2}$  &	$1.38^{+0.16}_{-0.29}$   & 1.04   & 0.400 \\
UGC 11583         & 4 & $8.32^{+0.94}_{-1.8}$ &	$9.1^{+1.4}_{-2.6}$      & 0.046  & 0.176 \\
UGC 11819         & 2 & $22.6^{+0.75}_{-0.84}$ &$5.67\pm0.30$ 			   & 5.21	& 0.987 \\
\hline
\end{tabular}
\end{table*}

In Table~\ref{tab:results5} we show the results for the remaining 7 LSB galaxies,
reproduced with two excited states $i,j$. We report the fitting parameters,
$\rho_0^i$ and $\rho_0^j$ (the central densities for the states $i$ and $j$)
and the halo radius $R$, all the quantities $\pm 1\sigma$ errors.  We also
computed the mass in the highest energy level $M_j$ and in the ground state $M_1$
(for the galaxies displaying $j=1$), as well as the mass ratio $\eta=M_j/M_1$.
\citet{urena-bernal10} showed it is possible to have stable multistate
configurations of the Schr\"odinger-Poisson system for values of $\eta=M_j/M_1
\lesssim 1.3$, i.e. for this values in the simulations, the multistate halos do
not decay to the ground state. Therefore, from our results, we might argue that
the final configurations are stable where the ground state appears as a
dominant component. Further numerical simulations are needed to test the
validity of those results in the context of the mSFDM analytical model.

\begin{figure*}
    \begin{center}$
    \begin{array}{ccc}
\includegraphics[width=2.3in]{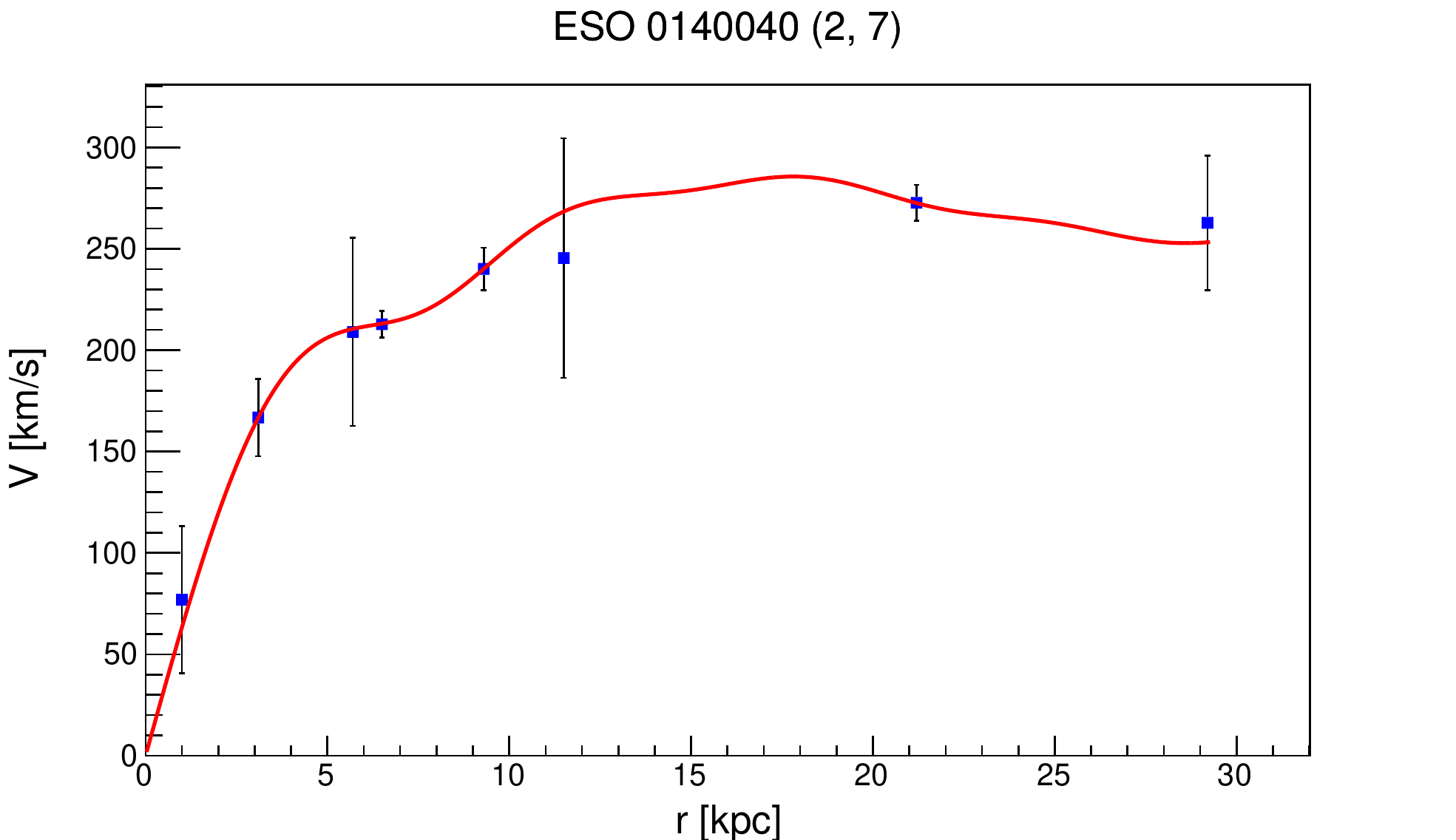} &
\includegraphics[width=2.3in]{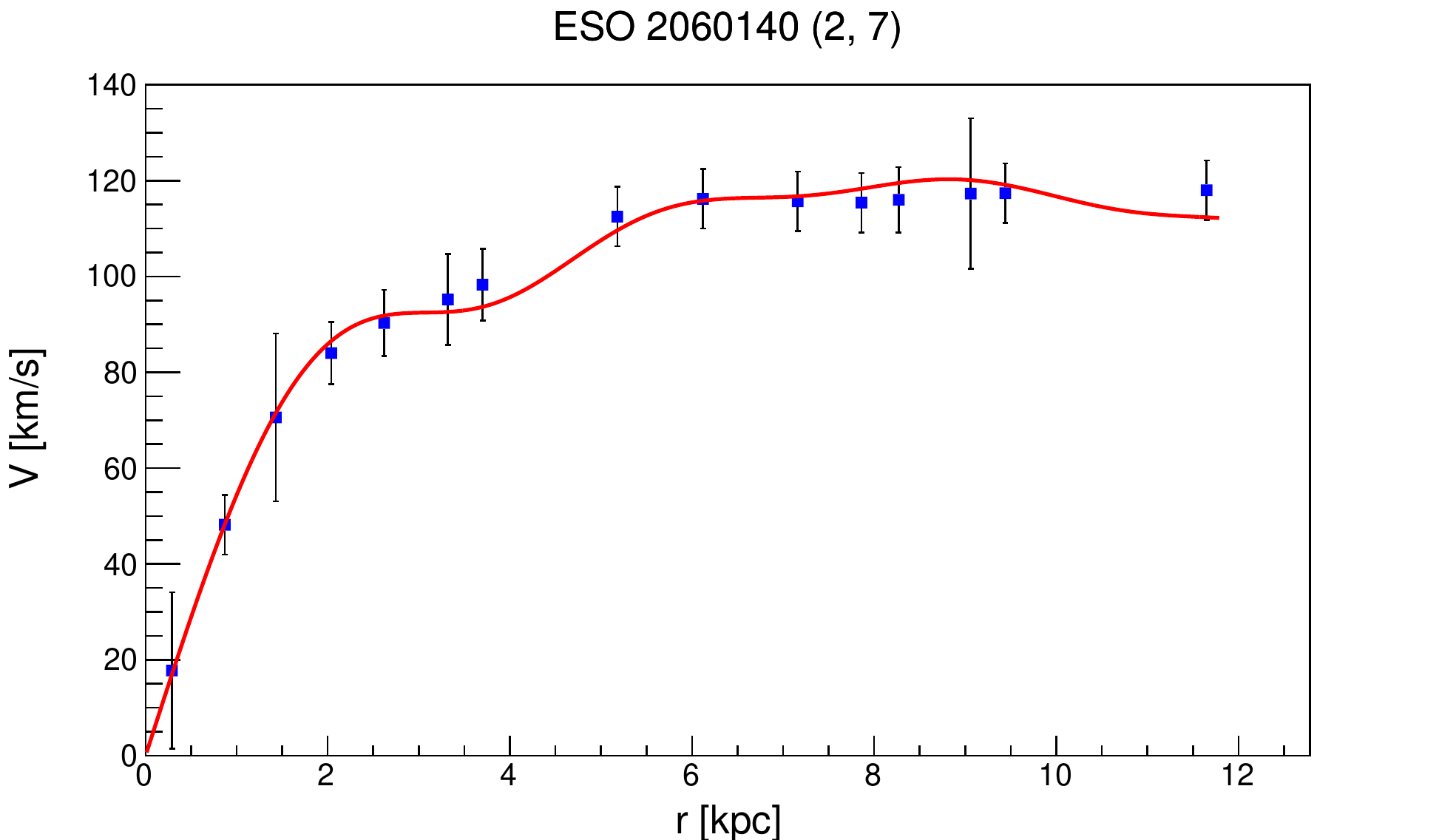} &
\includegraphics[width=2.3in]{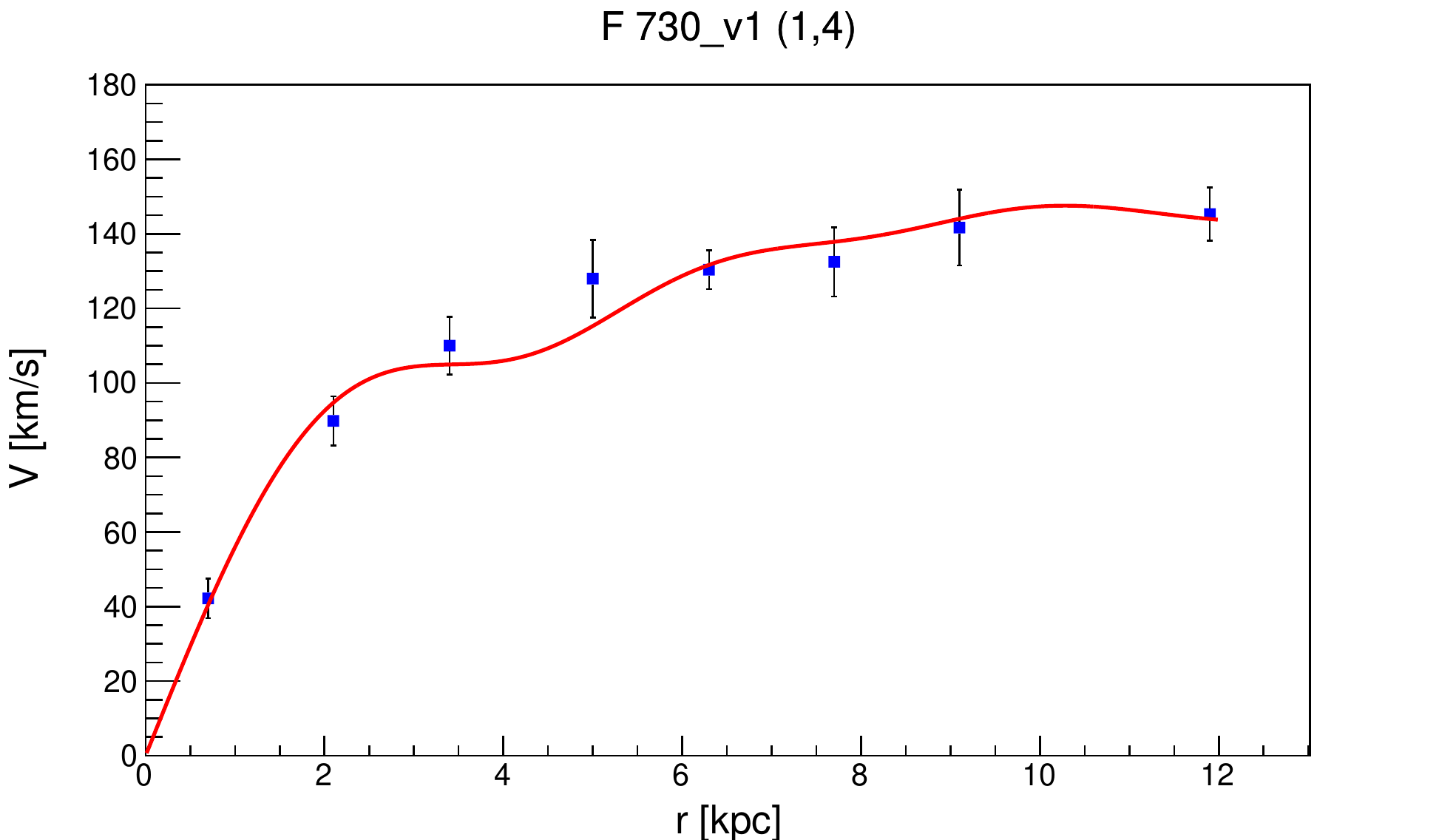} \\
\includegraphics[width=2.3in]{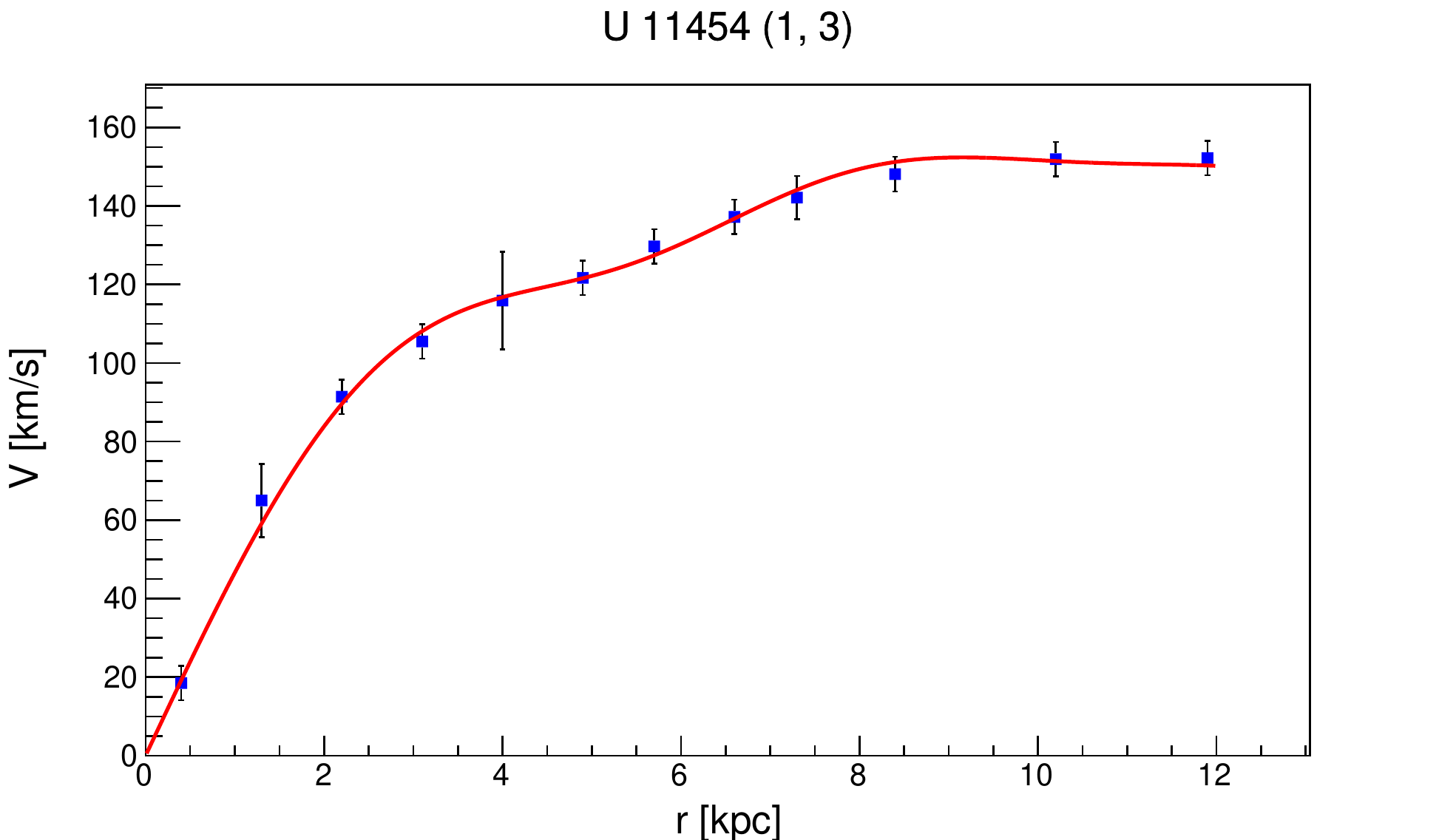} &
\includegraphics[width=2.3in]{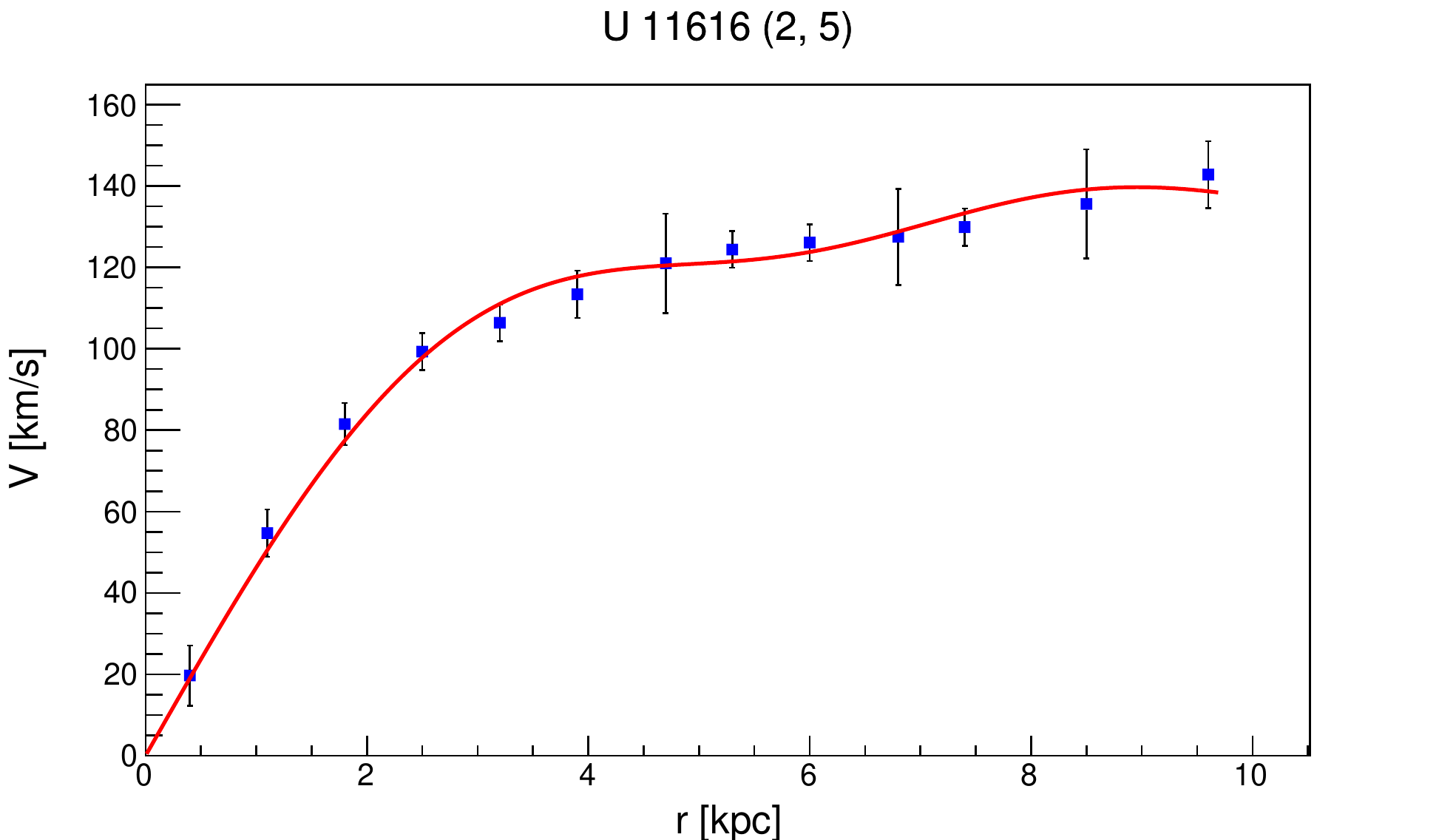} &
\includegraphics[width=2.3in]{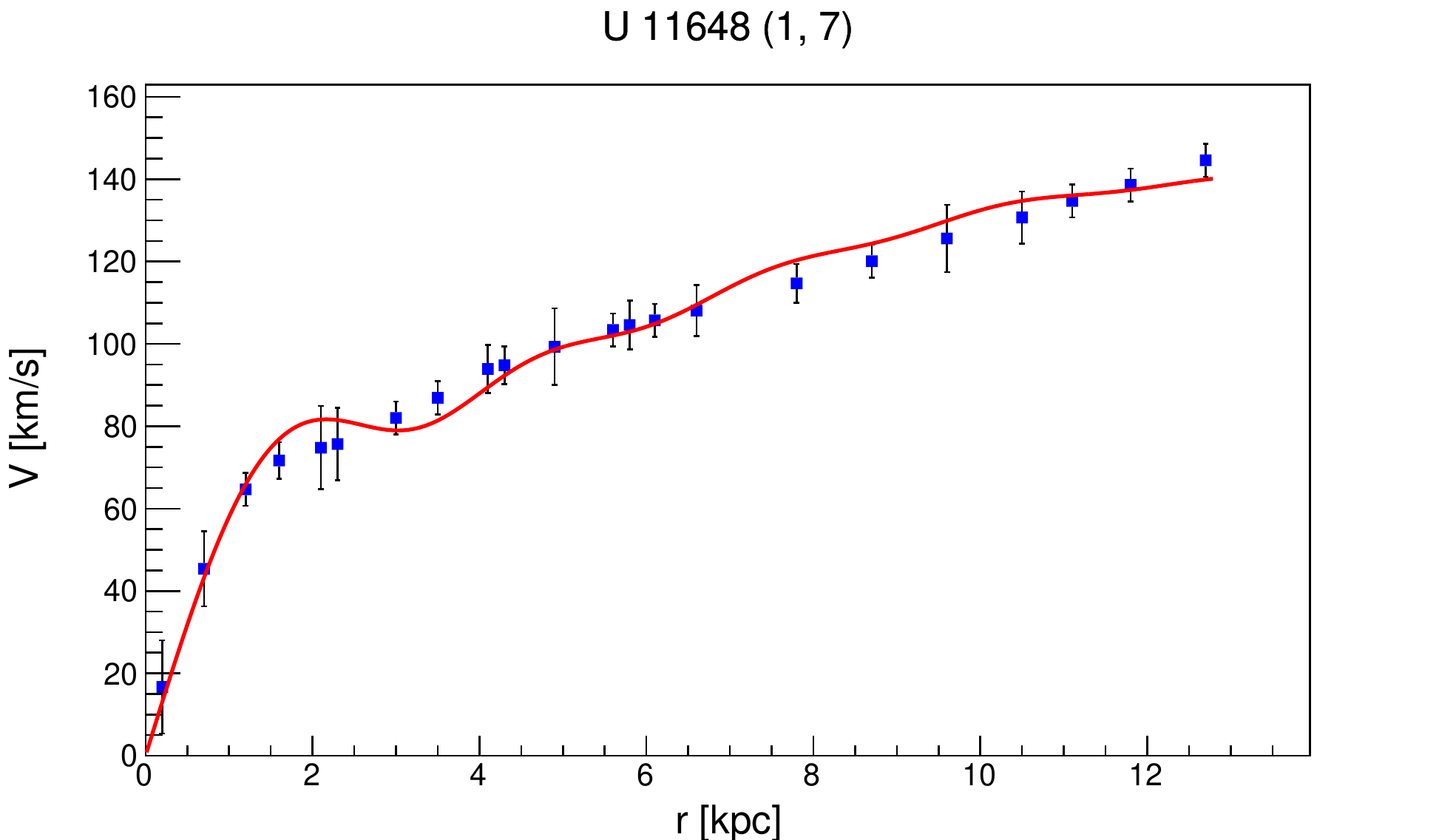} \\
\includegraphics[width=2.3in]{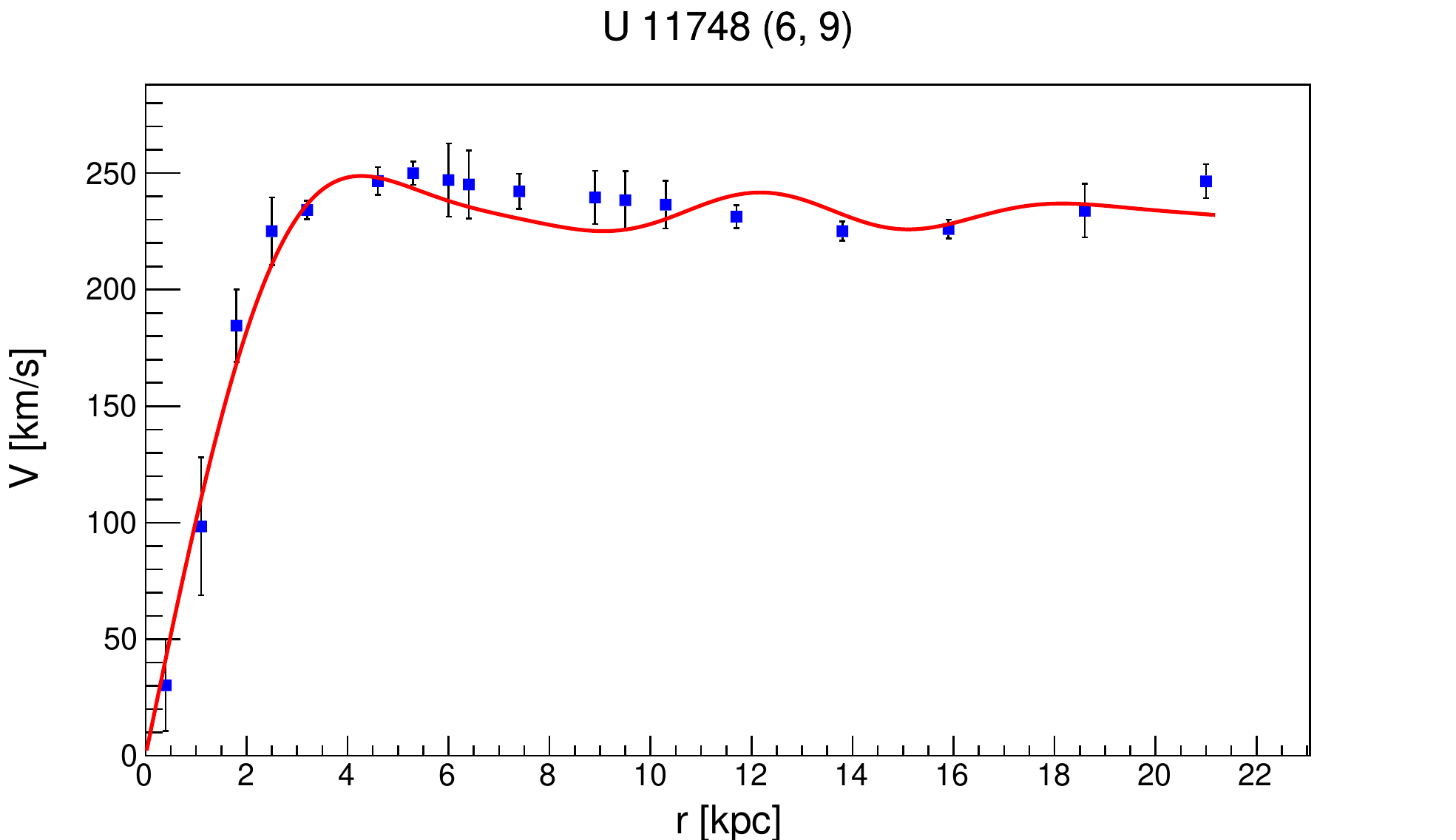}
 \end{array}$
\end{center}
\caption{Best-fit profiles for the high-resolution LSB galaxies reported in
Table~\ref{tab:results5} in the mSFDM model with two excited states $(i,j)$.}
\label{fig:SFDM5}
\end{figure*}

Additionally, there are 4 galaxies where the lowest dominant state is $i=2$ or
even $i=6$; in these systems, the total distribution in different excitation
levels must be the final product of galaxy formation processes and interactions
with the baryonic matter. Again, numerical simulations are needed to
investigate the stability of such multistate configurations without a dominant
ground state and the influence of the baryonic matter in the mSFDM halo.
With the halos formed by two excitation states, we found a very good agreement
with the observational velocities for the galaxies, except for UGC 11748. For
this galaxy the fit might be improved by introducing a third excited state.
Fig.~\ref{fig:SFDM5} shows the resulting best-fit rotation curves for the
parameters reported in Table~\ref{tab:results5}.

\begin{table*}
\caption{Multistate SFDM with two excited states in high-resolution LSB
galaxies. In this Table we show the fitting parameters $R$, $\rho_0^i$ and
$\rho_0^j$ for two excited states $i,j$, $\pm1\sigma$ errors from the MCMC
method used, in the mSFDM model~\eqref{densitytotal} for 7 high-resolution LSB
galaxies in \citet{deBlok:2001}. We report the resulting DM masses, $M_i$ and
$M_j$, for each state, the mass ratio $\eta=M_j/M_1$ (where the ground state
$j=1$ appears in the fit) and $\chi^2_\mathrm{red}$ errors from the fitting
method.}
\label{tab:results5}
\begin{tabular}{lcccccccc}
    \hline
     Galaxy & $i,j$ & $R$ & $\rho_0^i$ & $\rho_0^j$ & $M_{i}$ & $M_{j}$ & $\eta=M_j/M_1$ & $\chi^2_\mathrm{red}$ \\
     & & $(\mathrm{kpc})$ & $(10^{-2}M_\odot/\mathrm{pc}^3)$ & $(10^{-2}M_\odot/\mathrm{pc}^3)$ & $(10^{10}M_\odot)$ & $(10^{10}M_\odot)$ & & \\
\hline
ESO 014-0040   & 2,7 & $48.1^{+2.6}_{-4.4}$  &	$2.93^{+0.71}_{-0.89}$ & $19.3^{+3.1}_{-3.7}$	& 26.9  &	16.0	& ---	& 0.206 \\
ESO 206-0140   & 2,7 & $24.3^{+1.8}_{-2.5}$  &	$1.85^{+0.37}_{-0.53}$ & $16.6^{+2.1}_{-3.0}$	& 2.11  &	1.43	& --- & 0.275 \\
F730-V1        & 1,4 & $15.6^{+1.3}_{-2.6}$  &	$1.89^{+0.49}_{-0.69}$ & $17.4^{+2.3}_{-3.3}$	& 4.20  &	1.97	& 0.469	& 1.28 \\
UGC 11454      & 1,3 & $14.56^{+0.72}_{-1.1}$&	$2.40 \pm 0.37$		   & $10.4 \pm 1.1$			& 4.54 	&	1.81	& 0.399	& 0.322 \\
UGC 11616      & 2,5 & $25.8^{+2.9}_{-2.5}$  &	$2.07^{+0.42}_{-0.72}$ & $10.62^{+0.95}_{-1.2}$	& 2.56  &	1.84	& --- & 0.597 \\
UGC 11648      & 1,7 & $21.1^{+1.8}_{-2.4}$  &	$1.07^{+0.12}_{-0.19}$ & $20.5^{+2.6}_{-2.9}$   & 4.48	&	1.45	& 0.324	& 0.698 \\
UGC 11748      & 6,9 & $41.2^{+1.2}_{-1.6}$  &	$28.0 \pm 5.8$		   & $31.9 \pm 7.8$			& 17.3  &	9.11	& --- & 1.68 \\
\hline
\end{tabular}
\end{table*}

\subsubsection{NGC galaxies with photometric data}
\label{subsubsec:sparc-2}

  In Table~\ref{tab:SFDM1} we present the results for the multistate
SFDM model, for the 3 representative SPARC galaxies in \citet{McGaugh:2016}
(NGC 7814, 6503, 3741) and the 3 sample galaxies analyzed in
\citet{Robles:2013} (NGC 1003, data from SPARC; NGC 1560, data from
\citet{deBlok:2001}; NGC 6946, data from \citet{McGaugh:2005}). We present
separately the DM-only fits and the DM+baryons fits including the baryonic data
to reproduce the rotation curves. For these galaxies, it was not possible to
fit the data with one state only, and we report the two states-combinations.
For the halos formed with two states we report: $\rho_0^i$, $\rho_0^j$ and $R$,
all $\pm1\sigma$ errors from the MCMC method. We also report the total masses
$M_i$ and $M_j$, and the resulting mass ratios $\eta=M_j/M_1$ (where the ground
state appears as a dominant component). For the DM+baryons fits we obtained 4
galaxies with $j=1$ and 2 with excited states only. In the case of the galaxies
with a ground state, the values of the mass ratios are $\eta\lesssim 1.3$,
consistent with the stability condition mentioned before, found through
numerical simulations of multistate configurations \citep{urena-bernal10}.

\begin{figure*}
    \begin{center}
\includegraphics[width=0.5\textwidth]{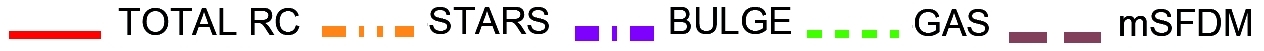}
    $\begin{array}{ccc}
\includegraphics[width=2.3in]{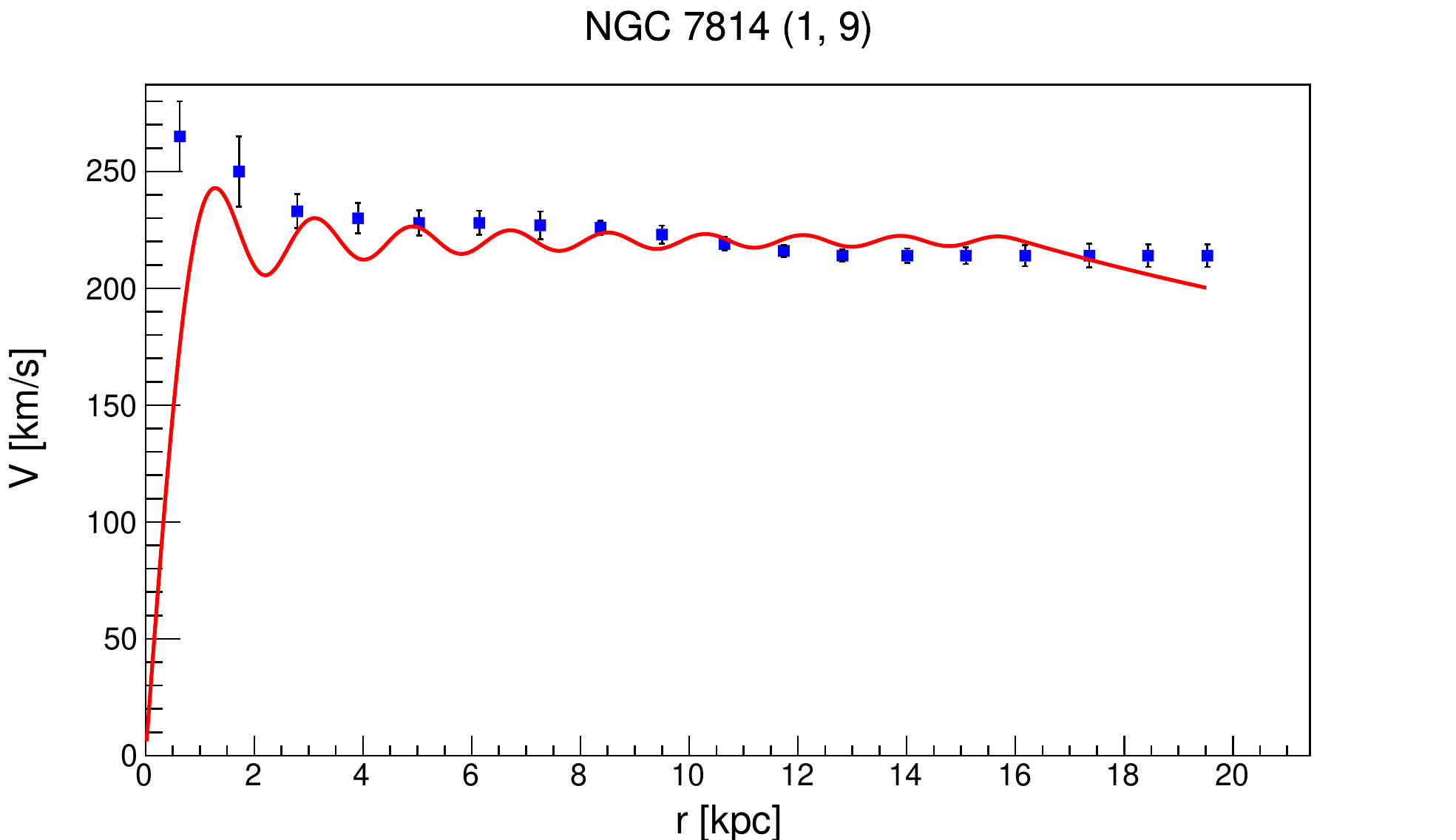} &
\includegraphics[width=2.3in]{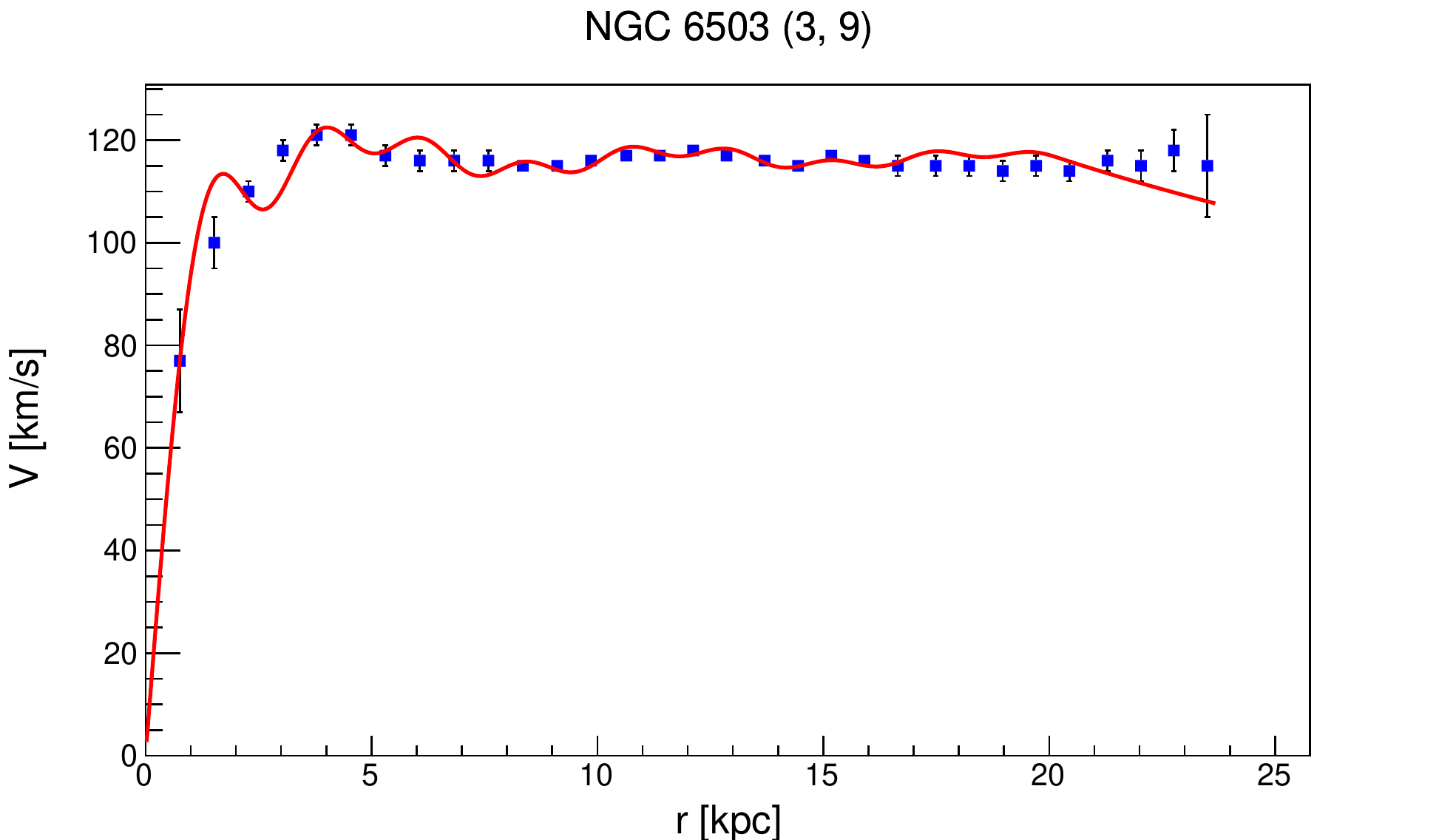} &
\includegraphics[width=2.3in]{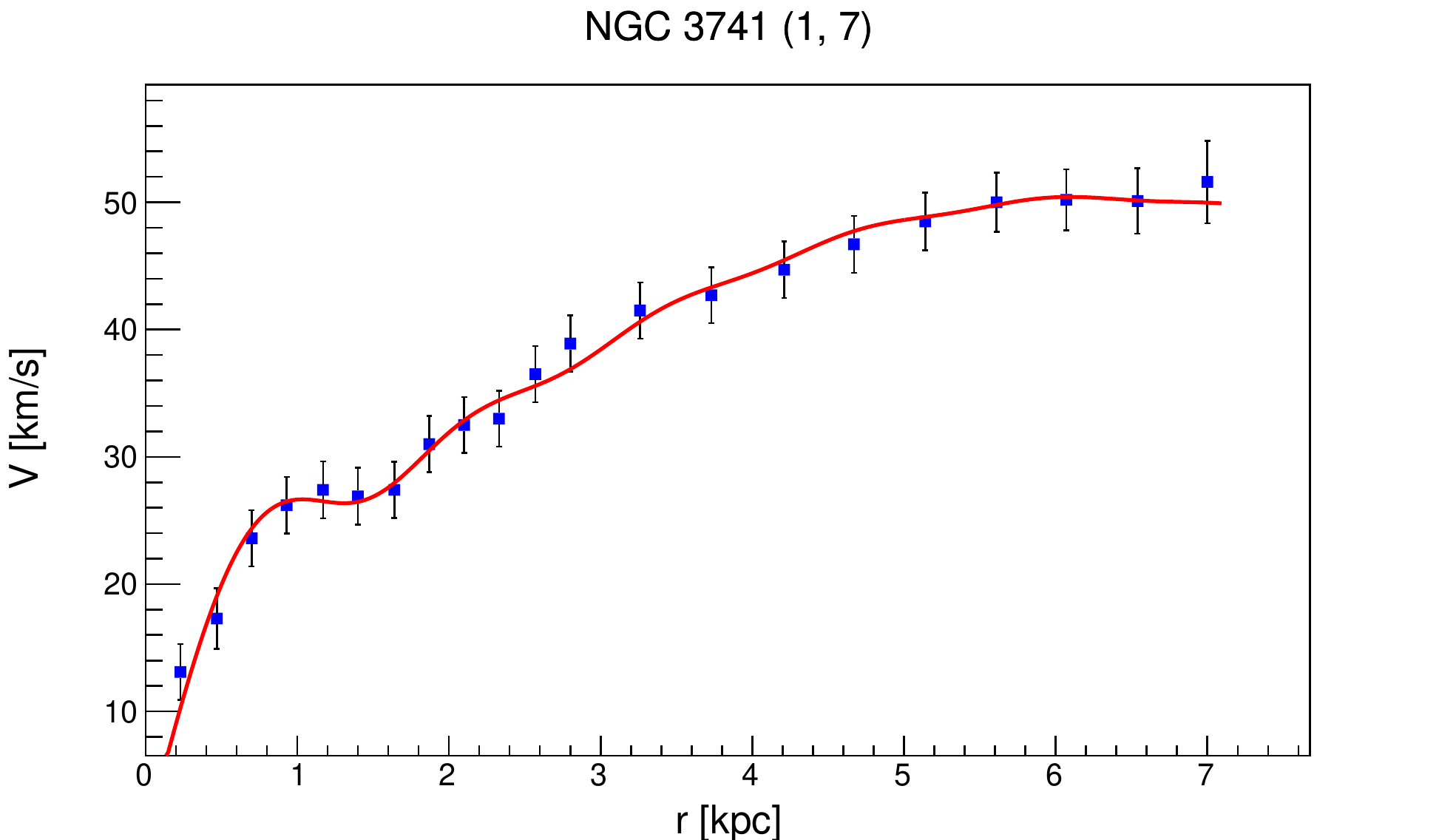} \\
\includegraphics[width=2.3in]{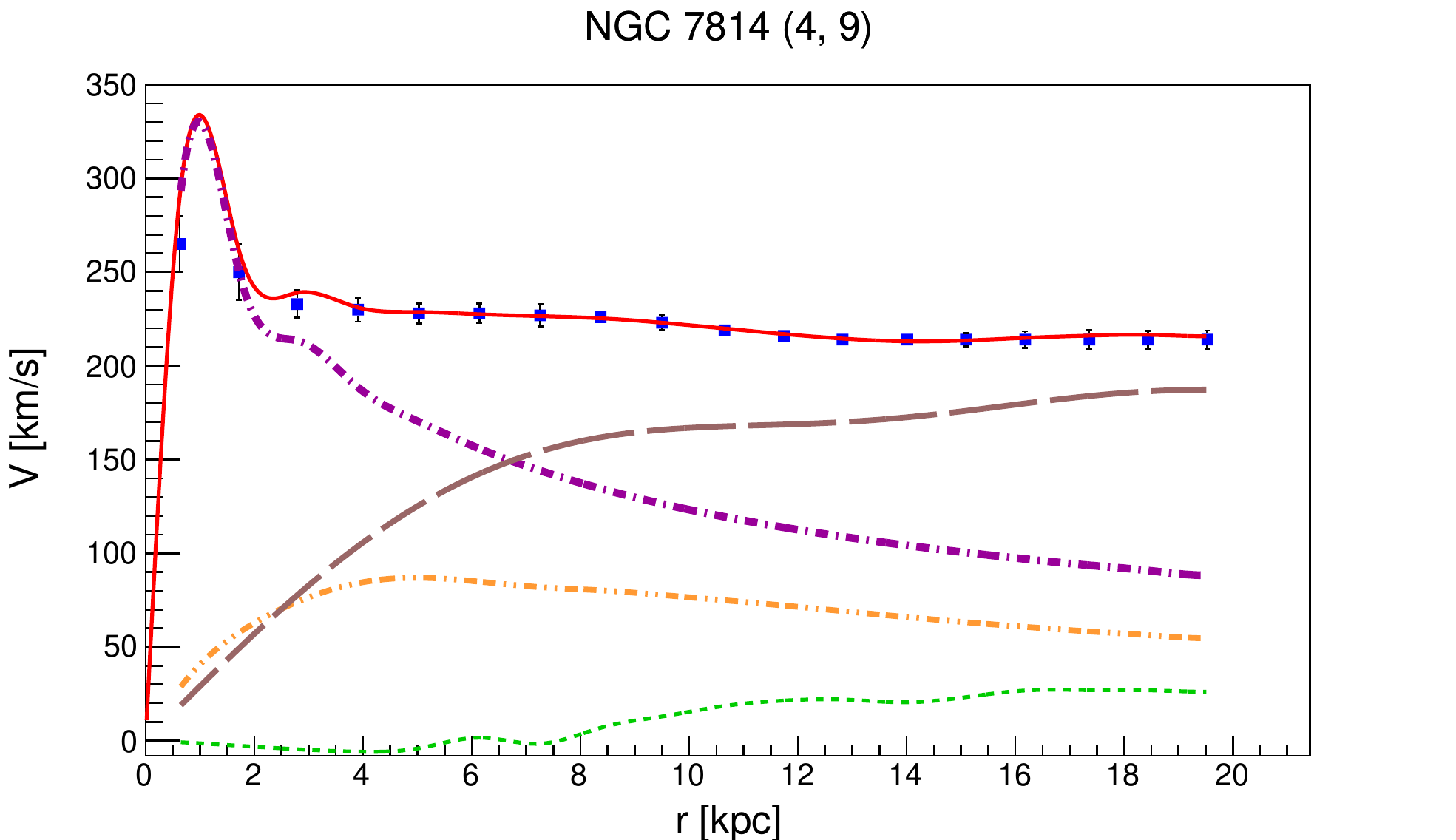} &
\includegraphics[width=2.3in]{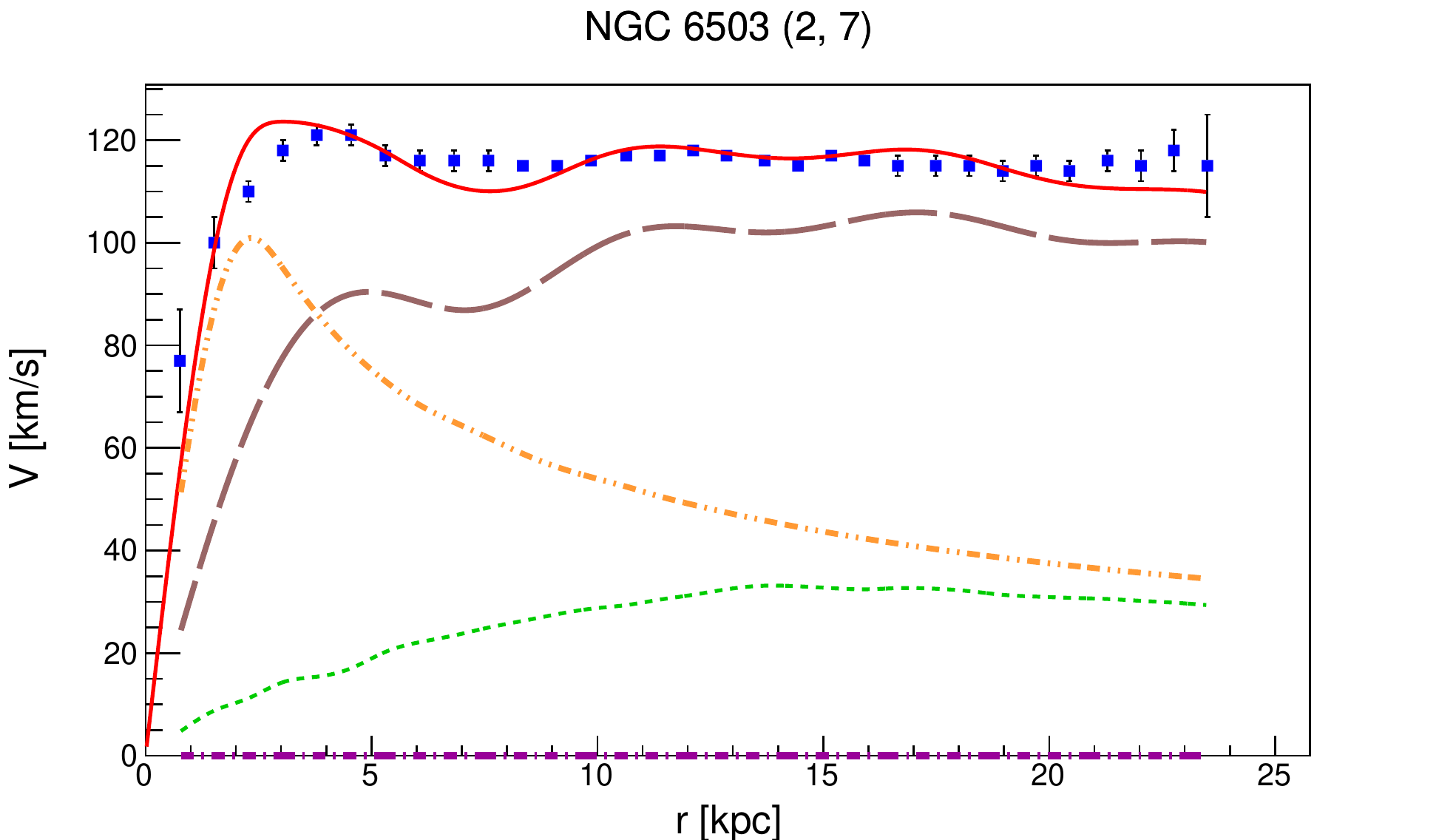} &
\includegraphics[width=2.3in]{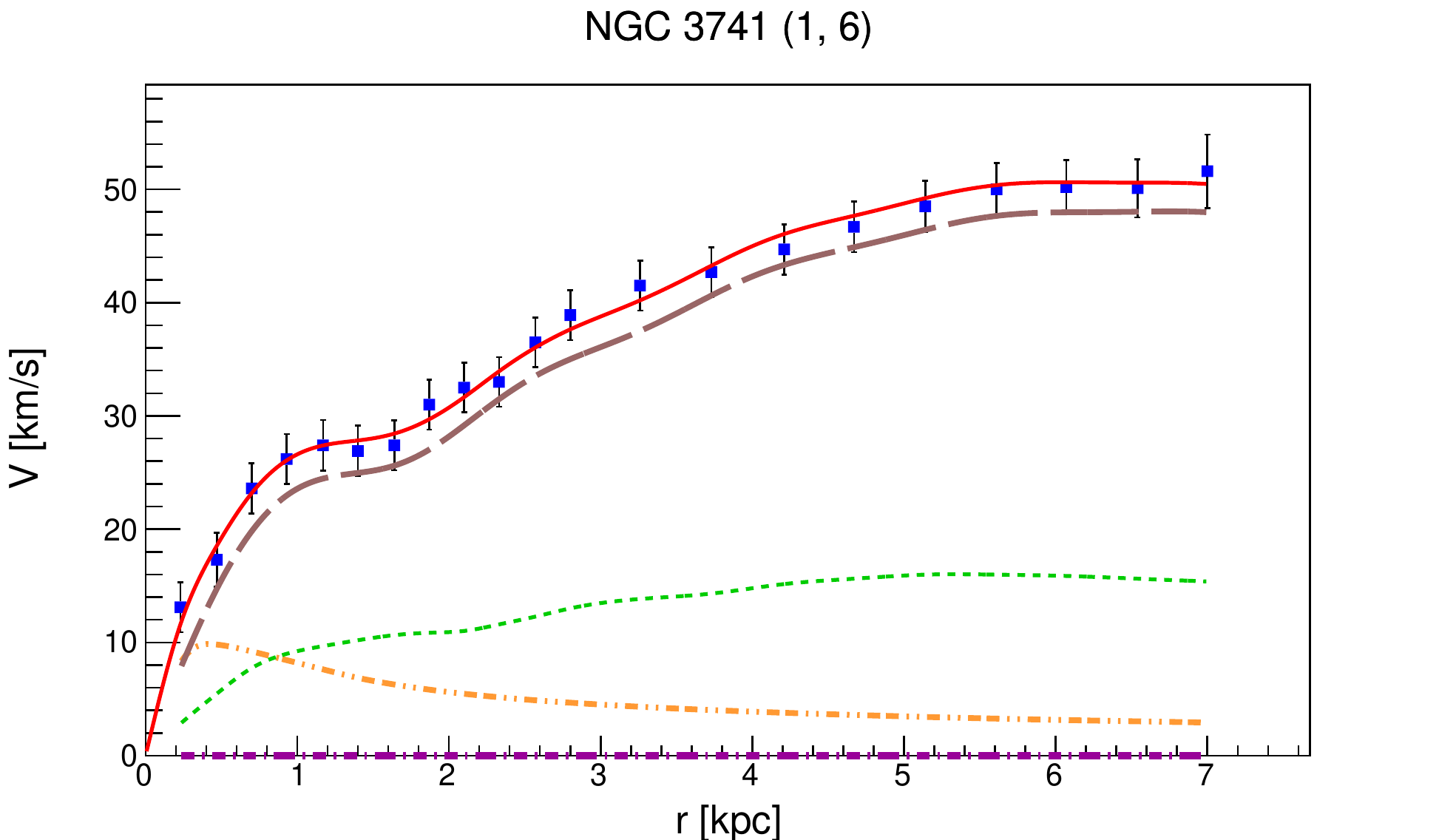} \\
\\
\\
\includegraphics[width=2.3in]{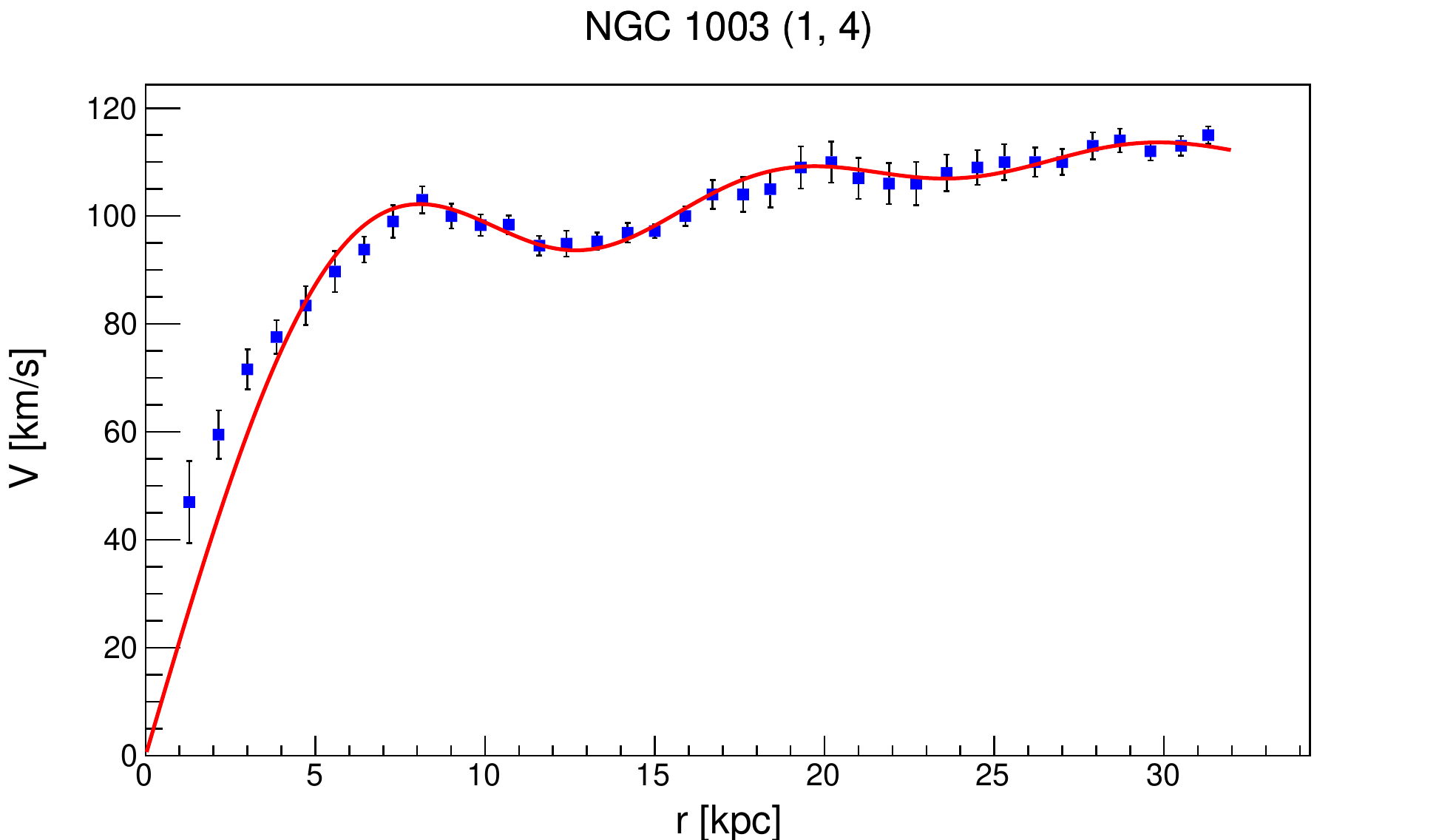} &
\includegraphics[width=2.3in]{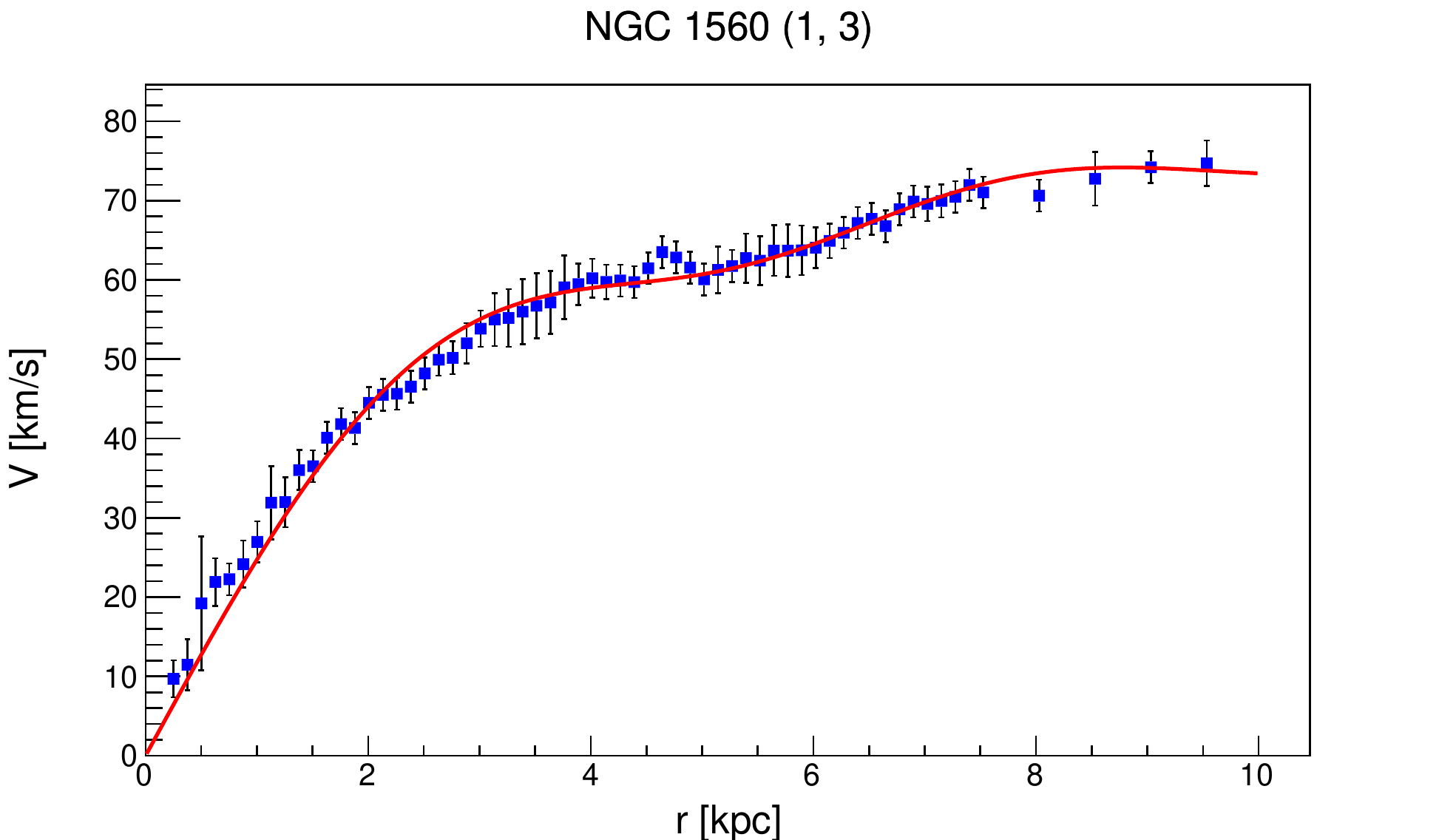} &
\includegraphics[width=2.3in]{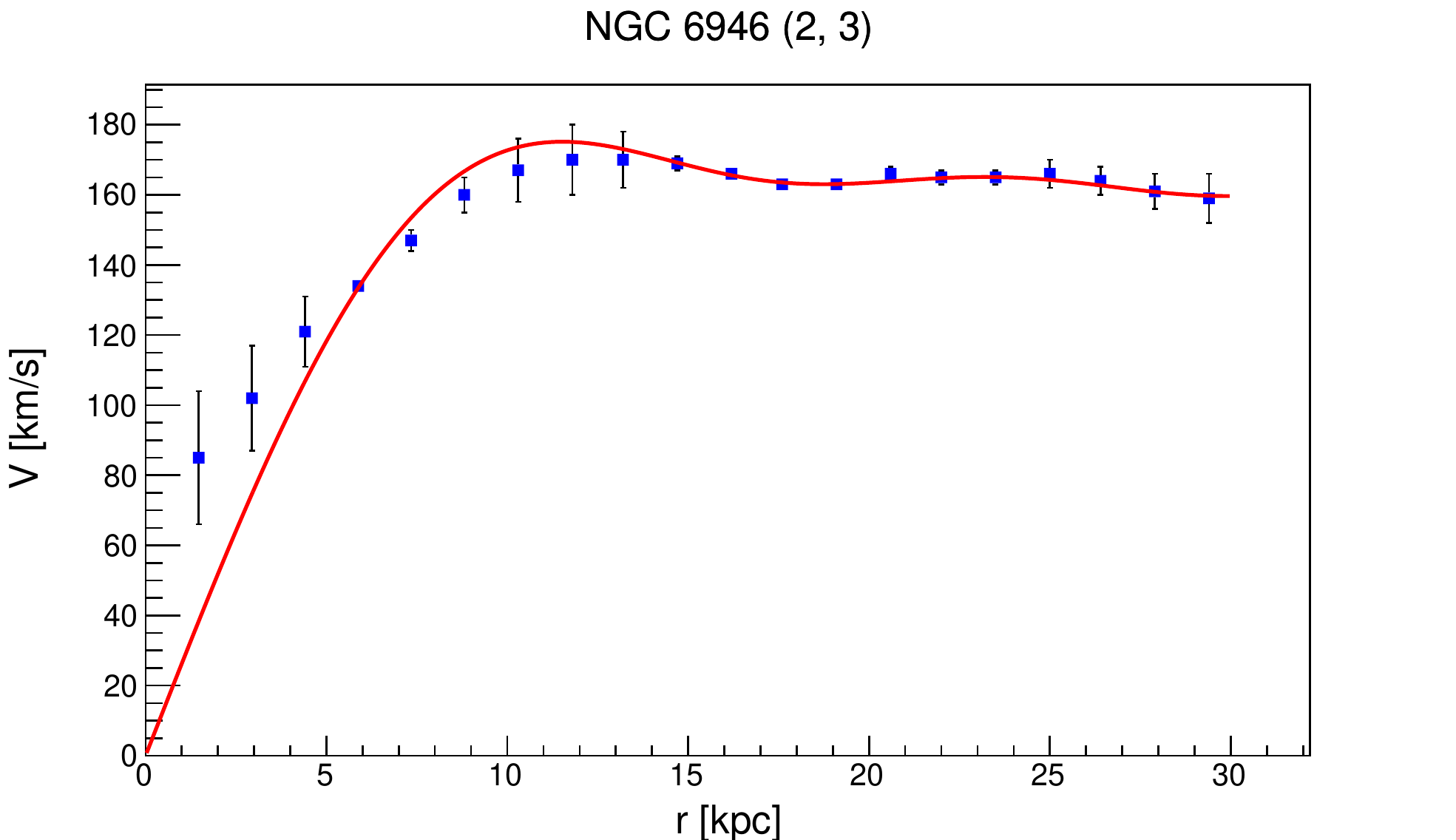} \\
\includegraphics[width=2.3in]{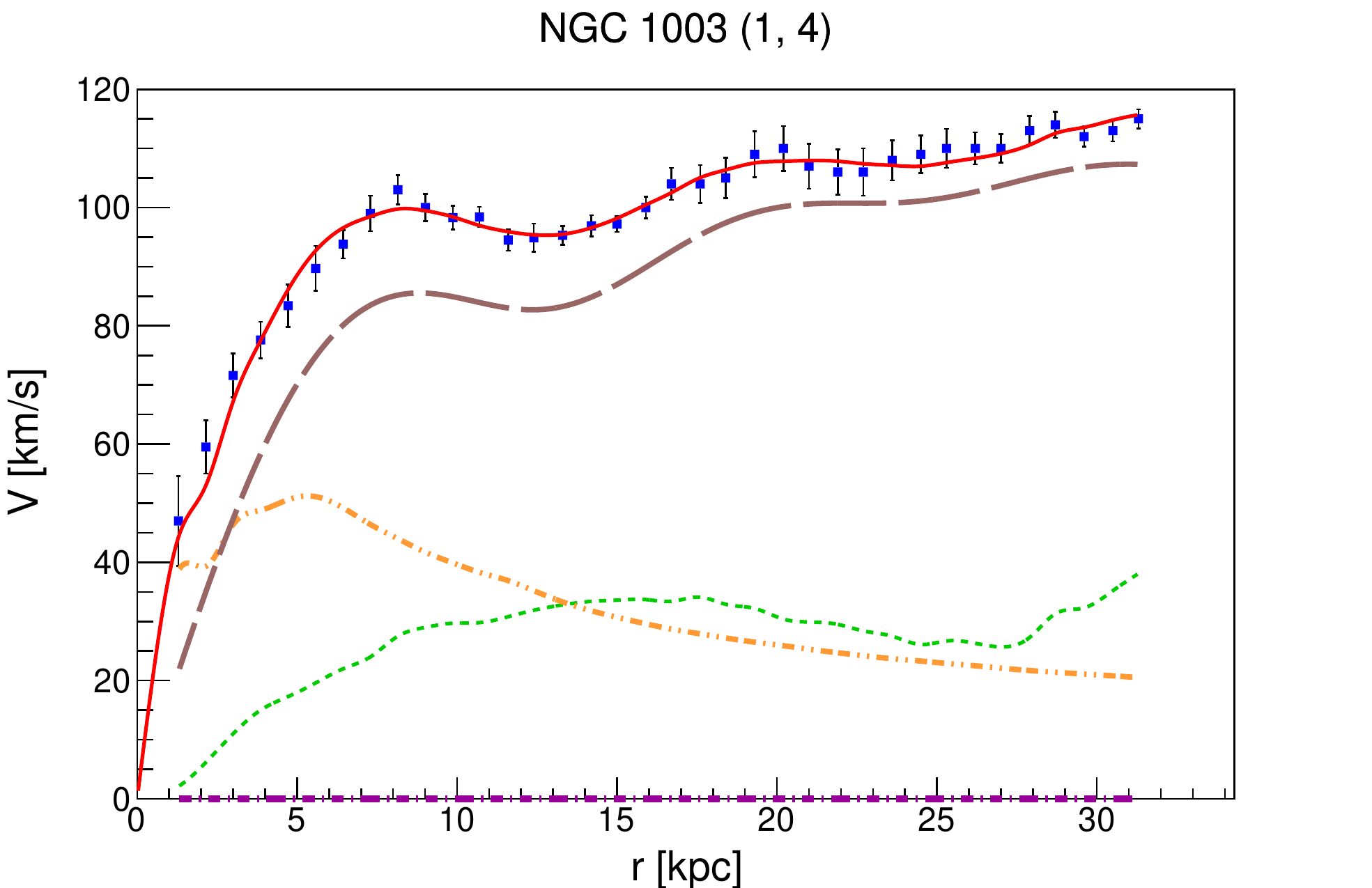} &
\includegraphics[width=2.3in]{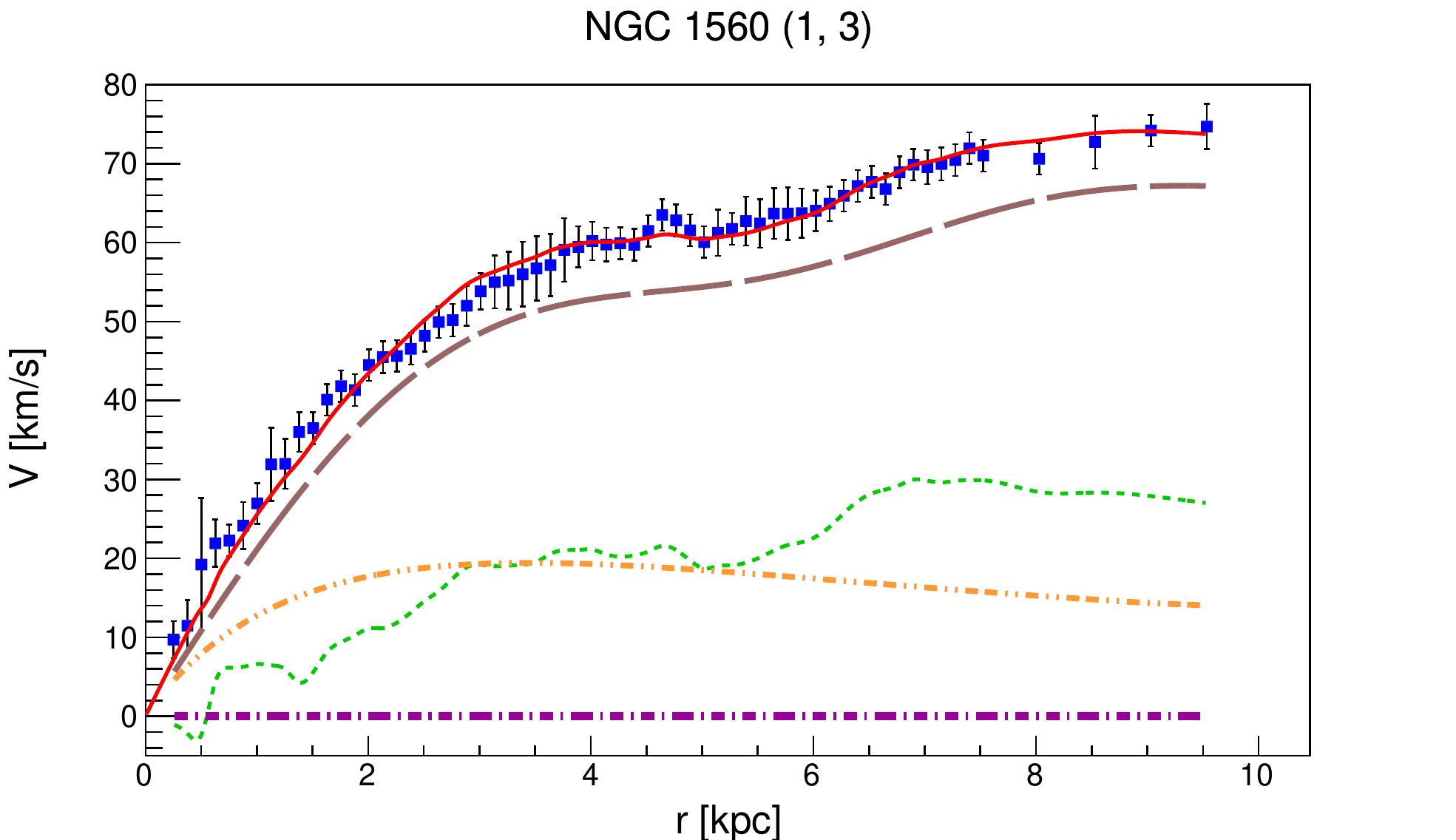} &
\includegraphics[width=2.3in]{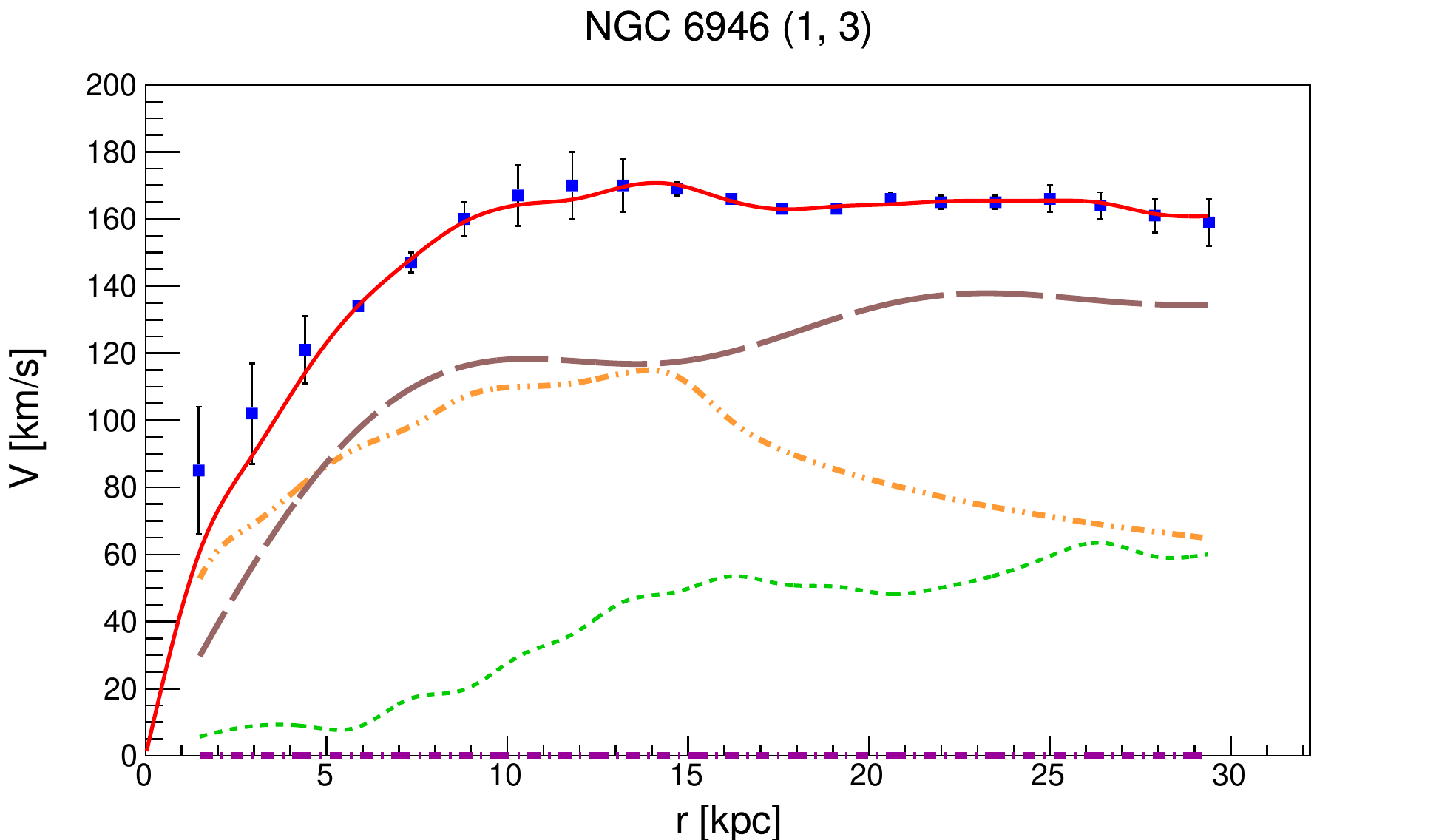}
 \end{array}$
    \end{center}
   \caption{Best fits for the NGC galaxies with photometric data in the
   mSFDM model with two excited states $(i,j)$, for DM-only and
   DM+baryons contribution. The corresponding fitting parameters are shown in
   Table~\ref{tab:SFDM1}.}
    \label{fig:SFDM1}
\end{figure*}

Fig.~\ref{fig:SFDM1} shows the resulting profiles with these fitting
parameters, without and taking into account the baryonic information.
For NGC 7814, the bulge-dominated galaxy, once the photometric information is
included, the model reproduces very well the observations. For NGC 6503, the
disk-dominated galaxy, the best fit fails at reproducing some regions, even
taking into account the baryonic data; for this galaxy we used three excited
states. For NGC 3741, the gas-dominated galaxy, and for NGC 1560, both fits are
very good. For NGC 1003 and 6946 the inner regions are reproduced once the
baryonic information is included. For NGC 1003, the mSFDM model reproduces very
well the `wiggles' or oscillations, which are not reproduced by the baryonic
components \citep[see the previous results in][]{Robles:2013}, neither the
soliton+NFW profile (see Subsection~\ref{subsubsec:sparc-1}).

\begin{table*}
\caption{Multistate SFDM with two excited states in NGC galaxies with
photometric data. The top panel shows the DM-only fits and the bottom panel the
DM+baryons analysis, both for the same sample galaxies with high-resolution
photometric information. Both panels show the fitting parameters $R$, $\rho_0^i$
and $\rho_0^j$ for two excited states $i,j$, $\pm1\sigma$ errors from the MCMC
method used, in the multistate SFDM model~\eqref{densitytotal}. We also report the
resulting DM masses, $M_i$ and $M_j$, for each state, the mass ratio $\eta
= M_j/M_1$ and $\chi^2_\mathrm{red}$ errors from the fitting method.}
\label{tab:SFDM1}
\begin{tabular}{lcccccccc}
    \hline
    \multicolumn{9}{c}{DM-only fits} \\
    \hline
     Galaxy & $i,j$ & $R$ & $\rho_0^i$ & $\rho_0^j$ & $M_{i}$ & $M_{j}$ & $\eta=M_j/M_1$ & $\chi^2_\mathrm{red}$ \\
     & & $(\mathrm{kpc})$ & $(10^{-2}M_\odot/\mathrm{pc}^3)$ & $(10^{-2}M_\odot/\mathrm{pc}^3)$ & $(10^{10}M_\odot)$ & $(10^{10}M_\odot)$ & & \\
\hline
NGC 7814	&  9 &  $11.25^{+0.10}_{-0.12}$	&  ---						&  $1122^{+24}_{-22}$	&  --- 		& 22.0	&	---	 	& 31.7\\
NGC 6503	&  3,9 &  $19.30^{+0.54}_{-0.43}$	&  $3.79^{+0.55}_{-0.65}$	&  $73.5^{+5.2}_{-6.5}$	&  2.429 	&  5.08	&	---	& 2.35\\
NGC 3741	&  1,7 &  $9.04^{+0.49}_{-0.76}$	&  $0.774 \pm 0.096$		&  $10.7^{+1.2}_{-1.3}$	&  0.334 	& 0.0783&	0.234	& 0.33\\
NGC 1003	&  1,4 &  $43.41 \pm 0.61$			&  $0.0717 \pm 0.0045$		&  $2.470 \pm 0.078$	&  3.279 	& 6.01	&	1.83	& 1.46\\
NGC 1560	&  1,3 &  $13.88 \pm 0.46$			&  $0.574^{+0.035}_{-0.042}$&  $3.10 \pm 0.13$		&  0.814	& 0.391	&	0.48	& 0.61\\
NGC 6946	&  2,3 &  $42.28 \pm 0.90$			&  $0.89 \pm 0.12$			&  $2.96 \pm 0.12$		&  6.901 	& 10.6	&	---	& 1.40\\
\hline
    \multicolumn{9}{c}{DM+baryons fits}\\
\hline
NGC 7814	&  4,9 &  $103.9^{+5.8}_{-7.0}$		&  $0.88^{+0.11}_{-0.15}$	&  $3.85^{+0.36}_{-0.48}$	&  8.818  & 6.902  &	---	 & 0.388 	\\
NGC 6503	&  2,7 &  $44.26 \pm 0.88$			&  $0.318 \pm 0.024$		&  $5.26 \pm 0.20$			&  2.194  & 3.270  &	---	 & 4.05 	\\
NGC 3741	&  1,6 &  $8.99^{+0.52}_{-0.75}$	&  $0.730^{+0.085}_{-0.10}$	&  $5.75^{+0.75}_{-0.89}$	&  0.306  & 0.059  &	0.19	 & 0.209 	\\
NGC 1003	&  1,4 &  $44.84^{+0.88}_{-1.2}$	&  $0.0912 \pm 0.0050$		&  $1.488^{+0.077}_{-0.068}$&  4.442  & 3.933  &	0.885    & 0.527 	\\
NGC 1560	&  1,3 &  $14.77 \pm 0.61$			&  $0.417^{+0.030}_{-0.037}$&  $2.23 \pm 0.12$			&  0.661  & 0.338  &	0.511    & 0.488 	\\
NGC 6946	&  1,3 &  $38.19 \pm 0.99$			&  $0.199 \pm 0.015$		&  $2.035 \pm 0.079$		&  6.559  & 5.775  &	0.88     & 0.374 	\\
\hline
\end{tabular}
\end{table*}

\begin{figure*}
    \begin{center}$
    \begin{array}{ccc}
\includegraphics[width=2.5in]{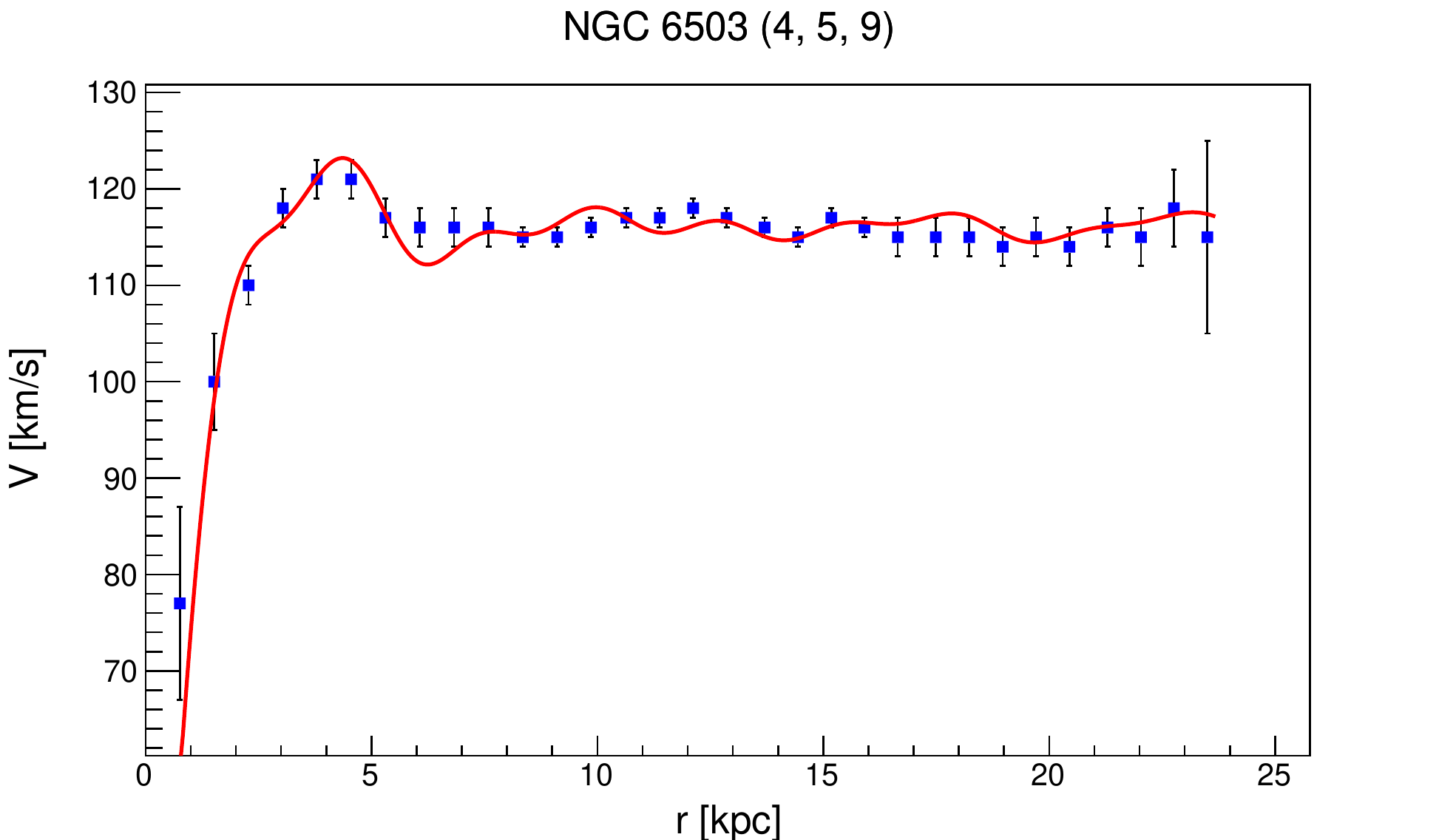} &
\includegraphics[width=2.5in]{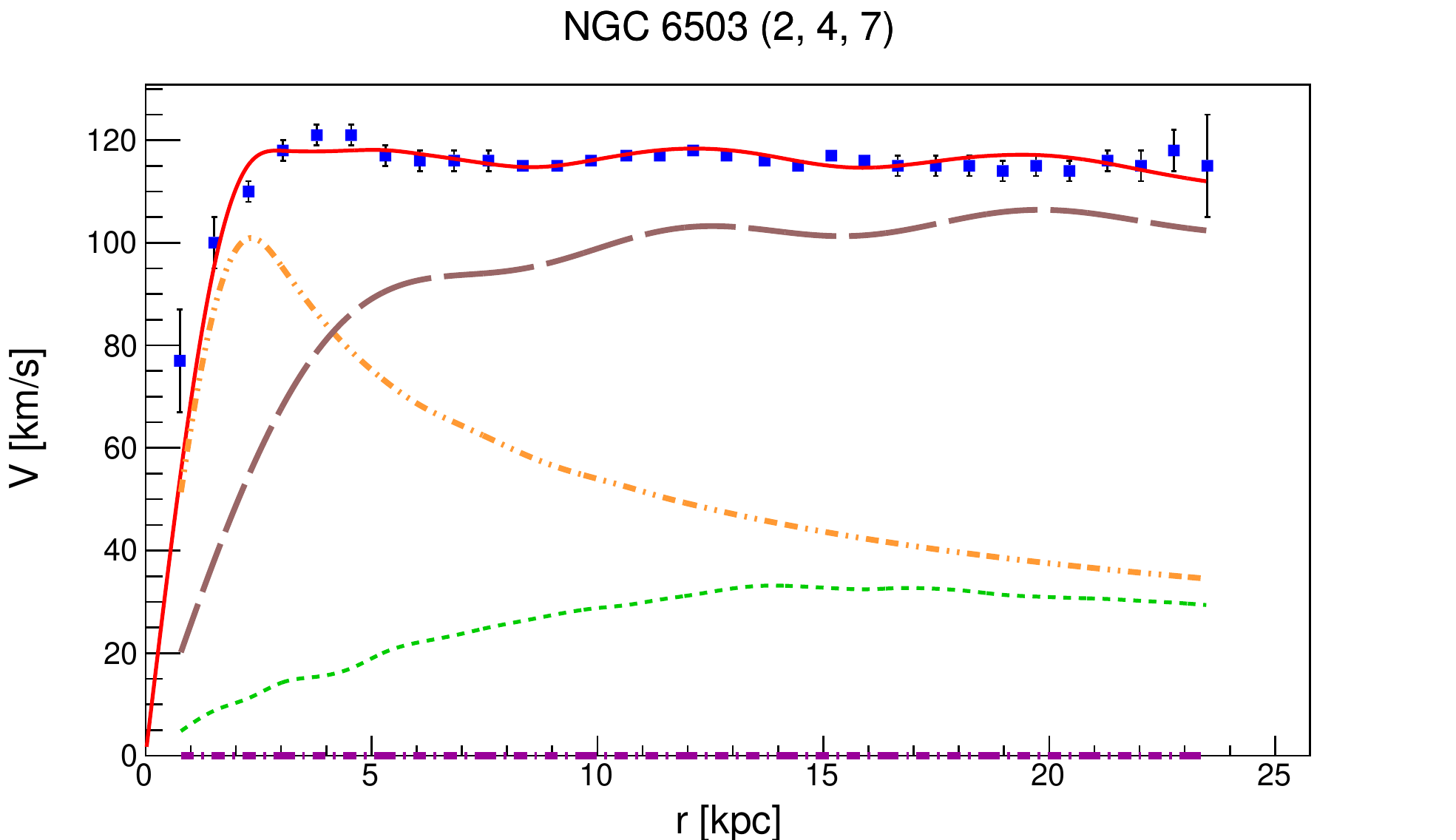}
 \end{array}$
    \end{center}
   \caption{The same as Fig.~\ref{fig:SFDM1} for NGC 6503 with three excited
   states $(i,j,k)$. The fitting parameters are reported in Table~\ref{tab:SFDM3}.}
\label{SFDM:ngc6503}
\end{figure*}

  In Table~\ref{tab:SFDM3}, we present the best-fit three states configuration
for NGC 6503, for the data without and with photometry, and the resulting
rotation curves are in Fig.~\ref{SFDM:ngc6503}. For the other galaxies, the
fittings might be improved by introducing a third excited state, but is
not necessary to introduce more degrees of freedom since two excited
states are well enough. It is worth noting that in these cases the other
intermediate states (including the ground state) might be present in the mSFDM
halo, but their contribution to the total density profile is negligible with
respect to the dominant ones.

\begin{table*}
\caption{Multistate SFDM with three excited states in NGC 6503. The top panel
shows the DM-only fit and the bottom panel the DM+baryons analysis. Both panels
show the fitting parameters $R$, $\rho_0^i$, $\rho_0^j$ and $\rho_0^k$ for
three excited states $i,j,k$, $\pm1\sigma$ errors from the MCMC method. We also
report the resulting DM masses, $M_i$, $M_j$ and $M_k$, for each state and
$\chi^2_\mathrm{red}$ errors from the fitting method.}
\label{tab:SFDM3}
\begin{tabular}{lccccccccc}
    \hline
    \multicolumn{10}{c}{DM-only fit} \\
    \hline
     Galaxy & $i,j,k$ & $R$ & $\rho_0^i$ & $\rho_0^j$ & $\rho_0^k$ & $M_{i}$ & $M_{j}$ & $M_{k}$ & $\chi^2_\mathrm{red}$ \\
     & & $(\mathrm{kpc})$ & $(10^{-2}M_\odot/\mathrm{pc}^3)$ & $(10^{-2}M_\odot/\mathrm{pc}^3)$ & $(10^{-2}M_\odot/\mathrm{pc}^3)$ & $(10^{10}M_\odot)$ & $(10^{10}M_\odot)$ & $(10^{10}M_\odot)$ & \\
\hline
NGC 6503	& 4,5,9 & 24.2 $\pm$ 0.48 & 4.37$\pm$ 0.63 & 5.56 $\pm$ 0.87 & 27.98 $\pm$ 3.3 & 2.456 & 2.000 & 3.082 & 1.27	\\
\hline
    \multicolumn{10}{c}{DM+baryons fit} \\
    \hline
NGC 6503	&  2,4,7 &	 $50.0^{+2.0}_{-2.5}$  &  $0.197 \pm 0.032$  &  $0.671^{+0.066}_{-0.082}$  &  $2.89^{+0.24}_{-0.29}$  & 1.957  & 1.659	&  2.103	& 1.54 	\\
\hline
\end{tabular}
\end{table*}

\begin{figure*}
	\includegraphics[width=0.49\textwidth]{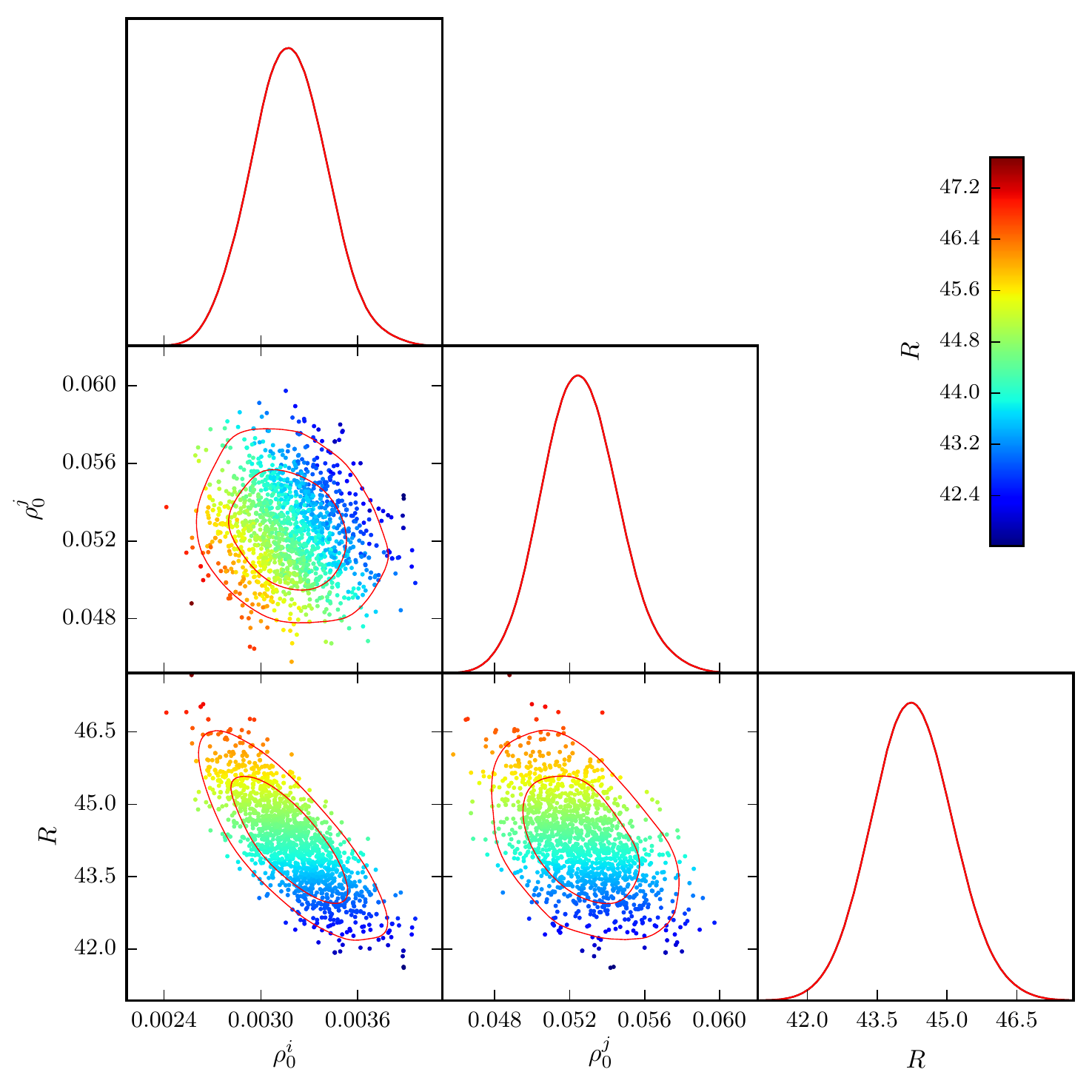}
    \includegraphics[width=0.49\textwidth]{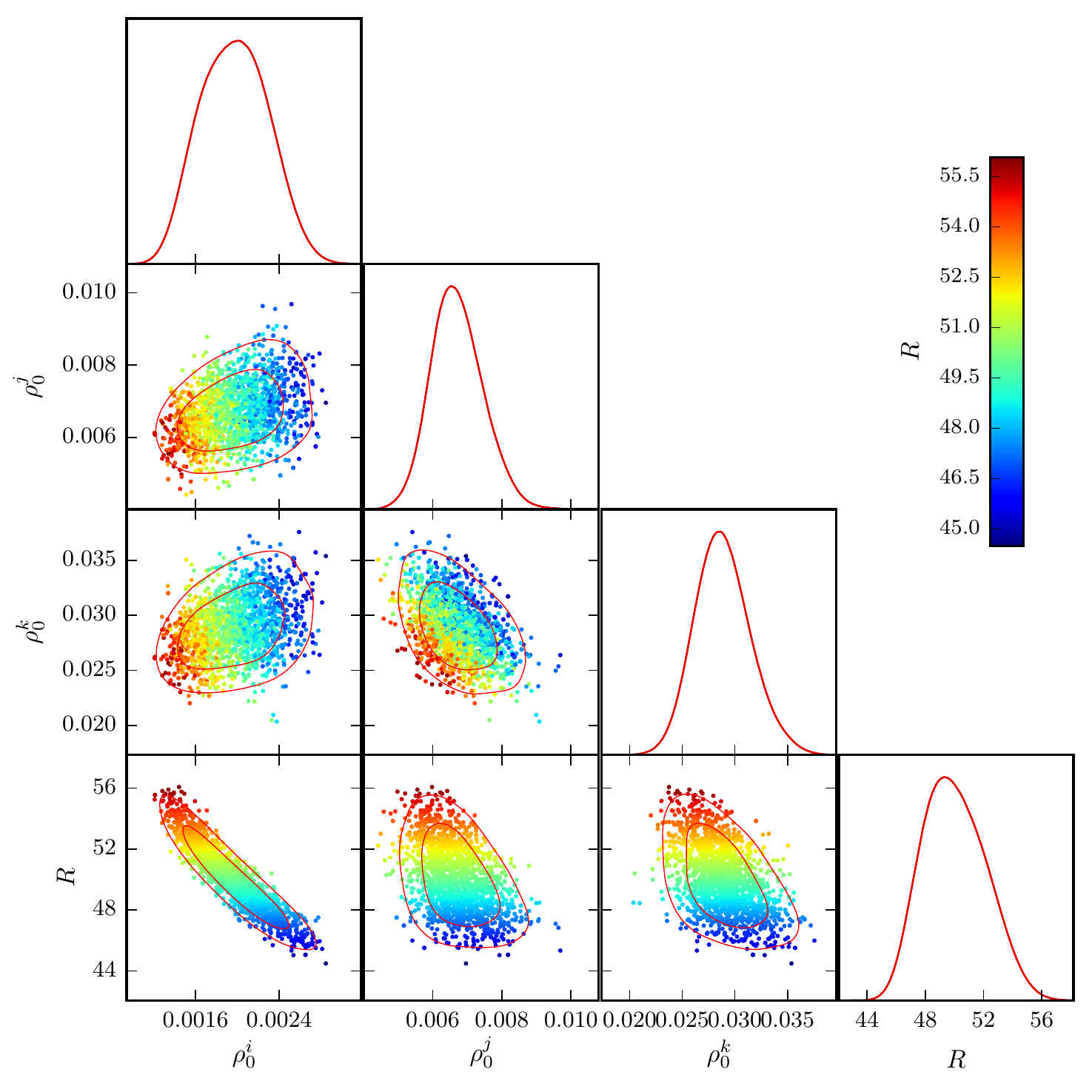}
\caption{Posterior distributions for the parameters $\rho_0^i(M_\odot/
\mathrm{pc}^3)$, $\rho_0^j(M_\odot/\mathrm{pc}^3)$ and $R(\mathrm{kpc})$, with
$R$ as the scatter plot, for the two (left panel) and three (right panel)
states configurations for NGC 6503.}
\label{ellipses:ngc6503}
\end{figure*}

Finally, Fig.~\ref{ellipses:ngc6503} illustrates the posterior distributions
and $1\sigma$ and $2\sigma$ contours obtained from the MCMC method for NGC
6503, with two and three excited states. The parameter $R$ is the scatter plot.

%%%%%%%%%%%%%%%% CONCLUSIONS %%%%%%%%%%%%%%%%
\section{Discussion and conclusions}
\label{sec:Conclusions}

  From the cosmological FDM simulations it is found that the galaxy halos are
formed by prominent dense cores, that approximate well the soliton-like
solutions of the Schr\"odinger-Poisson equation. However, the soliton profile
alone is unable to reproduce the rotation curves, in particular of large
galaxies. It is in this case where is necessary to introduce a more general
profile than the soliton one in order to reproduce the observations at the
outer radii of large systems. An ad hoc approach smoothly matches the soliton
solution with a NFW profile (soliton+NFW) \citep{Schive:2014dra,Schive:2014hza}.
On the other side, a theoretically motivated alternative that has been proposed
in the literature is to include excited states in the SFDM halo (multistate
SFDM) coming from thermal excitations in the SF potential \citep{Robles:2013},
that presents a soliton core-like behavior at the innermost radii and `wiggles'
or oscillations at the outer radii in the density profile and rotation curves
(see Fig.~\ref{plot-densities}). In this article we take both alternatives, and
apply the soliton+NFW profile and the mSFDM model to reproduce the rotation
curves of 18 LSB galaxies and 6 NGC galaxies with high-resolution photometric
data.

  In the first place, we apply the soliton+NFW profile with the continuity
condition between both profiles at the transition radius $r_\epsilon$. For all
the galaxies, from the individual fitting analysis leaving vary freely the four
parameters of the model (boson mass $m_\psi$, soliton core radius $r_c$,
transition radius $r_\epsilon$ and characteristic NFW radius $r_s$), we found
that the resulting boson mass is in the range $0.212 \leq m_\psi/10^{-23}
\mathrm{eV} \leq 27.0$, the core radius within $0.326 \leq r_c/\mathrm{kpc} \leq
8.96$, and the ratio between the transition and core radii in $0.46 \leq
r_\epsilon/r_c \leq 1.94$ (see Tables~\ref{tab:results2}
and~\ref{tab:results1}). However, as the boson mass must be a constant of the
FDM model, we performed a combined analysis of the 18 LSB galaxies for a single
boson mass for all the galaxies, and obtained $m_\psi = 0.554\times 10^{-23}
\mathrm{eV}$, with core radius in the range $1.57 \leq r_c/\mathrm{kpc} \leq
5.82$, and ratio between the transition and core radii in the interval $0.15
\leq r_\epsilon/r_c \leq 2.3$ (see Table~\ref{tab:LSB-comb}). We point
out that we are not reporting the confidence interval for the boson mass in
this analysis since the combined computation time was too expensive. However,
we believe that the confidence level is not too large based on the boson mass
errors reported in Table~\ref{tab:results2} for the individual analysis, and
also given the value of $\chi^2_\mathrm{red}=1.208$ obtained in the combined
analysis.

  Remarkably, in both the individual and the combined analysis, in order to fit
the rotation curves of the galaxies, we obtained different results from the
expected $r_\epsilon > 3r_c$ in the FDM cosmological simulations \citep{Schive:2014dra}.
Furthermore, in the combined analysis we obtained a best-fit mass $m_\psi <
10^{-23}\mathrm{eV}$, that is in conflict with the stringent cosmological
constraints \citep{Bozek:2014,Sarkar:2016}, setting up a possible tension
between the FDM model and the galaxy rotation curves. As it is known, in the
SFDM there is a sharp break in the matter power spectrum leading to a natural
suppression of substructure below a scale dependent on the boson mass, $k \sim
m_\psi^{1/3}$. The cosmological implications for such light boson 
masses ($m_\psi < 10^{-23}\mathrm{eV}$) would imply less substructure than
observed. For these galaxies, we expected that the range in core radius will
decrease as the boson mass increases to have $m_\psi>10^{-23} \mathrm{eV}$,
once the cosmological restriction was imposed; however, if we restrict the
fitting methods to obtain $m_\psi \geq 10^{-23}\mathrm{eV}$ it is not possible
to fit all the rotation curves together, so we did not include those results in
the paper.

\begin{figure}
   \begin{center}
   \includegraphics[width=3.5in]{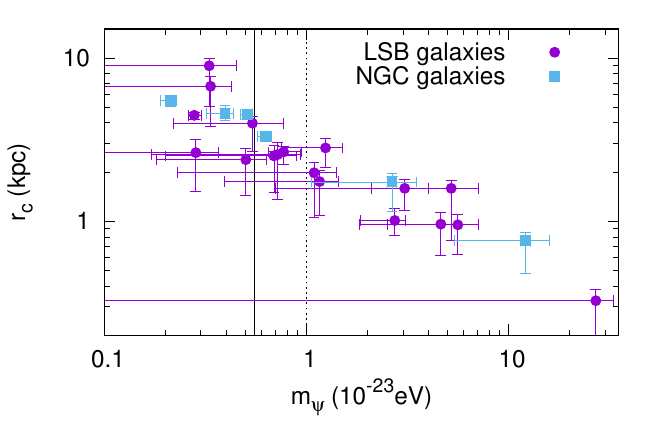}
   \end{center}
   \caption{Distribution of resulting core radii $r_c$ vs. boson masses $m_\psi$
   from the soliton+NFW density profile in the FDM model, for the individual
   analysis of the 18 LSB and 6 NGC (including baryonic information) galaxies,
   $\pm 1 \sigma$ errors from the MCMC method. The vertical dashed line at
   $m_\psi=10^{-23}\mathrm{eV}$ shows the cosmological constraint
   \citep{Bozek:2014,Sarkar:2016} for the minimum mass required for the
   ultra-light boson. The vertical solid line at $m_\psi=0.554 \times 10^{-23}
   \mathrm{eV}$ shows the result for the boson mass from the combined analysis
   of the 18 LSB galaxies.}
    \label{mpsi-rc}
\end{figure}

\begin{figure}
   \begin{center}
   \includegraphics[width=3.5in]{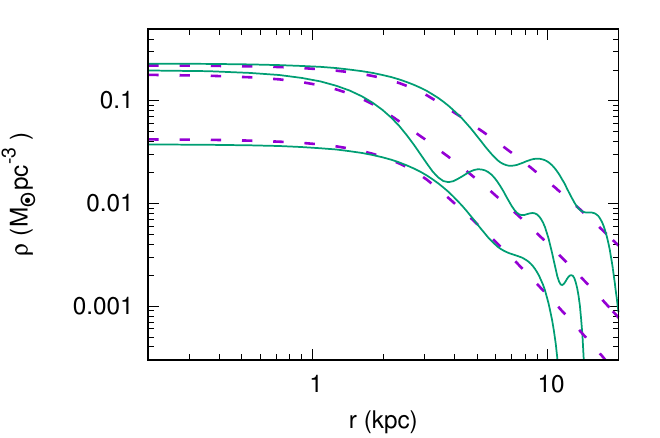}
   \end{center}
   \caption{Resulting density profiles for three sample LSB galaxies. The
   dashed lines show the soliton+NFW profiles and the solid lines the
   corresponding mSFDM fits. The overlap between both profiles is evident for
   the galaxies, showing the core-like behavior at the innermost radii, the
   `wiggles' or oscillations around the NFW-region, inherent to the multistate
   profile, and the steeper slope at the outermost radii of the mSFDM model
   compared to the NFW profile. Every galaxy shows different oscillations
   depending on its particular multistate configuration.}
   \label{plot-densities}
\end{figure}

  In Fig.~\ref{mpsi-rc}, we show the resulting distribution of core radii $r_c$
vs. boson masses $m_\psi$ for all the galaxies in the individual analysis,
$\pm 1 \sigma$ errors from the MCMC method. The solid line in the figure, at
$m_\psi = 0.554\times 10^{-23}\mathrm{eV}$, shows the result for the boson mass
from the combined analysis with the LSB galaxies, and the dashed line shows the
minimum ultra-light boson mass required from the cosmological constraints.
Notice the correlation between both parameters: for large galaxies the core
radius is larger and the resulting boson mass is smaller, and vice versa, for
smaller galaxies the core radius is smaller and the boson mass larger. Such
correlation has been discussed very recently in \citet{Urena-Robles:2017},
including dSph galaxies' results (whose $R_\mathrm{max}\sim 0.5-2 \ \mathrm{kpc}$),
for which the estimated boson mass from the Jeans analysis of the classical
dSphs is $m_\psi = 1.79 \times 10^{-22} \mathrm{eV}$ \citep{Chen-Schive:2016},
and from a detailed analysis using kinematic mock data of Fornax and Sculptor
is $m_\psi = 2.4 \times 10^{-22} \mathrm{eV}$ \citep{Gonzalez-Marsh:2016}, two
orders of magnitude larger than the value obtained from the combined analysis
of the LSB galaxies in this work. In our case, the galaxies span a wide range
in sizes ($R_\mathrm{max} \sim 1-30 \ \mathrm{kpc}$), thus there is large
scattering for both $r_c$ and $m_\psi$, and it is expected that for large
galaxies the resulting boson mass will be smaller than $10^{-23}\mathrm{eV}$.

  As noted in \citet{Urena-Robles:2017}, this could happen because the parameters
of the FDM model, including the boson mass and soliton core radius, are allowed
to vary freely, thus the value of $m_\psi$ will depend on the properties of the
sample and the fitting methods will not necessarily imply the FDM model is
wrong in these systems. This kind of fitting methods could not provide a
reliable determination of the boson mass $m_\psi$, and this could be a possible
explanation of the variety of results coming from different astronomical
observations (see Subsection~\ref{section:Schive+NFW}). Based on the mass
discrepancy-acceleration relation \citep{McGaugh:2016,Lelli-McGaugh:2016b},
\citet{Urena-Robles:2017} proposed there is a universal central surface density
for any DM profile, leading to a correlation between its central density and
scale radius \citep[see also][]{Garcia-Aspeitia:2015}. In the SFDM model, this
implies that all the galaxy halos should have a universal soliton structure at
the galaxy centers, with a constant boson mass $m_\psi = 1.2 \times 10^{-21}
\mathrm{eV}$, that is in agreement with the severe cosmological constraints
($m_\psi > 10^{-23}\mathrm{eV}$). This universal soliton should have a core
radius $r_c \approx 0.3 \ \mathrm{kpc}$ and total mass $M_s = 1.8 \times 10^7
M_\odot$. Those authors analyzed the complete soliton+NFW profile and found the
results are in agreement with the universal soliton profile solution with a
constant boson mass $m_\psi \sim 10^{-21}\mathrm{eV}$, with surface densities
close to the soliton-only results and $r_c \sim 0.3 \ \mathrm{kpc}$. These
results suggest the rotation curves of a diverse sample of galaxies might be
fitted with this large mass, leaving open the possibility to alleviate the
tension between the soliton+NFW profile and the large galaxies observations.
However, from our results in this article, we found instead $m_\psi < 10^{-23}
\mathrm{eV}$. A more detailed study of large galaxies including the universal
soliton hypothesis, is needed.

  On the other hand, we apply the exact mSFDM model, that includes
self-interactions and multistate solutions coming from thermal excitations in
the SF potential, as a realistic SFDM halo. In this work, we find that the
model is consistent with the rotation curves for both LSB and NGC galaxies with
photometric information. For one excitation state in equation~\eqref{densitytotal}
(with the excited state $i$, the radius of the halo $R$ and the central density
$\rho_0^i$ as free parameters), the $\chi^2_\mathrm{red}$ errors are smaller
than those of the soliton+NFW profile for 6 of the 11 LSB galaxies fitted,
almost the same for 4 and larger for 1. In this case, it was possible to fit
the galaxies with the ground state only ($j=1$) for almost the half of the
sample. For combinations of two excited states, the $\chi^2_\mathrm{red}$
errors are smaller for the mSFDM profile compared to the soliton+NFW results
for the other 7 LSB galaxies, and for 5 of the 6 NGCs with photometric data.
For these fits, the ground state is present in more than a half of the samples.
Finally, in order to improve the fit for NGC~6503, we used three excited
states. Notice the freedom in choosing the excitation states $i,j$ in
order to fit the observations within the mSFDM model; in principle, the excitation
state $j$ can take any value $j=1,2,3,...$, however, we preferred lower 
states since, in this scenario, it is expected that the mSFDM halos evolve from
BEC configurations with all bosons in the ground state, thus it would be easier
to reach the lower energy levels first.

  For the resulting two-states-configurations in the multistate SFDM model, we
assume the stability threshold by \citet{urena-bernal10}, where for multistate
systems where the mass ratios $\eta=M_j/M_1\lesssim1.3$, with respect to the
ground state $M_1=M_\mathrm{mSFDM}(j=1)$, the excited state $j$ does not decay
to the ground state, i.e. is a stable multistate configuration. In our fits, we
obtained 3 of 7 LSB galaxies and 4 of 6 NGC galaxies with baryonic data with a
dominant ground state $j$$=$$1$; for these multistate galaxies we argue that
they might be stable systems since the mass ratios satisfy such numerical
restriction. However, further numerical simulations including the baryonic
components are needed to investigate the stability of the multistate
configurations and the influence of the baryonic matter in the mSFDM halo.
Moreover, for the galaxies with excited states only ($j>1$), the final
distribution in different excitation levels might be the final product of
galaxy formation processes and interactions with the baryonic matter, and
numerical simulations are required in order to investigate the stability of
such configurations.

  Finally, Fig.~\ref{plot-densities} shows the resulting soliton+NFW and mSFDM
density profiles for three sample LSB galaxies. There is an evident overlap
between both profiles, showing the core-like behavior at the innermost radii,
the oscillations at intermediate radii (inherent to the mSFDM model) around the
$r^{-3}$ NFW decline, but a steeper slope of the mSFDM with respect to the NFW
decline at the outermost radii, defining more compact mSFDM halos due to the
election of the radius $R$. Such overlap for the outer regions was already
discussed in \citet{Bernal:2016}, from the results of fitting the mSFDM model
and the NFW profile to the X-ray observations of clusters of galaxies, noticing
the core-like behavior that the mSFDM profile should share with the soliton
core density. Such overlap of the NFW and SFDM profiles was also seen in recent
simulations of mergers of FDM halos in the ground state \citep{Schive:2014dra,
Schwabe:2016,Mocz-Robles:2017}. Furthermore, we notice the similar behavior of
the oscillations of the mSFDM around the NFW profile and the numerical
simulations of solitonic core mergers in \citet{Schwabe:2016}. The final merger
configurations display new `excited' solitonic cores. One possible
interpretation of such excited cores is that those are the mSFDM solutions at
the outer radii after the merger simulations, as a realistic scenario where the
SF does not remain in the soliton DM cores exclusively, after galaxy and galaxy
clusters formation processes, showing excitation states corresponding to
multistate solutions.

  To conclude, we obtained the ultra-light boson mass required to fit the
rotation curves of LSB and NGC galaxies in the context of the FDM model. The
combined analysis of the LSB galaxies gives the value $m_\psi=0.554 \times
10^{-23}\mathrm{eV}$, setting up a tension of the model with the cosmological
constraints for the minimum boson mass required to reproduce the observed
substructure. As an interesting alternative, the analytic mSFDM solution is
successful in reproducing the observations of dwarf and larger galaxies and
galaxy clusters, in particular the wiggles or oscillations observed in some
systems. Even when, at this point, with the mSFDM model it is not possible to
constrain the boson properties, the ultra-light mass $m_\psi$ and the
self-interaction $\lambda$, we show that this model is a theoretically
motivated framework for realistic SFDM halos, and as an analytic solution can
be used, at least, as a useful fitting-function in several astrophysical
systems and along with, or as a substitute of the empirical 
soliton+NFW profile.

  Unfortunately, in the analysis we performed here, we are in fact neglecting
processes in the galaxies that are somehow important, like supernovae
explosions, star formation, stellar winds, etc. This processes give some
uncertainties to the observations compared with the SFDM profiles, and it is
one of the main reasons we are not able to fit the boson mass with better
accuracy with these kind of fitting methods.

%%%%%%%%%%%%%%%% Acknowledgements %%%%%%%%%%%%%%%%
\section{Acknowledgements}

  We gratefully acknowledge Stacy McGaugh for providing us the observational data
used in this article and for helpful comments and discussions about his work in
previous publications. We thank Luis Arturo Ure\~na-L\'opez and Victor Robles
who provided insights and suggestions for the final results in this article,
and the referee who reviewed carefully the previous version and help us to
strengthen the results in the final version. This work was partially supported
by CONACyT M\'exico under grants CB-2011 No. 166212, CB-2014-01 No. 240512,
Project No. 269652 and Fronteras Project 281; Xiuhcoatl and Abacus clusters at
Cinvestav, IPN; I0101/131/07 C-234/07 of the Instituto Avanzado de
Cosmolog\'{\i}a (IAC) collaboration (http://www.iac.edu.mx).

\appendix
\section{Soliton+NFW mass profile}
\label{appendix}

  For the soliton+NFW profile~\eqref{eq:rho-schive} the density is restricted
to be continuous at the transition radius $r_\epsilon$ \citep{Schive:2014dra,
Marsh-Pop:2015}:
\begin{equation}
	\rho_\mathrm{sol} (r_\epsilon) = \rho_\mathrm{NFW} (r_\epsilon) .
\label{cont-condition}
\end{equation}
With this condition, we have four free parameters, $\rho_c$, $r_c$, $r_\epsilon$
from the soliton density~\eqref{eq:rho-core} and $r_s$ from the NFW
profile~\eqref{nfw-rho}. The NFW density parameter $\rho_s$ can be written as
\begin{equation}
	\rho_s(\rho_c,r_c,r_\epsilon,r_s) = \rho_c \frac{(r_\epsilon/r_s)
	(1+r_\epsilon/r_s)^2}{[1+0.091(r_\epsilon/r_c)^2]^8} ,
\end{equation}
where $\rho_c := 1.9 (m_\psi/10^{-23}\mathrm{eV})^{-2}(r_c/\mathrm{kpc})^{-4}$.

The total soliton+NFW mass is given by
\begin{equation}
	M_{\mathrm{FDM}}(r) = \begin{cases} M_\mathrm{sol}(r), & \mbox{if } r \leq
	r_\epsilon ; \\
	M_\mathrm{sol}(r_\epsilon) - M_\mathrm{NFW}(r_\epsilon) + M_\mathrm{NFW}(r), &
	\mbox{if } r > r_\epsilon ; \end{cases}
\end{equation}
where the core mass function is \citep{Chen-Schive:2016}
\begin{eqnarray}
	&&M_\mathrm{sol}(r) = \frac{4.2 \times 10^6 M_\odot}{(m_\psi/10^{-23}\mathrm{eV})^2
	(r_c/\mathrm{kpc})} \frac{1}{(1+a^2)^7} \nonumber \\ &&\left[ 3465 a^{13} + 23100
	a^{11} + 65373 a^9 + 101376 a^7 + 92323 a^5 \right. \nonumber \\ && \left.
	+ 48580 a^3 - 3465 a + 3465 (1+a^2)^7 \arctan{a} \right] ,
\end{eqnarray}
with $a:=0.301 (r/r_c)$, and the NFW mass profile is given by \citep{Navarro:1997}
\begin{equation}
	M_\mathrm{NFW}(r) = 4 \pi \rho_s r_s^3 \left[ \ln \left( 1 + \frac{r}{r_s}
	\right) - \frac{r/r_s}{1+r/r_s} \right] .
\end{equation}

%%%%%%%%%%%%%%%%%%%%%%%%%%%%%%%%%%%%%%%%%%%%%%%%%%%%%%%%%%%%%
\section{Soliton+NFW analysis with differentiability condition}
\label{appendix2}

  Additionally to the performed analysis of the FDM model, we explored the
soliton+NFW density profile~\eqref{eq:rho-schive} to be continuous
(equation~\eqref{cont-condition}) and differentiable at the transition radius
$r_\epsilon$:
\begin{equation}
	\rho'_\mathrm{sol} (r_\epsilon) = \rho'_\mathrm{NFW} (r_\epsilon) .
\end{equation}
With these conditions, the two NFW parameters $r_s$ and $\rho_s$ can be written in
terms of the other three free parameters, $\rho_c$, $r_c$ and $r_\epsilon$, as
\begin{eqnarray}
	r_s(r_c,r_\epsilon)&=&\left( \frac{b-3}{1-b} \right) r_\epsilon , \\
	\rho_s(\rho_c,r_c,r_\epsilon)&=&\frac{1-b}{(b-3)^3} \frac{4
	\rho_c}{[1+0.091 (r_\epsilon / r_c)^2
	]^8} , 
\end{eqnarray}
where $b:= 1.456 (r_\epsilon / r_c)^2 / [1+0.091(r_\epsilon / r_c)^2]$.

In Table~\ref{tab:results2-diff}, we report the results of the individual
analysis for the 18 LSB galaxies with three free parameters, $\rho_c$, $r_c$
and $r_\epsilon$, and the resulting boson mass $m_\psi$, all the quantities $\pm
1\sigma$ errors from the MCMC method, also the ratio $r_\epsilon/r_c$ and
$\chi^2_\mathrm{red}$ errors for each galaxy. Table~\ref{tab:LSB-comb-diff}
shows the results of the combined analysis for the same LSB galaxies.

\begin{table*}
\caption{In this Table we show the resulting parameters $\rho_c$, $r_c$ and
$r_\epsilon$ for the FDM model with the soliton+NFW density
profile~\eqref{eq:rho-schive} with the differentiability condition at the
transition radius $r_\epsilon$, and the resulting boson mass $m_\psi$, all the
quantities $\pm 1\sigma$ errors from the MCMC method used. We also show the
resulting $r_{200}$ from the NFW halo radius, concentration parameter $c$ and
$\chi^2_\mathrm{red}$ errors for the 18 high-resolution LSB galaxies in
\citet{deBlok:2001}.}
\label{tab:results2-diff}
\begin{tabular}{lcccccccc}
    \hline
    Galaxy	&	$\rho_c$	&	$r_c$	&	$r_\epsilon$	&	$r_\epsilon/r_c$	&	$m_\psi$	&	$r_{200}$	&	$c$	&	$\chi^2_\mathrm{red}$ \\
    &	$(10^{-2} M_\odot/\mathrm{pc}^3)$	&	$(\mathrm{kpc})$	&	$(\mathrm{kpc})$	&	&	$(10^{-23}\mathrm{eV})$	&		$(\mathrm{kpc})$	&	&	\\
\hline
ESO 014-0040 & $22.2^{+4.4}_{-9.1}$ 	& $3.12\pm 0.76$ 			& $3.23\pm 0.92$ 		 	 & 1.04	& $0.352^{+0.063}_{-0.16}$ 	& 254 & 21.0 & 0.658	\\
ESO 084-0411 & $0.588^{+0.068}_{-0.15}$	& $4.90^{+1.5}_{-0.99}$ 	& $4.0^{+1.1}_{-1.6}$ 	 	 & 0.816& $0.986^{+0.011}_{-0.58}$ 	& --- & ---  & 0.092	\\
ESO 120-0211 & $2.66^{+0.72}_{-0.91}$ 	& $1.01^{+0.17}_{-0.27}$	& $1.44^{+0.35}_{-0.78}$ 	 & 1.43	& $9.6^{+1.6}_{-3.9}$ 		& 21.5& 44.5 & 0.074	\\
ESO 187-0510 & $4.66^{+0.92}_{-1.4}$ 	& $1.20^{+0.27}_{-0.34}$	& $1.46^{+0.36}_{-0.84}$ 	 & 1.22	& $5.28^{+0.87}_{-2.7}$ 	& 37.7& 21.1 & 0.101	\\
ESO 206-0140 & $17.2^{+3.1}_{-4.8}$ 	& $1.65^{+0.31}_{-0.36}$	& $1.80^{+0.39}_{-0.56}$ 	 & 1.09	& $1.34^{+0.25}_{-0.49}$ 	& 106 & 23.1 & 0.133	\\
ESO 302-0120 & $4.23^{+0.68}_{-1.2}$ 	& $2.66\pm 0.57$ 			& $3.28^{+0.89}_{-1.4}$  	 & 1.23	& $1.09^{+0.14}_{-0.48}$ 	& 79.0& 21.6 & 0.032	\\
ESO 305-0090 & $2.65^{+0.44}_{-0.99}$ 	& $2.26^{+0.68}_{-0.89}$	& $2.30^{+0.77}_{-1.4}$  	 & 1.02	& $2.589^{+0.020}_{-1.9}$ 	& --- & ---  & 0.076	\\
ESO 425-0180 & $4.04^{+0.67}_{-2.2}$ 	& $3.7\pm 1.1$ 				& $3.42^{+1.2}_{-0.83}$  	 & 0.924& $0.653^{+0.053}_{-0.35}$ 	& 251 & 6.14 & 0.509	\\
ESO 488-0490 & $8.9^{+1.4}_{-2.5}$ 		& $1.98\pm 0.45$ 			& $2.24^{+0.52}_{-1.0}$  	 & 1.13	& $1.37^{+0.22}_{-0.62}$ 	& 90.9& 20.4 & 0.131	\\
F730-V1      & $18.1^{+3.5}_{-5.0}$ 	& $1.84^{+0.34}_{-0.46}$	& $1.95^{+0.39}_{-0.67}$ 	 & 1.06	& $1.08^{+0.22}_{-0.44}$ 	& 130 & 21.2 & 0.329	\\
UGC 4115     & $15.2^{+2.0}_{-4.3}$ 	& $1.06^{+0.14}_{-0.66}$	&$1.037^{+0.075}_{-0.79}$	 & 0.978& $6.5^{+1.1}_{-5.8}$ 		& 91.1& 14.4 & 0.132	\\
UGC 11454    & $15.7^{+3.0}_{-4.0}$ 	& $1.96^{+0.39}_{-0.44}$	& $1.96^{+0.44}_{-0.55}$ 	 & 1.00	& $1.0^{+0.21}_{-0.38}$ 	& 157 & 16.0 & 0.364	\\
UGC 11557    & $1.91^{+0.18}_{-0.62}$ 	& $3.7^{+1.4}_{-1.2}$ 		& $2.77^{+0.94}_{-1.1}$  	 & 0.749& $1.152^{-0.072}_{-0.83}$ 	& --- & ---  & 0.12 	\\
UGC 11583    & $9.5^{+2.2}_{-2.6}$ 		& $0.86^{+0.11}_{-0.26}$	& $1.45^{+0.25}_{-1.0}$  	 & 1.69	& $7.2^{+1.9}_{-3.0}$ 		& --- & ---  & 0.097	\\
UGC 11616    & $16.3^{+1.9}_{-3.2}$ 	& $1.93\pm 0.28$ 			& $2.10^{+0.38}_{-0.49}$ 	 & 1.09	& $0.97^{+0.13}_{-0.27}$ 	& 122 & 22.4 & 0.189	\\
UGC 11648    & $226\pm 25$ 				&$0.0858^{+0.010}_{-0.0093}$&$0.0737^{+0.0089}_{-0.0080}$& 0.859& $128^{+20}_{-30}$ 		& 184 & 8.38 & 1.097	\\
UGC 11748    & $97^{+13}_{-23}$ 		& $1.70^{+0.23}_{-0.21}$	& $2.14\pm 0.35$ 			 & 1.26	& $0.502^{+0.056}_{-0.098}$ & 160 & 77.0 & 1.978	\\
UGC 11819    & $7.10^{+0.40}_{-0.96}$ 	& $3.92^{+0.51}_{-0.26}$	& $4.9\pm 1.1$ 				 & 1.25	& $0.351^{+0.017}_{-0.077}$ & 139 & 28.0 & 0.28 	\\
\hline
\end{tabular}
\end{table*}

\begin{table*}
\caption{Soliton+NFW density profile -- Combined analysis in high-resolution
LSB galaxies. In this Table we show the resulting fitting parameters $\rho_c$,
$r_c$, $r_\epsilon$ and $r_s$ for the FDM model with the soliton+NFW density
profile~\eqref{eq:rho-schive} with the differentiability condition, and the
resulting boson mass $m_\psi=0.554\times 10^{-23}\mathrm{eV}$. The combined
analysis was performed minimizing the $\chi^2$ errors and we obtained
$\chi_\mathrm{red}^2=1.208$. We also show the resulting NFW density parameter
$\rho_s$, the ratio $r_\epsilon/r_c$, the $r_{200}$ radius from the NFW halo,
concentration parameter $c$ and total DM mass $M_{200}=M_\mathrm{FDM}(r_{200})$,
for the 18 high-resolution LSB galaxies in \citet{deBlok:2001}.}
\label{tab:LSB-comb-diff}
\begin{tabular}{lcccccccccc}
\hline
    Galaxy	&	$\rho_c$	&	$r_c$	&	$r_\epsilon$	&	$r_\epsilon/r_c$	&	$m_\psi$	&	$r_s$	&	$\rho_s$	&	$r_{200}$	&	$c$	&	$M_{200}$ \\
    	&	$(10^{-2} M_\odot/\mathrm{pc}^3)$	&	$(\mathrm{kpc})$	&	$(\mathrm{kpc})$	&	&	$(10^{-23}\mathrm{eV})$	&	$(\mathrm{kpc})$	&($10^{-2} M_\odot/\mathrm{pc}^3$)	&	$(\mathrm{kpc})$	&	&	($10^{12} M_\odot$)	\\
\hline
ESO 014-0040 & 2.61	& 0.855 & 0.767 & 0.897 & 2.024 & 16.0  & 2.61  & 289  & 18.1  & 2.66  	\\
ESO 084-0411 & 0.0020& 2.593 & 2.227 & 0.859 & 2.024 & 696   & 0.0020 & 412  & 0.592 & 7.76  	\\
ESO 120-0211 & 1.99	& 2.654 & 3.440 & 1.296 & 2.024 & 2.69  & 1.99  & 43.8 & 16.3  & 0.009 	\\
ESO 187-0510 & 192	& 2.013 & 3.018 & 1.499 & 2.024 & 0.499 & 192   & 45.7 & 91.7  & 0.009 	\\
ESO 206-0140 & 3.15	& 1.183 & 1.173 & 0.992 & 2.024 & 6.27  & 3.15  & 122  & 19.5  & 0.201 	\\
ESO 302-0120 & 0.549& 1.654 & 1.584 & 0.958 & 2.024 & 12.0  & 0.549  & 117  & 9.70  & 0.176 	\\
ESO 305-0090 & 0.110& 2.032 & 1.848 & 0.909 & 2.024 & 29.0  & 0.110  & 142  & 4.90  & 0.317 	\\
ESO 425-0180 & 0.0945& 1.537 & 1.341 & 0.873 & 2.024 & 71.6  & 0.0945 & 328  & 4.59  & 3.92  	\\
ESO 488-0490 & 0.577& 1.429 & 1.318 & 0.922 & 2.024 & 16.3  & 0.577 	& 162  & 9.90  & 0.467 	\\
F730-V1      & 1.85	& 1.143 & 1.073 & 0.939 & 2.024 & 10.4  & 1.85  & 164  & 15.8  & 0.484 	\\
UGC 4115     & 7.17	& 1.387 & 1.601 & 1.155 & 2.024 & 2.78  & 7.17  & 74.5 & 26.8  & 0.045 	\\
UGC 11454    & 1.21	& 1.161 & 1.065 & 0.917 & 2.024 & 14.4  & 1.21  & 192  & 13.4  & 0.780 	\\
UGC 11557    & 0.0049& 1.940 & 1.665 & 0.858 & 2.024 & 662   & 0.0049 & 693  & 1.05  & 36.8  	\\
UGC 11583    & 3.59	& 1.692 & 1.974 & 1.167 & 2.024 & 3.20  & 3.59  & 65.6 & 20.5  & 0.031 	\\
UGC 11616    & 1.85	& 1.152 & 1.084 & 0.941 & 2.024 & 10.2  & 1.85  & 161  & 15.8  & 0.459 	\\
UGC 11648    & 0.533& 1.352 & 1.227 & 0.907 & 2.024 & 20.16 & 0.533  & 193  & 9.58  & 0.798 	\\
UGC 11748    & 54.2	& 0.581 & 0.576 & 0.992 & 2.024 & 3.07  & 54.2  & 177  & 57.5  & 0.610 	\\
UGC 11819    & 0.307& 1.235 & 1.084 & 0.878 & 2.024 & 43.0 	& 0.307  & 328  & 7.62  & 3.90  	\\
\hline
\end{tabular}
\end{table*}

  From Table~\ref{tab:results2-diff} it is noticeable that, in general,
$r_c$ is of the order of the transition radius $r_\epsilon$. This means that
the soliton contribution is overlapped with the NFW halo (as we ask for a
smooth transition between both profiles), and from the best fits, the values of
$r_\epsilon$ do not correspond to $r_\epsilon>3r_c$ \citep{Schive:2014dra}. The
transition from the soliton to the NFW profile is smoother in this case, and
that behavior is maintained as in the continuity condition analysis (see
Section~\ref{subsec:Schive+NFW}). From the combined analysis of the 18 LSB
galaxies, we obtained an ultra-light mass $m_\psi = 2.024 \times 10^{-23}
\mathrm{eV}$, that is in agreement with the cosmological constraints
\citep{Bozek:2014,Sarkar:2016}, unlike the results with the continuity
condition only.

\begin{figure}
   \begin{center}
   \includegraphics[width=0.5\textwidth]{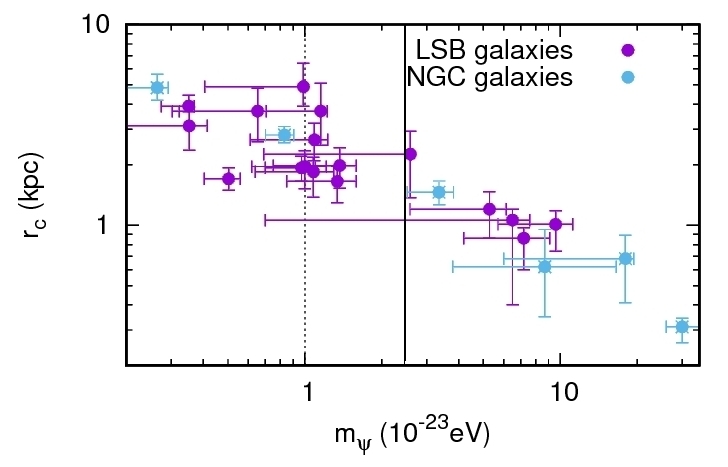}
   \end{center}
   \caption{Distribution of resulting core radii $r_c$ vs. boson masses $m_\psi$
   from the soliton+NFW density profile in the FDM model with the
   differentiability condition. The data is from the individual analysis of the
   18 LSB and 6 NGC (including baryonic information) galaxies, $\pm 1 \sigma$
   errors from the MCMC method. The vertical dashed line at $m_\psi=10^{-23}
   \mathrm{eV}$ shows the cosmological constraint \citep{Bozek:2014,Sarkar:2016}.
   The vertical solid line at $m_\psi=2.024 \times 10^{-23}\mathrm{eV}$ shows
   the resulting boson mass from the combined analysis of the 18 LSB galaxies.}
   \label{mpsi-rc-diff}
\end{figure}

  Finally, in Table~\ref{tab:results1-diff} we show the results from the
individual analysis of the 6 NGC galaxies with photometric information. In this
case, the boson mass is in the range $0.264 \leq m_\psi/\mathrm{eV} \leq 30.0$,
core radius within $0.311 \leq r_c/\mathrm{kpc} \leq 4.84$ and ratio between
the transition and core radii in $0.882 \leq r_\epsilon \leq 1.157$.
Fig.~\ref{mpsi-rc-diff} shows the distribution of $r_c$ vs. $m_\psi$ for all
the LSB and NGC galaxies.

\begin{table*}
\caption{NGC galaxies with the soliton+NFW profile with differentiability
condition. In this Table we show the fitting parameters $\rho_c$, $r_c$ and
$r_\epsilon$, the resulting boson mass $m_\psi$, the concentration parameters $c$
and $r_{200}$ radii from the NFW profile, and the $\chi^2_\mathrm{red}$ errors
for the soliton+NFW profile, for the NGC galaxies with photometric
information. In the top panel we show the DM-only fits and in the bottom panel
the fits taking into account the baryonic contribution.}
\label{tab:results1-diff}
\begin{tabular}{lcccccccc}
    \hline
    \multicolumn{8}{c}{DM-only fits} \\
    \hline
    Galaxy	&	$\rho_c$	&	$r_c$	&	$r_\epsilon$	& $r_\epsilon/r_c$	&	$m_\psi$	&	$c$	&	$r_{200}$	&	$\chi^2_\mathrm{red}$ \\
    	&	$(10^{-2}M_\odot/\mathrm{pc}^3)$	&	$(\mathrm{kpc})$	&	$(\mathrm{kpc})$	&	& $(10^{-23}\mathrm{eV})$	&	&	$(\mathrm{kpc})$	&	\\
\hline
NGC7814  & $4730\pm 4.8$ & $0.0872^{+0.010}_{-0.0067}$ & $0.0774^{+0.0096}_{-0.0063}$ &  0.888 & $27.0^{+2.8}_{-5.5}$ & 69.425 & 147.32 & 5.541 \\
NGC6503  & $1430^{+190}_{-130}$ & $0.0464^{+0.0036}_{-0.0028}$ & $0.0402^{+0.0032}_{-0.0025}$ & 0.866 & $171\pm 20$ & 28.59 & 98.725 & 6.38 \\
NGC3741  & $133^{+13}_{-20}$ & $0.0367\pm 0.0050$ & $0.0315\pm 0.0043$ & 0.858 & $932^{+200}_{-300}$ & 6.032 & 72.329 & 0.646 \\
NGC1003  & $90^{+52}_{-24}$ & $0.154^{+0.046}_{-0.065}$ & $0.133^{+0.040}_{-0.056}$ & 0.864 & $69^{+70}_{-30}$ & 8.778 & 124.246 & 4.864 \\
NGC1560  & $5.63^{+0.64}_{-0.88}$ & $1.39\pm 0.18$ & $1.31\pm 0.20$ & 0.942 & $3.16^{+0.50}_{-0.78}$ & 7.983 & 95.974 & 0.395 \\
NGC6946  & $5.14^{+0.32}_{-0.47}$ & $4.92^{+0.36}_{-0.30}$ & $6.28\pm 0.70$ & 1.276 & $0.254^{+0.019}_{-0.029}$ & 27.294 & 150.211 & 1.616 \\
\hline
    \multicolumn{8}{c}{DM+baryons fits}\\
\hline
NGC7814  & $5.57^{+0.66}_{-1.4}$ & $4.84^{+0.80}_{-0.66}$ & $5.6\pm 1.2$ & 1.157 & $0.264^{+0.028}_{-0.070}$ & 18.492 & 177.278 & 0.86 \\ 
NGC6503  & $4.59^{+0.41}_{-0.68}$ & $2.81^{+0.29}_{-0.24}$ & $3.07^{+0.38}_{-0.34}$ & 1.093 & $0.831^{+0.072}_{-0.13}$ & 13.809 & 106.975 & 1.547 \\
NGC3741	 & $6.33^{+0.82}_{-2.9}$ & $0.68^{+0.21}_{-0.27}$ & $0.60^{+0.18}_{-0.25}$ & 0.882 & $18.0^{+1.4}_{-12}$ & 4.687 & 93.217 & 2.115 \\
NGC1003  & $23.1^{+6.1}_{-3.2}$ & $0.311^{+0.033}_{-0.052}$ & $0.269^{+0.029}_{-0.045}$ & 0.865 & $30.0^{+8}_{-4}$ & 5.062 & 135.755 & 2.93 \\
NGC1560  & $4.11^{+0.42}_{-0.65}$ & $1.46\pm 0.20$ & $1.38^{+0.20}_{-0.23}$ & 0.945 & $3.35^{+0.47}_{-0.83}$ & 7.113 & 86.89 & 0.234 \\
NGC6946  & $27^{+16}_{-12}$ & $0.62^{+0.33}_{-0.27}$ & $0.55^{+0.30}_{-0.24}$ & 0.887 & $8.7^{+7.9}_{-4.9}$ & 9.376 & 143.84 & 5.34 \\
\hline
\end{tabular}
\end{table*}

%%%%%%%%%%%%%%%% REFERENCES %%%%%%%%%%%%%%%%

\bibliographystyle{mnras}
\bibliography{articulo}

% Don't change these lines
\bsp	% typesetting comment
\label{lastpage}
\end{document}